\makeatletter

\newcommand{\LyX}{L\kern-.1667em\lower.25em\hbox{Y}\kern-.125emX\@}

\documentclass[saving]{disser}

\newcommand\printindex{\@input@{\jobname.ind}}

\makeindex
\newenvironment{lyxlist}[1]
  {\begin{list}{}
    {\settowidth{\labelwidth}{#1}
     \setlength{\leftmargin}{\labelwidth}
     \addtolength{\leftmargin}{\labelsep}
     \renewcommand{\makelabel}[1]{##1 \hfill}}}
  {\end{list}}

\include{fancybox}
\usepackage[sl]{caption}
\renewcommand{\captionfont}{\rm}
\usepackage{fancybox}
\usepackage{graphicx}
\usepackage{pifont}

\title{LINE EMISSION FROM STELLAR WINDS \\ IN ACTIVE GALACTIC NUCLEI}

\author {Jason A. Taylor}

\department {Laboratory of High Energy Astrophysics \\
             NASA Goddard Space Flight Center}

\advisor{   Dr. Demosthenes Kazanas}
\committee{Professor Jordan A. Goodman, Chair \\
                Professor Abolhassan Jawahery\\
                Dr. Timothy Kallman\\
                Professor Dennis Papadopoulos\\
                Professor Andrew S. Wilson}

\prefacefile{preface}

\dedication {\centering In memory of David Waylonis, who taught me 
the importance of uncovering the truth.}

\acknowledgements{\begin{flushleft}I thank my advisor, Demos Kazanas,
for his guidance over the past years.  Despite the fact that we did
not, at least until very recently, agree on some of the fundamental
scientific issues raised by our research, I have benefited from Demos'
abilities, and in particular his knack at getting to the bottom of
complex questions as fast they can be dished out.  Even when his
office was as jammed as a subway during rush hour, he would always
manage to make time for me.  I am indebted to the many people at the
NASA/GSFC who took time out of their busy schedules to answer my
numerous questions.  These people include Mike Harris, Damien Audley,
Glen Piner, Mike Corcoran, and Richard Mushotzky.  I thank Andrew
Wilson, Tim Kallman, Dennis Papadopoulos, Jordan Goodman, and
Abolhassan Jawahery for participating at my defense.  I especially
thank Andrew, Tim, and Dennis for uncovering errors and omissions.  I
also thank Ivan Hubeny for letting me use his excellent partial
redistribution line transfer program.  I thank Dimitris Christodoulou
for his assistance in obtaining much needed supercomputer time.  I am
grateful to Chris Shrader for his offering me a position during the
past two summers (when TAs are difficult to find).  I thank the USNA
Department of Physics for providing an atmosphere conducive towards my
research.  I also thank my parents.  In addition to creating me, they
gave me a computer which is eight times faster than the fastest one I
could get my hands on at NASA.  Finally, I am grateful to Rebecca
Zeltinger for her support and understanding while this dissertation
was being written.  \\ \vspace{0.5in} The initial phase of this work
was supported by NASA grant NCC-5-54.\end{flushleft}}

\abstract{This dissertation presents synthetic spectra and response
functions of the red giant stellar line emission model of active
galactic nuclei (e.g., Kazanas 1989).  Our results agree with the
fundamental line emission characteristics of active galactic nuclei
within the model uncertainties if the following new assumptions are
made: 1) the mean stellar mass loss rates decrease with distance from
the black hole, and 2) the mean ionization parameters are lower than
those postulated in Kazanas (1989).  For models with enhanced mass
loss, the zero-intensity-full-widths of the line profiles are
proportional to the black hole mass to the power of 1/3.  This
scaling relation suggests that the black hole masses of NLS1s
(narrow-line Seyfert 1s) are relatively low.  Models with enhanced
mass loss also predict minimum line/continuum delays that are
proportional to the zero-intensity-full-widths of the profiles.
Because of their high column densities, these models yield
triangle-shaped response functions, which are not generally observed.
On the other hand, models without enhanced mass loss yield
line-continuum delays that are proportional to the square root of the
continuum luminosity.  This prediction appears to agree with results
from reverberation mapping campaigns.

If the intercloud (interstellar) medium densities are high enough, the
winds are ``comet shaped,'' with the shock fronts having higher
densities than the cloud ``tails.''  In this case, the densities in
the ionized (inverse Str\"omgren regions) of the outbound clouds are
lower than those of the inbound clouds.  For models in which an
accretion disk occults the broad line region, the broadest line
emission and absorption profile components of lines similar to C IV, N
V, and O VI are redshifted.  Conversely, the narrowest emission and
absorption profile components are blueshifted.  The shifts of the
Lyman alpha profile components are much smaller.  One particularly
interesting prediction of the nonspherical wind models is that their C
IV red wings respond faster than their blue wings, as has been
observed (e.g., Done \& Krolik 1996).  These same models, however,
yield opposite results for the C III] line, such that the C III] {\it
blue} wings respond first.  For this reason, measurements of the
velocity dependence of the C III] profile response could be used to
test the viability of nonspherical stellar wind line emission models.}

\hasfigures{true}
\hastables{true}
\hascopyright{true}
\renewcommand\bibname{References}
\newcommand\indexnamekludge{Phrase Index}
\renewcommand\indexname{}

\def\s{\sigma}

\def\lesssim{\mathrel{\hbox{\rlap{\hbox{\lower5pt\hbox{$\sim$}}}\hbox{$<$}}}}
\def\gtrsim{\mathrel{\hbox{\rlap{\hbox{\lower5pt\hbox{$\sim$}}}\hbox{$>$}}}}
\def\lessim{\lesssim}
\def\lsim{\lessim}
\def\gsim{\gtrsim}
\def\la{\lessim}
\def\ga{\gtrsim}

\def\keywords{\def\acomment}
\def\affil{\def\acomment}
\def\tableline{\hline}
\def\tablecomments{}
\def\altaffiltext#1#2{}

\def \s{\begin{equation}}
\def \e{\end{equation}}
\def \labep{\label}

\def\infootbegplain{}
\def\infootendplain{}

\let\rfn\footnote                          

\newcommand{\qfn}[1]{\rfn{\it{Question} I would like your opinion on:\it{#1}}}               
\newcommand{\pfn}[1]{\rfn{Private comment to myself (please ignore this):#1}}                      
\renewcommand{\footnote}[1]{\rfn{Private comment to myself (please ignore this):#1}}
\def\home{.}
\def\figdirprefix{\home/scon/}
\def\pfn{\def\acomment}
\def\qfn{\def\acomment}
\def\footnote{\def\acomment}

\usepackage[section]{placeins}
\let\subsectionoriginal\subsection
\def\subsection{\FloatBarrier\subsectionoriginal}
\def\fig#1#2{\begin{figure}[htbp]
 \centering
 \ovalbox{\includegraphics[width=5.525in,height=4.in]{\figdirprefix#1.eps}}
 \caption{\setlength{\baselineskip}{.2in} #2}
 \label{#1}
 \end{figure}}
\def\fignobox#1#2{\begin{figure}[htbp]
 \centering
 \includegraphics[width=5.525in,height=4.in]{\figdirprefix#1.eps}
 \caption{\setlength{\baselineskip}{.2in} #2}
 \label{#1}
 \end{figure}}
\def\Fig#1#2{\fig{#1}{#2}Figure~\ref{#1}}
\def\Fignobox#1#2{\fignobox{#1}{#2}Figure~\ref{#1}}

\makeatother

\begin{document}

\makefrontmatter
\spacing{1.5}

\chapter{Introduction
}

\begin{quote}
\spacing{1.0}``What you have to do, if you get caught in this gumption trap
of value rigidity, is slow down---you're going to have to slow down anyway whether
you want to or not---but slow down deliberately and go over ground that you've
been over before to see if the things you thought were important were really
important and to . . . well . . . just \emph{stare} at the machine. There's
nothing wrong with that. Just live with it for a while. Watch it the way you
watch a line when fishing and before long, as sure as you live, you'll get a
little nibble, a little fact asking in a timid, humble way if you're interested
in it. That's the way the world keeps on happening. Be interested in it.

At first try to understand this new fact not so much in terms of your big problem
as for its own sake. That problem may not be as big as you think it is. And
that fact may not be as small as you think it is. It may not be the fact you
want but at least you should be very sure of that before you send the fact away.
Often before you send it away you will discover it has friends who are right
next to it and are watching to see what your response is. Among the friends
may be the exact fact you are looking for.

After a while you may find that the nibbles you get are more interesting than
your original purpose of fixing the machine. When that happens you've reached
a kind of point of arrival. Then you're no longer strictly a motorcycle mechanic,
you're also a motorcycle scientist, and you've completely conquered the gumption
trap of value rigidity.

...

I can just see somebody asking with great frustration, `Yes, but which facts
do you fish for? There's got to be more to it than that.'

But the answer is that if you know which facts you're fishing for you're no
longer fishing. You've caught them.''

~~~~~~~~~~~~~~~~~~~~~~~~~~~~~~~~~~~~~~~~~~~~~~~~~~~~~~~~~~~~~~~~---\emph{Robert
M. Pirsig}
\end{quote}

\section{Background}

\spacing{1.5}Galaxies are composed of stars. Most of these stars emit Plank
continua with myriad absorption lines. It is, therefore, understandable that
the spectra of galaxies also have absorption lines. But in 1908, before the
distances to ``spiral nebulae'' had even been determined, Edward A. Fath discovered
line \emph{emission} from the center of the nearby spiral nebula M77, more commonly
known as NGC 1068 (Fath 1909). NGC 1068 is now categorized as an ``active''
galaxy. To this day, there is still no consensus on what produces the lines
emitted by NGC 1068 and similar active galaxies.

The centers of $  \approx 46  $\% of all galaxies are active in that they emit
at least some form of line emission or nonthermal continuum radiation (Ho 1996).
A picture of the active galaxy NGC 5548 taken by the \emph{Hubble Space Telescope}
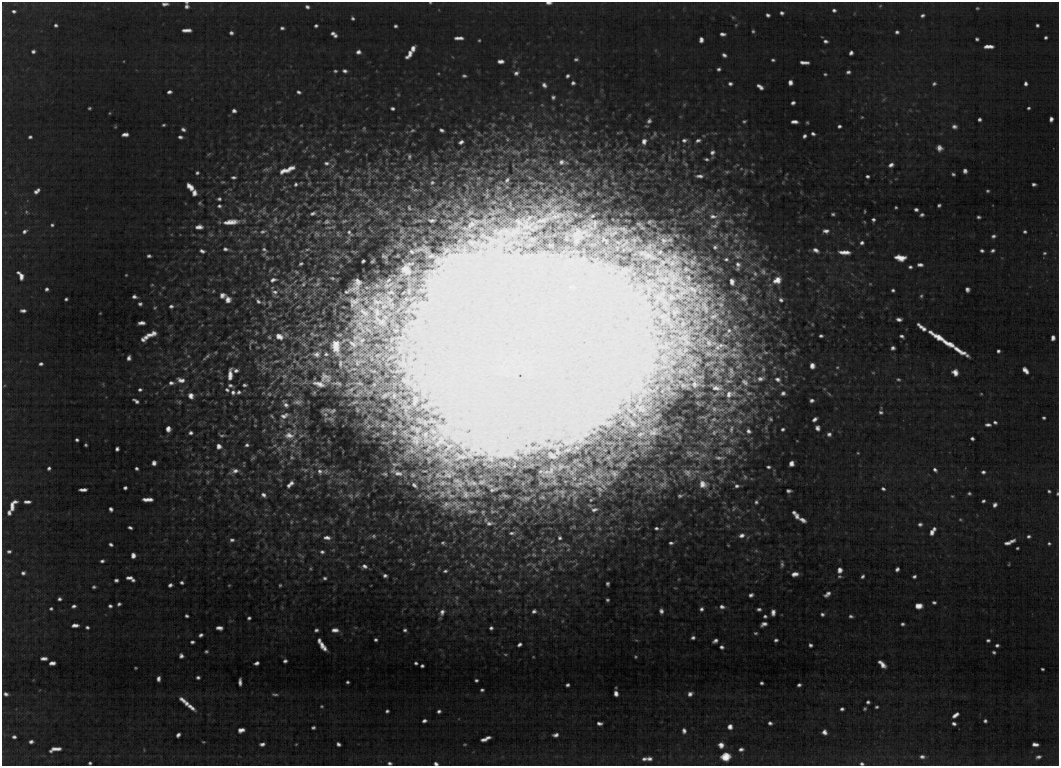
(\emph{HST}) WFPC2 camera is shown in \Fignobox{thesispics/clyde}{WFPC2 image
of active spiral galaxy NGC 5548. This galaxy has a redshift of $  z=0.0174  $,
placing it $  7\times 10^{7}  $ pc ($  2.3\times 10^{8}  $ light-years or
$  2.2\times 10^{26}  $ cm) away from us if a Hubble constant of 75 km s$  ^{-1}  $
Mpc$  ^{-1}  $ (where Mpc is an abbreviation for 10$  ^{6}  $ parsecs) is
assumed and all non-cosmological motions are neglected.}. The unresolved central
continuum source in NGC 5548 is brighter in the ultraviolet (UV) than the entire
remainder of the galaxy. The UV spectrum of the entire galaxy is, therefore,
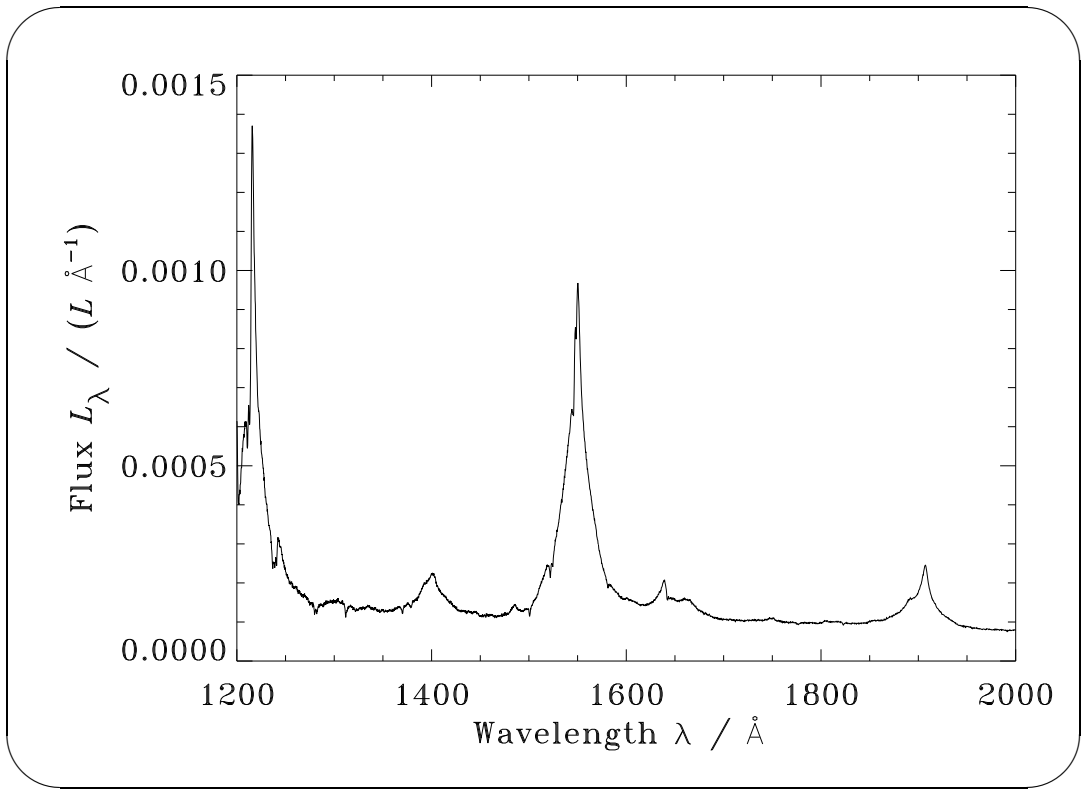
quite similar to the central source. This spectrum is shown in \Fig{thesispics/5548spec}{Time-averaged
UV spectrum of AGN NGC 5548 obtained from the \emph{HST} FOS. Data courtesy
of K. Korista.}. The majority of AGNs have only narrow (equivalent Doppler shifts
of $  c\Delta \lambda /\lambda _{0}\lsim   $ 1000 km s$  ^{-1}  $) lines superimposed
upon a nonthermal continuum. NGC 5548 is one of the $  \sim   $20\% of AGNs
(Ho 1996) that is also a Seyfert 1, with larger line widths of $  \Delta \lambda /\lambda _{0}\sim 0.02  $.
Seyfert 1s and their brighter counterparts called quasars (also known as ``QSOs'')
are the primary topic of this dissertation. They are AGNs with line emission
profiles that are very broad (equivalent Doppler shifts of $  \sim   $ 5000
km s$  ^{-1}  $), but their profiles may also contain the narrower profile
components that are more commonly observed. In Seyfert 1s, the broad profile
components are readily apparent in the UV spectral region due to their large
equivalent widths\rfn{The equivalent width represents the strength of a line
relative to the continuum. It can be defined as $  F_{l}/F_{{\rm c}\lambda }  $,
where $  F_{l}  $ is the total continuum-subtracted line flux and $  F_{{\rm c}\lambda }  $
is the continuum flux per unit wavelength.} of $  \sim 10  $ \AA$    $ ($  1.0  $ \AA$  \equiv 10^{-10}\textrm{ m}  $).
Several other AGN categories have been invented. For instance, AGNs that are
extremely bright in the radio band are known as radio-loud AGNs. This category
constitutes 10\% of all Seyfert 1s. One such radio-loud AGN is M87, shown in
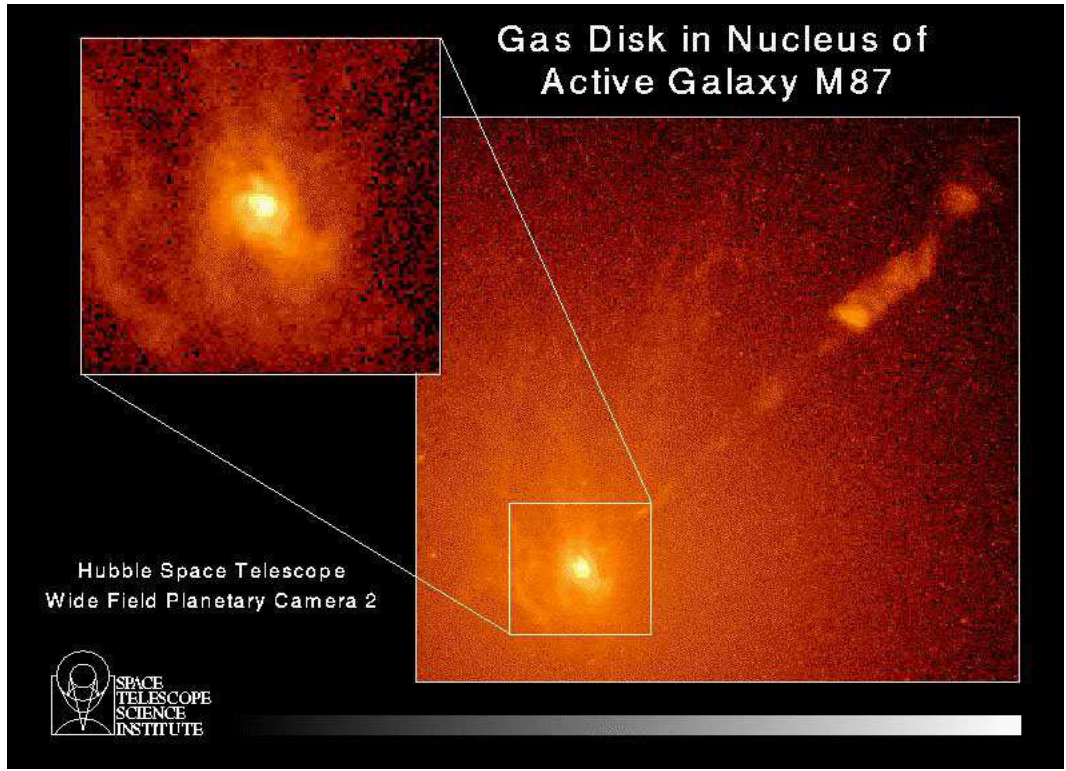
Figures \ref{thesispics/M87Disk} and \ref{thesispics/M87Plot}.\fignobox{thesispics/M87Disk}{WFPC2
image of supermassive active elliptical galaxy M87, center of the Virgo, the
nearest cluster of galaxies. Upper right: the near side of a jet protruding
perpendicularly out of a dusty disk toward us from the continuum source. Photo
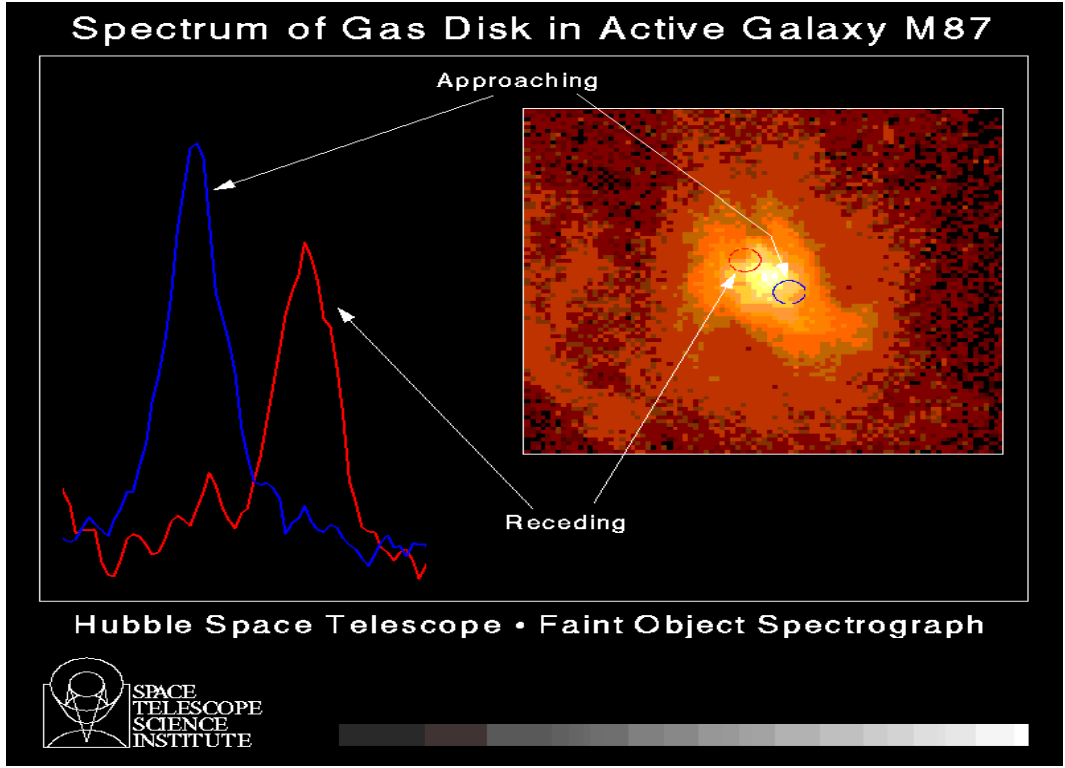
courtesy of STScI.}\fignobox{thesispics/M87Plot}{Right: blowup of central disk
in M87. Left: superimposed spectra of two regions of the disk illustrating Doppler
shifts of the {[}O {\footnotesize III}{]} $  \lambda \lambda   $4959, 5007
doublet. Photo courtesy of STScI.}

Although 90 years have elapsed since Fath's original publication, the mystery
of what produces the narrow and broad line emission remains unsolved. Recent
space-based observations have literally shed new light on the problem by making
UV spectra available for the first time. This has fueled research; 14\% of all
1995 \emph{Astrophysical Journal} papers mention ``AGN'' in their abstract.

Before discussing the Seyfert 1 and QSO AGN models that have been proposed to
date, let us first review what is generally suspected about AGNs. This review
will provide perspective and reduce the chance that we box ourselves into a
narrow corner of model parameter space. Because there is so little consensus
in the field, let us begin \emph{at} the beginning, with the suspected formation
of AGNs and their host galaxies.

Before galaxies formed, there were density and velocity fluctuations left over
from the big bang. As gravitational attraction pulled the higher density regions
together, protogalaxies with even higher densities began to form. These protogalaxies
had small velocity gradients induced by tidal torques from neighboring protogalaxies.
Because gas is inherently dissipative (both collisional and inelastic), its
orbits cannot cross. For this reason, the gas in the protogalaxies formed disks
that had axes aligned with the initial average angular momentum vectors. If
viscous forces can be neglected, the distance between protogalaxies is small
compared to the initial velocity perturbation scale length, and star formation
does not consume most of the gas prior to its collapse, then it can be shown
that conservation of angular momentum yields a post-collapse disk surface density
of 

\begin{equation}
\label{1.1}
\Sigma (r)={G\over r}\left( {\rho d_{{\rm g}}\over |\nabla \times {\bf {v}}|}\right) ^{2}.
\end{equation}
 In this equation, $  r  $ is the galactic radius, $  G  $ is the gravitational
constant, $  \rho   $ is the initial mean cosmological density near the protogalactic
center of mass, $  d_{{\rm g}}  $\rfn{In this dissertation, roman-typefaced
subscripts represent abbreviations of descriptive words while italic-typefaced
subscripts represent variables.}is the initial distance between protogalaxies,
and $  \nabla \times {\bf {v}}  $ is the curl of the initial velocity function
of the gas near the protogalactic center of mass. Relations similar to equation
(\ref{1.1}) were first derived by Mestel (1963, 1965) and have since been verified
by numerical calculations (e.g., Fall \& Efstathiou 1980). Though it is currently
a matter of debate as to what the dominant contributor of mass in galaxies is,
the surface density of equation (\ref{1.1}) yields a flat rotation curve and
a divergent density at $  r=0  $. But an initial assumption made in deriving
equation (\ref{1.1}) was that viscous forces could be neglected. Since the
importance of viscosity increases sharply with density, equation (\ref{1.1})
\emph{must} be violated near $  r=0  $. Such viscosity would force the formation
of and the rapid accretion onto a compact object such as a supermassive black
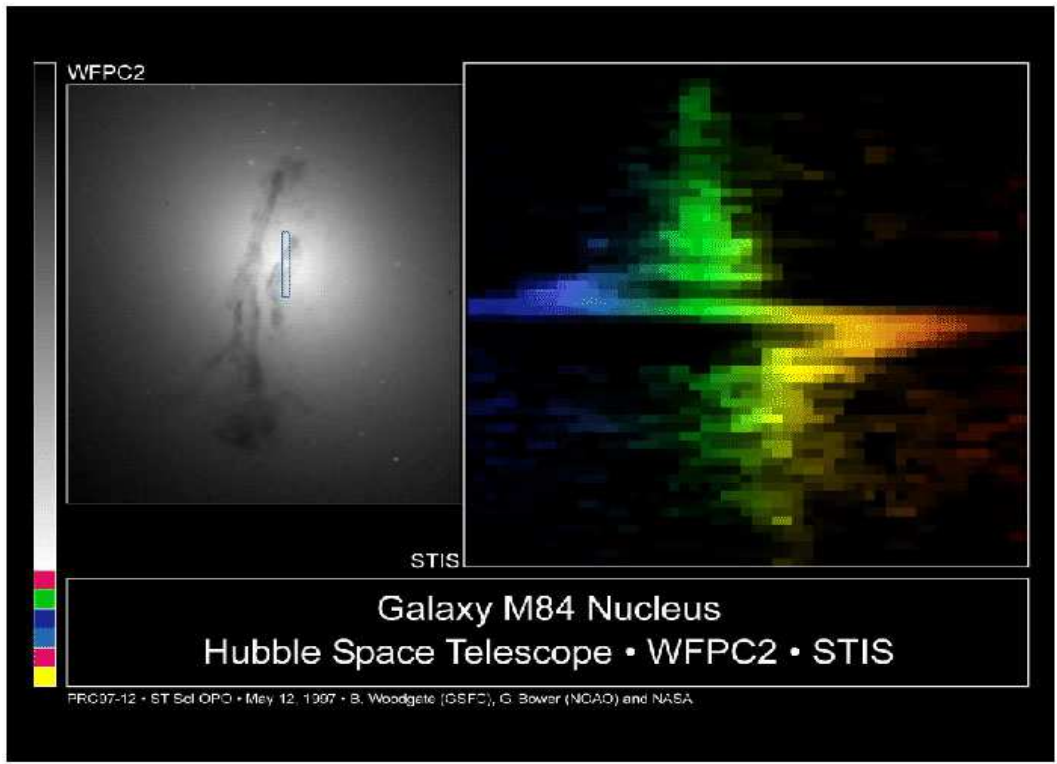
hole not unlike the one suspected of residing in M84 (see \Fignobox{thesispics/M84}{Left:
photograph of nearby galaxy M84. Right, vertical axis: position within central
boxed region in photograph on left. Right, horizontal axis: equivalent line-of-sight
velocity of line peaks in position-dependent spectrum within boxed region. The
``S'' shape is suggestive of a supermassive black hole. Photo courtesy of
STScI.}). The gravitational potential energy that must be released for this
accretion process to occur is sufficient to power the continuum energy released
by the high-redshift quasars we see today.

\section{A Critique of AGN Line Emission Models}

Though the accretion disk theory of AGN continuum emission currently has little
``competition,'' there is no consensus on the AGN line emission models. The
following is a an incomplete list of the models that have been proposed:

\begin{itemize}
\item Non-Doppler line broadening (e.g., Raine \& Smith 1981, Kallman \& Krolik 1986)
\item Two-phase, pressure-equilibrium clouds in radial flows or chaotic motions (e.g.;
Wolfe 1974; McCray 1979; Krolik, McKee, \& Tarter 1981)
\item Accretion disks with central tori or coronas (e.g., Collin-Souffrin 1987, Eracleous
\& Halpern 1994)
\item Hydromagnetically driven outflows from accretion disks (e.g., Emmering, Blandford,
\& Shlosman 1992; Cassidy \& Raine 1993)
\item Outflowing disk winds (Murray et al. 1995)
\item Stellar winds (e.g., Edwards 1980, Norman \& Scoville 1988, Kazanas 1989, Alexander
\& Netzer 1994)
\item Tidally disrupted stars (e.g., Roos 1992)
\item Supernovae remnants (e.g., Aretxaga, Fernandes, \& Terlevich 1997)
\end{itemize}
The next several subsections comment on the viabilities of some of these models.

\subsection{Non-Doppler Broadening?\label{non-doppler}}

Before a species as questionable as ``rapidly moving broad line region cloud''
(hereafter, ``BLR cloud'') is considered seriously in order to describe AGN
line emission, one should first estimate the importance of the non-Doppler broadening
mechanisms. Such estimates are also useful in estimating the minimal number
of clouds that some AGN models require. As discussed in \S\S~ 2.1 \& 1.3, this
is because AGN emission profiles are observed to be extremely smooth.\pfn{{[}look
for the word stark in {*}.chi documents.{]}}

There are several different potential sources of line broadening in AGNs. Let
us discuss some of the most important. The existence of a broad, yet relatively
weak component of the C {\small III}{]} $  \lambda   $1909 inter-combination
line (where the number after ``$  \lambda   $'' represents the transition
wavelength in units of \AA) suggests electron densities for the BLR edge surfaces
(as defined by the positions in the line-emitting region where the optical depths
to the observer are $  \sim   $3/4) of the BLR emission gas (more specifically,
a plasma) of $  \sim 10^{11}  $ cm$  ^{-3}  $ (e.g. Ferland et al. 1992).
For the lines AGNs are observed to emit, such densities are too low for Stark
broadening (also referred to as pressure broadening) to yield profiles as broad
as 5000 km s$  ^{-1}  $. Another argument against Stark broadening as being
the dominant source of the line broadening in AGNs is the overall similarity
between the various lines. Stark broadening would generally predict a different
profile width for each line. The opposite trend is observed (see, e.g., Laor
et al. 1994).\pfn{ {[}this was from section on Finally, there are other possible
explanations for the double peaked line profiles that have been observed in
addition to the ones described by Eracleous \& Halpern (1994). That the the
$  H_{\alpha }  $ line in these objects was found to have average widths twice
as large as in the rest of the sample prompts the question of whether transient
micro-electric fields, perhaps from $  {\textbf {v}}x{\textbf {B}}  $ terms
involving motion of uncharged clouds in magnetic fields, could be causing stark
splitting to occur in some of these objects for the $  H_{\alpha }  $ line,
which is relatively sensitive to the effect. However, the required magnetic
field would need to be $  \sim 10^{5}  $ gauss before such effects would be
dominate in determining the shape of the line profile. This number is too high
by most standards even for these rare radio-loud objects. Incidentally, answering
this question would be straightforward by comparing the $  H_{\alpha }  $ peak
separation in these objects to the peak separation of a line such as $  H_{\beta }  $,
which would be almost 1/2 for pure stark effects. {[}Nevertheless, that some
of the 12 objects detected were responding to continuum bursts that caused the
wings to flair . . ask mike.{]}}

One non-Doppler broadening mechanism which does predict similar profile shapes
for different transitions is electron scattering (e.g., Raine \& Smith 1981,
Kallman \& Krolik 1986). Electron scattering becomes important for electron
column densities greater than $  \sim \sigma _{{\rm T}}^{-1}=1.5\times 10^{24}  $
cm$  ^{-2}  $, where $  \sigma _{{T}}  $ is the Thomson cross section. 

One important feature of AGNs is that their continua luminosities vary as a
function of time. The lines are also observed to vary. Because the gas emitting
the lines is expected to be heated by the continuum, this line variation is
expected. The lines, however, take a finite time to respond to the changes in
the continuum. Line/continuum delays are discussed in much more detail in \S~2.2.
Electron scattering models predict a minimum variability time for both the lines
and the continua of $  \tau \sim r_{{\rm BLR}}/c  $, where $  c  $ is the
speed of light and $  r_{{\rm BLR}}  $ is the frequency-dependent radius where
the electron scattering optical depth to the observer is 3/4. The continuum-subtracted
line emission does not appear to have the same minimum time scales as their
underlying continua at the same frequency. In fact, the UV lines are observed
to take approximately 2-6 times longer to vary than the UV continua (see Fig.
\ref{thesispics/alldata}). Thus, electron scattering is probably not the dominant
source of line broadening in AGNs.

For these reasons, non-Doppler line broadening is unlikely to be the dominant
broadening mechanism for most of the strong and broad AGN UV lines.

\subsection{Clouds in Pressure Equilibrium With a Hot, Intercloud Medium?}

Many papers published between 1985 and 1992 refer to the ``standard model''
of AGN line emission. This standard model is the pressure-equilibrium model
developed by Wolfe (1974), McCray (1979), and Krolik, McKee, \& Tarter (1981).
The model assumes cool, high-density regions called clouds which have temperatures
of $  \sim   $ 10$  ^{4}  $ K embedded in a much larger, hot, $  \sim   $
10$  ^{8}  $ K, intercloud medium. Because line emission becomes a relatively
inefficient coolant at high enough temperatures, the dependence of the equilibrium
temperature upon ionization permits the two phases to exist in adjacent pressure
equilibrium. The clouds and their confining medium are then assumed to be in
either radial or chaotic motions at velocities of $  v\sim c\Delta \lambda /\lambda _{0}  $,
where $  \Delta \lambda   $ is the approximate line width and $  \lambda _{0}  $
is the laboratory wavelength of the transition. An essential element of the
pressure-equilibrium model is that it assumes a very low filling factor. That
is, most of the BLR region is assumed to be of very low density. This is compatible
with several empirical features of AGNs, including the fact that the continua
(emitted by accretion disks around the black holes) are able to vary much faster
than the lines (emitted by the clouds which are much farther away from the black
hole). Another feature of the pressure-equilibrium cloud model is its compatibility
with several of the observed line ratios. This agreement indicates that the
plasma emitting the lines is photoionized primarily by the continuum source
and is not in LTE (local thermal equilibrium).

One can distinguish between two quite different types of pressure-equilibrium
cloud models. The original pressure-equilibrium cloud model assumed that the
inter-cloud medium is co-moving with the clouds. Several authors have, however,
also assumed clouds that move within the intercloud medium. Let us consider
each of these types of models in turn. 

For models in which the clouds are co-moving with the intercloud medium, the
kinematic structure of the intercloud medium is important. As mentioned previously,
gas is dissipative. As a result, unless special external forces such as magnetic
fields act upon it, any local velocity gradients within an intercloud medium
would quickly die out. If magnetic forces are important, however, the clouds
are probably confined by the magnetic fields, not the intercloud medium pressure.
This would violate the fundamental assumption of the gas pressure-equilibrium
cloud model.\pfn{Models invoking magnetic confinement are discussed in \S~\ref{murray}.}
Thus, the intercloud medium in the pressure-equilibrium model either should
fall into the accretion disk or be in radial motion. Line emission from disks
is discussed in \S\S~\ref{accretiondisks}-\ref{murray}. Radial motion is ruled
out in objects like NGC 4151 by red versus blue wing variability studies (see,
e.g., Maoz et al. 1991). AGNs are noted for their conformity over extreme variations
in parameter space, so it is unlikely that this object is an exception. 

But this is just one of many problems with the co-moving pressure-equilibrium
model. The accretion efficiency can be estimated from the observed AGN luminosities
and the mass flux. The mass flux is a simple function of the required radial
velocities of the gas in such models (which must be roughly the same as the
observed line widths) and the estimated density of the line-emitting gas (see
\S~1.2.1). The mass flux rates for the co-moving inter-cloud models are over
$  10^{3}  $ $  M_{\odot }  $ yr$  ^{-1}  $ (Kallman et al. 1993), where
``$  M_{\odot }  $'' denotes the solar mass unit. This mass flux is simply
too large for most concepts involving AGNs to make sense; it implies accretion
efficiencies of $  10^{-5}-10^{-7}  $. This efficiency is not only much less
than the typically assumed efficiency of an accretion disk $  (\sim 0.3mc^{2})  $,
but is even less than the efficiency of fusion $  (\sim 10^{-3}mc^{2})  $.
For these reasons, pressure-equilibrium models in which the clouds are co-moving
with the intercloud medium are poor BLR candidates. 

Let us now consider pressure-equilibrium models in which the clouds are \emph{not}
co-moving with the intercloud medium. It is now believed that the clouds in
these models would be disrupted by instabilities at an especially high rate.
Updated estimates of the temperatures of the intercloud medium temperatures
are only $  \sim 10^{7}  $ K; the implied inter-cloud densities are high enough
that the ram pressure at the leading edge of a cloud should cause rapid breakup
of the clouds due to Rayleigh-Taylor instabilities, unless the medium is co-moving
with the clouds (Mathews \& Blumenthal 1977; Allen 1984; Mathews \& Ferland
1987). Krinsky \& Puetter (1992) showed that the outer edges of even the co-moving
clouds are unstable to a thermal instability which grows on times scales of
$  \sim   $10$  ^{3}  $ s, which corresponds to evaporation times scales of
$  \gsim 10  $ years. Krinsky \& Puetter (1992) also found that the clouds
are dynamically unstable to trapped Ly$  \alpha   $ radiation with growth times
of $  10^{6}  $ s. If the covering factor is $  \sim   $0.1, these numbers
imply $  10^{6}  $ cloud births per year. It is questionable that the proposed
cloud formation mechanisms (e.g., Eilek \& Caroff 1979; Beltrametti 1981; Krolik
1988; Emmering, Blandford, \& Shlosman 1992) would be able to compete with these
high disruption rates, especially without the co-moving assumption. Even if
cloud generation did not require conditions different from those in the BLR
and radial models somehow worked, how exactly what the clouds are and why they
would be created with the required velocities is unclear. 

But perhaps the most severe problem with the non-co-moving pressure-equilibrium
model is simply that the clouds would slow down in $  \sim   $1.0 days as they
transfer their momenta to the intercloud medium (Taylor 1996). The energy released
by this process if the intercloud medium pressure were actually as high as the
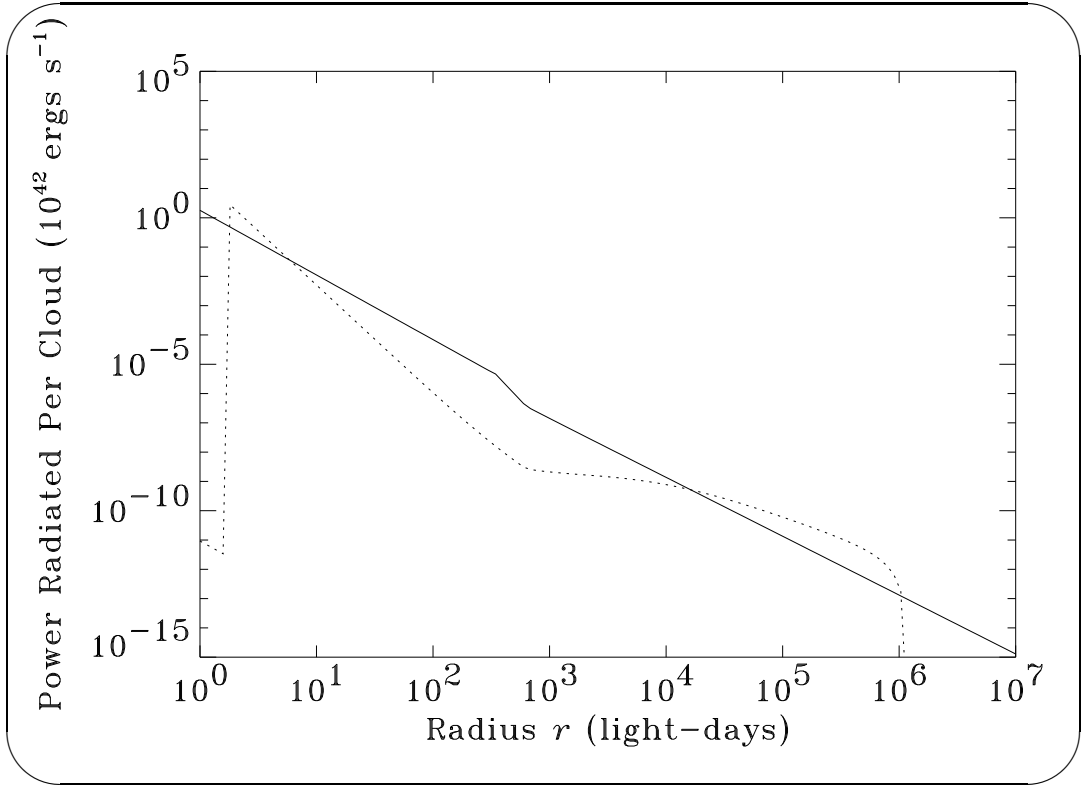
cloud pressure would be quite high. This is shown quantitatively in \Fig{model2/shockpower}{Solid
line: approximate power received per stellar wind (or cloud) for model 2 by
continuum heating. These results are dependent upon the characteristics and
parameters of model 2, which is a sub-Eddington AGN model discussed in detail
in Chapter 4. Dotted line: the approximate power $  P_{\rm d}  $ due to the
effects of drag. This calculation assumes that $  P_{\rm d}=A\rho _{\rm HIM}<v^{2}>^{3/2}  $,
where $  A  $ is the cloud cross sectional area, $  \rho _{\rm HIM}  $ is
the mass density of the intercloud medium, $  <v^{2}>  $ is the squared velocity
dispersion of the clouds. It also assumes that the inter-cloud temperature is
$  T_{\rm HIM}=10^{7}  $ K, the mean molecular weight per HIM particle is 1.3,
and the inter-cloud pressure is the same as the cloud pressure. For radii below
$  \sim   $1 light-days and greater than $  \sim   $ 10$  ^{6}  $ light days,
neither $  P_{\rm d}  $ nor $  <v^{2}>  $ are defined for model 2 because
there are no line-emitting objects in this region.}\qfn{Tim, check if this 1.3
\# is 100\% ok.}. For this reason, drag was at one point even proposed to power
AGN line emission. An argument against this idea (and the non-co-moving pressure-equilibrium
model as well) is that the lines of some AGNs would probably be unable to respond
substantially to changes in the continuum. This predicament is, therefore, similar
to that of electron scattering (\S~\ref{non-doppler}) in that it is largely
incompatible with the observed characteristics of line variability. 

In summary, though the pressure-equilibrium cloud model is able to match most
of the line ratios reasonably well, its other problems appear to make it untenable.

\subsection{Modified Accretion Disks?\label{accretiondisks}}

There are several different types of accretion disk AGN BLR models. Most assume
that the accretion disks producing the continua also produce the UV and optical
lines. Before discussing the BLR models of modified accretion disks, let us
first review the standard Shakura \& Sunyaev (1973) accretion disk theory and
why accretion disks were not originally expected to be important sources of
AGN line emission. 

Because the kinetic energy of matter in Keplerian motion must be -1/2 of the
gravitational potential energy, the heating per unit surface area due to viscosity
in an accretion disk is required to scale as $  r^{-3}  $, where $  r  $ is
the distance from the black hole. This function has a very steep slope. The
slope is so steep, in fact, that most of the emission would be emitted near
the innermost emission region, which is generally assumed to be $  \sim 3r_{{\rm S}}  $,
where $  r_{{\rm S}}  $ is the Schwarzschild radius. The temperatures on the
$  z=0  $ plane inside accretion disks are slightly more uncertain, but if
we assume each annulus to emit similar to a blackbody, we obtain (see, e.g.,
Pringle 1981) 
\begin{equation}
\label{temperatureofadisk}
T\sim 3.5\times 10^{6}(r/r_{{\rm S}})^{-3/4}\, \textrm{K}
\end{equation}
assuming accretion at 10\% of the Eddington limit and a black hole mass of 10$  ^{8}      M_{\odot }  $.
Thus, the region of continuum emission from a normal AGN accretion disk depends
upon the frequency, with the regions emitting optical continuum radiation being
much farther out than the regions near the Schwarzschild radius responsible
for the ``big, blue'' $  \nu F_{\nu }  $ bump. Nevertheless, equation (\ref{temperatureofadisk})
predicts that even the optical continuum of typical AGNs should be emitted from
a region less than several light days from the continuum source. Partial support
of equation (\ref{temperatureofadisk}) is provided by time-sampled AGN data
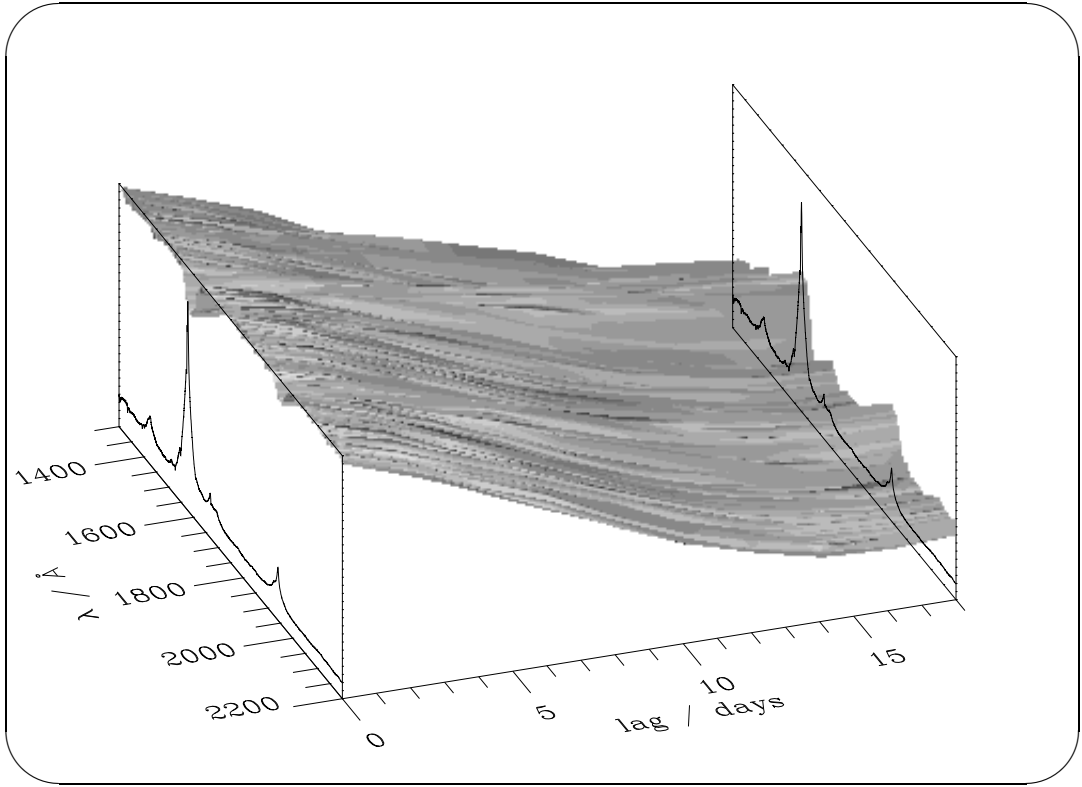
such as that shown in \Fig{thesispics/alldata}{The $  \tau \geq 0  $ frequency-dependent
cross correlation function between the 1135-1180 \AA ~de-redshifted continuum
and the UV spectral region of NGC 5548, an $  M_{{\rm h}}\la 10^{8}  $ AGN.
This plot was made using the AGN watch (Alloin et al. 1994) \emph{HST} variability
campaign data described in Korista et al. (1995).}. These data show that the
lines respond on a distinctly different, and in particular longer, time scale
than the underlying continuum. They also show that the $  \lambda   $2200:1160
\AA  ~continuum emission variability time lag, which according to equation (\ref{temperatureofadisk})
should be about twice as long as the underlying $  \lambda   $1300 \AA ~variability
time scale, is still too short to be detected with this data, which has a sampling
rate of 3 days. Models which would emit substantial UV continua from equivalent
distances farther than this are clearly ruled out.

For various reasons, the line-emitting gas in AGNs is generally expected to
have temperatures of $  T\sim 2\times 10^{4}  $ K. So, for ordinary disk models
without, e.g., a hot yet optically thick coronal region above the disk, the
region with the temperature appropriate for the UV line emission similar to
that observed in AGNs is at a distance from the black hole of 
\begin{equation}
\label{disktemperature}
r_{\rm BLR}\sim \left( \frac{3.5\times 10^{6}\, \textrm{K}}{T}\right) ^{4/3}r_{\rm S}\sim 10^{3}r_{\rm S},
\end{equation}
 assuming the same parameters as before. Given the steep $  r^{-3}  $ falloff
in the surface brightness, the above result implies that normal accretion disks
around supermassive black holes would only emit a small fraction of their flux
as UV emission lines. Also, since the majority of the energy in a disk is deposited
inside the high-density regions of the disk rather than the outside surface,
the lines from an ordinary accretion disk may actually be in absorption (like
stellar lines) rather than emission.

Let us now consider a few of the AGN disk-like emission models that have been
proposed. These models are different than the Shakura \& Sunyaev (1973) accretion
disks. Eracleous \& Halpern (1994) employ a model based on that described in
Collin-Souffrin (1987). Depending on whether or not a narrow component was added,
their models yielded double- or triple-peaked AGN H$  \alpha   $ and H$  \beta   $
line emission profiles. (Apparently, fits for other lines were not attempted.)
Their models have a free parameter which signifies the inner edge of the line-emitting
region of the disk. By treating this parameter as free, they are effectively
permitting the line emissivity to be a free parameter as well. They are, therefore,
bypassing the results implied and associated with equation (\ref{disktemperature}).
Though equation (\ref{disktemperature}) has a large variety of systematic uncertainties
associated with it, it is probably not justifiable to assume they are infinite. 

The primary reasons given by Eracleous \& Halpern (1994) for deviating from
the Shakura \& Sunyaev (1973) model is the possibility that the inner region
of the accretion disk is bloated in the form of a torus. If such a torus were
hot and high enough, it would heat the outer regions of the disk. However, the
heating per unit area that the outer regions of the disk would receive due to
such a hot torus again falls as $  r^{-3}  $, like the Shakura \& Sunyaev (1973)
thin disk. Thus, provided the radius of this torus is small enough, the dependence
of the heating function upon $  r  $ is the same as that due to viscosity. 

If, on the other hand, the radius of the torus is made large enough, there is
indeed a new effective inner radius of the accretion disk. However, such large-radius
tori models would probably have difficultly simultaneously matching the data
for various lines. This is because the response functions are different for
each line. In particular, the high-ionization lines have much shorter lags than
the low-ionization lines. In such a model, how could the Mg {\small II} $  \lambda   $
2798 line, which is generally stronger than N {\small V} $  \lambda   $ 1240
line, have a lag that is a approximately a decade larger (e.g., Krolik et al.
1991; Horne, Welsh, \& Peterson 1991)? (Unless the radius of the inner edge
of the accretion disk changes for each lines, each line should have a similar
response function.) This issue is important because the equivalent widths of
the broad lines appear to be correlated. Thus, if the large-radius tori models
are intended only to explain the Balmer/Fe {\small II} lines, they appear to
suffer from a ``conspiracy problem.'' (This problem is analogous to, for instance,
that of certain theories of dark matter which have difficulty explaining why
the disk and halos of spiral galaxies have the same rotational velocities.)
So there would appear to be problems with this disk model for AGN line emission
regardless of the size of a possible torus. 

In addition to the inner radius parameter, Eracleous \& Halpern (1994) also
employed a local ``turbulent broadening'' parameter. This was assumed to be
820-8200 km s$  ^{-1}  $ in their fits. Without a clear explanation of what
could cause such large turbulences (and why they would not be damped from viscous
forces), their model is largely incomplete. 

Even with the free parameters, only 8 of the 94 objects they examined could
be fit with their model. That this fraction is less than unity is yet another important problem with
this model.

Additional potential problems with disk line emission models concern the variability
of individual lines. The observed delays of lines are generally longer than
what would be expected from disk models (Eracleous \& Halpern 1993). Moreover,
the \emph{relative} differences between the delays of the profile cores and
wings (clearly apparent in Figure \ref{thesispics/alldata}) do not appear to
be compatible with disk models (Eracleous \& Halpern 1993). Finally, observed
profiles respond without the near perfect symmetry that ordinary accretion disk
models (in particular, those that have azimuthal symmetry) would predict. It
has been proposed that disk ``hot spots'' might accommodate these observed
asymmetries. However, the response asymmetries do not appear to vary with epoch
or object. For instance, the C {\small IV} red wings consistently respond faster
than the blue wings (Gaskell 1997). Hot spots would probably yield a more random
response behavior. 

It has been suggested that the problems with traditional accretion disks in
producing line emission might be eliminated by allowing external irradiation,
such as that given off by a jet (see, e.g., Osterbrock 1993 and references therein).
However, the continuum emission from a jet would be severely beamed away from
the disk due to relativistic effects. Also, even if relativistic effects could
be ignored, it is straightforward to show that such illumination again would
result in the steep $  r^{-3}  $ heating falloff. In light of these problems,
it appears unlikely that these models produce the UV AGN lines emission. Also,
the required geometry of these models seems somewhat contrived in that it apparently
does not stem any published analytic or numerical calculations.

\subsection{Outflowing Disk Winds?\label{murray}}

In the Murray et al. (1995) AGN model, the accretion disk surrounding the black
hole has a radiatively accelerated outflowing wind similar to that in O-stars.
This wind is also assumed to have a sheer velocity of the order of the radial
velocity. This model is discussed in more detail in Appendix \ref{adpbal}, where
it is shown pictorially in Figure \ref{figure2}. Unlike the Eracleous \& Halpern
(1994) model, the relevant region of the Murray et al. (1995) disk is assumed
to have concave flaring. One of the best features of this model is that it attempts
to explain the broad line self-absorption that several AGNs exhibit without
resorting to any special new class of absorption clouds. 

Despite some of the amicable qualities of the model, there appear to be many
difficulties with it as well. Perhaps the most widely discussed problem is that
the X-rays may inhibit the wind. This is because the wind is presumably accelerated
by line trapping and the observed X-ray flux produced by accretion disks, though
lower than the UV flux, is still high enough to fully ionize the lighter, abundant
ions. If this occurs, the force obtained from line trapping is greatly reduced.

But there are other problems that may be even more severe. For instance, the
region between the continuum source and the inner edge of the UV disk does not
appear to have been accounted for in a self-consistent fashion. The continuum
radiation $  F_{{\rm c}}  $ in the Shakura \& Sunyaev (1973) thin disk is emitted
anisotropically, with $  F_{{\rm c}}={\bf \hat{r}}\cdot {\bf \hat{z}}L/(2\pi r^{2})  $,
where ``\^{}'' denotes unit vectors, $  {\bf r}  $ is the radius vector,
$  L  $ is the luminosity, and $  {\bf \hat{z}}  $ is the unit vector of the
disk axis. This is because the column density in the plane of the disk is high
enough that the disk is assumed to insulate most of itself from the hot, innermost
region and the apparent solid angle subtended by the bright portion of the disk
is lower for off-axis observers. In other words, each ring in the disk radiates
all of its own heat. In the calculations of the Murray et al. (1995) model,
the local continuum flux in the region of interest (where the winds would originate)
was evidently assumed to be $  F_{{\rm c}}=L/(4\pi r^{2})  $. The high-column
density region between the continuum source and the UV section of the disk appears
to have been treated as a vacuum. Ideally, their models would account for the
self-absorption of the disk and a nonspherical inner disk continuum flux. Because
$  {\bf \hat{r}}\cdot {\bf \hat{z}}\ll 1  $ at the inner edge of the hypothesized
line-emitting region of the disk, such accounting would probably result in much
less UV line emission. However, because the Murray et al. (1995) model invokes
substantial concave flaring even this is unclear.

The mere use of an inner edge makes the Murray et al. (1995) disks similar to
the disks employed by Eracleous \& Halpern (1994). This is because both models
assume this somewhat artificially induced, innermost section of the disk far
outside the Schwarzschild radius. The justification for the use of an inner
edge is that the wind could not start closer to the continuum source center
because the ionization state of the gas would be too high for line trapping
to be important. In contrast, models of AGN disks which include the effects
of magnetic fields (generally discussed to explain the existence of jets) appear
to produce the strongest outward radial velocities much farther in, at $  r\sim 3r_{\rm S}  $.
For a given AGN, at least one of these results must be incorrect. Since the
radio-loud AGNs with jets appear quite similar to other classes of AGNs, it
is unlikely that their lines are produced in a fundamentally different way than
the lines in other AGNs.

Another potential problem with the Murray et al. (1995) model is that viscosity
might not have been included in a self-consistent fashion. Disks produce their
heat because of viscosity, so viscosity cannot be ignored entirely in disk models.
The outbound gas in the Murray et al. (1995) model is assumed to follow thin
streamlines that reside just on top of a stationary surface of the disk below.
Since this outbound gas is required to travel at speeds of 10,000 km s$  ^{-1}  $
or more, the velocity gradients in the outbound line-emitting gas could easily
be much higher than those within the disk itself. Thus, it is possible that
viscosity, had it been accounted for in a more complete rendition of the model,
would be the dominant force on the ions of the outbound gas. If this is true,
the line-emitting gas might never flow outwards in the first place. 

In addition to the above concerns regarding the physics put into the Murray
et al. (1995) model, there are also several problems with the predictions of
the model. For instance, the model predicts that the profile shape is a strong
function of optical depth. Therefore, since each line transition has a different
oscillator strength, each line profile in the Murray et al. (1995) model is
predicted to be different. This poses the same problem discussed earlier regarding
Stark broadening. Thus, due to the six decade difference in the oscillator strengths
of the C {\small IV} and C {\small III}{]} lines, for example, this model would
probably have difficulty in fitting the observed profile similarities. 

One alleged success of the Murray et al. model (1995) is that it purports to
explain why the BALQSOs are generally heavily absorbed in the X-rays. However,
most AGN models with UV BAL absorption will also have X-ray absorption. This
is simply because warm enough gas with substantial optical depths in UV lines
generally also has strong X-ray absorption edges. So this ``success'' should
be expected for \emph{all} BALQSO models with warm enough absorbing material,
not merely the Murray et al. (1995) model. \qfn{ASK TIM TO CHECK THIS PARAGRAPH}In
summary, the Murray et al. (1995) model does not appear be very compelling.

\subsection{Ordinary Stars as the Invisible Cores of AGN Clouds?}

The next subsections discuss models in which the BLR clouds are actually stars
or stellar winds. Before we discuss them in detail, some general comments about
the the history of AGN models are perhaps in order. The pressure-confined, two-phase
equilibrium cloud model appears to have been founded on the inferred physical
conditions (density, temperature, and ionization) of the line-emitting gas as
derived from the observed line ratios. In some sense, it could be claimed that
this cloud model ``bypassed'' the scientific method. This is because the scientific
method (at least in its original form) states that the data used to test a hypothesis
should be analyzed only after the testable conclusions of the hypothesis have
been developed. By building models for the purposes of fitting our data, we
can violate the scientific method, especially in non-interactive situations
like these. When calculations suggested that the AGN clouds were optically thick
to their emission lines (e.g., Kwan \& Krolik 1979, 1981; Weisheit, Shields,
\& Tarter 1981), the various line diagnostic techniques yielded little information
about the density near the center of the clouds or, for that matter, the upper
limit to the column density. For this reason, line ratios obtained from clouds
that are winds surrounding stars are quite close to those of the standard two-phase,
pressure-equilibrium clouds (\S~4.2); in both cases low-density gas is subjected
to ionizing AGN continuum radiation. While cloud models with and without dense
stellar cores describe many of the primary observations well, models with dense
cores are more immune to the effects of drag, instability, and dissipation. This
distinction is probably the single most important feature of all stellar AGN
line emission models.\rfn{The following hypothetical and humorous example helps
to elucidate this point: a hard-core astrophysicist goes to the local pub, has
too much to drink, and passes out. A friend takes the astrophysicist to his
home, which happens to be located near a freeway. The astrophysicist wakes up
in the middle of the night and looks out the window. He sees flashing red and
white lights in the distance and wonders what they are. Not knowing any better,
and still being partially drunk, he decides to take an optical spectrum of the
lights. He sees Doppler-shifted line emission. He concludes that clumps of ionized
halogen gas are somehow moving around in the atmosphere. He calculates the mass,
density, temperature, and dissipation time of the clumps and becomes puzzled
why the clumps fail to slow down. Of course, he is just seeing lights from cars
in traffic! Cars with headlights can make more sense than headlights alone,
even though some observers cannot distinguish between the two.} 

A main sequence star which reprocesses continuum radiation at its photosphere
is arguably the simplest AGN cloud line emission model one could envision. External
heating would change the boundary condition of the intensity in the inward directions
at the surface, which would normally be zero. The most obvious change to the
stars would be an increase of the photospheric temperature required to bring
the stellar surface back into thermal equilibrium. At least in a limited number
of binary accreting systems, such changes in spectral type have been observed
(e.g., Hutchings et al. 1979). 

The result of this heating could indeed be line emission from the stellar surfaces.
For a line to be in emission, the temperature as a function of depth must be
such that the source function decreases near $  \tau \simeq 0.7  $. Though
this process depends upon the specifics of the line transition in question,
it generally occurs when the heated photosphere is slightly hotter than the
temperature of the unheated photosphere obtained without chromospheric heating
from the star itself (see, e.g., Hubeny 1994). Thus, AGN continua could produce
line emission from stars, even stars later than spectral type O or B (which
interestingly emit lines without external heating).

The position-dependent geometrical covering factor for this type of AGN model
can be estimated by
\begin{equation}
\label{coveringfactor}
\Omega (r)=1-\exp \left( -\int _{0}^{r}dr'\sum _{t}n_{t}(r')A_{t}(r')\right) ,
\end{equation}
 where $  r  $ is the distance from the black hole, $  A_{t}  $ is the cross-sectional
area of stars of type $  t  $, and $  n_{t}  $ is the density of such stars.
With the above definition, the geometrical covering factor represents the maximum
fraction of continuum light that could be affected by the clouds; the exponential
factor serves to prevent this fraction from being greater than unity. The geometrical
covering factor is also the absorption coefficient for light emitted at $  r_{\rm t}\bf {\hat{r}}  $
and received at $  r\bf {\hat{r}}  $ assuming that each cloud is completely
opaque. An estimate of $  n_{t}  $ near a massive black hole is provided in
Murphy, Cohn, \& Durisen (1991). The highest density of stars that occurred
anywhere during the evolution of their cluster model $  2B  $ was $  10^{11}  $
pc$  ^{-3}  $ at $  10^{-4}  $ pc from the black hole $  10^{6}  $ years
into its evolution for the 0.3 solar mass stars. Upon extending this to $  10^{-2}  $
pc with a slope of $  r^{-0.5}  $ and assuming $  A_{t}=\pi (0.3\cdot R_{\odot })^{2}  $,
the above equation yields a broad line covering factor of only $  \sim 10^{-8}  $;
the contribution toward the broad line emission would appear to be negligible
for this model.

Due to this fundamental problem, it seems unlikely that normal stellar surfaces
produce the line emission that is observed in AGNs.

\subsection{Tidally Disrupted Stars?}

Stars near a supermassive black hole are subject to strong tidal accelerations.
At the tidal radius these accelerations are by definition equal in magnitude
to the normal stellar surface gravity. Stars near enough to the central black
hole can therefore be ``tidally disrupted.'' It has been proposed that this
resulting stellar disruption is the source of BLR clouds in AGNs (Lacy et al.
1982; Rees 1982, 1988; Sanders 1984; Roos 1992). 

The importance of the effect (i.e., the associated covering factor) was estimated
in Roos (1992). This paper states that only $  0.1\, M_{\odot }  $ stellar
masses need to be tidally disrupted per year to obtain the observed line emission.
This value, however, is much higher than the $  10^{-2}\, M_{\odot }  $ yr$  ^{-1}  $
maximum loss-cone rate that occurred in the system with a $  5\cdot 10^{8}\, M_{\odot }  $
supermassive black hole calculated by Murphy, Cohn, \& Durisen (1991). Murphy
et al. also showed, at least for high stellar densities (which is the case of
interest here), that the mass lost by stellar collisions and stellar evolution
to the BLR interstellar medium actually dominates that of tidal disruption by
about two orders of magnitude. Moreover, the phase space available for stars
to be disrupted, yet not accreted, can be very small. If the black hole mass
is greater than $  \sim 10^{8}\, M_{\odot }  $ (which would not be unusual
for a QSO) the vast majority of stars are presumably accreted whole (Hills 1975;
Rees 1990). This could be a problem with this model because these objects do
of course emit lines, though the associated covering factors are somewhat lower
those of the lower luminosity Seyfert 1s.

If we assume the existence of an accretion disk which occults the broad line
region (Taylor 1994), the tidal disruption model described in Roos (1992) may
have additional problems. In particular, since the BLR gas associated with a
stellar disruption is not emitted until after the star reaches pericenter, for
AGNs in which the remnants dissipate rapidly enough the line emission would
be blueshifted in all but the edge-on-disk AGNs. Three of the more obvious such
remnant dissipation mechanisms include the accretion disk itself (which remnant
clouds would collide with), momentum exchange with the inter-stellar medium,
and evaporation of remnant gas to the hot ($  T\gg 10^{5}  $ K) phase. Incidentally,
none of these effects were accounted for in the simulations by Evans \& Kochanek
(1989), and only evaporation was accounted for in Roos (1992). These blueshifts
would cause the profiles to have shift/width ratios much higher than the values
of $  \la   $30\% that are typically observed.

In particular, the post-disruption velocity of ejected remnant components is
(e.g., Roos 1992)
\[
\simeq 5000(M_{\rm h}/10^{6}\, M_{\odot })^{1/6}\, \textrm{km s}^{-1},\]
where $  M_{\rm h}  $ is the mass of the black hole. Even if we reduce this
velocity by a factor $  \sim 2  $ to account for random disk orientations,
the line-emitting material farthest from the continuum source would appear to
be extremely blueshifted in this model. This does not appear to be consistent
with the observed characteristics of AGN variability (such as the nearly unshifted
profiles peaks shown in Figure \ref{thesispics/alldata}), though the blue sides
of the C {\small IV} $  \lambda   $1550 wings do respond slightly slower than
the red wings.

Finally, the return time for bound remnants is only $  \sim .03\, (M_{\rm h}/10^{6}\, M_{\odot })^{1/2}  $
(e.g., Rees 1990). This should be approximately the same as the disk-collision
time scale. If we assume that the smooth, nearly time-independent profiles imply
that there are at least 10 line-emitting clouds at any given time, the disruption
rate would be $  \ga   $300 yr$  ^{-1}  $. This lower limit is incompatible
with some models. 

Tidal disruption is interesting, and it will be discussed in more detail elsewhere.
However, the above results imply that the model described in Roos (1992) may
have difficulty producing the observed, essentially non-transient AGN line intensities
and low shift/width profile shapes.

\subsection{Heated Stars?\label{Heated stars?}}

In 1980, A. Edwards proposed that AGN continuum radiation would enhance stellar
mass losses. He asserted that the additional mass would be accelerated by the
continuum to form gaseous, comet-like plumes, which would then emit the BLR
line radiation (Edwards 1980). The effect of heating upon stellar wind strength
has been studied in moderate detail for AGNs in Voit \& Shull (1988) and for
X-ray binaries in Basko \& Sunyaev (1973), London et al. (1981), London \& Flannery
(1982), Tavani \& London (1993), and Banit \& Shaham (1992). One of the reasons
the studies on binaries were performed is because enhanced mass loss rates of
heated stars have also been proposed to explain the difference in formation
rates between low-mass X-ray binaries and low-mass binary pulsars (e.g., Kulkarni
\& Narayan 1988). The influence of heating upon stellar structure has been studied
for the same reason in Podsiadlowski (1991), Harpaz \& Rappaport (1991), Frank
et al. (1992), Hameury et al. (1993) and several others. Tout et al. (1989)
studied it within the context of AGNs.

To estimate the significance of the external heating, most of these studies
begin by comparing it to normal stellar cooling. The two are approximately equal
when 
\begin{equation}
\label{starheating=cooling}
\alpha \pi R_{*}^{2}\frac{L}{4\pi r^{2}}=\sigma T^{4}_{*}4\pi R_{*}^{2},
\end{equation}
where $  \alpha   $ is the fraction of radiation that penetrates the wind and
corona of the star, $  R_{*}  $ is the radius of the star, $  r  $ is the
distance from the continuum source, $  L  $ is the continuum luminosity, $  T_{*}  $
is the temperature of the photosphere, and $  \sigma   $ is the Steffan-Boltzmann
constant. For a normal AGN spectrum with a ``big, blue bump'' and stellar
winds that are highly ionized, one expects $  \alpha \simeq 1  $ (see also
Appendix \ref{Ostararea}). For a Seyfert-class AGN with a fiducial luminosity
of $  L=3\cdot 10^{44}  $ ergs sec$  ^{-1}  $ and a BLR size of $  r=3.5  $
light-days, the flux is $  3\times 10^{11}  $ ergs s$  ^{-1}  $, and the above
equality holds for $  T_{*}=6050  $ K. An H-R diagram reveals that red dwarf
main sequence stars and red giants can be affected in the suspected BLR.

But the BLR appears to span a decade in radius, so stars hotter than this and
nearer to the continuum source could also be affected. The stars nearest to
the continuum source are presumably at their tidal radius, which is 
\begin{equation}
\label{tidalradiusequation}
r_{\rm t}\equiv R_{*}(2M_{\rm h}/M_{*})^{1/3}.
\end{equation}
 For NGC 5548, this is also near the ``inner edge'' of the BLR covering function
( \S~\ref{tidal section}). Approximating the main-sequence stellar luminosity
as $  L_{*}=L_{\odot }(M_{*}/M_{\odot })^{a}  $, we obtain 

\[
M_{*}\la M_{\odot }\left( {\alpha L/L_{\odot }\over 4\cdot (r/r_{\rm t})^{2}(2M_{\rm h}/M_{\odot })^{2/3}}\right) ^{1/(a-2/3)}.\]
 Adopting $  a=3  $ and a black hole mass of $  5\cdot 10^{7}M_{\odot }  $,
this yields $  M_{*}\la 130M_{\odot }  $ at $  r=r_{\rm t}  $. These parameters,
which are typical of the objects that have been studied with reverberation mapping,
show that heating can affect all main sequence stars. If one assumes Eddington
luminosities, which are a few orders of magnitude larger than the above luminosity,
this result is strengthened by a substantial amount (Edwards 1980).

Tavani \& London (1993) found that the mass lost due to such external heating
is proportional to the ratio of the heating to the wind power, 
\begin{equation}
\label{tavaniandlondon}
\dot{M}_{*}\frac{GM_{*}}{R_{*}}=\epsilon \frac{L\pi R_{*}^{2}}{4\pi r^{2}}
\end{equation}
 where  $  \epsilon \sim 10^{-3}-10^{-1}  $ for various values of the heating-cooling
parameter. For solar parameters and the above fiducial AGN luminosity and BLR
size, this yields a mass loss of $  \epsilon \cdot 4\times 10^{-8}\, M_{\odot }  $
yr$  ^{-1}  $. The models of London, McCray, \& Auer (1981) yield lower mass
loss estimates. They found mass loss enhancements of only $  10^{-9}-10^{-7}\, M_{\odot }  $
yr$  ^{-1}  $ assuming a local continuum flux of $  10^{13}  $ ergs s$  ^{-1}  $,
which is $  \sim 10^{2}  $ times greater than the suspected continuum flux
heating to which the BLR clouds in AGNs are exposed. Even for these extreme
values of the local continuum fluxes, such mass losses are too weak to yield
substantial BLR covering factors. The models of Voit \& Shull (1988) for red
giants and red supergiants showed similar results. 

But there are two even more severe problems with assuming that radiatively excited
winds in AGNs cause the BLR line emission. First, if the winds indeed reprocess
continuum radiation into line emission, then they must also \emph{shield} the
stellar chromosphere from most of the continuum heating that would occur. In
other words, $  \alpha \simeq 0  $. (This is not strictly true for the X-rays,
but their luminosity in quasars is usually much less than that of the UV emission.)
Second, the Baldwin Effect (Baldwin 1977) suggests that the line emission goes
\emph{down} when the luminosity increases. The opposite would be naively expected
for this model. So the existence of the Baldwin Effect would appear to pose
a problem for models which assume radiatively excited winds (see also Taylor
1994).

Whether or not external heating substantially alters the stellar mass loss rates,
it probably does influence stellar structure and evolution. The questions of
interest here are whether or not the affective areas of the stars might increase,
if the number density of red giants becomes enhanced in AGNs, what the time
scales are for such increases to occur, etc. External radiation should significantly
increase the size of the convective layer of a star in order to maintain the
required heat losses from fusion (e.g., Edwards 1980). The steady-state analysis by Tout et al. (1989) showed that stars with convective
envelopes should, in order to satisfy the virial theorem, expand and cool upon
a prolonged increase in external effective temperature. While the luminosity
of the star increases, its fusion luminosity decreases, which slows down the
evolutionary process. The evolutionary end points in this situation need not
be neutron stars or white dwarfs, but rather can be evaporating ``passive stars\char`\"{}
that simply reprocess the exposing radiation. Tout et al. (1989) showed that
the evolutionary slowdown, coupled with the enhanced mass-loss rate that the
few late-type stars would have, should decrease the overall number of red giant
stars. 

The situation is perhaps best summed up by Harpaz \& Rappaport (1991, 1995).
Upon initial exposure to external heating, the photosphere of a star heats up
within hours to reradiate the majority of the additional heating. The accompanying
pressure increase in the photosphere takes the system out of its previous equilibrium,
and it expands. The outer portion of the convective envelope is no longer able
to operate under the temperature inversion; the energy of the star instead goes
into expanding the size of the convective envelope. Time-dependent plots of
surface temperature given by Antona \& Ergma (1993) and the results of Harpaz
\& Rappaport (1991) seem to indicate that the surface remains out of equilibrium
for a brief time of only $  \sim 10^{2}-10^{4}  $ years. The convective layer
as a whole adjusts to equilibrium on a much longer time scale that is somewhere
between the Kelvin-Helmholtz time scale of the entire star and that of just
its convective envelope, which is $  \sim 10^{7}-10^{9}  $ years for stars
of respective mass of 0.8 and 0.1 years. For the AGN case, the time scales of
interest can be much less than a year.

While radiative stars with masses greater than $  1.5  $ $  M_{\odot }  $
are able to reradiate the additional heat, convective stars with a mass of less
than $  1.5  $ $  M_{\odot }  $ will, after the initial exposure, expand,
powered by the internal fusion heating in the stars. This expansion will lower
the temperature near the center of the star, which eventually reduces fusion.
As described by Tout et al. (1989), the stars with convective envelopes expand
and cool upon a prolonged increase in the external temperature in order to satisfy
the virial theorem. While the radius increases by only a factor of 1.2 for a
1.0 solar mass star subjected to a flux of $  10^{11}  $ ergs s$  ^{-1}  $,
the radius of a $  0.2  $ solar mass star should increase by a factor of 3.0
if subjected to a flux of $  10^{12}  $ ergs s$  ^{-1}  $ (Podsiadlowski 1991;
Hameury et al. 1993). The $  0.1\, M_{\odot }  $ mass stars, which are the
most numerous in normal clusters, probably can expand even more, but their evolution
is difficult to follow (Antona \& Ergma 1993). Although the observed luminosity
of such line-emitting irradiated stars increases, their luminosity due to fusion
decreases. But since these stars have such smaller radii to begin with, their
cross-sectional areas (which of course scale as $  R^{2}_{*}  $) should be
relatively small even with any bloating. 

The debate of whether or not radiatively excited stellar winds in AGNs could
be strong enough to account for the observed line emission is far from over.
Wind formation, even in normal stars, remains poorly understood. But if the
above calculations are correct, radiatively excited stellar winds are too weak
to produce the observed AGN line emission.

\section{The Stellar Wind AGN Line Emission Model\label{start-of-demos-intro}}

The stellar wind line emission AGN cloud model was pioneered by Penston (1985,
1988), Scoville \& Norman (1988), Norman \& Scoville (1988), and Kazanas (1989).
In this model, the clouds are winds emitted from red giants or supergiants (Scoville
\& Norman 1988; Kazanas 1989; cf Kwan, Cheng, \& Zongwei 1992).\pfn{either ``bloated''
main sequence stars (Penston 1988) or, alternatively, }

This proposal is interesting for the following reasons:

\begin{itemize}
\item It provides a precise description of what the line emitting clouds are.
\item It explains why the ionization parameter is similar for most AGNs. The similarity
occurs because the clouds are stratified and the ionizing radiation ``evaporates''
the outer high-ionization/low-density parts of the wind. 
\item It provides a decisive answer to the issue of the dynamics of the line emitting
clouds in favor of virial motions. This relatively old prediction is now supported
by recent AGN Watch consortium data (e.g., Korista et al. 1995; Done \& Krolik
1996).
\item It dispenses with the need for a high-density hot intercloud medium. Such a
medium is a necessary ingredient of other AGN cloud models in order to provide
the pressure necessary to confine and preserve these clouds over dynamical time
scales. (These time scales are longer than the cloud expansion times.) 
\item It supports the notion that the covering factor of the continuum source by the
clouds in a specific AGN decreases with increasing continuum luminosity (Kazanas
1989). This is because the density at the wind edges increases with increasing
local continuum flux assuming all other parameters of each stellar wind are
held constant. This makes the line-emitting region of the stellar winds shrink
as the continuum luminosity increases. This is a possible explanation for the
``intrinsic Baldwin effects'' for the UV emission lines such as C {\small IV}
$  \lambda   $1550. These effects have been observed in a number of AGNs (e.g.,
Kinney, Rivolo, \& Koratkar 1990). 
\end{itemize}
Following the original set of ideas on the nature of the AGN line emitting clouds,
Alexander \& Netzer (1994, 1997) (hereafter, ``AN94'' and ``AN97'') explored
this model from the point of view of the emission line ratios. These authors
treated the radiative transfer and line emission problem in the case of a ``bloated
star'' exposed to the ionizing radiation emitted by the continuum source. They
also presented the line ratios of the most prominent AGN lines. Their models
covered a large range of parameter space (base density and velocity as well
as their functional dependence on their distance from the star) of the stellar
winds which replace the AGN clouds. (Hereafter, the words ``cloud'' and ``wind''
are treated synonymously.) Within the observational uncertainties, they found
that the continuum shape they assumed did not affect their results.

One finding of AN94 was that the line emission spectrum depends mainly on the
conditions at the boundary of the line emitting wind rather than on its entire
structure. Like Kazanas (1989), AN94 initially assumed that the size of a cloud
is determined by the Compton temperature of the continuum AGN radiation. AN94
found that the ionization parameter at this boundary is sufficiently high to
produce much more broad, high-ionization, forbidden lines (such as {[}Fe XI{]}
$  \lambda   $7892 and {[}Ne {\small V}{]} $  \lambda   $3426) than is observed.
To reduce this unwanted emission, they found that they could either artificially
reduce the temperature of this layer, or terminate the clouds not by Comptonization
but by some other mechanism. In particular, if an upper limit to the mass of
each wind is imposed, then the ionization parameter becomes small enough to
suppress the unwanted forbidden line emission. 

They also examined the efficiency of line emission in AGNs within this model.
They found, as expected from previous studies, that the slowest, densest winds
provide the most favorable conditions for simulating the line emission in AGNs
from the point of view of both efficiency and line ratios. Their results indicate
that the slow and decelerating winds yield density gradients that maximize the
line emission.

Additionally, they found that the models with the most successful line emission
properties require the smallest number of clouds in order to account for the
observed line emission in terms of source covering. In particular, AN94 found
that models with the slowest, densest winds require less than $  5\times 10^{4}  $
supergiants within the inner $  1/3  $ pc. On the one hand, this low number
alleviates the constraints of Begelman \& Sikora (1991) on collision rates and
accretion rates onto the black hole. On the other hand, Arav et al. (1997, 1998)
show that it is ruled out by the observed smoothness of the profiles if the
terminal wind velocities of the supergiants are less than $  \sim   $100 km
s$  ^{-1}  $. Arav et al. (1997, 1998) also show that if the clouds are only
thermally broadened, there are at least $  \sim 3\times 10^{7}  $ of them.
The wind velocity we assume in most of our models is only 10 km s$  ^{-1}  $
(rather than the 0.5 km s$  ^{-1}  $ assumed by AN94), so this issue is less
important to our models. We also employ more clouds than AN94; for nearly all
of the models we show in Chapter 4, we assume $  3.2\times 10^{6}  $ supergiants
inside 20 light-days (the BLR) and $  3.0\times 10^{7}  $ supergiants in total,
which is compatible (though just barely) with the Arav et al. (1997, 1998) results.
Moreover, the model discussed in Appendix \ref{shiftsap} has internal velocity
broadening of $  \sim   $1000 km s$  ^{-1}  $ and clearly passes the profile
smoothness constraint. 

AN94 also found systematic deficiencies of the ratios Mg {\scriptsize II}/Ly$  \alpha   $,
N {\scriptsize V}/Ly$  \alpha   $. The former requires a lower value for the
ionization parameter, while the latter a higher value. Clearly, more complicated
models are necessary if one is to account for all line systematics in AGN.

Our approach is different from that of AN94. Although we also use photoionization
calculations to determine the ratios of the emission lines, our scheme is much
more approximate in calculating the detailed line emission than that of AN94.
Instead of the continuous multi-zone cloud of AN94, we use a two zone approximation
which is described in \S~\ref{lineefficiencies}. We avoid the arbitrariness of the radial wind velocity profiles used by AN94
by fixing our profile, but we allow the ionization parameter at the edge of
a cloud to be a function of the distance from the continuum source. Moreover,
our scope is much broader. For example, we account for the dynamics of the clouds
(presumed to be stars) in the combined gravitational field of both the black
hole and the stellar cluster, whose constituents are the giants which produce
the AGN line emission. 

In Chapter 2 we derive the equations employed to compute the various kinematic
quantities associated with our models, such as the line profiles and response
functions. In Chapter 3 we describe the approximations we made in order to compute
these quantities. In Chapter 4 we provide and discuss the results of our basic
model centered around a black hole mass of $  M_{\rm h}=3\times 10^{7}\, M_{\odot }  $,
a cluster mass of $  M_{\rm c}=5\times 10^{8}\, M_{\odot }  $, and a continuum
luminosity of $  L=3\times 10^{43}  $ ergs s$  ^{-1}  $. In Chapter 5 the
results of our study are discussed and the conclusions are drawn.

\chapter{Fundamental Assumptions
}

The AGN wind model consists of a point-like continuum source located at the
center of a stellar cluster. The winds of the red giants and supergiants in
the cluster reprocess part of the continuum radiation into lines. These winds
thus play the role of the line-emitting clouds. The dynamics of the stars are
determined by the combined gravitational field of the black hole and cluster.
The line emission from the stellar winds is dictated by the physics of radiative
transfer. The relevant observables, such as the line profiles or response functions,
can be computed by integrating over the phase space distribution function $  f  $
of the stars and their winds. In this section, we provide the expressions we
use to calculate these quantities.

\section{The Line Profiles\label{lineprofilessection} 
}

We denote $  L_{l*}(t,{\bf {r,\hat{s}}})  $ as the apparent luminosity in line
$  l  $ at time $  t  $ of an individual cloud/stellar wind at position \textbf{$  r  $}
when viewed from position {\scriptsize V}ector $  {\bf {D}}\equiv {\bf {r+s}}  $
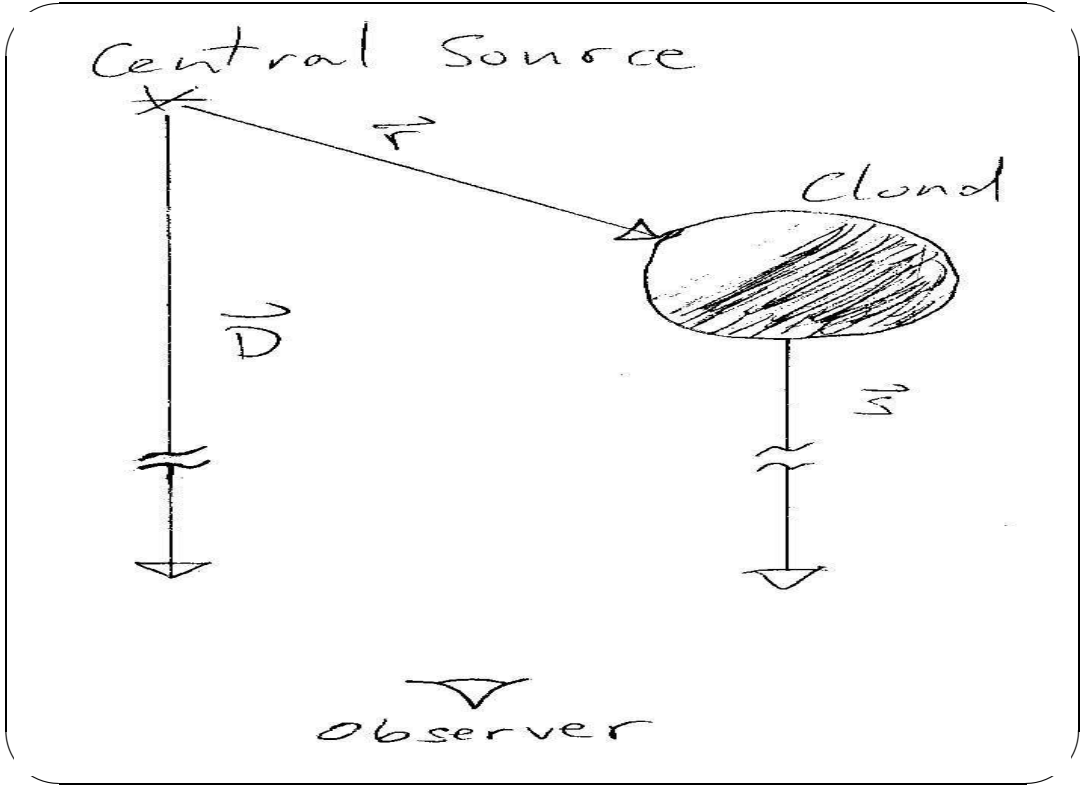
(see \Fig{thesispics/focus}{Definitions of $  {\bf D}  $, $  {\bf r}  $, and
$  {\bf s}  $.}\pfn{ This figure is not drawn to scale; because $  D\gg r  $,
$  {\bf D}  $ and $  {\bf s}  $ are essentially parallel.}). Here we take
$  {\bf \hat{s}}  $ as the unit vector of $  {\bf s}  $. As will be discussed
in \S~3.1, $  L_{l*}  $ depends upon the position of the star and time-dependent
luminosity of the central source $  L  $. The line profile $  F_{l}(t,v_{\rm D})  $
is the contribution to the line flux by the objects that have a line-of-sight
velocity $  v_{\rm D}  $. As shown in Blandford \& McKee (1982), for example,
this profile can be expressed as 
\begin{equation}
\label{2}
F_{l}(t,v_{\rm D})=\int d^{3}rd^{3}v{L_{l*}(t-s/c,{\bf r,\hat{s}})\over 4\pi s^{2}}f\delta (v_{\rm D}+{\bf v\cdot \hat{s}}),
\end{equation}
 where $  {\bf v}  $ is the cloud velocity vector and $  f  $ is the stellar
phase space distribution function. In the above equation, the delta function
in velocity serves to select the clouds with line-of-sight velocity $  v_{\rm D}  $
under the simplifying assumption that we can ignore the intrinsic line widths
of the clouds. 

Strictly speaking, equation (\ref{2}) is only valid under special conditions.
For instance, as discussed in Appendix \ref{equilibriumiscrossing}, the cloud
equilibrium times must be much less than the times for clouds to traverse the
line-emitting region. As shown in Appendix \ref{equilibriumiscrossing}, the
cloud crossing times are generally more than a year, so this assumption is probably
valid for many models. A more questionable assumption of equation (\ref{2})
is that the line emissions from the clouds do not have any explicit dependences
upon the directions of the velocity vectors. The validity of this assumption
is discussed in Appendix \ref{shiftsap}. Another assumption of equation (\ref{2})
is that the number of clouds must be much larger than a fraction of the ratio
of the entire line width to the line widths of the individual clouds. This is
so that the emitted line profiles are smooth (in accordance with observations).
As discussed in \S~\ref{start-of-demos-intro}, this assumption appears to be
reasonable in emission, at least for most of our models. For reasons discussed
in Appendix \ref{adpbal}, however, it is probably invalid in absorption. In
fact, equation (\ref{2}) is inapplicable for systems with significant BLR line
or continuum absorption. In the spirit of simplicity, and because we are primarily
interested in just the emission characteristics, we will hereafter assume that
each of these conditions is met. 

We also assume that $  f  $ is not an explicit function of the velocity vector
\textbf{$  {\bf v}  $}. Specifically, we assume that $  f  $ and $  L_{l*}  $
are the same on the $  \theta _{v}=\cos ^{-1}(v_{\rm D}/v)  $ and $  \theta _{v}=\pi -\cos ^{-1}(v_{\rm D}/v)  $
cones of velocity phase space. As a result, the profiles and their responses
of our AGN wind models are symmetric in velocity space. These assumptions permit
the delta function in equation (\ref{2}) to remove the remaining integration
over velocity. The time-averaged line profile thus becomes simply an integration
over the number density of the reprocessing objects.

\section{The Linearized Response Functions 
}

As mentioned earlier in this and the previous chapter, most AGN line emission
models consist of a point continuum source that ionizes the surrounding clouds.
These clouds then respond to changes in the ionizing flux through changes in
the associated line emission. Since the continuum sources in AGN are highly
time-dependent, these models predict that the line emission should also be time-dependent.
However, since it is believed that these clouds are located at distances much
larger than the size of the continuum source, the lines should be delayed compared
to continuum flux. These delays can then be used to probe the distribution and
radially dependent properties of the line-emitting clouds. 

If the system responds linearly, the time-dependent luminosity emitted in line
$  l  $ can be written as an integral over the time history of the continuum
source luminosity (e.g., Blandford \& McKee 1982):

\begin{equation}
\label{3}
F_{l}(t)=\int _{0}^{\infty }dt^{\prime }L(t-t^{\prime })\Psi _{l}(t^{\prime }).
\end{equation}
 In this equation, $  \Psi _{l}  $ denotes the response function of the system.
In the linear regime it indicates the response in time to an impulsive input
at $  t=0  $. As discussed in the previous subsection, variability in the luminosity
of the continuum source results in variation of the total line flux; equation
(\ref{3}) provides a means of calculating the response of the system to changes
in the continuum as a function of both the time and the velocity across the
line profile, i.e. to compute the time-dependent line emission flux given the
history of the continuum light curve. 

The procedure employing response functions has also been called ``echo mapping.''
This is because it is similar to the procedure used with sonar and other active
measurement systems. With sonar, the input is a ``ping'' given off by a submarine,
for instance. The delay of the echo from this ping provides the distance of
nearby underwater objects. If there are enough objects underwater, the echo
is a smooth function. This is the response function. However, in an AGN there
are no well-defined pings. Rather, there is only a smooth time-dependent continuum
light curve. Provided the system is linear and one has obtained enough data,
the response function obtained upon deconvolution of the data is unique and
the same that would be obtained if AGNs pinged. In this linear case, an analysis
of line and continuum light curves could be used to constrain the spatial distribution
of clouds around the black hole. For instance, as shown in Appendix \ref{localdelays},
an optically thin spherically distributed shell of linearly responding clouds
would have a hat-shaped response function. 

There are, however, indications that AGN systems \emph{are} nonlinear (e.g.,
Kinney et al. 1990, Maoz 1992). Because most AGN line emission models predict
nonlinear line emissivity and cloud area functions, this nonlinearity is actually
what one would expect. In the nonlinear case, linear response functions do not
represent the system well. In particular, even with a hypothetical infinite
quantity of error-free data, the response functions obtained upon deconvolution
would be highly dependent upon the shape of the continuum light curve in the
data set. 

To avoid this problem, one could resort to fully nonlinear model fitting. Unfortunately,
as discussed in Appendix \ref{localdelays}\pfn{ include other appendix if time
permits}, this would be very computationally expensive. 

As an alternative to abandoning response functions altogether, one can simply
include the next term in the Taylor expansion of the nonlinear response. The
procedure for doing this is described in detail in Appendix \ref{linearization}.
This leads to the introduction of the following additional parameters: the average
of the input, the average of the output, and the ``gain'' of the output. The
gain of the flux of a line is defined as its logarithmic derivative with respect
to the continuum flux. It simply tells the extent to which the system is nonlinear.
Thus, a system with gain of unity acts linearly, for example, while a system
with a gain of 2 has an output response amplitude that is twice that of the
amplitude input to the system. 

Equation (\ref{3}) modified to include these extra parameters becomes

\begin{equation}
\label{2.3}
F_{l}(t)=<F_{l}>\left( 1+\int _{0}^{\infty }dt^{\prime }\left[ {L(t-t^{\prime })\over <L>}-1\right] {\hat{\Psi }_{l}}(t^{\prime })\right) .
\end{equation}
 In this equation and throughout this dissertation, $  <x>  $ represents the
average over the entire data set of quantity $  x  $. The new input for the
system the above equation represents is the expression in square brackets, which
is simply the fractional fluctuation of the continuum about the mean. We have
used the symbol ``$  \bigwedge   $'' above $  \Psi _{l}  $ to indicate two
things: 

\begin{enumerate}
\item That we are dealing with the first order in the Taylor expansion of a nonlinear
system rather than a truly linear system. The output of a linear system is proportional
to the input. Because of the term of unity in equation (\ref{2.3}), this does
not occur. However, it does occur if the $  \bigwedge   $'ed variables are
taken as the actual input and output of the system. Thus, for the $  \bigwedge   $'ed
variables, the analysis is fully linear. We hereafter adopt the term ``linearized''
to denote the transformation of variables to their deviations about their means
such that tools like equation (\ref{3}) can be employed. 
\item That we are using a ``normalization-independent'' response function with units
of inverse time, such that the transfer function (which is just the Fourier
transform of the response function) and gain are dimensionless. We will, however,
adhere to the standard convention that the ``$  \bigwedge   $'' symbol denotes
only the unit vector when it is above a vector. 
\end{enumerate}
The normalization-independent response function in equation (\ref{2.3}) can
also be obtained by applying the procedure described in described in Appendix
\ref{linearization}. Applying this procedure to equation (\ref{2}) yields
\begin{equation}
\label{linearizedrf}
{\hat{\Psi }_{l}(t)}=\int d^{3}rd^{3}v{<L_{l*}(t,{\bf {r,\hat{s}}})>\over <L>4\pi s^{2}}\left\{ \eta (L_{l*}|L)\delta \left( t-{{\bf {r}}\over c}\cdot ({\bf {\hat{r}-\hat{D}}})\right) \right\} f.
\end{equation}
This equation is used to obtain the response functions shown in Chapter 4. The
bracketed factor is the linearized response function of an individual cloud
due to variations in the observed continuum luminosity. The factor of $  \eta (L_{l*}|L)  $
is the \emph{asymptotic} gain of an individual stellar wind line flux $  L_{l*}  $
due to small variations in the continuum luminosity $  L  $ about its local
average. The asymptotic gain is the normalization-independent transfer function
at an excitation frequency of zero. This gain is dependent upon the local continuum
flux (among other things), which is of course position-dependent. 

Strictly speaking, equation (\ref{linearizedrf}) is merely the \emph{spatial}
response function. If the cloud equilibrium times are comparable to the light
crossing times, the response function of the system is the convolution of the
spatial response function with the \emph{local} response functions of the clouds.
For simplicity, we will hereafter make the ``fast cloud'' assumption, which
is that the local response functions are delta functions in lag, i.e., that
the local response time is much less than the light crossing time. In other
words, we assume that we can ignore the details of the time-dependent responses
of the clouds and concern ourselves only with the asymptotic gains of their
line emission fluxes. The validity of this assumption is discussed in Appendix
\ref{localdelays}. This assumption yields fewer parameters because any frequency
dependence of the cloud gains is ignored; only the asymptotic gains are employed. 

The beauty of the above expansions is that, for a given nonlinear model and
continuum average, we can now fully compute the response function that would
be obtained for hypothetical infinitesimal continuum variations. Unlike the
fully linear response function, this ``linearized'' response function is unique
for a given nonlinear model\rfn{Strictly speaking, there are nonlinear models
for which this is not true. In particular, it is not true for models that violate
the fast clouds assumption and in which the sign of the time-derivative of the
local continuum flux is an important physical variable.} and average continuum.
Moreover, it yields the important physical information about the positions of
the clouds in the nonlinear system. The main drawback is that it cannot tell
us the absolute number of clouds along the iso-delay surface, but rather only
the number relative to the other clouds in the system.

One complexity of our model is its very large spatial extent. This yields an
extended tail in the response functions. Study of response function tails is
systematically difficult because the duration of such monitoring campaigns must
be several times that of the light travel time of the region one intends to
probe. The recent AGN monitoring campaigns, lasting only a month or so, may
have only probed a small fraction of the nonzero response function of the system.
Because our model is well defined across its entire spatial extent, we are able
to use equation (\ref{linearizedrf}) to compute the response function at arbitrarily
large time delays.

\section{The Velocity-Resolved Linearized Response Functions 
}

In the previous subsection, we obtained the profile-integrated response functions
for nonlinear models. With the higher resolution data recently made available,
we can also accurately measure time-dependent line profile shapes. In order
to employ this data to constrain models, we can simply extend the results of
the previous subsection into velocity/wavelength space. 

The velocity-resolved normalization-independent response function is just
\begin{equation}
\label{linearizedvrrf}
{\hat{\Psi }_{l}(t,v_{\rm D})}=\int d^{3}rd^{3}v\frac{<L_{l*}(t,{\bf {r,\hat{s}}})>}{<L>4\pi s^{2}}\left\{ \eta (L_{l*}|L)\delta \left( t-{{\bf {r}}\over c}\cdot ({\bf {{\hat{r}}-{\hat{D}}}})\right) \right\} f\delta (v_{\rm D}+{\bf {v\cdot {\hat{D}}}}).
\end{equation}
This equation gives the response of the cloud system to changes in the continuum
flux at a specific velocity \emph{$  v_{\rm D}  $.} It is different from equation
(\ref{linearizedrf}) only by the addition of a delta function in {\scriptsize V}elocity.
This delta function isolates just the clouds moving at a line-of-sight velocity
$  v_{\rm D}  $. We employ this equation to produce the velocity-resolved response
functions shown in Chapter 4.

\chapter{Approximations for the Cloud Luminosity \protect$  L_{l*}\protect  $ and the
Distribution Function \protect$  f\protect  $ 
}

In Chapter 2 we showed general expressions of the line profiles and response
functions associated with a generic system of ``clouds'' which reprocess continuum
radiation into lines. The details of the model are contained in the precise
forms of the apparent line luminosity per reprocessing object $  L_{l*}  $
and the distribution function of reprocessing objects $  f  $. In this chapter,
we discuss the assumptions and methods we used in calculating these quantities
for the models presented in Chapter 4.

\section{Approximations of the Cloud Luminosity \protect$  L_{l*}\protect  $
}

We assume that the angular distribution of the continuum radiation is isotropic
and that there is no intervening absorption. However, as we shall see shortly,
the mean column densities of the clouds can be very high. We, therefore, do
not assume that they emit isotropically. An expression compatible with these
assumptions for the apparent line luminosity of an individual reprocessing stellar
wind for an observer at $  {\bf {D}}\equiv {\bf {r+s}}  $ is

\begin{equation}
\label{3.1}
L_{l*}(t,{\bf {r,\hat{s}}})={L\over 4\pi r^{2}}A\epsilon _{l}(1+\epsilon _{{\rm {A}}l}\bf {\hat{r}}\cdot {\bf {\hat{s}}}).
\end{equation}
 In the above equation, $  A  $ is the area per cloud (which in general depends
on the time-retarded continuum luminosity and the position of the cloud), $  \epsilon _{l}  $
is the efficiency for converting continuum into line radiation, and $  -1\le \epsilon _{{\rm {A}}l}\lsim 0  $
is the line-dependent anisotropy factor, also called the beaming factor. The
anisotropy factor, which incidentally has nothing to do with the wind area,
accounts in a simple fashion for the line emission that is beamed back towards
the continuum source. Such beaming is especially important for lines in which
the clouds are very optically thick, such as Ly$  \alpha   $. According to
this prescription, such a line would have $  \epsilon _{{\rm {A}}l}\simeq -1  $.
In this case, equations (\ref{linearizedrf}) and (\ref{linearizedvrrf}) would
yield a response function that is proportional to the time lag $  \tau   $
for small enough values of $  \tau   $. Such a response function would therefore
be equal to zero at $  \tau =0  $. This would occur because the clouds that
would be both nearest to the observer and along the zero-delay line of sight
towards the observer would emit all of their line emission back towards the
continuum source. Conversely, if $  \epsilon _{{\rm {A}}l}\simeq 0  $, the
response function would, at least for simple geometries, be a monotonically
decreasing function of lag. 

Linearization of equation (\ref{3.1}) yields (see eq. {[}\ref{localdelays3.24}{]})

\begin{equation}
\label{gain}
\eta (L_{l*}|L)=1+\eta (A|L)+\eta (\epsilon _{l}|L)+\frac{<\epsilon _{{\rm A}l}>\bf {\hat{r}}\cdot {\bf {\hat{s}}}}{1+<\epsilon _{{\rm A}l}>\bf {\hat{r}}\cdot {\bf {\hat{s}}}}\eta (\epsilon _{{\rm A}l}|L).
\end{equation}
 The second and third terms of this equation are discussed, respectively, in
\S~\ref{area} and \S~\ref{lineefficiencies}. Because the column densities are
so high for nearly all of the models we compute, $  \epsilon _{{\rm A}l}  $
is insensitive to the continuum flux and $  \eta (\epsilon _{{\rm A}l}|L)  $
is negligible. In our calculations we therefore ignore the last (fourth) term
of this equation.

Equation (\ref{3.1}) states that the line emission from a wind is dependent
upon the cloud area $  A  $, the line efficiency $  \epsilon _{l}  $, and
the anisotropy factor $  \epsilon _{{\rm {A}}l}  $. In the following, we discuss
in turn the assumptions and approximations made for each of these functions.

\subsection{Approximations for the Cloud Area \protect$  A\protect  $\label{area} 
}

Following Kazanas (1989), we assume that the cooler stars such as the red giants
have winds that slowly emanate from the stars. The properties of an individual
wind are a strong function of the value of the local continuum flux to which
it is exposed. Since the luminosity of a given AGN varies much less than $  r^{-2}  $
(a factor of $  \sim 2  $ verses a factor of $  \sim   $100), these wind characteristics
are primarily a function of the distance from the black hole. There are at least
four different regions interest:  

\begin{itemize}
\item A region in which the winds are optically thick to the UV continuum. The winds
terminate at distances from the red giants where the ionization parameters at
the wind edges would be greater than $  \Xi ^{*}_{\rm {c}}  $. In this dissertation,
$  \Xi \equiv F_{{\rm c}}/cP  $ (where $  F_{{\rm c}}  $ is the normal component
of the local continuum flux vector between 1 and 1000 Ry and $  P  $ is the
gas pressure at the wind edge) is the pressure ionization parameter and $  \Xi ^{*}_{\rm {c}}  $
is the critical pressure ionization parameter in Krolik, McKee, \& Tarter (1981)
above which the temperature rises to the ``hot phase'' of $  T\sim   $10$  ^{8}  $
K. As a result of this wind edge condition, the effective area of the reprocessing
portion of a cloud $  A  $ decreases with increasing local continuum flux.
In particular, Kazanas (1989) obtained for clouds in this region
\begin{equation}
\label{windradius}
R_{\rm {w}}\simeq 7\times 10^{13}r_{17}\left( \frac{\dot{M}_{-6}}{v_{\infty 10}L_{44}}\right) ^{1/2}\, \textrm{cm},
\end{equation}
where $  r_{17}  $ is the distance to the cloud in units of $  10^{17}  $
cm, $  {M}_{-6}  $ is the wind mass loss in units of $  10^{-6}  $~$  M      _{\odot }  $
per year, $  v_{\infty 10}  $ is the wind terminal velocity in units of 10
km s$  ^{-1}  $, and $  L_{44}  $ is the bolometric continuum luminosity measured
in $  10^{44}  $ erg s$  ^{-1}  $, which is a typical luminosity for the Seyfert
class of AGNs. 
\item A region where the clouds are optically thin to UV continuum radiation. The
above functions yield a cloud size to ionizing region fraction of $  R_{\rm w}/d_{\rm {S}}\propto N_{\rm c}/\Xi \propto (L/r^{2})^{-1/2}  $,
where $  d_{\rm {S}}  $ is the depth of the inverse Str\"{o}mgren region and
$  N_{\rm c}  $ is the column density. Thus, the clouds are recombination-limited
and optically thin to UV continuum radiation at large enough distances from
the black hole. In this second possible region, the wind boundary conditions
are less important, and the effective area shrinks down to the cross-sectional
area of the inverse Str\"{o}mgren region, which is straightforward to calculate.
The result is $  A_{\rm eff}\propto (L/r^{2})^{-2/3}  $ for the clouds in this
distant region, where $  A_{\rm eff}  $ denotes an effective area different
from the physical wind area $  A  $. This area function maintains a constant
$  N_{\rm c}/\Xi   $, where the cloud edge parameters now represent an average
over the line-emitting section of the wind. Both of the area expressions for
these first two possible regions are approximations to the one obtained upon
introducing an upper cutoff to the integral in equation (4) of Scoville \& Norman
(1988). 
\item A region very near to the black hole in which the stellar winds are approximately
the same sizes as the photospheres. This region exists because the winds can
become fully ``stripped'' if the local continuum flux is high enough.
\item A region of low column density clouds far from the black hole. Far enough from
the black hole, the mean column density associated with the cool region of the
winds decreases at least as fast as $  r^{-2/3}  $. Thus, as the distance from
the continuum source increases, the mean column density would decrease without
limit as the winds became arbitrarily large and tenuous. Though the wind cloud
model does not ``need'' an external hot medium to pressure confine the clouds,
such a medium is bound to exist, and in certain models is even necessary to
fuel the continuum source. Because the stars are accelerated by the black hole
to extremely high speeds and the intercloud medium is probably stationary, ram
pressure must affect the more tenuous elements of the wind. As discussed in
Appendix \ref{cloudshape}, this is important for high enough values of the inter-cloud
density. We do not attempt a detailed calculation accounting for such effects,
as it would be beyond the scope of this dissertation. We do, however, impose
upper limits to the sizes of the clouds in the models by employing a free parameter
$  N_{\rm cmin}  $ to denote the minimum permissible mean column density. By
adjusting this parameter, we are able to avoid models with winds that would
be unrealistically large and tenuous. In this fourth and final possible region,
we therefore assume $  N_{\rm c}=N_{\rm cmin}  $. 
\end{itemize}
In practice, we calculate the line emission from the actual column densities
of the geometrical wind and do not use the effective area function $  A_{\rm eff}  $.
For almost all of the models we calculate, $  N_{\rm cmin}  $ is high enough
that the outermost region in which the winds are optically thin is never actually
realized, especially where the covering is substantial. Thus, the errors introduced
by ignoring the high-pressure region associated with $  A_{\rm eff}  $ are
probably small. As a result, most of our models have at most only three, successively
more distant regions with at most two boundaries. We denote these two boundaries
by the parameters $  r_{12}  $ and $  r_{23}  $.

In the simplest case, the cloud properties at these three boundaries would be
non-differentiable, with their radial dependences changing suddenly. This would
result in discontinuous gains. It is unlikely, however, that the stars in AGNs
are identical. Thus, in more realistic models with several different stellar
types, the actual boundaries would probably be smooth without any discontinuities.
To account for this, our gain computer code smoothes the relevant properties
of clouds located near a boundary. 

While we assume spherical symmetry in calculating the radius and particle density
of the wind, we attempt to account for suspected asymmetry due to continuum
radiation pressure. We do this by increasing the final area by a factor of four
(Kazanas 1989).

In Kazanas (1989), the radial velocity of the winds was assumed to be independent
of the distance from the stellar surface. In this paper, we attempt a more accurate
analysis. We\qfn{{[}refs here, perhaps tavani and mahalas even{]}} assume that
the wind velocity $  v  $ as a function of the distance $  R  $ from the center
of the star is

\begin{equation}
\label{3.2}
v(R)=v_{\infty }\left( 1-{R_{*}\over R}\right) ^{-1/2},
\end{equation}
 where $  R_{*}  $ is the radius of the red giant, which is $  1.5\times 10^{13}  $
cm for the models presented here. For most of our models $  R_{\rm w}\gg R_{*}  $
and the particular exponent of equation (\ref{3.2}) (assumed in our work to
be -1/2) is not particularly important. This is because the form of equation
(\ref{3.2}) yields a sudden decrease in the wind velocity near the surface
of the star. Since we assume 
\[
\dot{M}=4\pi R^{2}v\rho ,\]
where $  \rho   $ is the mass density of the wind, the pressure of the wind
is a steep function of $  R  $ only for $  R\sim R_{*}  $. The ionized, line-emitting,
inverse Str\"{o}mgren region is much smaller than $  R  $ in most of our models,
so even iso-velocity wind models yield similar results. The exceptions to this
occur at very high fluxes (very near to the continuum source) in our models
without mass loss enhancement. In this situation, ionization parameters as low
as $  \Xi ^{*}_{\rm {c}}  $ are only obtained at very high densities very close
to the stellar surface. This results in $  R_{\rm w}\simeq R_{*}  $ independent
of the precise value of the very high local continuum flux. In other words,
$  R_{*}  $ is a lower limit to the effective size of the line-emitting area
of the stellar wind. 

Under the above assumptions, the functional dependence of the line-emitting
cross-sectional area of the wind upon the local continuum flux level is straightforward
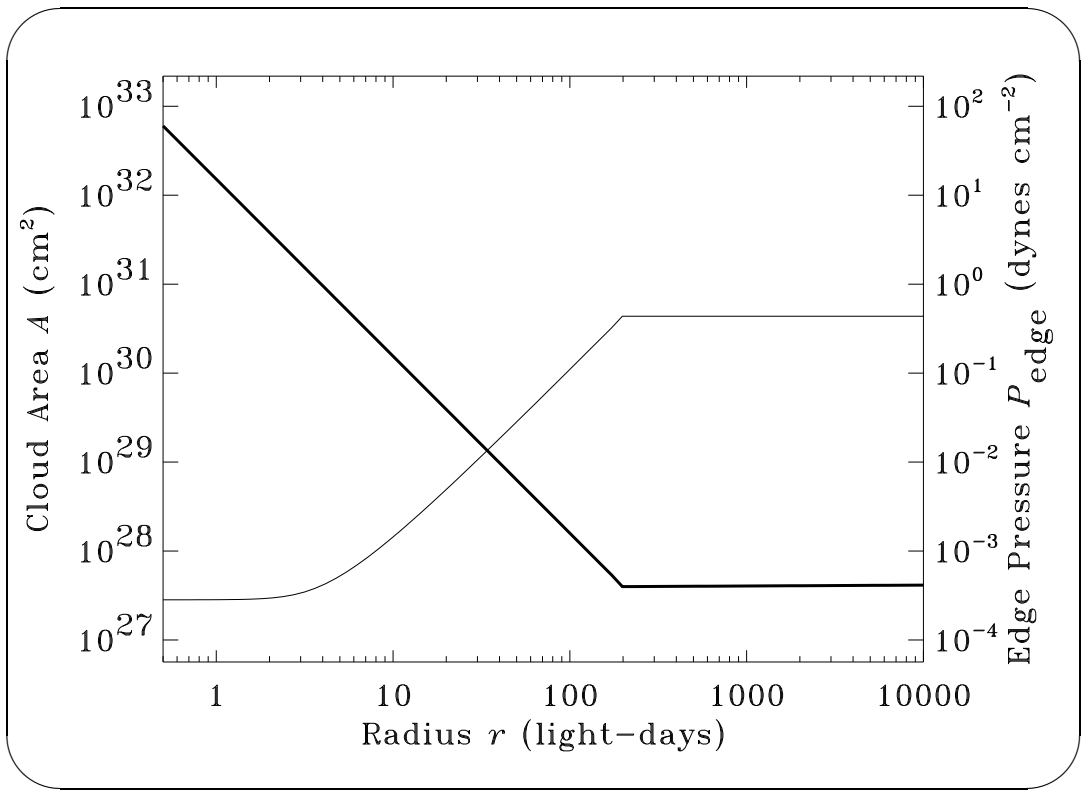
to calculate. This area is shown for model 1 in \Fig{model1/area}{Solid thin
line, left vertical axis: wind area $  A  $ as a function of distance from
the black hole $  r  $ (horizontal axis) for model 1. Solid thick line, right
vertical axis: the pressure $  P  $ at the wind edge.}. The bolometric luminosity
of model 1 is $  1.0\times 10^{44}  $ ergs s$  ^{-1}  $. All other parameters
of model 1 are the same as the fiducial ones assumed in Kazanas (1989). Specifically,
the wind edge ionization parameter is $  \Xi =10  $, the giant star mass loss
rate is $  \dot{M}=10^{-6}\, M_{\odot }  $ yr$  ^{-1}  $, and the terminal
wind velocity is $  v_{\infty }=10  $ km s$  ^{-1}  $. Note that near the
black hole (at relatively high values of the local continuum flux), the area
is essentially the size of the stellar envelope. Conversely, far away from the
black hole (at relatively low values of the local continuum flux), the area
attains its upper limit in size. For this model, $  N_{\rm {cmin}}=8\times 10^{21}  $
cm$  ^{-2}  $, which yields a maximum wind cross-sectional area of $  4.8\times 10^{-7}  $
pc$  ^{2}  $ ($  4.6\times 10^{30}  $ cm$  ^{2}  $).

Though the above parameters define the cloud area function of model 1, some
of their values are now questionable. For instance, Kazanas (1989) assumed $  \Xi =\Xi ^{*}_{\rm c}=10  $
for the outer edge of the cool zone for all winds; the ionization parameter
was independent of the local continuum flux and the edge pressure obeyed $  P\propto (L/r^{2})^{1.0}  $.
However, near the wind edge boundary there is a strong temperature and ionization
gradient which defines the conduction zone between hot and cool plasma. In the
pressure-equilibrium model, these regions are assumed to be in pressure equilibrium.\rfn{It
is precisely this assumed pressure equilibrium which prompts us to employ $  \Xi   $
as our ionization parameter rather than $  U  $, which currently appears more
frequently in AGN papers.} Because of the difficulties that would be involved,
essentially all computations of line emission from pressure-equilibrium clouds
ignore this relatively high ionization region. One difficulty is that the plasma
in the conduction zone would not even be in static equilibrium; because line
cooling becomes unimportant at high enough temperatures and ionization parameters,
this plasma would be heated up to the Compton temperature on the thermal instability
time scale. This time scale is inversely proportional to the position-dependent
difference between the heating and cooling rates from photoionization, Compton
scattering, line emission, and thermal Bremsstrahlung.\qfn{{[}Please check this
demos.{]}} This position-dependent heating rate function in turn would depend
upon the structure of the local temperature and ionization parameter functions
in the conduction zone. The precise ionization parameter to impose at the cool
edge of our simple models in order to best account for the conduction zone and
the evaporation within it is not clear to us. However, near and above the Krolik,
McKee, \& Tarter (1981) $  \Xi =\Xi _{\rm {c}}^{*}\simeq 10  $ limit, the thermal
instability is very strong. Therefore, plasma with ionization parameters near
this limit should rapidly evaporate and be physically thin. For this reason,
the value of the ionization parameter at the edge of the cool zone which best
approximates the complex situation is probably much smaller than this. In the
stellar wind case of interest here, the evaporation is balanced by mass loss
rate (as in eq. {[}\ref{tavaniandlondon}{]}) and pressure equilibrium is not
required. Thus, for the case of interest here, the location the conduction zone
and the ionization parameter at the cloud edge are additionally a function of
the mass loss rate. In contrast, the two-phase, pressure-equilibrium clouds,
which have no reservoir of cool gas, eventually would evaporate completely (Krinsky
\& Puetter 1992). 

To account for the required thinness of any high ionization layers on the clouds,
Taylor (1994) assumed that the location of the wind edge cannot be in equilibrium
if the efficiency of C {\footnotesize IV} (one of the most important cooling
lines) has a gain of less than -1.0 due to changes in the ionization parameter. With
this assumption, the first and third terms in equation (\ref{gain}) cancel
each another for models in which $  P\propto L/r^{2}  $.\pfn{XXX Put in dissertation
only if time: As Appendix xx shows, a} A side benefit of doing this was that
it resulted in asymptotic gains of Ly$  \alpha   $ and C {\footnotesize IV}
that agree with the observations by Krolik et al. (1991) that indicate, among
other things, an intrinsic Baldwin effect in NGC 5548.

In this dissertation, we make no such attempt to compute the precise location
and ionizational structure of the conduction zone. We do, however, attempt to
account for the possibility of flux-dependent edge conditions by considering
models in which $  P\propto (L/r^{2})^{s/2}  $, where $  s  $ is a free parameter.
This prescription is slightly different from that of Rees, Netzer, \& Ferland
(1989), who assumed the cloud pressures scale with distance according to $  P\propto r^{-s}  $.
Rees, Netzer, \& Ferland (1989), however, did not compute response functions.
Thus, in that paper it was irrelevant as to whether the cloud pressure in their
models was a direct function of position or merely an indirect function via
the radial dependence of the local continuum flux. In the models shown in Chapter
4, we assume the latter.

An additional assumption in question concerns the mass loss. It has been suggested
that the AGN continuum is strong enough to influence some of the cooler stars
in the vicinity of the continuum source (e.g., Edwards 1980, Penston 1988, Norman
\& Scoville 1988, Tout et al. 1989; cf Voit \& Shull 1988). Problems with this
assumption are discussed in \S~\ref{Heated stars?}. A detailed analysis of the
effect of continuum heating upon the structure of stellar winds is beyond the
scope of our work. Nevertheless, for completeness we also compute models which
follow the prescription of Norman \& Scoville (1988), who suggested that the
mass loss rate per star would be a function of position, with $  \dot{M}\propto r^{\delta }  $.
We assume a luminosity dependence and adopt $  \dot{M}\propto (L/r^{2})^{\alpha (r)}  $,
where the radial dependence in $  \alpha (r)  $ is only employed to prevent
$  \dot{M}<\dot{M}_{\rm min}  $, where $  \dot{M}_{\rm min}  $ is the mass
loss of stars outside of the line-emitting region. The addition of the $  \alpha   $
parameter permits us to consider models somewhat different from that of Kazanas
(1989). If $  \alpha >0  $, the effective area is a weaker function of the
local continuum flux than if $  \alpha =0  $. For this reason, $  r_{12}  $
and $  r_{23}  $ can be undefined for some $  \alpha >0  $ models. For models
with winds that are radiatively excited, one expects $  \alpha \sim 1  $ (\S~\ref{Heated stars?}).

Employing the above parameters yields an asymptotic area gain (which, as before,
is simply $  d\textrm{ln}A/d\textrm{ln}L  $) of 
\begin{equation}
\label{areagain}
\eta (A|L)=\left\{ \begin{array}{cc}
0 & r\lsim r_{12}\\
\alpha (r)-s/2 & r_{12}\lsim r\lsim r_{23}\\
0 & r\gsim r_{23}
\end{array},\right. 
\end{equation}
 where, as discussed in the beginning of this subsection, $  r_{12}  $ and
$  r_{23}  $ denote the boundaries between the ``bleached,'' optically thick,
and column density-limited regions. Equations (\ref{gain}) and (\ref{areagain})
are used to compute the linearized response functions that are shown in Chapter
4 for our models. For comparison, the parameters in Kazanas (1989) yield $  \eta (A|L)=-1.0  $.

\subsection{Approximations for the Line Efficiencies \protect$  \epsilon _{l}\protect  $
and the Anisotropy Factors \protect$  \epsilon _{{\rm {A}}l}\protect  $ \label{lineefficiencies}
}

Equation (\ref{3.1}) for the cloud line luminosity $  L_{l*}  $ also contains
the line efficiencies $  \epsilon _{l}  $ and the anisotropy factors $  \epsilon _{{\rm {A}}l}  $.
The line efficiency of a stellar wind is harder to compute than the efficiency
of a traditional cloud such as the ones considered by Kwan \& Krolik (1979).
This is partly because the fundamental properties of the line emitting gas within
a wind can be a function of the ``impact parameter'' $  p  $, which is defined
to be the perpendicular distance from the radius vector $  {\bf r}  $ that
extends from the black hole to the star. In Alexander \& Netzer (1994), the
line efficiency was calculated by partitioning the clouds/winds into shells
of various impact parameters. One-dimensional radiative transfer was then applied
to each shell.

Our approach in determining $  \epsilon _{l}  $ is more modest; we merely partition
each wind of cross-sectional area $  A  $ into two sections of equal area:
a ``face-on region'' and an ``angled region.'' The face-on region represents
the inner cylindrically shaped region of a cloud with low impact parameters.
This region is assumed to have a pressure at the edge of the cloud of $  P=F/(c\Xi )  $,
where $  c  $ is the speed of light and $  F  $ is simply $  L/(4\pi r^{2})  $
rather than the slightly more complicated normal component of the flux vector.
The angled region represents the outer portions of the wind that have higher
impact parameters. In the angled region, the normals of the tangential surfaces
of the wind edge are not parallel to the continuum flux, but are rather offset
by a mean angle of $  \theta   $. We assume that the edge pressure in this
region is $  P=F/(c\Xi \cos \theta )  $, which should be valid for optically
thick winds. We then simply adopt $  \theta =\pi /4  $.

For this prescription to be accurate, the size of the ionized, line-emitting,
inverse Str\"{o}mgren region must be much smaller than the size of the winds.
In other words, as stated above, the winds must be highly optically thick to
most of the continuum. In this case, line emission occurs from only a thin (relative
to the scaling size of the pressure gradient) outer skin, and only the edge
conditions (including the incidence angle of the flux with respect to the normal
of the wind surface $  \theta   $) are important. (Conversely, if a significant
portion of a wind is optically thin, the edge conditions become relatively unimportant.)
Assuming the size of the ionized region is approximately $  d_{{\rm Ly}\alpha }\simeq F_{\rm c}/(\textrm{Ry}\cdot n_{\rm H}\alpha _{\rm B})  $
(where $  n_{\rm H}  $ is hydrogen density, $  \alpha _{B}  $ is the Menzel-Baker
case B hydrogen recombination coefficient, and 1.0 Ry=13.6 eV), the winds become
thin when the column density to ionization parameter ratio $  N_{\rm c}/\Xi   $
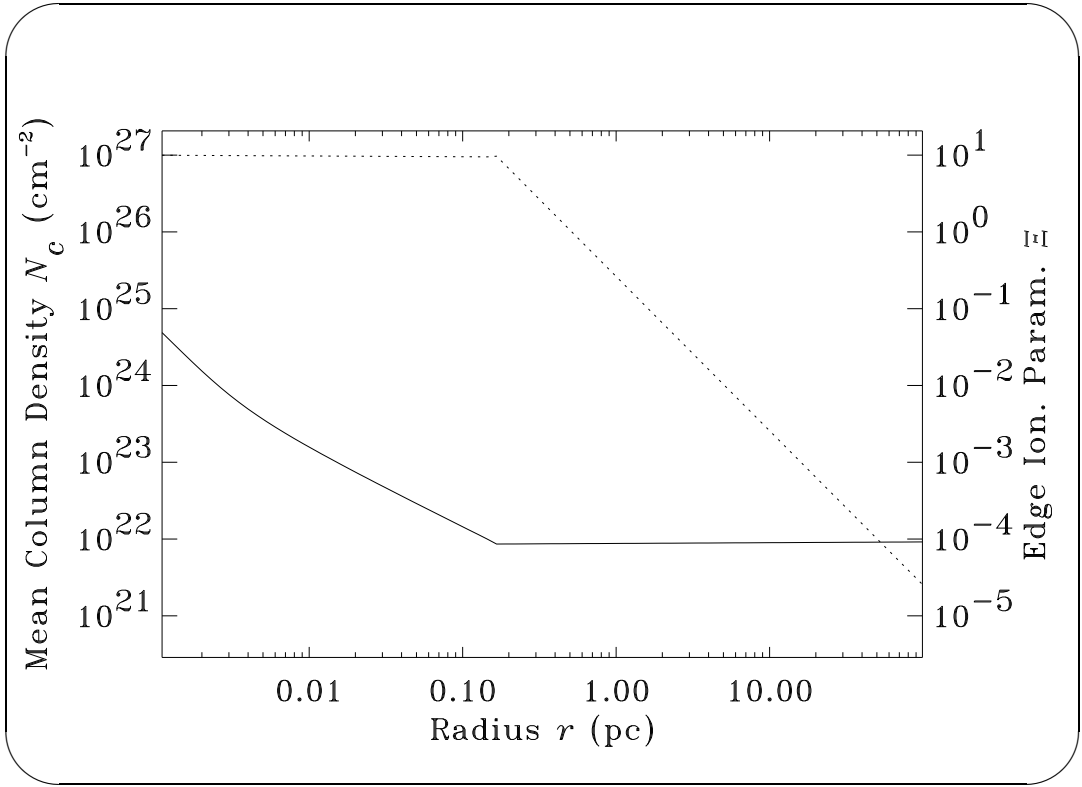
exceeds $  \simeq 10^{22}  $ cm$  ^{-2}  $. In \Fig{model1/N_c+Xi}{Solid line,
left axis: the mean column density as a function of radius from the continuum
source $  r  $ (horizontal axis) for model 1. In the broad line region, the
column density drops off faster than 1/$  r  $ because the radial velocity
of these winds at the edge is significantly less than the terminal velocity.
Dotted line, right axis: the ionization parameter $  \Xi   $. For this model,
the ionization parameter is constant throughout most of the broad line region
with $  \Xi =\Xi ^{*}_{\rm {c}}\simeq 10  $. Beyond $  0.15  $pc, the mean
column density reaches its minimum permitted value, the edge pressure becomes
independent of $  r,  $ and the effective ionization parameter is assumed to
be inversely proportional to the local continuum flux.}, the mean column density
and ionization parameter of  the winds for model 1 are shown. The column density
to ionization parameter ratio is generally above 10$  ^{23}  $ cm$  ^{-2}  $,
especially in the broad line region. Therefore, calculations accounting for
pressure gradients within clouds are probably not necessary for most of the
regimes of parameter space we are concerned with, and our approximations should
be valid.\clearpage

We, therefore, treat the two regions of the cloud as being independent and assume
that the line efficiency of each region is a simple function of only the edge
conditions (e.g., the pressure and ionization parameter) and the column density,
which we assume is less in the angled region by a factor of $  \sqrt{2}  $.
The line efficiency of each region was calculated via three-dimensional interpolation
of results obtained using the radiative transfer code XSTAR version 1.20 (e.g.,
Kallman 1995)\pfn{XSTAR is a code similar to CLOUDY and ION. As the manual states,
``XSTAR is a command-driven, interactive, computer program for calculating
the physical conditions and emission spectra of photoionized gases. It may be
applied in a wide variety of astrophysical contexts. Stripped to essentials,
its job may be described simply: A spherical gas shell surrounding a central
source of ionizing radiation absorbs some of this radiation and re-radiates
it in other portions of the spectrum; XSTAR computes the effects on the gas
of absorbing this energy, and the spectrum of re-radiated light. In many cases
other sources (or sinks) of heat may exist, for example, mechanical compression
or expansion, or cosmic ray scattering. XSTAR permits consideration of these
effects as well. The user supplies the shape and strength of the incident continuum,
the elemental abundances in the gas, its density or pressure, and its thickness;
the code can be directed to return any of a large number of derived quantities,
including (but not limited to) the ionization balance and temperature, opacity
tables, and emitted line and continuum fluxes. The solution divides into several
distinct parts: transfer of the incident radiation into the cloud; calculation
of the temperature, ionization, and excited state populations (the last only
for selected atoms and ions) at each point in the cloud; and transfer of the
emitted radiation out of the cloud.''}. The grid we constructed was large;
the ionization parameter, pressure, and column density were calculated, respectively,
at 10, 6, and 6 values (for a total of 360 node points). In accordance with
the edge conditions being regulated by the ionization parameter (rather than
the density), we assumed the pressure is constant throughout each region.  

Unlike line profiles, linearized response functions contain a factor of the
line emissivity gain $  d\textrm{ln}L_{l*}/d\textrm{ln}L  $ (e.g., Goad et
al. 1993, Taylor 1996). Because the expression for the gain contains derivatives
of the emissivity and because the resolution of the grid we used was so high,
the calculated gain was very sensitive to the systematic error in the calculated
line emissivity. In order to obtain gains that did not have numerous discontinuities,
we modified the step size routine of XSTAR. XSTAR normally calculates the step
sizes such that changes in continuum optical depth at each energy is at most
1.0 from one slab to the next. The routine we used is simpler, yet is more computationally
expensive. It sets the step size within each region to be inversely proportional
to the heating rate in the previously calculated zone. This increases the number
of steps taken near the cloud ``face'' and decreases the number of steps taken
in the cooler (and less energetically important) portions of the clouds. The
actual number of steps within each cloud was then adjusted such that the flux
from each of the 32 spectral regions that were calculated deviated by 5\% or
less from a run at much higher resolution. For ionization parameters above $  \sim 1  $,
the modified step routine affects some of the high ionization lines, such as
He II $  \lambda   $1640 \AA~and N {\scriptsize V} $  \lambda   $1243 \AA. In
addition, the modified routine yields smoother line emissivity functions (as
determined from viewing two-dimensional cuts of their three-dimensional structure)
and relatively continuous gains. This may be due to the relative insensitivity
of our step routine to highly position-dependent continuum opacities at high
ionization parameters (which might be caused by, for instance, the He II $  \lambda 228  $ \AA~
(54 eV) absorption edge).  

XSTAR was also used to obtain the anisotropy factor $  \epsilon _{{\rm {A}}l}  $
of each region. For simplicity, we assume the angled region has the same anisotropy
factor as the face-on region. Because of the very high column densities we assumed,
we found $  \epsilon _{{\rm {A}}l}\sim -1  $ for most of the lines and regimes
of parameter space we explored.

In Kazanas (1989), the ionization parameter at the edge of the clouds $  \Xi   $
was fixed, so the asymptotic gain of the line emissivity due to variations in
the continuum $  \eta (\epsilon _{l}|L)  $ was simply $  \eta (\epsilon _{l}|P)  $.
In other words, only the dependence of the emissivity upon the pressure affected
the line ratios. Employing the more general parameterization described in \S~3.1.1,
this becomes 

\[
\eta (\epsilon _{l}|L)=\left\{ \begin{array}{cc}
\eta (\epsilon _{l}|P)(1-\frac{s}{2})\eta (\epsilon _{l}|\Xi )+\frac{s}{2}\eta (\epsilon _{l}|P)+\frac{1}{2}\left( \alpha (r)+\frac{s}{2}\right) \eta (\epsilon _{l}|N_{\rm c}) & r\lsim r_{23}\\
\eta (\epsilon _{l}|\Xi ) & r\gsim r_{23}
\end{array}.\right. \]
 The upper expression above is valid for winds near the continuum source at
$  r\lsim r_{23}  $. Without any mass loss enhancement, these winds would shrink
when the continuum brightens to maintain a relatively constant ionization parameter
(assuming $  s>0  $). The lower expression is valid for winds that are optically
thin or that have a truncated column density. For nearly all of the models we
calculate, the third, right-most term in the upper expression for the $  r\lsim r_{23}  $
region is approximately an order of magnitude smaller than the first two terms.
This is because the column densities of our models are so high. We therefore
ignore this term in our calculations. Standard interpolation algorithms were
used to solve the other terms. As mentioned in \S~\ref{area}, we smoothed $  \eta (\epsilon _{l}|L)  $
near the $  r_{23}  $ boundary to prevent unphysical discontinuities in $  \eta (\epsilon _{l}|L)  $
as well as to avoid potential numerical complications.

\section{Approximations for the Phase-Space Distribution Function \protect$  f\protect  $
\label{f}
}

In addition to $  L_{l*}  $, knowledge of the cloud phase space distribution
function $  f  $ is necessary in order to determine the line profiles (eq.
{[}\ref{2}{]}) and response functions (eqs. {[}\ref{linearizedrf}{]} \& {[}\ref{linearizedvrrf}{]})
of the models under consideration. In this subsection we discuss $  f  $ and
our assumptions about it.

The stellar distribution functions in AGNs are not well known. One might hope
to try to base the stellar distribution functions upon available observations.
Unfortunately, the measurement of the distribution function in real systems
is difficult since one measures only moments of it, such as the surface brightness
profile. From this, one can infer the density as a function of radius. The velocity
information that does exist for some systems which are thought to harbor ``dead''
black holes at their centers is usually just the velocity dispersion alone,
which is not sufficient to provide the stellar distribution function but rather
only imposes quantitative constraints on it. For galaxies which contain active
nuclei, such measurements are very difficult since the emission near the center
is dominated by the point-like source of the continuum.  

Fortunately, if we assume that the clusters are in steady state in most regions,
we can employ the Jeans theorem. This states that distribution functions of
such systems are dependent only upon integration constants of the equations
of motion, such as the angular momentum and energy per unit mass

\begin{equation}
E=\phi (r)+\frac{v^{2}}{2}\lsim 0.
\end{equation}
 In this equation, $  \phi   $ is the gravitational potential. If we assume
that the entire system (in which the supergiants constitute a small fraction
of the total mass) is spherically symmetric, then this gravitational potential
is 

\begin{equation}
\label{3.5}
\phi (r)=-\left\{ \frac{GM_{\rm h}}{r}+4\pi GM_{\rm {rat}}\left( {1\over r}\int ^{r}_{0}dr'n_{*}(r')r'^{2}+\int ^{\infty }_{r}dr'n_{*}(r')r'\right) \right\} ,
\end{equation}
 where $  M_{\rm h}  $ is the mass of the supermassive black hole and $  M_{\rm {rat}}  $
is the mass density divided by the giant star number density $  n_{*}  $. For
simplicity, we assume here that $  M_{\rm rat}  $ is independent of radius.
This assumption is clearly not ideal; in the models of Murphy, Cohn, \& Durisen
(1991), the density functions have a strong mass and radial dependence, especially
for $  r_{\rm t}\ll r\ll r_{\rm c}  $ (where, as before, $  r_{\rm t}  $ is
the tidal radius and $  r_{\rm c}  $ is the core or gravitational radius of
the system). However, this assumption has an important feature of permitting
the distribution function to be determined purely from the surface brightness
functions. These surface brightness functions have been measured in nearby galactic
nuclei, though only for $  r      \gsim 0.1  $ pc.

Inside $  r\sim 0.1  $ pc, constraints upon $  f\textrm{ }  $ can be imposed
by employing theoretical results, such as those obtained with Murphy, Cohn,
\& Durisen's Fokker-Plank code. Note that their models are highly input dependent
for the densities beyond about one parsec, where the stars do not interact fast
enough to be independent of the somewhat arbitrary initial distribution that
was assumed. On the other hand, their results for radial distances less than
a parsec show that collisions become important when the stellar density is above
$  \sim 10^{7}  $ M$  _{\odot }  $ pc$  ^{-3}  $. Murphy, Cohn, \& Durisen
(1991) assumed that tidal disruption imposes the boundary condition that the
density of giant stars $  n_{*}(r)  $ is zero for $  r<r_{\rm t}  $ and increases
with radius for $  r\sim r_{\rm t}  $. It is, however, straightforward to show
that were $  f  $ only a function of $  E  $ everywhere, $  n_{*}\propto r^{-1/2}  $
is the flattest density function possible for a collisionless system in which
the black hole provides the dominate contribution toward the potential. For
this reason, the system is unlikely to be in steady state at radii significantly
below a parsec (near the tidal radius). Thus, in this region, we permit $  f  $
to be an explicit function of $  r  $.

Despite this additional freedom granted to $  f  $, the relation $  n_{*}=\int d^{3}vf  $
can be inverted nonetheless. If we assume that $  f(E,r)=0  $ for $  E>0  $
(that is, no stars are escaping from the system), we obtain (see also Binney
\& Tremaine 1987)
\begin{equation}
\label{df-eq.}
f(E,r)={\Theta (r-r_{1/2})\over 2\sqrt{2}\pi ^{2}}\int ^{\infty }_{\phi ^{-1}(E)}{dn_{*}\over dr}{dr\over \sqrt{\phi (r)-E}}
\end{equation}
 where $  r_{1/2}>r_{\rm t}  $ is the distance from the black hole inside of
which the stellar densities are low and the logarithmic slope of $  n_{*}  $
is greater than -1/2.\pfn{The density cannot fall faster than this If the density
of stars is so high that stellar collisions dominate the evolution of the system,
the distribution function at the energies density profile is a power law, with
$  n_{*}\propto r^{-1/2}  $. ???? for {[}look at other version of dissertation{]}
densities such that the stellar collision time scale equals the age of the system.
The distance from the black hole where the distribution function in such a system
is zero for energies higher than those corresponding to the radii at which the
density profile law deviates from 1/2.} Under these assumptions, $  f  $ is
a function of $  n_{*}  $, $  r_{1/2}  $, $  r_{\rm t}  $, $  M_{\rm {rat}}  $,
and the black hole mass $  M_{\rm {h}}  $. A benefit of this method of determining
$  f  $ (the use of eq. {[}\ref{df-eq.}{]}) is that it has the freedom to
match observations; there does not seem to be only one global class of density
distributions for galactic centers dependent upon a few canonical momenta, but
rather, each galactic center has its own particular stellar density slopes and
scales (e.g., Lauer 1995).

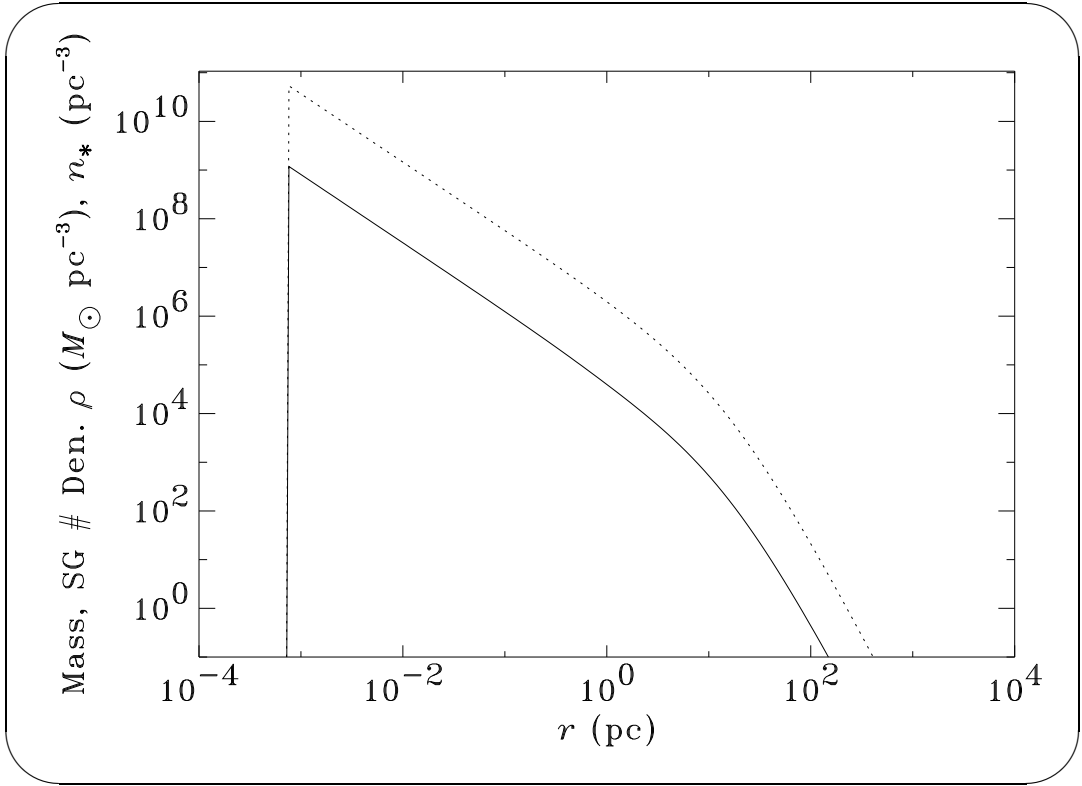
The mass and giant number densities assumed for model 1 are shown in \Fig{model1/density}{Dotted
line: the mass density $  \rho   $ as a function of radius from the continuum
source $  r  $ (horizontal axis) for model 1. Solid line: supergiant number
density $  n_{*}  $.}. The upper curve is the mass distribution, which includes
all types of stars. It is used to obtain the gravitational potential. The lower
curve represents the density of stars with reprocessing winds. The mass of the
black hole for this model is $  3\times 10^{7}M_{\odot }  $. The form of the
stellar density we have assumed is similar to that in Tremaine et al. (1994),
with  

\begin{equation}
\label{Inputdensity}
n_{*}(r)={C_{*}\theta (r-r_{\rm t})\over r^{3-\eta }\left[ 1+(r/r_{\rm c})\right] ^{1+\eta }},
\end{equation}
 where $  C_{*}  $ is a constant, $  r_{\rm t}  $ is the tidal radius, $  \theta   $
is the step function, $  \eta   $ is a free parameter that dictates the density
slope, and $  r_{\rm c}  $ is the core radius of the cluster. For $  r<r_{\rm t}  $,
we assumed $  n_{*}=0  $ in order to satisfy the tidal disruption boundary
condition. The tidal radius of model 1 was taken to be 0.90 light days. However,
in order to test our code, we computed some models in which $  r_{\rm t}  $
was artificially set to be 0. The distribution functions we obtained for these
hypothetical models agreed with those published in Tremaine et al. (1994).\pfn{This
was determined from scaling the radius at which the densities peaked in Fig.
6\emph{e} of Murphy, Cohn, \& Durisen (1991) for their model 2B assuming a black
hole mass of 5.5$  \times 10^{8}\, M_{\odot }  $. However, Though this black
hole mass is the value tabulated in their table 3, the actual black hole mass
associated with Fig. 6\emph{e} is probably much higher. For comparison, the
tidal radius of a 1 $  M_{\odot }  $ solar red giant mass star would have been
a more accurate starting point. This is because the precise black hole mass
associated with Fig. 2\emph{e} of Murphy, Cohn, \& Durisen's model 2B was not
given. Also, they assumed and because listed of model 2B is not specified If
$  M_{*}=1.0      M_{\odot }  $, For the parameters we assumed in \S~\ref{area},
with this is 2.55 light-days. However, we adopted the models assume the density
need not be ?? the radius where the density begins to}

Since the stars of interest in this study are the stars near the black hole,
one might be tempted to ignore the available observations, which are generally
of stars a parsec or more from the central source. However, stars beyond approximately
one parsec are important for at least four reasons:

\begin{enumerate}
\item They affect the strength of the narrow line profile components. 
\item They affect the ``tails'' of the response function. 
\item They govern the depth (of which the energy and correspondingly $  f  $ is a
function) of the potential well in which the nucleus resides. This depth governs
the width of the narrow line profile components. 
\item They may, in some as yet unknown way, correspond to the various AGN characteristics,
such as the covering function or the BLR/NLR line emission component ratios
(see, e.g., Malkan, Gorjian, \& Tam 1998). 
\end{enumerate}
~

For model 1, the local maximum mass density of stars is $  5.4\times 10^{10}  $
stellar masses per cubic parsec just outside the tidal radius of 7.6$  \times 10^{-4}  $
pc or 0.90 light-days. Though this density is higher than typical measurements
of stellar densities in inactive galactic nuclei, we employ it for the following
reasons: 

\begin{itemize}
\item Inactive galaxies probably have lower stellar densities than AGNs. The stellar
density observed for M32 (Lauer et al. 1995) is a modest factor of $  \simeq   $5
lower than the density of model 1 when scaled from $  r=r_{\rm t}  $ assuming
an $  r^{-2}  $ power law and a mass to light ratio of unity.\rfn{See, however,
Appendix \ref{darkma} for potential pitfalls with this common assumption.} Observations
by Eckart et al. (1993) of the Galactic Center were fit with an isothermal-like
model (i.e., $  n_{*}\propto 1/\{1+[r/r_{\rm c}]^{2}\}  $) that has a core
radius of $  r_{\rm c}=  $.015$  \pm   $.005 pc and a central stellar density
of 10$  ^{7-8}  $ M$  _{\odot }  $ pc$  ^{-3}  $. This density is a much
larger factor of $  \sim 10^{3}  $ lower than the mass densities of model 1. 
\item The total stellar mass within 1.0 pc from the center is $  1.6\times 10^{7}M_{\odot }  $.
This is 55\% of the black hole mass, which is within the range suspected for
nearby inactive galactic nuclei. Thus models with stellar densities lower than
model 1 might have $  M_{\rm c}/M_{\rm h}  $ ratios that are significantly
below those typically observed.
\end{itemize}
~

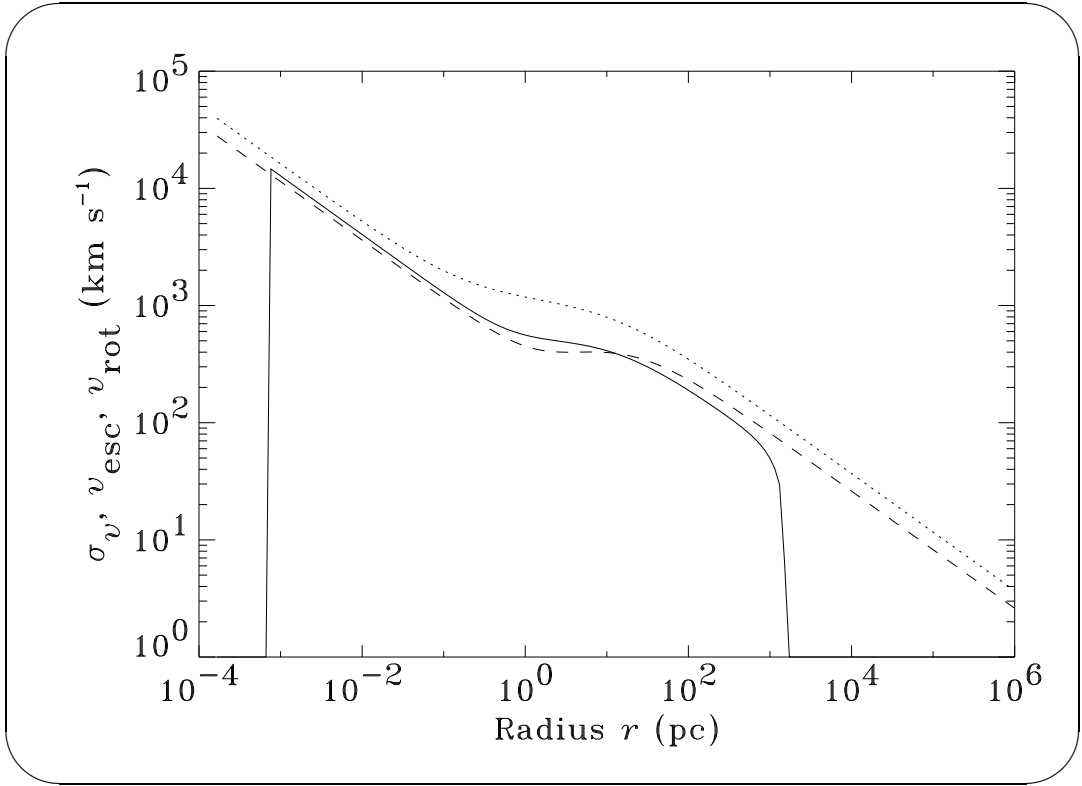
The velocity dispersion function of model 1 is shown in \Fig{model1/velocity}{Velocity
dispersion $  \sigma _{v}  $ (solid curve) of giant stars as a function of
$  r  $. The dotted curve is the escape velocity, while the dashed curve is
the circular orbital velocity.}. The black hole dominates the gravitational
potential up to $  \sim   $1 pc, at which point the contribution from other
matter (e.g., stars, remnants, and gas) becomes more important. At radii between
$  \sim   $1 pc and the core radius (which for model 1 is 9.6 pc) the velocity
dispersion remains relatively constant. This occurs because the potential is
nearly independent of radius in this region. At the edge of the cluster, the
velocity dispersion and mass density become zero. This is a consequence of assuming
that no stars ``evaporate'' from the cluster or, equivalently, that $  f(E>0)=0  $.

The energy dependence as calculated from equation (\ref{df-eq.}) of the distribution
function $  f  $ of the stars at distances greater than $  r_{1/2}  $ is shown
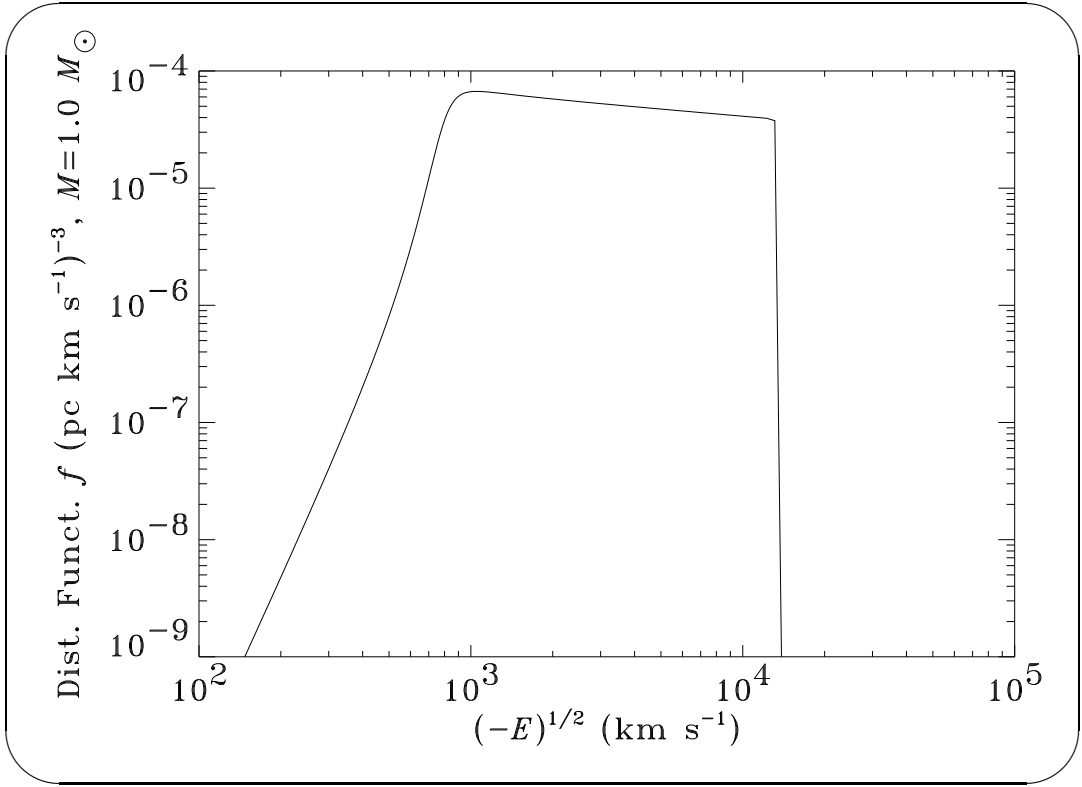
in \Fig{model1/f}{The distribution function $  f(E,r)  $ obtained from eq.
(\ref{df-eq.}) as a function of the energy per unit mass $  E  $ as expressed
in units of equivalent velocity $  (-E)^{1/2}  $. The bend in the curve occurs
because of the change in the density slope at $  r\sim r_{\rm c}  $. The energy
at which the distribution function suddenly drops to zero corresponds to the
potential energy at $  r=r_{1/2}  $, the radius where the density slope becomes
less steep than $  r^{-1/2}  $. For the models we present, this radius is set
equal to the tidal radius.}. The upper cutoff at highly negative energies occurs
at $  E=\phi (r_{1/2})  $. The distribution function at lower energies yields
the $  r^{-4}  $ drop in density that we assume in all of our models. Most
of the models we present in the next chapter assume the distribution function
shown in Figure \ref{model1/f}. 

Upon making the assumptions discussed in this chapter for $  L_{*l}  $ and
$  f  $, we computed the observable characteristics (e.g., the line profiles
and response functions) of several different stellar wind AGN line emission
models. We present these results in the next chapter.

\chapter{Results 
}

In the previous chapter, we discuss the theoretical assumptions of model 1.
In \S~4.1, we show the observable consequences of making these theoretical assumptions,
such as the line profiles and response functions that result. In \S~4.2, we
present model 2, which is more compatible with typical AGN data and NGC 5548
in particular. In \S \S~\ref{iontopratio}-\ref{s}, we demonstrate how some
of the model parameters regulate the line ratios. In \S \S~\ref{luminosity}-\ref{mdot_min},
we demonstrate how the parameters characterizing the stellar winds affect properties
such as the ``generic line profile'' strength and shape. Finally, in \S \S~\ref{eta}-\ref{m_core},
we present some of the more model-specific predictions of stellar wind AGN line
emission model---those concerning the stellar distribution function.

\section{Model 1: A Theoretically Motivated Model \label{model1}
}

As mentioned in \S~3, the parameters of model 1 are taken primarily from the
papers of Kazanas (1989); Murphy, Cohn, \& Durisen (1991); Tremaine et al. (1995);
and Lauer et al. (1995) without regard to their consequences upon the resulting
AGN line emission. Not surprisingly, we found that this approach yielded an
initial model that had severe shortcomings. Though we have subsequently produced
several models which surmount these shortcomings, we present model 1 first,
as it illustrates the fundamental problems that occur if certain changes are
not made to the simple stellar wind cloud model.

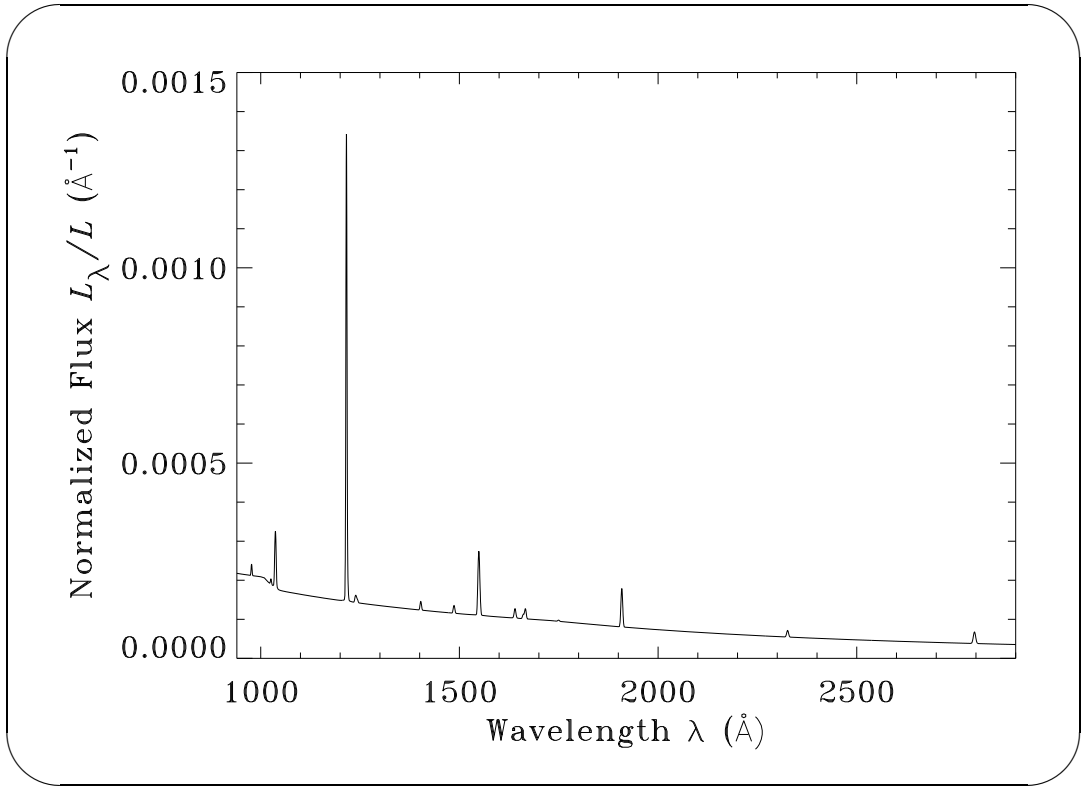
\Fig{model1/spectrum}{The UV spectrum of model 1 plotted in units of \AA$  ^{-1}  $.
This spectrum is scaled such that the integral of the continuum over all wavelengths
is unity for each plot. This scaling can help make differences between the various
spectra more apparent when the continuum luminosity is varied.} shows the UV
portion of the spectrum that results for model 1 for the parameters discussed
in \S~3. Solar abundances (e.g., Zombeck 1990) were assumed with the exception
of the carbon abundance, which was assumed to be 0.57 solar (2.0$  \times 10^{-4}  $).
The synthetic spectrum shown in Figure \ref{model1/spectrum} includes 37 separate transitions of the strongest \textit{observed} ultraviolet lines.
However, the ionization parameter in the broad line region of model 1 is very
high, with $  \Xi =10  $ or $  3.0\le U\le 3.2  $ (where $  U\equiv \phi /(cn_{\rm {H}})  $
and $  \phi   $ is the photon flux above 1 Ry) compared to values commonly
used to one-zone models ($  \Xi \sim 0.2  $; $  U\sim 0.03  $). Therefore,
the synthetic spectrum shown in Figure \ref{model1/spectrum}, which does not
include lines such as S {\footnotesize VI} $  \lambda   $933 \AA, is probably
not accurate, especially below the Lyman limit. 

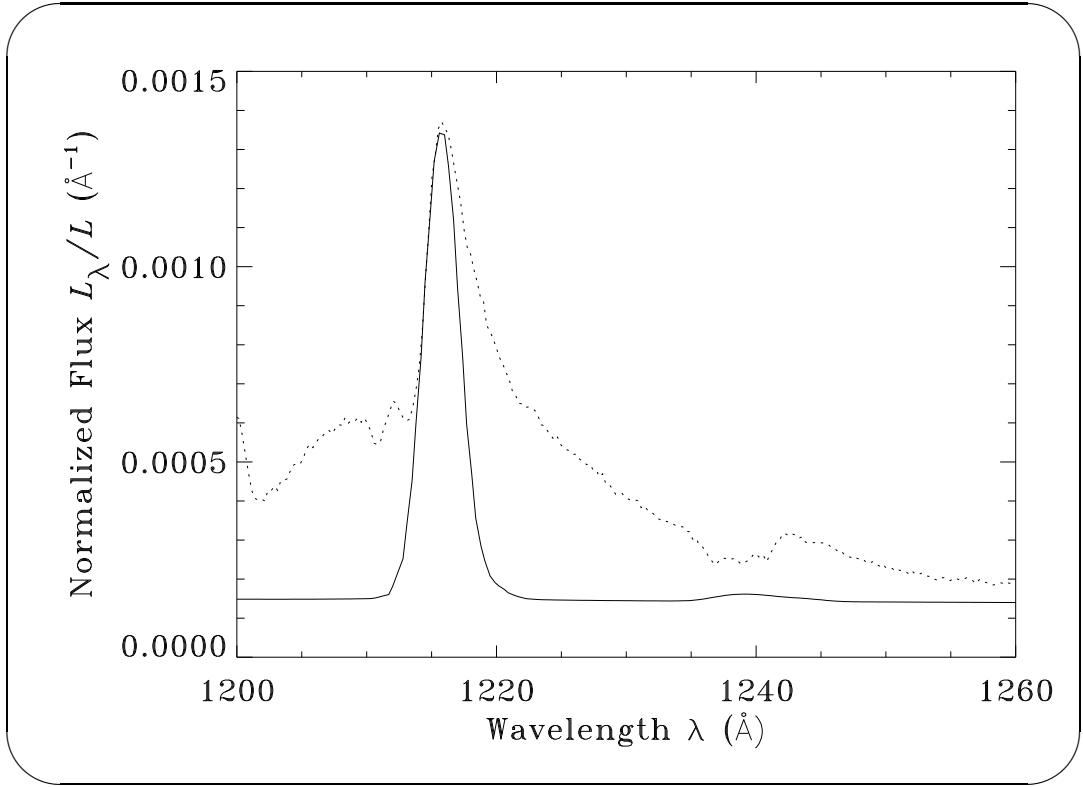
\Fig{model1/Lya}{Solid curve: the lines near and including Ly$  \alpha   $
for model 1. For comparison, the relevant de-redshifted portion of the mean
NGC 5548 spectrum is also shown as the dotted curve.} shows just the blended
Ly$  \alpha   $ complex, which includes He II $  \lambda   $1216 \AA. The
Ly$  \alpha   $ line is relatively insensitive to several of the cloud parameters,
such as the edge ionization parameter. Therefore, comparison of this profile
to observations serves as a useful diagnostic of the applicability of the covering
function of a model, which, for these models, is the radial dependence of the
product of the giant number density $  n_{*}  $ and the wind area $  A  $.
The covering function generally would have an additional dependence upon the
mean column density function. This dependence, however, is quite weak for the
wind models presented here because their column densities are so high.

For comparison, we also show the UV spectrum of NGC 5548 averaged over 39 HST
observations (see Korista et al. 1995). We chose this single object not because
of any particular features it has, but rather because of it has been so well
studied. We hope our detailed comparisons with a single object, rather than
with comparisons to a ``generic'' averaged composite, reduces the chance that
we deceive ourselves into thinking that a model is plausible when in actuality
nothing like it would be observed. This is important because several of the
parameters in our models are probably correlated. For instance, while both the
luminosities and the equivalent widths of C {\scriptsize IV} $  \lambda   $
1549 vary considerably in AGNs, relatively few $  L>10^{45}  $ ergs s$  ^{-1}  $
AGNs have large C {\scriptsize IV} equivalent widths (Baldwin 1977). Because
these two variables are anti-correlated, a model with a high luminosity and
a high C {\scriptsize IV} equivalent width would probably not exist. The parameters
of such a model could, however, still be within a few standard deviations of
the mean of a composite spectrum. By comparing most of our theoretical models
to one specific, yet not unusual AGN, we prevent ourselves from falling prey
to this statistical illusion. 

There are at least three features of model 1 that are apparent upon comparing
the spectra: 

\begin{enumerate}
\item The overall geometrical covering factor of model 1 (determined purely from the
wind area and cloud density functions) is only 0.04. Here we define the ``radial
geometrical covering factor'' to be the absorption coefficient for light emitted
at $  r_{\rm t}\bf {\hat{r}}  $ and received at $  r\bf {\hat{r}}  $; were
each of the stellar winds opaque, 
\begin{equation}
\label{eq. for covering function}
\Omega (r)=1-\exp \left( -\int _{0}^{r}dr'n_{*}(r')A(r')\right) .
\end{equation}
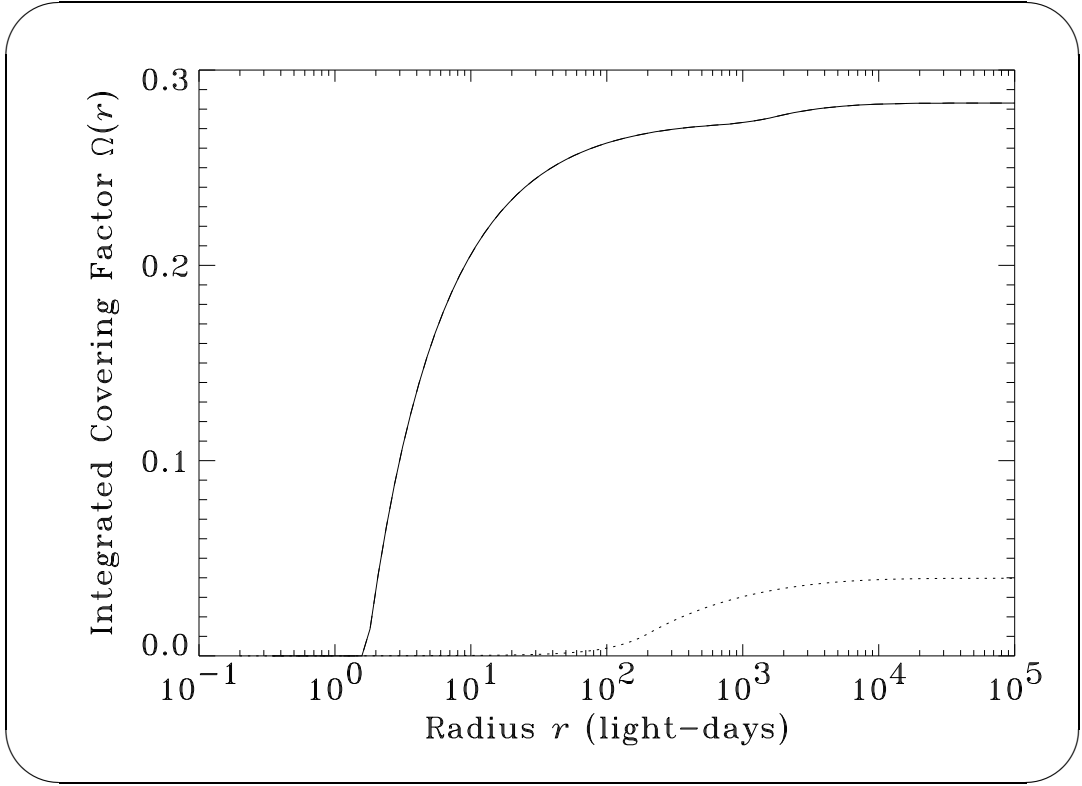
This function is shown in \Fig{model1/coveringfactor}{Dotted line: radial geometrical
covering factor $  \Omega (r)  $ as a function of $  r  $ for model 1. Solid
line: $  \Omega (r)  $ for model 2. For model 1, the covering factor in the
broad line region is small, with most of the emission being due to winds of
stars at distances greater than 100 light-days.}. In the ``optically thick
region,'' where $  r_{12}\lsim r\lsim r_{23}  $, $  R_{\rm w}\gg R_{*}  $,
and $  r_{\rm t}\lsim r\lsim r_{\rm c}  $, the slope of the covering function
according to equations (\ref{areagain}) and (\ref{Inputdensity}) is
\begin{equation}
\label{coveringslope}
\sim -(2\alpha -s+3-\eta ).
\end{equation}
At large $  r  $, the covering function for model 1 is much less than the empirical
estimates for NGC 5548 (e.g., $  \Omega [{\infty }]  $=0.28 {[}\S~4.2{]}; $  \Omega [{\infty }]  $=$  \sim   $0.3
{[}Krolik et al. 1991{]}). 
\item Though the equivalent widths of the lines are too low for the effect to be clearly
apparent, the profiles are much more concave than the ``logarithmic'' ones
that are generally observed and that are also a characteristic of NGC 5548.
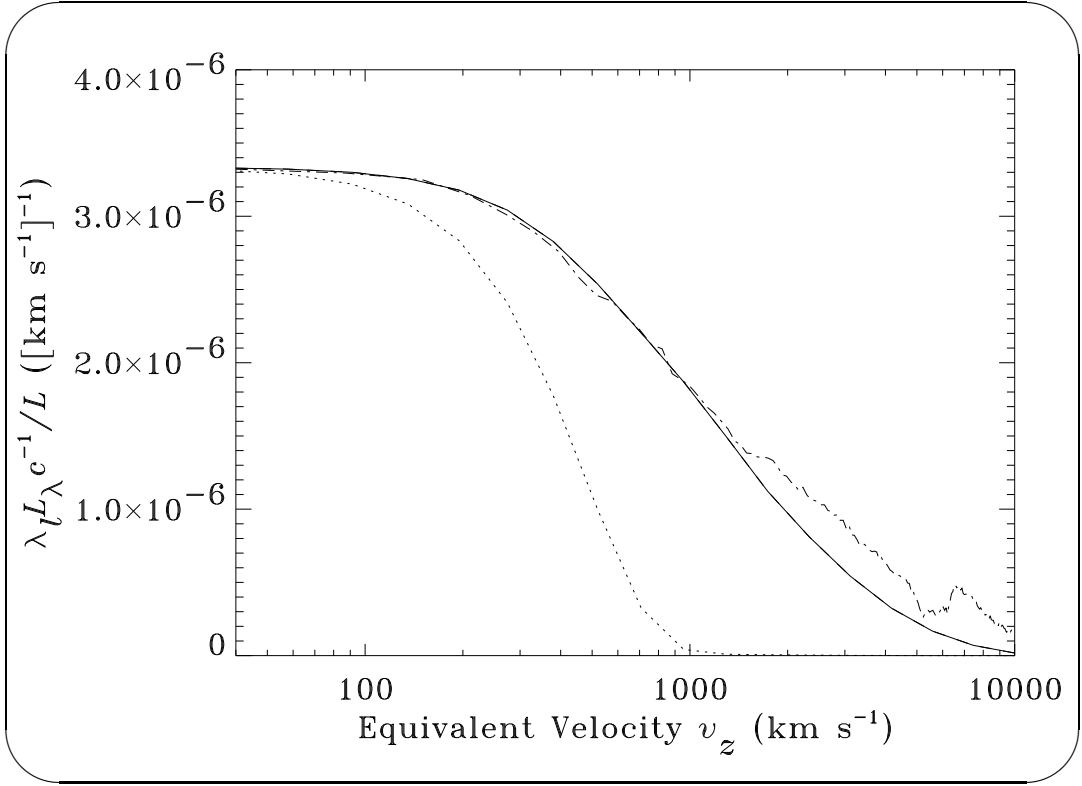
The discrepancy with the observed profile is especially apparent in \Fig{model1/Lyalog}{Dotted
curve: the redward side of the Ly$  \alpha   $ line profile for model 1 plotted
in log-linear coordinates. Solid curve: model 2, scaled to have the same peak
flux. Dot-dashed curve: the associated portion of the NGC 5548 spectrum, also
scaled. Note the drastic difference in profile shapes, with the spectrum of
model 1 having a much more concave and narrow profile than that of NGC 5548. Model
2, on the other hand, fits the observed Ly$  \alpha   $ shape quite well given
that our models do not account for absorption in the N {\scriptsize V} doublet.},
which shows the de-blended Ly$  \alpha   $ redward profile of model 1 in log-linear
coordinates. For comparison, we also show the redward side of the blended Ly$  \alpha   $/N~{\footnotesize V}
complex of NGC 5548.
\item The BLR line ratios are much different from those of NGC 5548. For instance,
largely independent of the particular BLR/NLR velocity cutoff that is assumed,
the broad component N {\footnotesize V}/C {\scriptsize IV} line ratio of model
1 is several times larger than that of the NGC 5548 spectrum shown in Figure
\ref{model1/spectrum}. This indicates that the value of the ionization parameter
in the BLR is smaller than that assumed by the model.
\end{enumerate}
Features (1) and (2) are the result of a relatively small BLR geometrical covering
function. The covering factor is determined purely by the product of the number
density of winds and their areas, $  n_{*}A  $. Therefore, increasing the number
density of supergiant stars in just the BLR might appear to solve several problems.
One way of doing this would be to assume that a stellar density falls significantly
faster than the $  r^{-1.4}  $ function employed by model 1 and observed in
inactive galactic nuclei (e.g., Lauer et al. 1995). In fact, as will be discussed
in \S \S~\ref{alpha} \& \ref{eta}, a function as steep as $  n_{*}\propto \sim r^{-4}  $
would be required. All of the galactic nuclei observed to date, however, have
a stellar density which falls off as $  r^{-2}  $ or \textit{slower}. Therefore,
we do not believe that drastically changing the slope parameter would result
in models that are fully self consistent. On the other hand, AN97 does assume
that the functional dependence of the ``bloated star'' density upon radius
is significantly steeper than the main sequence (i.e., mass) density. In Appendix
\ref{commentsonaandn}, we discuss in detail the evidence supporting this assumption.
We conclude that while there is some evidence for a weak mass density gradient,
modifying $  n_{*}  $ to have a gradient as steep as AN97 assumed is probably
implausible on both theoretical and empirical grounds.\clearpage

\section{Model 2: An Empirically Motivated Model\label{model2}}

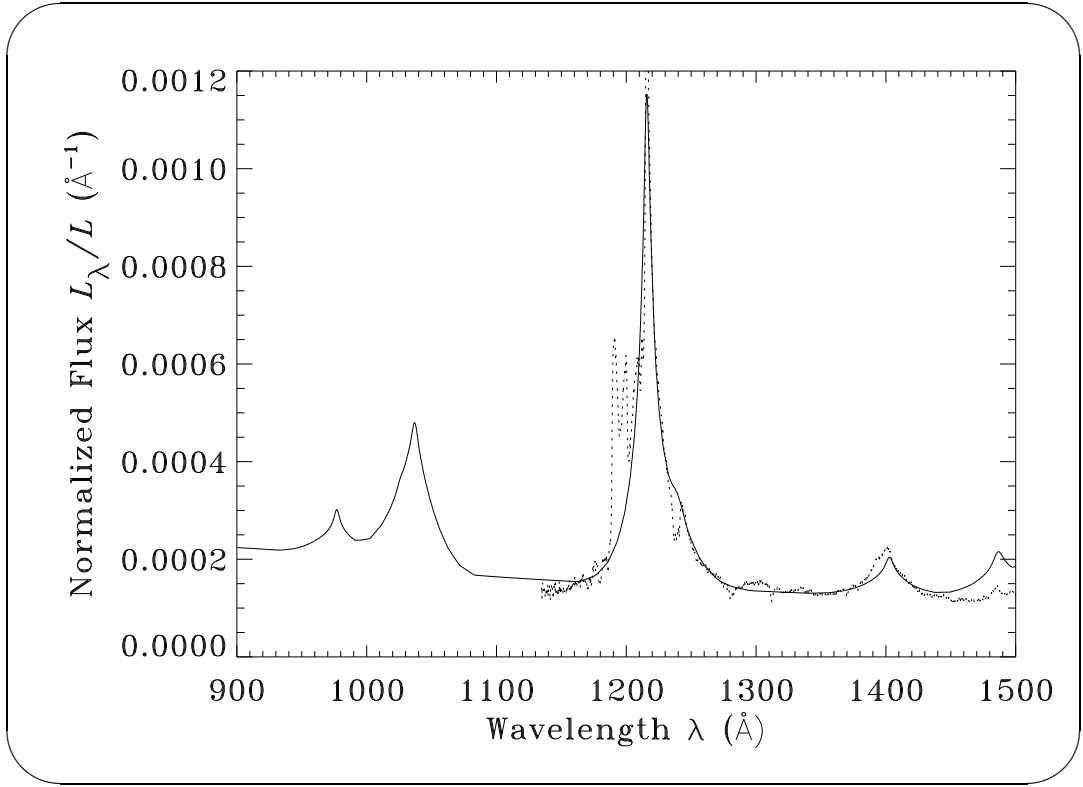
The UV spectrum of model 2 is shown in Figures \ref{model2/harduvb}\fig{model2/harduvb}{Solid
line: the ``hard UV'' spectrum of model 2. Dotted line: the spectrum of NGC
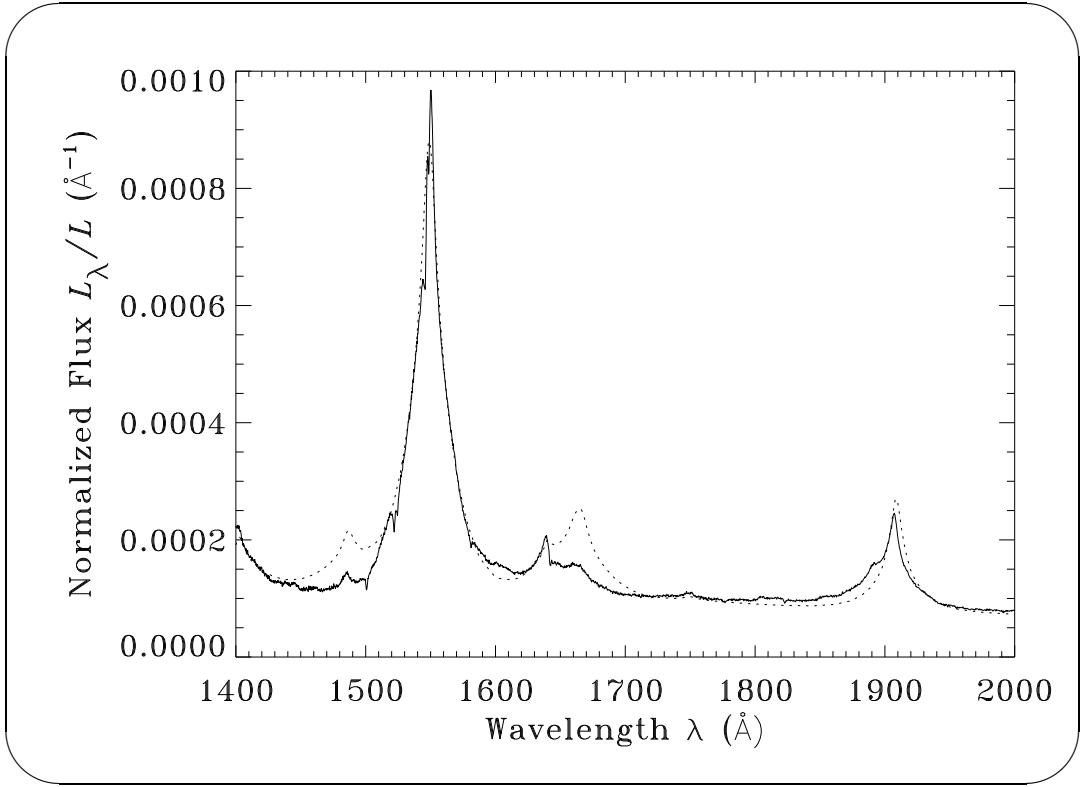
5548 averaged over 40 observations (Korista et al. 1995).} and \ref{model2/softuvb}\fig{model2/softuvb}{Solid
curve: the ``soft UV'' spectrum of model 2. Dotted curve: observed NGC 5548
spectrum.}. There are four differences between models 1 and 2. First, the mass
loss slope parameter $  \alpha   $ was changed from 0 for model 1 to 1.1 for
model 2. This makes the wind mass loss increase with increasing local continuum
flux instead of being constant. It also makes the cloud area relatively constant
(see Figure \ref{model3/area}). The resulting supergiant mass loss rate for
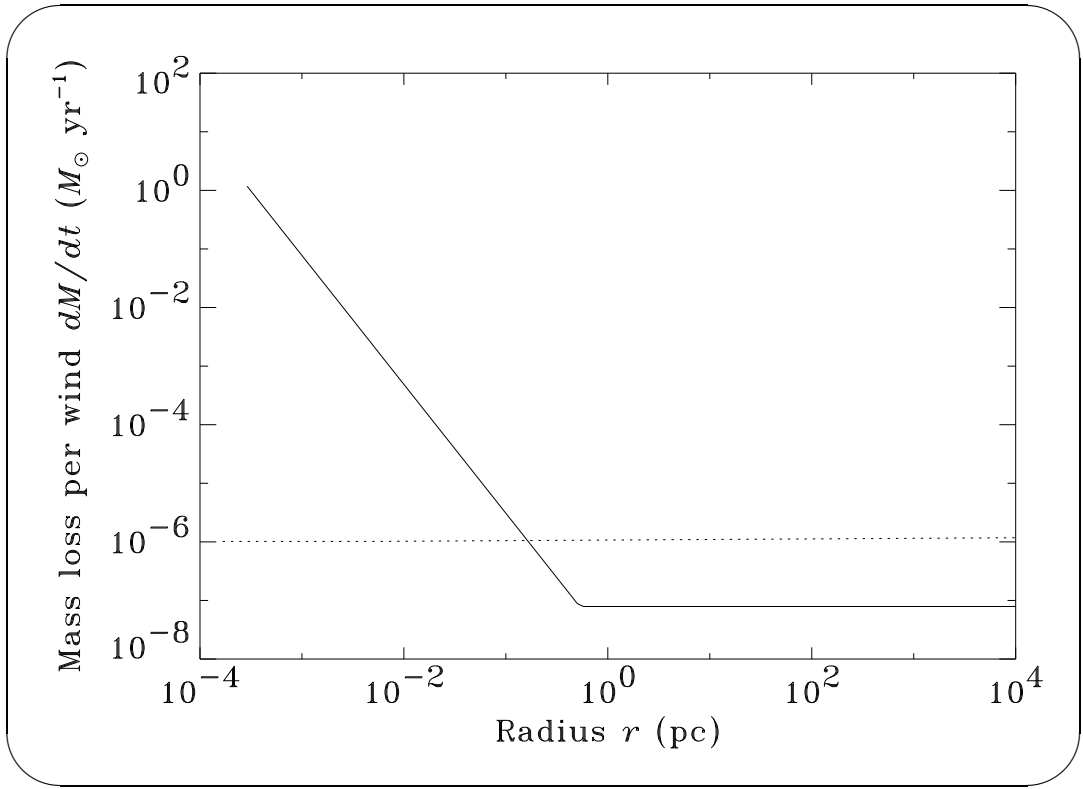
model 2 is shown in \Fig{model2/indivmass}{Solid line: mass loss rate per supergiant
wind for model 2. Dotted line: mass loss rate of model 1.}. This mass loss rate
falls as $  r^{-2.2}  $ for $  r\leq 650  $ light-days and is fixed at $  7.9\times 10^{-8}\, M_{\odot }  $
yr $  ^{-1}  $ for $  r>650  $ light-days.

A second difference between models 1 and 2 is the wind edge ionization parameter
function. For model 2, $  \Xi =1.0  $ for stellar winds near the tidal radius
(at the inner edge of the BLR of the model). This value of $  \Xi   $ is a
factor of 10 lower than that of model 1. The reduction is empirically necessary
if the Ly$  \alpha   $/O {\scriptsize VI} ratios are above $  \sim   $2.0.
(Laor et al. {[}1994{]} found ratios of 1.9-4.7.) This is a problem with the
model of Kazanas (1989), which without modification predicts an ionization parameter
of $  \sim   $10 at the boundary of the wind.  The ionization parameter of
model 1 was constant everywhere with $  s=2.0  $ (where, as mentioned previously,
$  P\propto [L/r^{2}]^{s/2}  $). For model 2, $  s=1.9  $. This slight difference
causes $  \Xi   $ to increase slowly with increasing local continuum flux.
Thus, regions of model 2 that are farther away from the continuum source have
lower wind edge ionization parameters.

The third and fourth differences between models 1 and 2 are the bolometric luminosity
and tidal radius. The time-averaged bolometric luminosity was reduced from $  1.0\times 10^{44}  $
ergs~s$  ^{-1}  $ for model 1 to $  3.2\times 10^{43}  $ ergs~s$  ^{-1}  $
for model 2.\pfn{DEMOS ASKS IF THIS IS CONSISTENT WITH OBSERVATIONS ... HERE BUT I TALK ABOUT THIS LATER so not doing it here} The
tidal radius was increased from $  7.6\times 10^{-4}  $ pc for model 1 to $  1.4\times 10^{-3}  $
pc for model 2. The parameters assumed in \S~\ref{area} for models 1 and 2
and the expression for the tidal radius (eq. {[}\ref{tidalradiusequation}{]})
would be compatible with a red giant stellar mass of $  M=2.9\, M_{\odot }  $.
For comparison, the tidal radius of a 1 $  M_{\odot }  $ mass star in models
1 and 2 is 2.1$  \times 10^{-3}  $ pc. 

The spectrum shown in Figures \ref{model2/harduvb} and \ref{model2/softuvb} is
in much better agreement with the mean AGN spectra compiled by Zheng et al.
(1997) than the spectrum of model 1. The spectrum of model 2 agrees reasonably
well in particular with that of NGC 5548. The Ly$  \alpha   $ profile shape
of model 2 is shown in Figure \ref{model1/Lyalog}, while the C {\footnotesize IV}
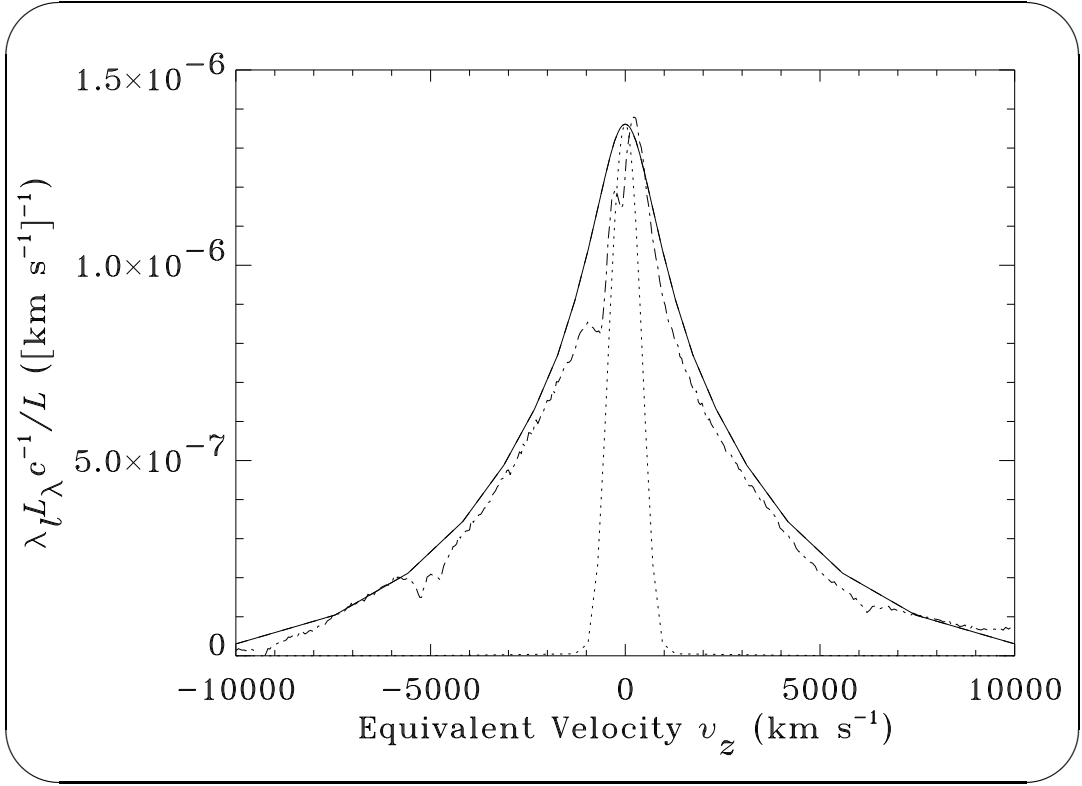
profile shape is shown in \Fig{model2/civ}{Solid line: the C {\scriptsize IV}
$  \lambda   $1548.20/1550.77 shape of model 2. Dotted line: same line of model
1, scaled to have the same peak flux. Dot-dashed line: NGC 5548, also scaled. Model
1 has a profile that is much narrower than observed. Model 2 lacks the blueshifted
absorption feature and apparent redshifted profile base, but fits the NGC 5548
profile well in other respects.}. The key difference in these profile shapes
stems from the higher BLR ($  r\leq 50  $ light-days) covering that model 2
has due to its enhanced wind area size. The effect this has upon the covering
function is shown in Figure \ref{model1/coveringfactor}.

Despite the good agreement in the strong lines, there is, nevertheless, poor
agreement between the spectrum of NGC 5548 and that of model 2 in several UV
lines. For instance, the N {\footnotesize IV} $  \lambda   $1486 continuum-subtracted
line peak is a factor of 3.7 times stronger in model 2 than in NGC 5548. Similarly,
the O {\footnotesize III}{]} $  \lambda   $1664 line flux is a factor of 2.5
times stronger than that of NGC 5548. In addition, like many other published
AGN models, the Ly$  \alpha   $/H$  \beta   $ ratio of model 2 is 20.8, which
is higher than the observed value of 12.6$  \pm 2.0  $ (e.g., Dumont, Collin-Souffrin,
\& Nazarova 1998). This problem persists despite the high densities and columns
we employ. As discussed in more detail below, several of these types of discrepancies
can be removed by invoking different model parameters. However, some discrepancies
persist even with many changes. These discrepancies, which we point out throughout
this chapter and summarize in Chapter 5, are important problems with the stellar
wind model.

The linearized response function for model 2 of the Ly$  \alpha   $ line flux
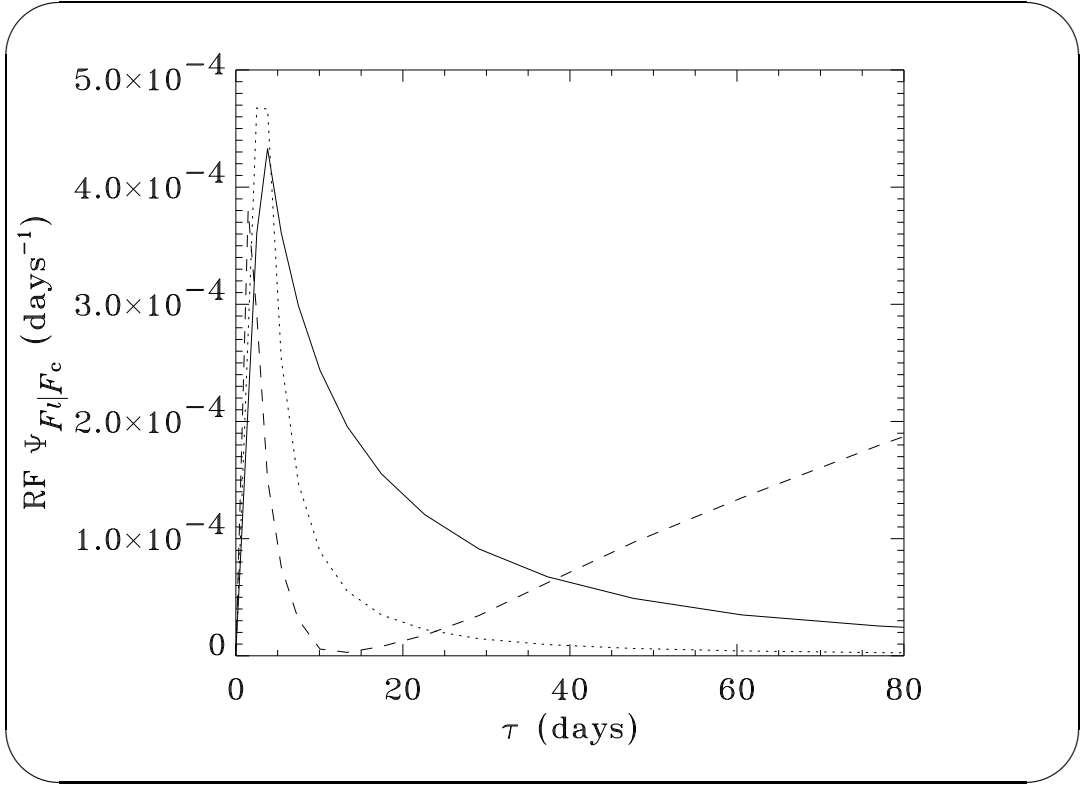
due to variations of the observed UV ionizing continuum is shown in \Fig{model2/Lya1-d}{Solid
line: the linearized response function of the Ly$  \alpha   $ line profile
due to variations in the UV continuum luminosity for model 2. Dotted line: the
response function of the ``summed line'' (the synthetic line for which $  \epsilon _{l}=1.0  $
everywhere) for model 2 scaled down by a factor of 10$  ^{2}  $ in order to
fit plot. Dashed line: the summed line response function for model 1 scaled
up by a factor of 10 in order to be visible in plot.}. For comparison, the response
function for model 1 is also shown. For model 1, the response function of Ly$  \alpha   $
is approximately zero at the intermediate delays of $  \sim   $10-15 days because
the cloud area gains are zero (with the area being inversely proportional to
the flux) at intermediate distances in the BLR. Conversely, the cloud sizes
are independent of the local continuum flux at both low and high radii, giving
the cloud areas of model 1 much larger gains (approaching unity) at high delays. Model
2, with $  \alpha >0  $, is relatively immune from these types of nonlinear
effects, which incidentally were not taken into account by AN97.

The response functions of model 2 peak at $  \sim   $3.3 days. In comparison,
Krolik \& Done (1995) obtained a peak of approximately $  7\pm 4  $ days for
C {\small IV} in NGC 5548. The uncertainty is primarily due to two things. First,
there is an ambiguity in the ideal deconvolution algorithm parameters. Second,
the sampling rate, which was 4 days, is poor. For these reasons, the peak could
be considered as an upper limit to the true peak that would be obtained from
a hypothetical infinitely high sampling rate. Incidentally, this also may be
the reason that the peak of the NGC 5548 C {\footnotesize IV} response function
obtained from using the 1989 IUE data set with a sampling interval of 4 days,
is approximately $  7\pm 4  $ days (Krolik \& Done 1995), while the 1993 data
set, with a sampling rate of 1 day, peaks at approximately $  0\pm 2  $ days
(Wanders et al. 1995; Done \& Krolik 1996).

Determining where the response function peaks reside is especially important
to wind models. This is because the innermost BLR emission radius is the well-defined
tidal radius of the stars (eq. {[}\ref{tidalradiusequation}{]}). The tidal radius
$  r_{\rm t}  $ regulates the size of the geometrical response function. This
function can have structure only on time scales greater than $  \sim 2r_{\rm t}/c  $,
or, for model 2, 3.3 days. In most other AGN cloud models, this innermost emission
region is a free parameter. Therefore, one way to test wind models is to measure
the shortest delay at which the response function has structure, or, as discussed
in Appendix \ref{veldtemp}, to measure the highest frequency at which the derivative
of the transfer function is nonzero. 

The particular nature that this structure takes depends highly upon the anisotropy
factor of the line $  \epsilon _{{\rm A}l}  $. As previously mentioned, the
column densities for model 2 are quite high compared to other BLR models (see
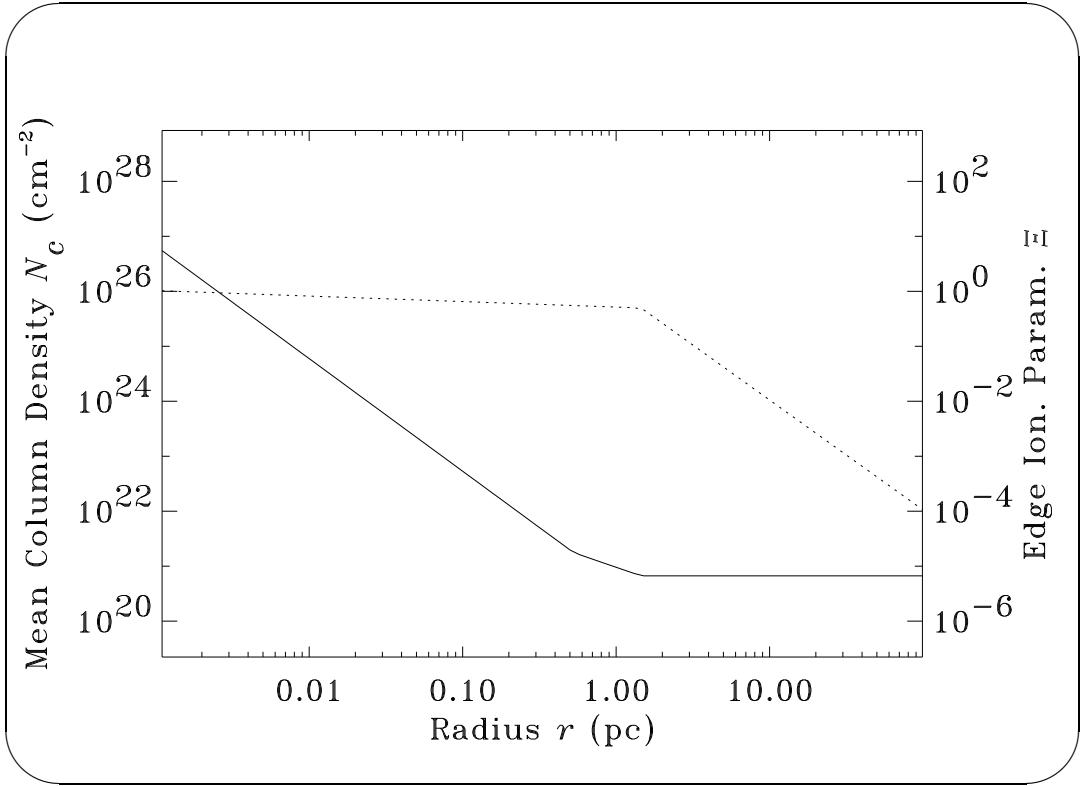
\Fig{model2/N_c+Xi}{Solid line, left axis: the mean column density for model
2. Dotted line, right axis: the ionization parameter $  \Xi   $.}). This makes
the clouds optically thick to Ly$  \alpha   $. Therefore, the winds at small
delays for a given radius (which are at small angles relative to the line of
sight) do not emit line radiation towards the observer, but rather mostly emit
back towards the continuum source. As a result, the Ly$  \alpha   $ response
function is approximately the shape of a rising triangle for $  0\leq \tau \leq 2r_{\rm t}/c  $.
This is shown in Figure \ref{model2/Lya1-d}. A more realistic model accounting
for a range of stellar radii and mass, each with a different tidal radius, would
have a much smoother response function. The effect of varying the tidal radius
is discussed in \S~\ref{tidal section}.

An additional characteristic of the Ly$  \alpha   $ response function (as well
as other functions that tend to peak near zero) is its ``FWHM,'' or full width
at half the maximum amplitude. The FWHM of the Ly$  \alpha   $ response function
for model 2 is 13 days. The FWHM obtained by Krolik \& Done (1995) was 28 days.
Thus, both the peak and the FWHM of the Ly$  \alpha   $ response function are
suggestive of the innermost BLR radius for model 2 being about a factor of 2
smaller than that appropriate for NGC 5548.

However, as will be discussed in more detail in \S~\ref{tidal section}, artificially
increasing the tidal radius of model 2 would not necessarily improve the overall
agreement with the observed response functions of NGC 5548. This is because
increasing $  r_{\rm t}  $ would worsen the fits to the C {\footnotesize IV}
response functions obtained using the 1993 data set. The C {\footnotesize IV}
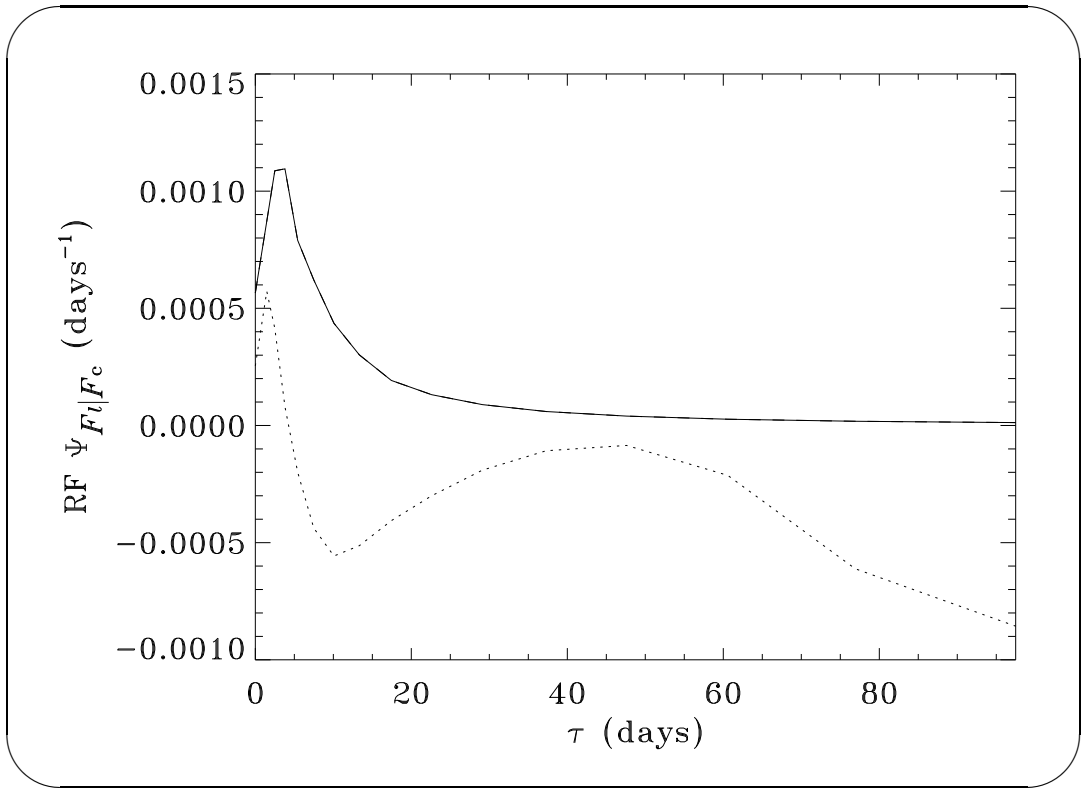
response function of model 2 is shown in \Fig{model2/civrf}{Solid line: the
linearized response function for the C {\footnotesize IV} line of model 2. Dotted
line: model 1, increased in scale by factor of 5000.}. It peaks at 3.3 days.
This is larger than the peak at $  0\pm \sim 2  $ days implied by the results
of Done \& Krolik (1996) and Wanders et al. (1995). Placing the clouds farther
away in model 2 would worsen this discrepancy.

The 1993 data set, in addition to being better sampled on the ``blue'' end
of the power spectrum, also has a higher signal to noise ratio than the 1989
\emph{IUE} data set---its energy resolution is a mere $  \pm 0.25  $~\AA. This
better resolution permitted the first measurements of ``velocity-resolved AGN
response functions,'' or, more correctly, velocity-\emph{integrated} response
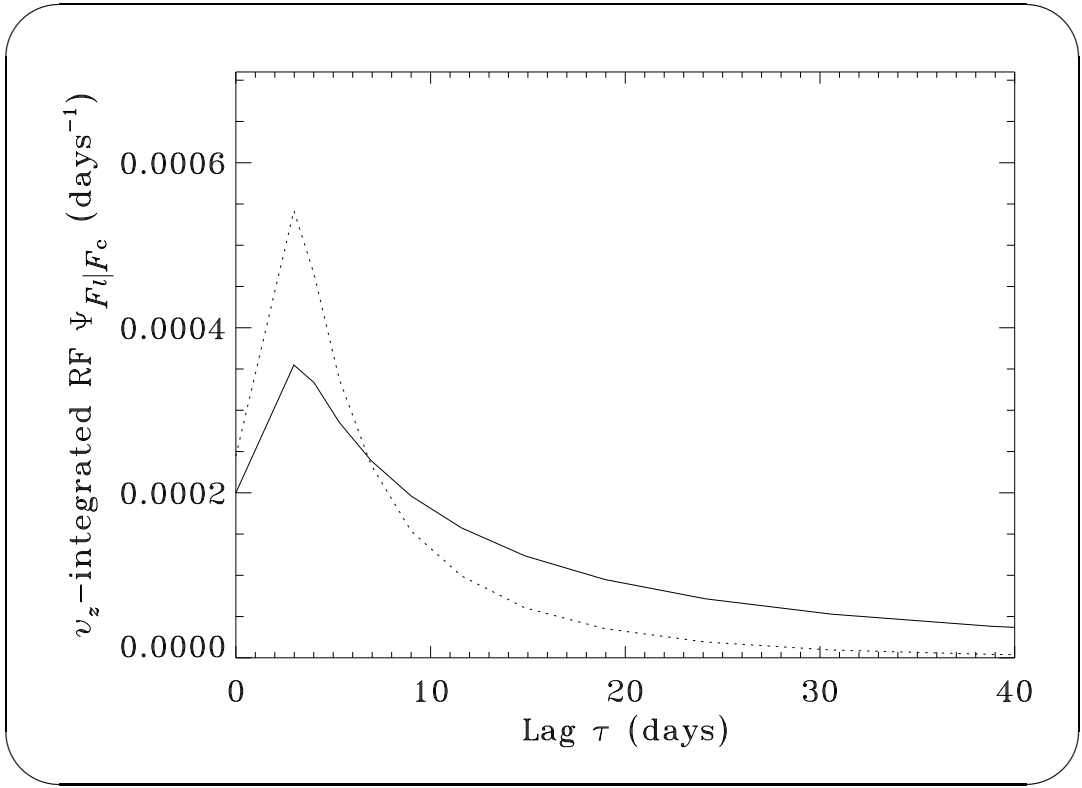
functions of profile components. \Fig{model2/civvel-intb}{The velocity-integrated
linearized response functions of the C {\scriptsize IV} line profile components
due to variations in the continuum luminosity for model 2. Dotted line: wing
component. Solid line: the response function for the C {\scriptsize IV} core
component. The peak of the wing response function is a factor of 1.8 higher
than the peak of the core response function.} shows the velocity-integrated
response functions of the ``wing'' and ``core'' C {\footnotesize IV} line
profile components for model 2. The velocity cutoff points between the wing
and core components were, respectively, $  2450  $ km~s$  ^{-1}      <|v_{z}|\leq 10,000  $
km s$  ^{-1}  $ and $  0  $ km~s$  ^{-1}      \leq |v_{z}|\leq 2450  $ km~s$  ^{-1}  $.
These are identical to those assumed by Wanders et al. (1995) and are very close
to those employed by Done \& Krolik (1995). Due to their simple rising triangle
shapes, the wing and core of the model 2 response functions peak at 3.3 days$  \approx 2r_{\rm t}/c  $.
The FWHM is 6 days for the wing component and 11 days for the core component.
In comparison, the C {\small IV} NGC 5548 response functions published by Wanders
et al. (1995) have a FWHM of 16 days for the red and blue cores, 10 days for
the red wing, and 24 days for the blue wing. These FWHMs are roughly 50\% larger
than those of model 2. However, these results employed the SOLA deconvolution
algorithm described by Pijpers \& Wanders (1994), which extracts response functions
rather than linearized response functions, with the difference being that the
former uses absolute variations (which incidentally are positive-definite) as
input while the latter uses variations about means (which can be negative).
Direct comparison with the results shown in \ref{model2/civvel-intb} is thus
difficult.

Done \& Krolik (1996) also obtained response functions of profile components,
but used the regularized inversion deconvolution algorithm (Krolik \& Done 1995).
They obtained a FWHM of 8 days for the response function of the red core, a
FWHM of $  \sim   $9 days for the blue core, a FWHM of $  \sim   $6 days for
the red wing, and a FWHM of $  \sim   $11 days for the blue wing. Model 2 agrees
much better with these results than the falling, nearly triangular SOLA ones.
They do resemble, however, the smoothed response function for the total C {\footnotesize IV}
emission obtained by Done \& Krolik (1996) (their Figure 3). Nevertheless, two
important problems remain:

\begin{enumerate}
\item Model 2 has a perfect red-blue symmetry. Model 2 is, therefore, not fully compatible
with NGC 5548 data. This is, in fact, a problem with each of the wind models
discussed in this section. Incidentally, it not a problem with the models discussed
in Appendix \ref{shiftsap}. 
\item The maximum amplitudes of the response functions $  \Psi _{F_{l}|F_{\rm c}}(0)  $
obtained by Wanders et al. (1995) and Done \& Krolik (1996) are roughly the
same for each C {\small IV} profile component. For model 2, however, the maximum
amplitude of the wing component is 1.6 times that of the core component. This
discrepancy could be removed by decreasing the velocity dispersion at the tidal
radius. As shown in \S \S~\ref{tidal section} \& \ref{m_h}, this decrease
in velocity dispersion can be achieved by decreasing $  M_{\rm h}/r_{\rm t}  $,
but only at the expense of making the C {\footnotesize IV} profile narrower
than observed. Ultimately, this problem arises because the velocity cut points
which define the wing and core components are fixed at $  \pm   $2540 km s$  ^{-1}  $,
despite the fact that the profile falls off rapidly at equivalent tangential
velocities beyond $  v_{z}\sim \sigma _{v}  $, where $  \sigma _{v}=[GM_{\rm h}/(3r_{\rm t})]^{1/2}  $
is the velocity dispersion of $  f  $ at $  r=r_{\rm t}  $. A velocity dispersion
of $  \sigma _{v}\ga 5000  $ km s$  ^{-1}  $ is required at $  r=r_{\rm t}  $
in order to obtain the semi-logarithmic profile out to 10,000 km s$  ^{-1}  $
as is observed. So by decreasing $  M_{\rm h}/r_{\rm t}  $, the wing/core ratio
could be reduced, but the width would be increased.\pfn{actually the local profile is not a guasssian and I need to say so}
\end{enumerate}

\section{Models 3 and 4: Varying the Edge Ionization to Pressure Ratio \protect$  \Xi /P\protect  $\label{iontopratio}}

Models 3 and 4 illustrate the effects of varying the ratio of the ionization
parameter at the wind edge $  \Xi   $ to the pressure $  P  $. Before we discuss
these models, let us state why the $  \Xi   $ parameter alone is not varied
and, more generally, how we probe the stellar wind AGN model parameter space. 

The parameters of AGN cloud models (including the red giant wind model) can
be grouped into two categories depending on which line emission characteristics
they affect most:

\begin{enumerate}
\item A category that consists of parameters which primarily affects the \emph{line
ratios}. In most AGN models the line ratios are regulated by the emissivities,
which are in turn three-dimensional functions of $  \Xi   $ (or $  U  $),
$  P  $ (or $  n_{\rm H}  $), and $  N_{\rm c}  $, where $  L/r^{2}\propto \Xi   $
and $  P\propto (L/r^{2})^{s/2}  $. As mentioned before, because of the very
high column densities that we assume, $  N_{\rm c}  $ is a relatively unimportant
parameter for most of the range of parameter space of interest. Also, $  L  $
is readily observable but time dependent. Thus, for a given observed value of
$  L  $, the $  \Xi   $ ratio could be adjusted, but not $  \Xi   $ or $  P  $
by itself. Conversely, as $  L  $ changes in an AGN with known a $  \Xi   $
ratio, both $  P  $ and $  \Xi   $ can vary, depending upon the value of $  s  $
assumed. This section as well as the next two sections explore the line ratio
parameter space by varying, respectively, the $  \Xi   $ ratio (with $  L  $
fixed) (\S~\ref{iontopratio}), $  L  $ (with $  \Xi   $ fixed but with changes
in $  \dot{M}  $ to counteract changes to the covering function that would
occur otherwise) (\S~\ref{L&P}), and $  s  $ (\S~\ref{s}). Thus, we probe
the structure of the emissivity function in a non-orthogonal fashion (similar
to that of Taylor {[}1994{]}). Because the underlying parameters are common
to most AGN line emission models, \S\S~\ref{iontopratio}-\ref{s} should be
of interest to the AGN community whether or not the AGN stellar wind model is
actually valid. 
\item A category of parameters that does not directly regulate line ratios but rather
primarily affects \emph{properties specific to the stellar wind AGN line emission
model}. Parameters in this category primarily affect the covering function,
or strength and shape of the ``generic line profile.'' As mentioned previously
(see Fig. \ref{model2/Lya1-d}), this generic or ``summed'' line emission flux
is defined to be the synthetic line resulting from a hypothetical line efficiency
of unity. (The summed line flux is equal to the incident continuum heating were
the clouds completely opaque.) The flux of the summed line is approximately
the difference between the spectrum and the continuum. The parameters in this
second category can be further subdivided into those which affect the stellar
winds and those which affect the stellar distribution function. In the former
(winds) subcategory are luminosity $  L  $ (without compensatory changes in
$  \dot{M}  $) (\S~\ref{luminosity}), the mass loss slope $  \alpha   $ (\S~\ref{alpha}),
the terminal stellar wind speed $  v_{\infty }  $ (\S~\ref{v_inf}), the ratio
of the stellar mass loss to the terminal wind velocity $  \dot{M}/v_{\infty }  $
(\S~\ref{dmdt/vinf}), the minimum column density $  N_{\rm {cmin}}  $ (\S~\ref{nc_min}),
and the minimum mass loss rate $  \dot{M}_{\rm {min}}  $ (\S~\ref{mdot_min}).
In the latter (distribution function) subcategory are the density slope $  3-\eta   $
(\S~\ref{eta}), the tidal radius $  r_{\rm t}  $ (\S~\ref{tidal section}),
the black hole mass $  M_{\rm h}  $ (\S~\ref{m_h}), and the cluster mass $  M_{\rm c}  $
(\S~\ref{m_core}). In \S \S~\ref{luminosity}-\ref{m_core} we discuss models
which illustrate how each of these parameters affect observable quantities.
\end{enumerate}
Before continuing, note that this partitioning of the model parameters into
a line ratio category and a summed line profile category is incomplete even
with some of the compensatory changes we make. This is because many parameters
affect both. For instance, the tidal radius $  r_{\rm t}  $ is classified as
being a summed line profile parameter. However, by lowering it, one increases
the O {\small VI}/C {\small IV} line ratio for $  s<2  $ models. This is because
the local O {\small VI}/C {\small IV} ratio is a decreasing function of the
distance from the continuum source. Similarly, if the ionization parameter (deemed
a ratio parameter) is increased enough, cooling via line emission decreases,
which of course would decrease the strength of the summed line. 

Models 3 and 4 are similar to model 2 yet have different $  \Xi   $ ratios.
As mentioned earlier, model 2 was given an ionization parameter of $  \Xi =1.0  $
so that it would be more compatible with observations. This is a full decade
lower than that of model 1. Models 3 and 4 have, respectively, an ionization
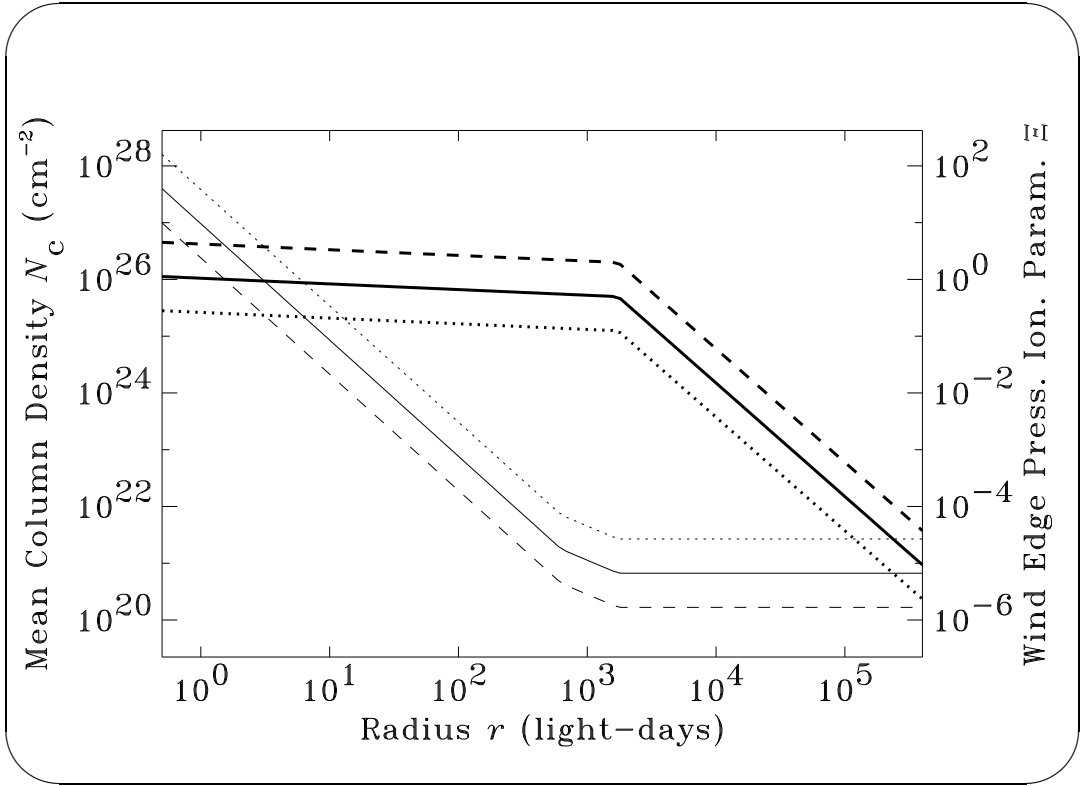
parameter of 0.25 and 4.0 in the BLR. This is shown in \Fig{model3/N_c+Xi}{Thin
lines, left axis: column density $  N_{\rm c}  $. Thick lines, right axis:
wind edge ionization parameter $  \Xi   $. Dotted lines: model 3. Dashed lines:
model 4. Solid lines: model 2, shown for comparison.}. Note that beyond $  1800  $
light days, $  N_{\rm {c}}=N_{\rm {cmin}}  $. In this region, the mass loss
is constant, and the edge pressure is nearly constant, so the cloud edge ionization
parameter falls as $  \sim r^{-2}  $.

One effect of increasing $  \Xi   $ is that it decreases the density at the
wind edge. For a given mass loss rate, this increases the radius of the cool
portion of the wind, which affects the summed line flux. Unfortunately, several
other parameters also affect the summed line flux. Therefore, in order to help
isolate just the effect of varying the ionization state of the line-emitting
gas without changing the covering factor at the same time, the mass loss rates
were also changed in models 3 and 4. Because the mass loss rate affects the
wind size (for a given column-density-limited wind), the minimum permissible
column density was additionally changed. These modified mass loss rates and
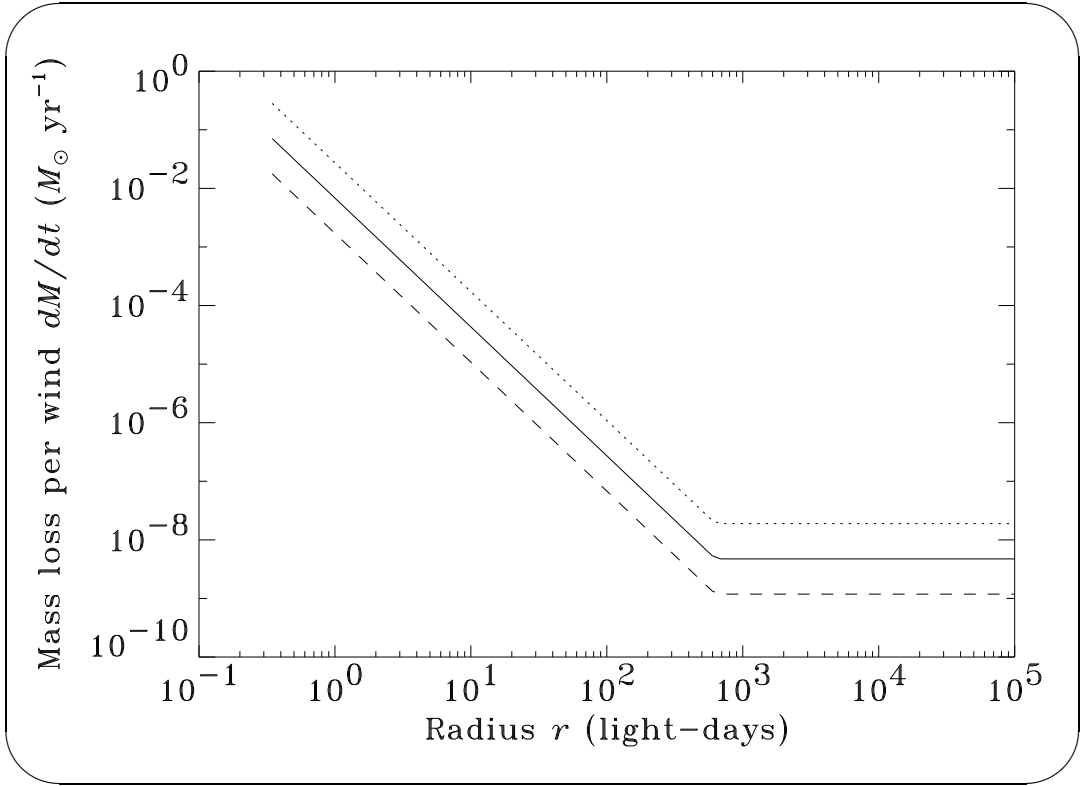
column densities for models 3 and 4 are shown, respectively, in Figures~\ref{model3/indivmass}\fig{model3/indivmass}{Dotted
line: the mass loss rate per star $  \dot{M}  $ as a function of radius from
the black hole $  r  $ for model 3. Dashed line: model 4. Solid line: model
2.} and \ref{model3/N_c+Xi}. The resulting effective wind areas for the supergiants
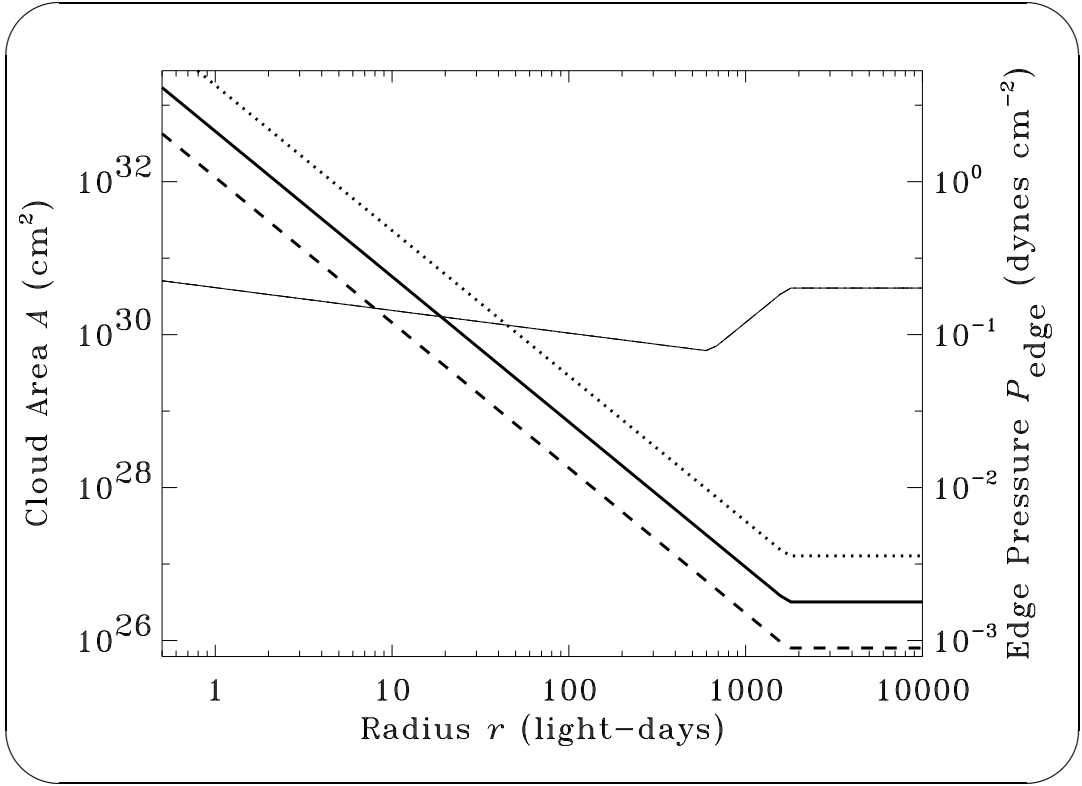
of models 3 and 4 are shown in \Fig{model3/area}{Solid thin line, left axis:
the cloud area $  A  $ as a function of radius $  r  $ for models 2, 3, and
4. (Due to the adjustments with the mass loss rate and the column density, the
area functions are identical for each model.) Thick dotted line, right axis:
the radial dependence of the pressure $  P_{\rm edge}  $ at the wind edge for
model 3. Thick dashed line: $  P_{\rm edge}  $ for model 4. Thick solid line:
$  P_{\rm edge}  $ for model 2.}.

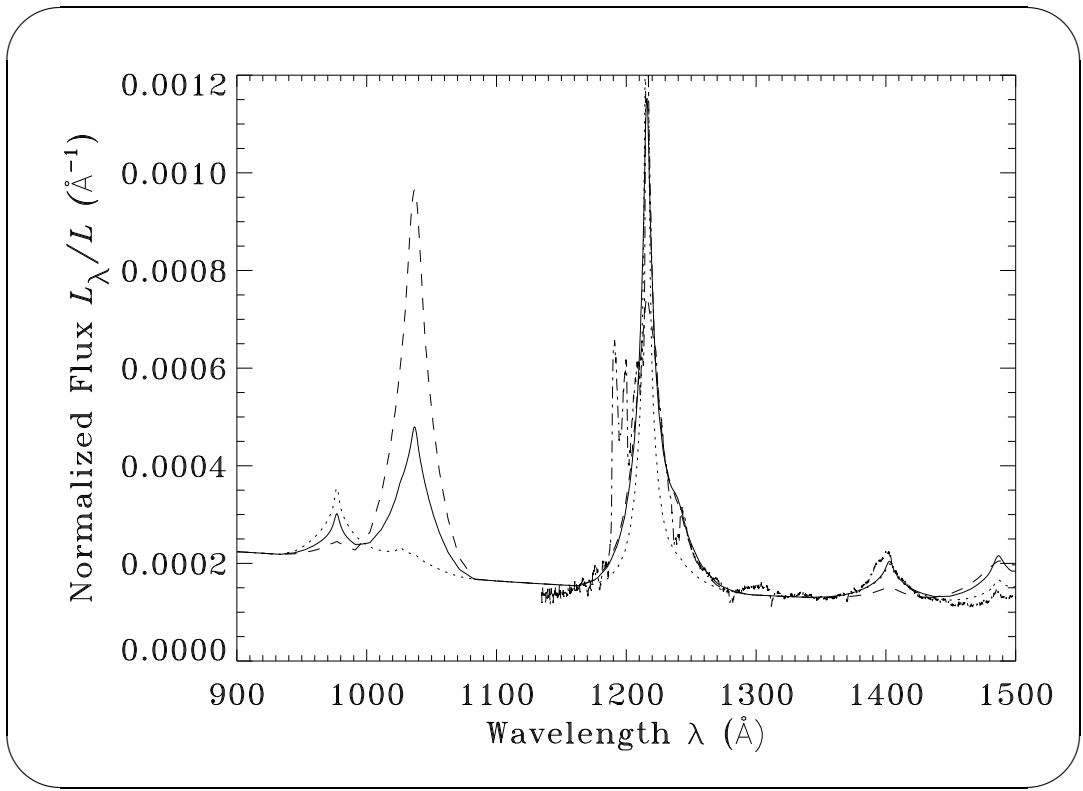
Spectra of models 3 and 4 are shown in Figures \ref{model3/harduv} and \ref{model3/softuv}.\fig{model3/harduv}{Dotted
line: hard UV spectrum of model 3. Dashed line: model 4. Solid line: model 2.
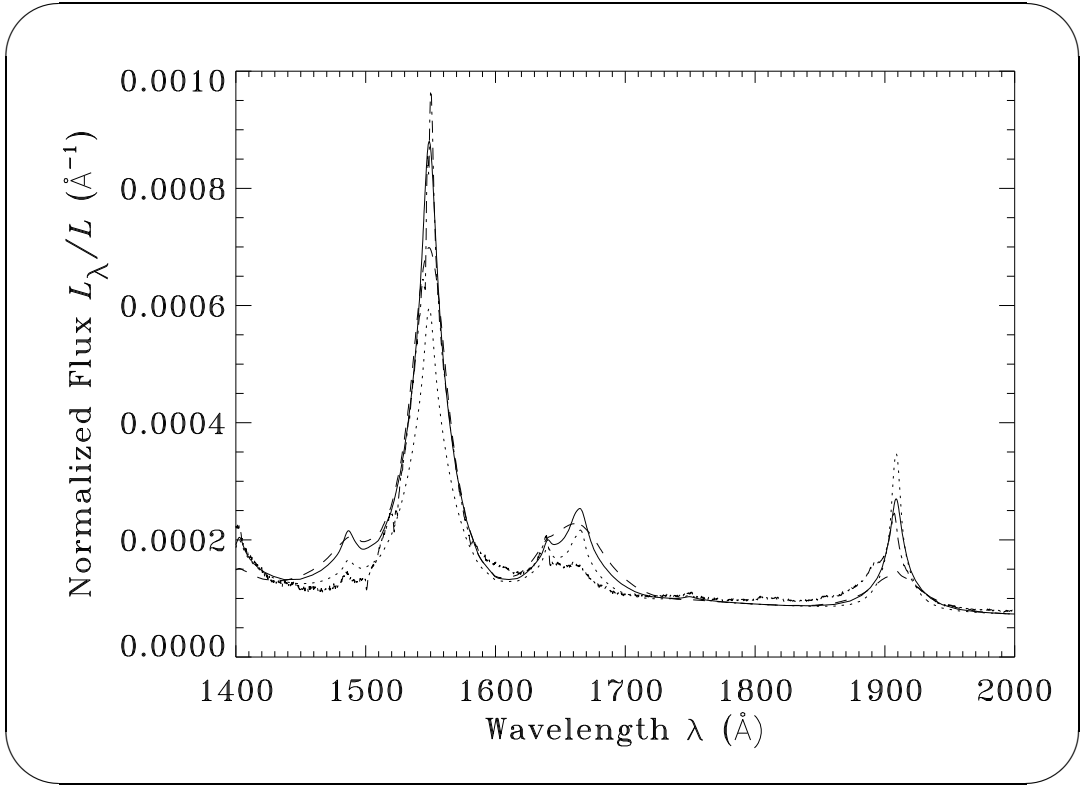
Dot-dashed line: NGC 5548.}\fig{model3/softuv}{Dotted line: soft UV spectrum
of model 3. Dashed line: model 4. Solid line: model 2. Dot-dashed line: NGC
5548.} As expected, model 3 ($  \Xi =0.25  $) has relatively low O{\scriptsize ~VI}
and N~{\scriptsize V} line fluxes. On the other hand, these lines are relatively
bright in model 4 ($  \Xi =4.0  $). Model 2, with its intermediate yet still
high ionization parameter of $  \Xi =1.0  $, is closest to matching the actual
N {\scriptsize V} line flux that was observed from NGC 5548 by \emph{HST}.

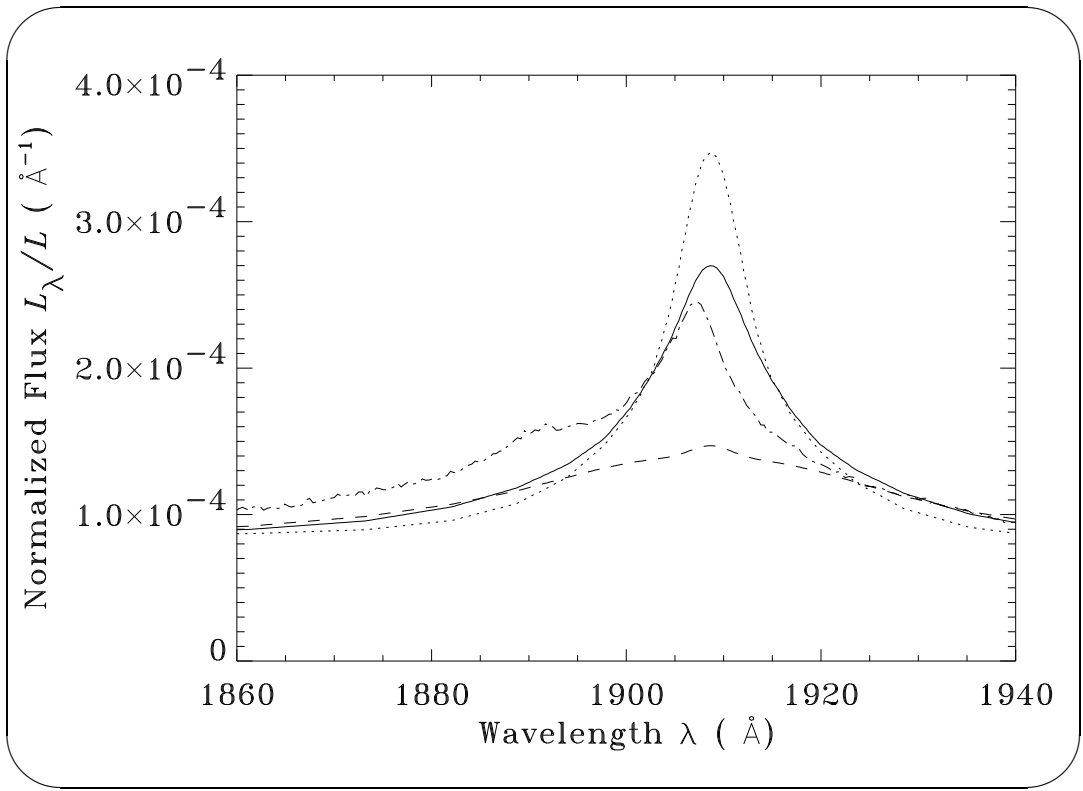
As \Fig{model3/ciii}{Dotted line: C {\footnotesize III}{]} spectral region of
model 3 ($  \Xi =0.25  $). Dashed line: model 4 ($  \Xi =4.0  $). Solid line:
model 2 ($  \Xi =1.0  $). Dot-dashed line: NGC 5548.} shows, the C {\small III}{]}
$  \lambda   $1909 \AA~profile shape is different for each of the three models.
This is, however, not true of its integrated intensity, which illustrates a
shortcoming of analysis based upon ratios alone. Model 4 ($  \Xi =4.0  $) has
the broadest and weakest C {\footnotesize III}{]} profile. Its ionization parameter
is very high everywhere, so the C {\footnotesize III}{]} flux is necessarily
low everywhere. However, because $  s=1.9  $ in each of these models and the
C {\small III}{]} inter-combination line is suppressed at high densities due
to quenching, the broader profile components are relatively weak. Model 4 has
the lowest density. It is therefore the most immune to this effect. As a result,
the high-velocity-dispersion clouds of model 4 have a relatively large C {\footnotesize III}{]}
flux, which is why this model has the broadest C {\footnotesize III}{]} profile.
Model 3, on the other hand, has a much lower ionization parameter that nearly
maximizes the mean fractional abundance of C$  ^{+2}  $ in the line-emitting
regions of its clouds, so it is relatively strong in C {\footnotesize III}{]}
on the whole. Model 3 also has the highest density, so it has the narrowest
of the three profiles. Model 2 comes closest to matching the actual C {\footnotesize III}{]}
profile that is observed in NGC 5548. 

Simultaneous comparison of other lines reveals important shortcomings that are
intrinsic to these red giant wind models. For instance, the NGC 5548 N {\footnotesize IV}
$  \lambda   $1486 and the O {\footnotesize III}{]} $  \lambda   $1664 line
fluxes are lower than all of the models shown. Model 3 ($  \Xi =0.25  $) comes
closest to matching the observed strength of these lines. However, model 3 fares
worst at matching the observed N {\footnotesize V} line strength. Therefore,
provided the other parameters remain constant, no single value of $  \Xi   $
can simultaneously match each of these line strengths.

In some cases, adjusting additional parameters can help line ratio problems
similar to this one. For instance, increasing the metallicity would help increase
the nitrogen line fluxes, which are relatively too low in model 3. However,
as is discussed in \S \S~\ref{model2}, \ref{L&P}, \& \ref{luminosity}, even
permitting each parameter to be freely adjustable would not yield complete agreement
with each NGC 5548 line flux.

\section{Models 4.4 and 4.8: Varying the Luminosity \protect$  L\protect  $, Pressure
\protect$  P\protect  $, and Mass Loss \protect$  \dot{M}\protect  $\label{L&P}
}

In the previous subsection, we presented models in which the wind edge pressure
and ionization parameter were varied while the luminosity was held constant.
In this section, we present models in which the continuum luminosity and cloud
pressures were changed, but the ionization parameter was held fixed.

Model 4.4 is similar to model 2 except that its bolometric luminosity is $  10^{43}  $
ergs s$  ^{-1}  $ instead of $  3.2\times 10^{43}  $ erg s$  ^{-1}  $. Model
4.8 has a bolometric luminosity of $  10^{44}  $ ergs s$  ^{-1}  $. The luminosity
parameter $  L  $ is similar to the $  \Xi /P  $ ratio discussed in the previous
subsection in that changing it also affects the covering function. In order
to help isolate just the effect of changing $  L  $ and $  P  $ upon the line
ratios, we again adjusted $  \dot{M}  $, $  \dot{M}_{\rm min}  $, and $  N_{\rm cmin}  $
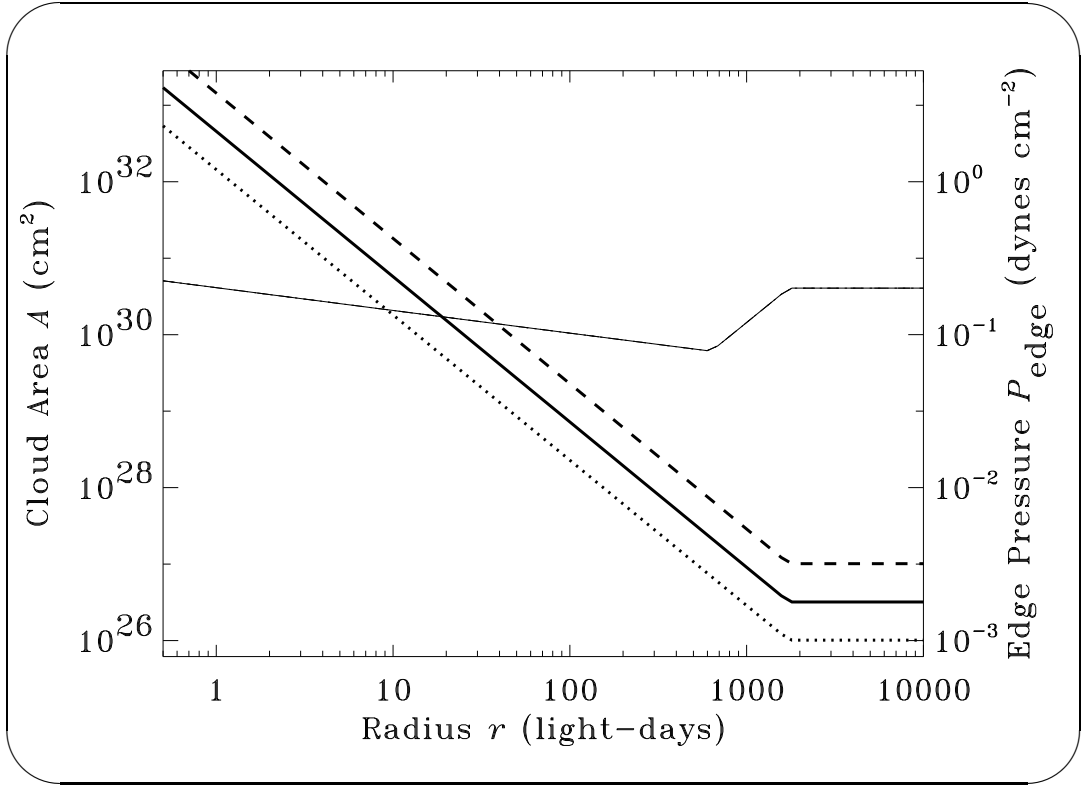
in models 4.4 and 4.8. This forced the area functions (shown in \Fig{model4.4/area}{Solid
line, left axis: wind area functions for models 4.4, 4.8, and 2. Thick lines,
right axis: wind edge pressures. Dotted line: model 4.4. Dashed line: model
4.8. Solid lines: model 2.}) and geometrical covering functions to be identical.
The full effects of varying the luminosity without any such adjustments are
discussed in \S~\ref{luminosity}. However, for model 2, $  s  $ is near $  2  $,
while $  \alpha   $ is near unity. Therefore, neither the cloud edge ionization
parameters nor the cloud area functions are a strong function of the local continuum
flux, and the models discussed here are very similar to those of \S~\ref{luminosity}. 

The UV spectra of models 4.4 and 4.8 are shown in Figures \ref{model4.4/harduv} and \ref{model4.4/softuv}.\fig{model4.4/harduv}{Dotted
line: model 4.4 ($  L=10^{43}  $ ergs s$  ^{-1}  $). Dashed line: model 6
($  L=10^{44}  $ ergs s$  ^{-1}  $). Solid line: model 2 ($  L=3.2\times 10^{43}  $
ergs s$  ^{-1}  $). Dot-dashed line: NGC 5548.}\fig{model4.4/softuv}{Dotted
line: model 4.4 ($  L=10^{43}  $ ergs s$  ^{-1}  $). Dashed line: model 6
($  L=10^{44}  $ ergs s$  ^{-1}  $). Solid line: model 2 ($  L=3.2\times 10^{43}  $
ergs s$  ^{-1}  $). Dot-dashed line: NGC 5548.} Because the geometrical covering
functions of the models are identical, the equivalent widths of the lines are
approximately the same in models 4.4 and 4.8. This is despite the factor of
10 variation in luminosity. 

Changing the luminosity does not have a discernible effect upon the absolute
C {\small III} $  \lambda   $977 \AA~line strength. This result is opposite
that of Rees, Netzer, \& Ferland (1989). The C {\small III} $  \lambda   $977 \AA~emission
characteristics obtained by Rees, Netzer, \& Ferland (1989) occurred because
of thermalization of strong lines that have high optical depths. This redistributes
the cooling to other lines including C {\small III} $  \lambda   $977. Our
results do not show such an increase in C {\small III} $  \lambda   $977 except
at high densities approaching $  10^{12}  $ cm$  ^{-3}  $, which is outside
the parameter space that XSTAR version 1.20 was designed to compute. Such high
density models are shown in \S~\ref{luminosity}.

On the other hand, the densities and local continuum fluxes of model 4.8 are
high enough to suppress the highest velocity components of the Ly$  \alpha   $
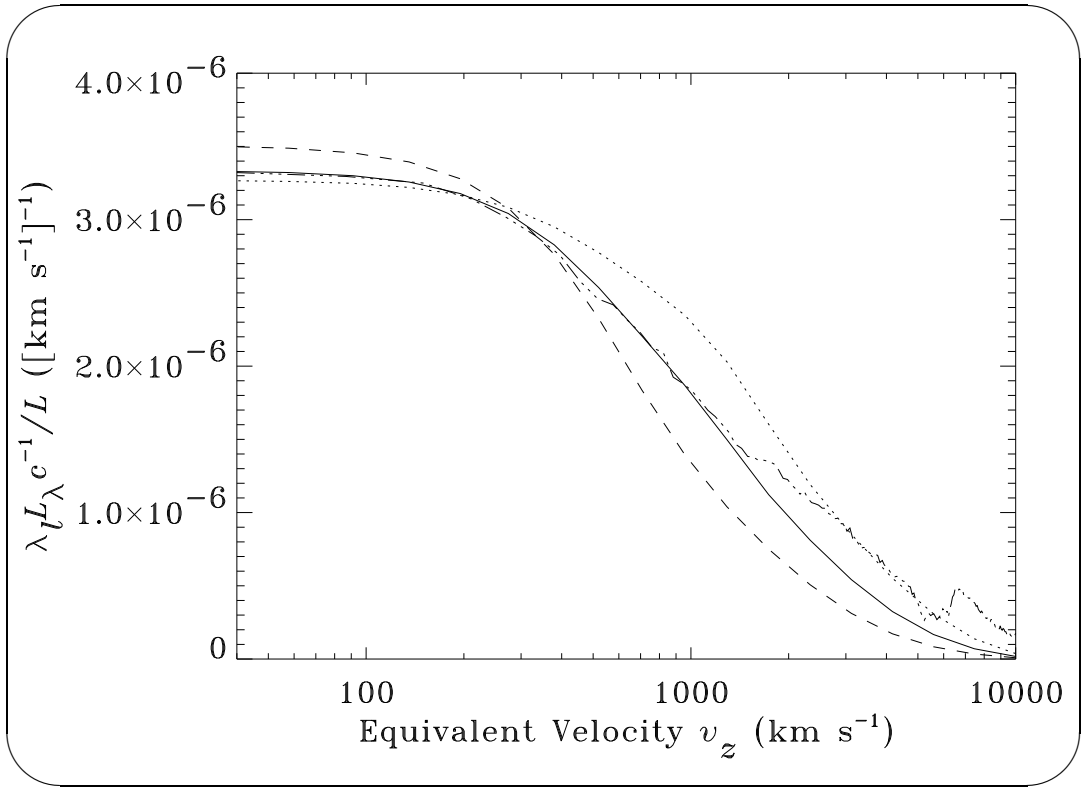
profile (shown at higher resolution in \Fig{model4.4/Lyaline}{Dotted line: Ly$  \alpha   $
line of model 4.4. Dashed line: Ly$  \alpha   $ line of model 4.8. Solid line:
Ly$  \alpha   $ line of model 2. Dot-dashed line: continuum subtracted NGC
5548 time-averaged Ly$  \alpha   $ spectral region, normalized to match the
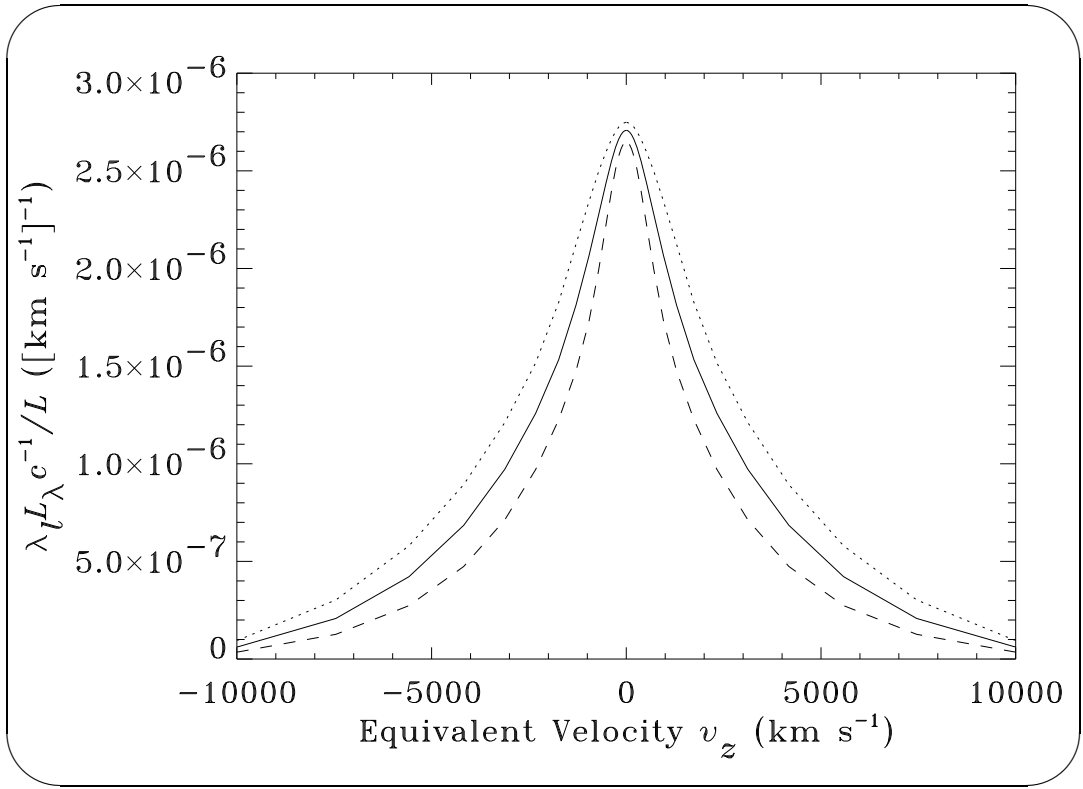
model 2 profile peak.}) and the C {\footnotesize IV} profile (shown in \Fig{model4.4/civ}{Dotted
line: C {\footnotesize IV} line of model 4.4. Dashed line: C {\footnotesize IV}
line of model 4.8.}). These lines behave in a fashion similar to C {\footnotesize III}{]}---the
line efficiencies in the high density models of the clouds nearest to the black
hole are suppressed. This results in these models having narrower profiles.

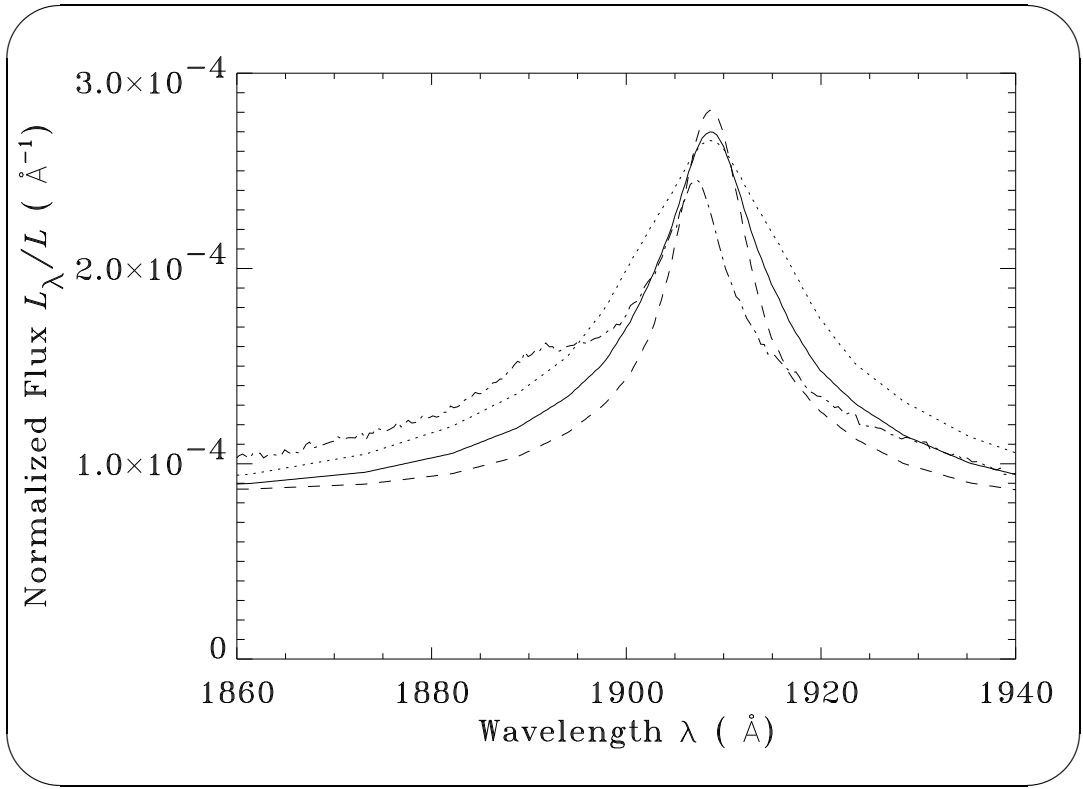
The behavior of the C {\small III}{]} profile itself is shown in \Fig{model4.4/ciii}{Dotted
line: C {\footnotesize III}{]} profile of model 4.4 ($  L=10^{43}  $ ergs s$  ^{-1}  $).
Dashed line: model 4.8 ($  L=10^{44}  $ ergs s$  ^{-1}  $). Solid line: model
2 ($  L=3.2\times 10^{43}  $ ergs s$  ^{-1}  $). Dot-dashed line: NGC 5548.}.
The dependence of the C {\footnotesize III}{]} profile shape upon luminosity
is perhaps the most prominent difference between models 4.4 and 4.8. This dependence
arises for the reasons mentioned in the previous subsection---the higher luminosity
model has higher densities, which yield less C {\small III}{]} emission. As
a result, the higher luminosity models have narrower profiles. 

Note that the broad and narrow components of the observed C {\small III}{]}
line blend appear to be shifted towards the blue. In contrast, the line shifts
of the synthetic spectra are zero (cf Appendix \ref{shiftsap}). Also, model
4.8 (and to a lesser extent the two other models) does not produce sufficient
broad C {\small III}{]} line emission. The time-averaged ionizing luminosity
of NGC 5548 and model 4.8 are approximately the same: $  \sim 4\times 10^{43}  $
ergs s$  ^{-1}  $. (For the continuum assumed, the bolometric luminosity of
each model is 2.61 times greater than 1-1000 Ry ionizing luminosity). Model
2, which has a luminosity that is a factor of $  \sim   $3 less than NGC 5548,
fairs better in this respect. However, the C {\small III}{]}/C {\small IV} line
ratio of the broad profile components is too low in all of our models in which
the luminosity was the same as the NGC 5548 value. This broad component C {\small III}{]}/C
{\small IV} line ratio problem was first discussed by Krolik et al. (1991).
The problem arises because the continuum flux in the BLR is so high for the
Hubble constant of $  H=65  $ km s$  ^{-1}  $ Mpc$  ^{-1}  $ that we assumed.
That is, because $  P=F/(c\Xi \cos \theta )  $, a decrease in the ionization
parameter results in an increase in the pressure, which increases collisional
de-excitation of C {\small III}{]}. Conversely, if the pressure is decreased,
the ionizational abundance of C$  ^{+2}  $ becomes too low. In this respect,
the stellar wind model has two advantages over some of the other models that
have been examined. First, the angled region of the optically thick wind effectively
has a lower continuum flux at the cloud edge. This permits it to have lower
densities and ionization parameters than the face-on cloud models that are traditionally
assumed by photoionization codes. Second, the optically thick winds effectively
have more than one single ionization parameter. This smoothes the radial dependence
of the total C {\small III}{]} emissivity function. Despite these features,
our models still exhibit the C {\small III}{]} deficiency. Incidentally, this
problem does not exist for the \emph{integrated} C {\small III}{]}/C {\small IV}
line ratio, which is an agreeable 6.8$  ^{-1}  $ in NGC 5548 (Krolik et al.
1991), 5.4$  ^{-1}  $ in model 2, and 5.2$  ^{-1}  $ in model 4.8. 

Dumont et al. (1998) state that the C {\small III}{]}/C {\small IV} problem
can be resolved simply by realizing that the response function of the C {\small III}{]}
line is more extended (suggesting emission from larger distances) than that
of the other lines. Caution should be exercised here, however, for two reasons.
First, the C {\small III}{]} line, though narrower than most of the other UV
lines, is still a broad line. Moving the emission covering function out would
only reduce the broad emission flux even further. (At issue should be only the
flux of the \emph{broader} C {\small III}{]} components.) Second, the gain of
the broad component of the C {\small III}{]} line is much less than that of
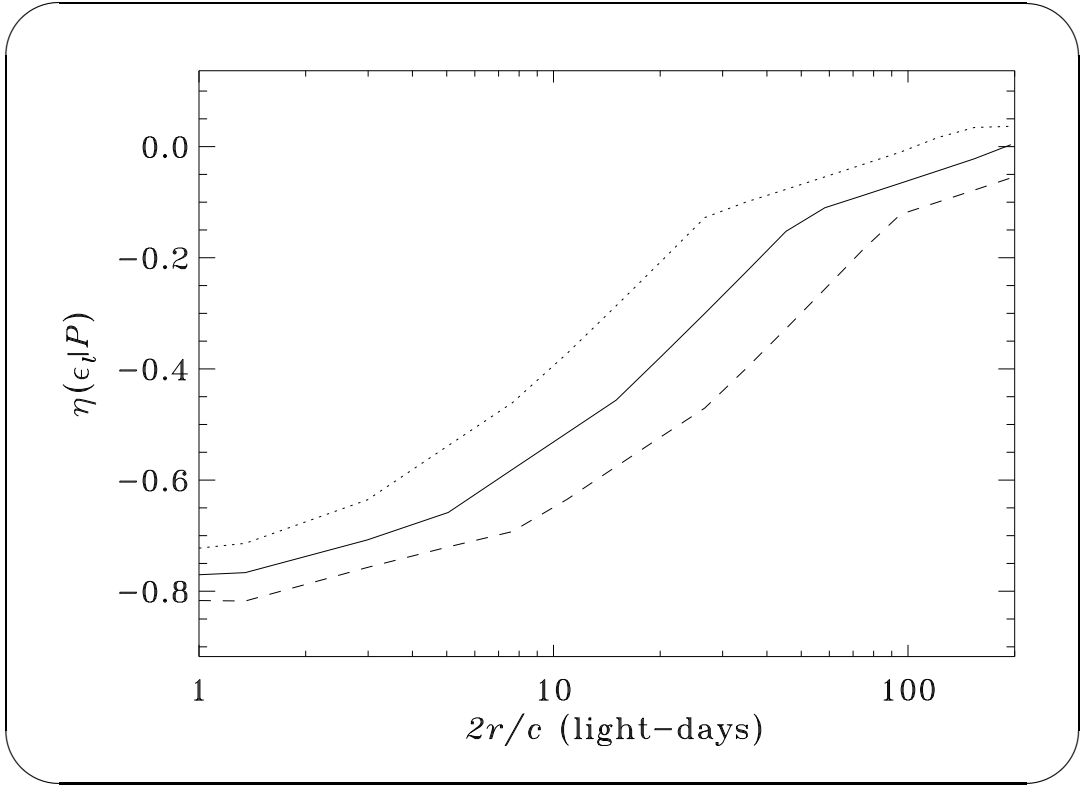
other UV lines (compare Figures \ref{model4.4/ciiietaP}\fig{model4.4/ciiietaP}{Dotted
line: the model-independent asymptotic gain of the C {\small III}{]} line with
respect to variations in the pressure as a function of $  2r/c  $ for model
4.4 ($  L=10^{43}  $ ergs s$  ^{-1}  $). Dashed line: model 4.8 ($  L=10^{44}  $
ergs s$  ^{-1}  $). Solid line: model 2 ($  L=3.2\times 10^{43}  $ ergs s$  ^{-1}  $).}
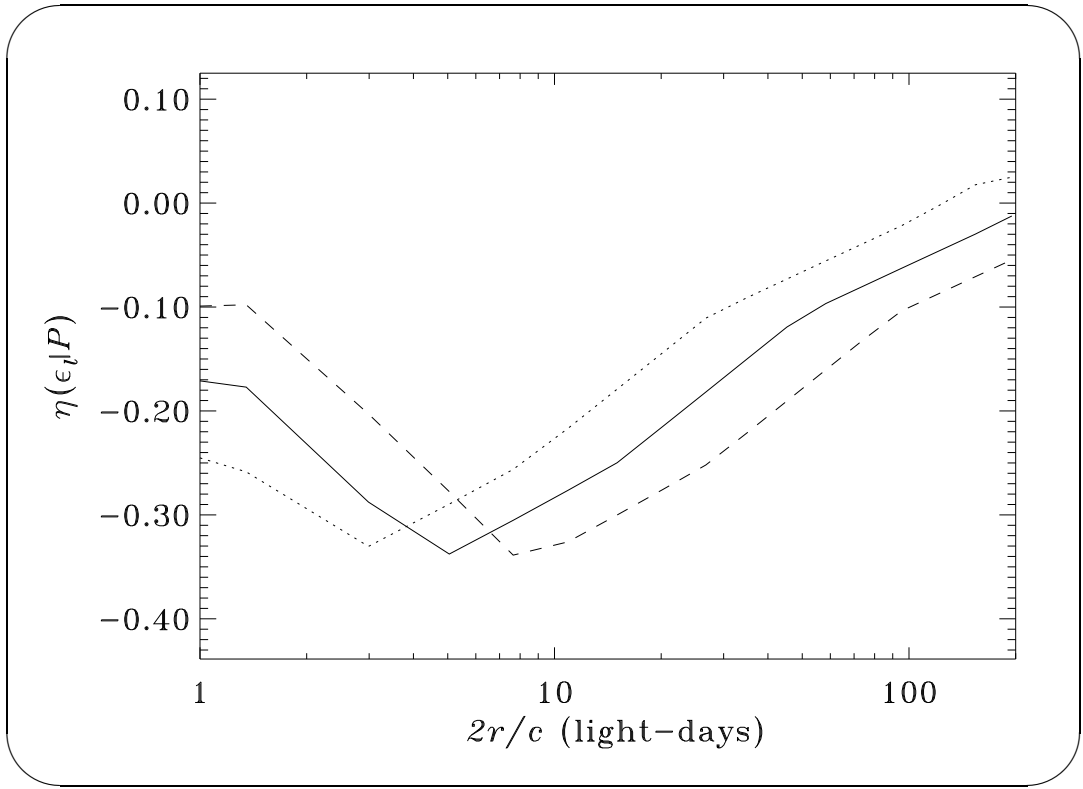
and \ref{model4.4/LyaetaP}\fig{model4.4/LyaetaP}{As in Figure \ref{model4.4/ciiietaP},
but for the Ly$  \alpha   $ line.}). For this reason, its response function
is more extended than its ``spatial response function'' that would exist if
the line responded linearly. Thus the large FWHM of the linearized response
function of this line does not necessarily translate to an extended emission
region. 

On the other hand, our models assume a much harder continuum than the one shown
in Dumont et al. (1998); our models may have a time-averaged ionizing continuum
luminosity that is too high. The issue is complicated by time variability. Fully
nonlinear time evolution models which fully account for such effects may someday
help resolve whether or not the C {\small III}{]}/C {\small IV} ratio problem
is significant.

\section{Models 5 and 6: Varying the Pressure Slope \protect$  s\protect  $\label{s}}

This subsection discusses models that have different values of the pressure
slope parameter $  s  $ (where $  P\propto [L/r^{2}]^{s/2}  $ and $  \Xi \propto [L/r^{2}]^{1-s/2}  $).
In model 5, $  s=1.0  $, $  \alpha =0.65  $, $  \dot{M}_{\rm min}=1.7\times 10^{-5}\, M_{\odot }  $
yr$  ^{-1}  $, and $  N_{\rm cmin}=1.4\times 10^{23}  $ cm$  ^{-2}  $. In
model 6, $  s=1.5  $, $  \alpha =0.90  $, $  \dot{M}_{\rm min}=8.7\times 10^{-7}\, M_{\odot }  $
yr$  ^{-1}  $, and $  N_{\rm cmin}=7.2\times 10^{21}  $ cm$  ^{-2}  $. As
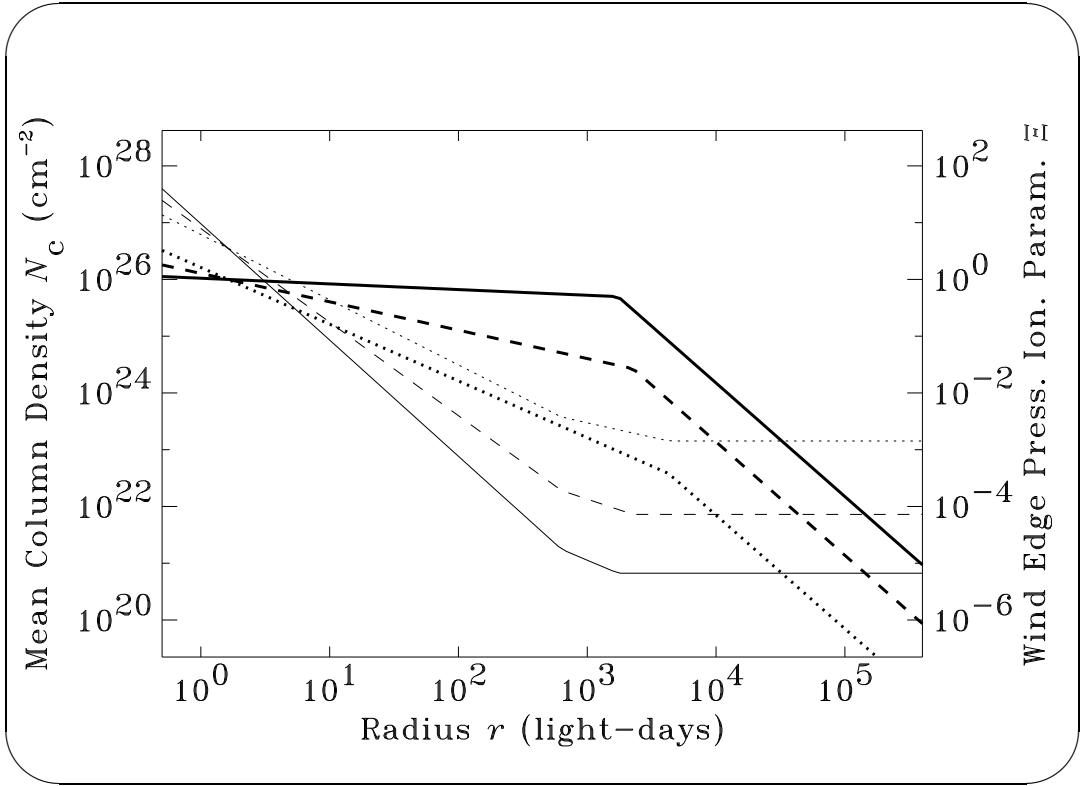
\Fig{model5/N_c+Xi}{Thin lines, left axis: column density $  N_{\rm c}  $.
Thick lines, right axis: wind edge ionization parameter $  \Xi   $. Dotted
lines: model 5 ($  s=1.0  $). Dashed lines: model 6 ($  s=1.5  $). Solid lines:
model 2.} shows, the pressure slope $  s  $ gives the dependence of the ionization
parameter upon position within the BLR. Thus, $  s  $ affects the line emissivities
and ratios. It also affects the summed line by changing the area function. This
changing of the covering function is not of interest here because it is discussed
in \S \S~\ref{luminosity}-\ref{m_core}. By adjusting $  \alpha   $, $  \dot{M}_{\rm min}  $,
and $  N_{\rm cmin}  $ in addition to $  s  $, we help isolate the effect
that changing $  s  $ has upon just the line ratios and emissivities. This
also keeps the approximation of the asymptotic area gain given by equation (\ref{areagain})
to be constant at +0.15. The wind area functions of models 4 and 5 are shown
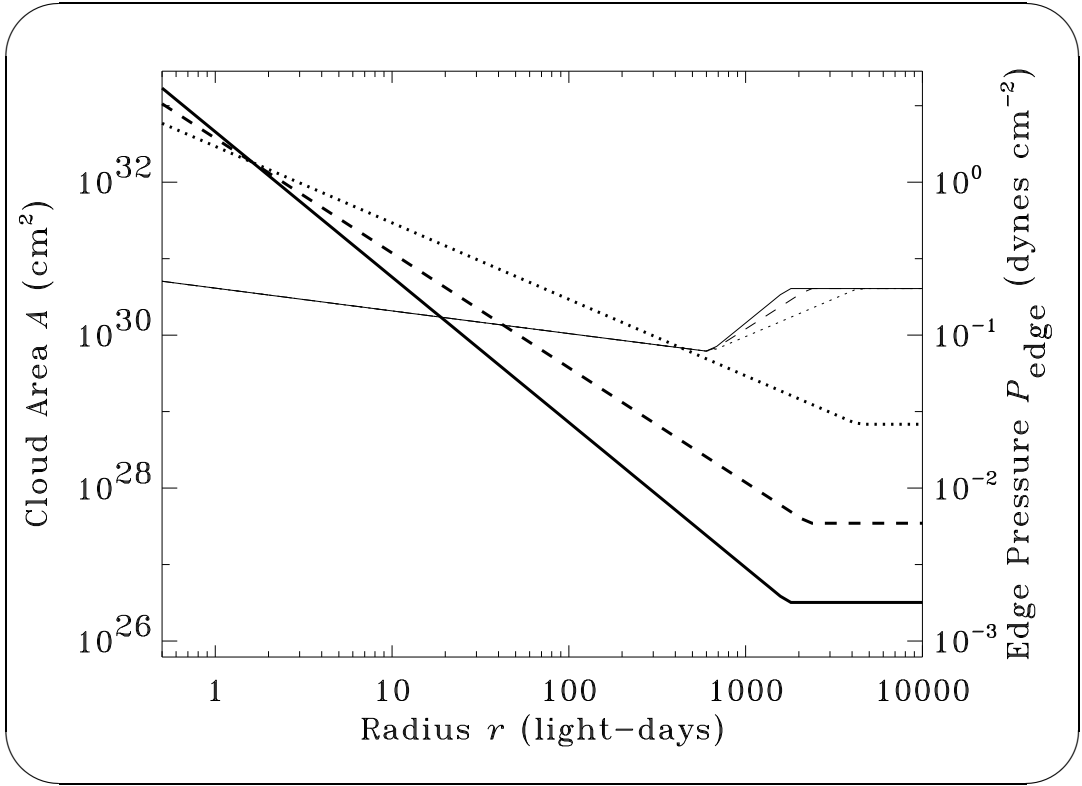
in \Fig{model5/area}{Thin lines, left axis: wind area functions. Thick lines,
right axis: wind edge pressures. Dotted lines: model 5. Dashed lines: model
6. Solid lines: model 2.}. Note that these adjustments are not totally complete;
in the relatively unimportant outer portion of the NLR the models have slightly
different area functions.

Spectra of models 5 and 6 are shown in Figures \ref{model5/harduv} and \ref{model5/softuv}.\fig{model5/harduv}{
Dotted line: model 5 ($  s=1.0  $). Dashed line: model 6 ($  s=1.5  $). Solid
line: model 2 ($  s=1.9  $).}\fig{model5/softuv}{Dotted line: model 5 ($  s=1.0  $).
Dashed line: model 6 ($  s=1.5  $). Solid line: model 2 ($  s=1.9  $).} The
intermediate-ionization lines of the spectrum of model 5 ($  s=1.0  $) are
broader than the low-ionization and recombination lines. This is especially
apparent for the C {\footnotesize IV}, C {\footnotesize III}{]}, and O {\footnotesize VI}
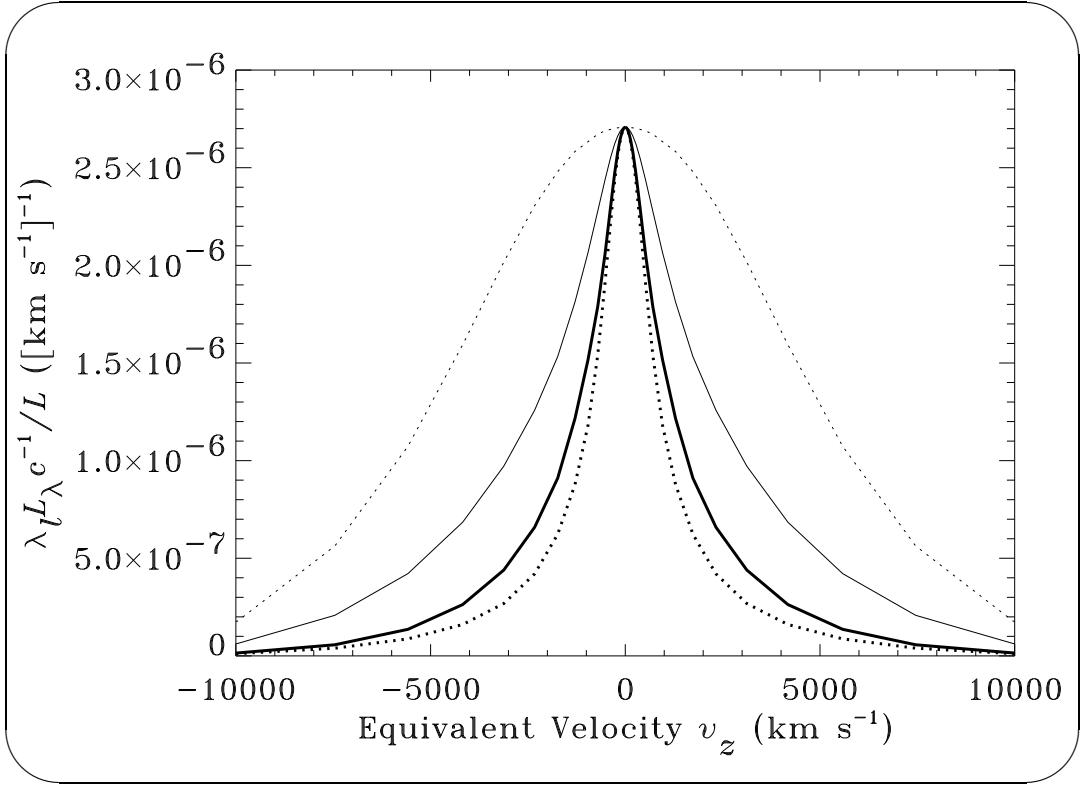
profiles, which are much broader than those of model 2. \Fig{model5/Lya+Civ}{Dotted
lines: model 5 ($  s=1.9  $) line profiles. Solid lines: model 2 ($  s=1.0  $)
line profiles. Thick lines: Ly$  \alpha   $. Thin lines: C {\footnotesize IV}.
The $  y  $-axis scale is applicable only for the C {\footnotesize IV} $  \lambda   $1548.2 \AA~line
flux of model 2; the other lines were rescaled. This plot confirms that the
$  s  $ parameter affects C {\small IV} much more than Ly$  \alpha   $.} shows
this effect more clearly. This broadening occurs if $  s<2.0  $ because the
ionization of low-velocity-dispersion clouds becomes too low for substantial
line emission to occur. Since Ly$  \alpha   $ is a recombination line, it is
relatively immune to this effect. These results imply $  s\ga 1.7  $. Comparable
results were obtained by Rees et al. (1989). On the other hand, several of the
model 5 ($  s=1.0  $) line shapes and fluxes are in closer agreement with NGC
5548 than model 2. This is particularly evident for the N {\footnotesize IV}
$  \lambda   $ 1486 and O {\footnotesize III}{]} $  \lambda   $1664 lines.\pfn{This
incompatibility is also discussed in Appendix xxx.} Thus, no value of $  s  $
would simultaneously fit all features of the NGC 5548 spectrum. This poses yet
another (admittedly minor) problem for these wind AGN models.

The response functions of models 5 and 6 also illustrate the effects of varying
the ionization gradient in the BLR. The response functions of C {\footnotesize IV}
for models 5 and 6 are shown in \Fig{model5/civrf}{C {\footnotesize IV} response
functions. Dotted line: model 5 ($  s=1.0  $). Dashed line: model 6 ($  s=1.5  $).
Solid line: model 2 ($  s=1.9  $).}. In accordance with $  \Xi \propto [L/r^{2}]^{1-s/2}  $,
the response drops off much more rapidly in the $  s=1.0  $ model than it does
in the $  s=1.9  $ model.\pfn{But even with $  s=1.9  $, the C {\footnotesize IV}
response function falls off too rapidly to match the SOLA-derived nonlinearized
C {\footnotesize IV} NGC 5548 response functions.} 

Many of the differences between the three response functions shown are caused
by differences in the gains. The radial dependences of the C {\small IV} line
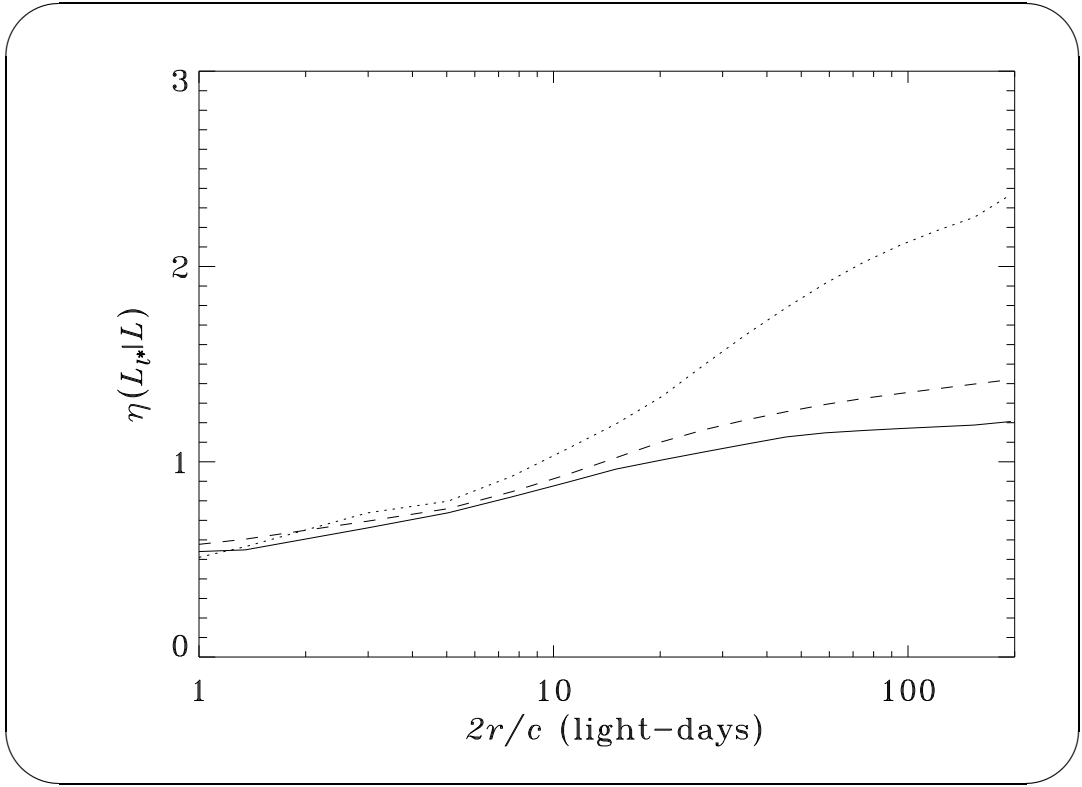
flux gains are in \Fig{model5/civgain}{The gains of the C {\footnotesize IV}
line emission from stellar winds due to variations of the continuum luminosity
as a function of radius $  r  $. Dotted line: model 5 ($  s=1.0  $). Dashed
line: model 6 ($  s=1.5  $). Solid line: model 2 ($  s=1.9  $).}. The two
key functions that regulate these gains are $  \eta (\epsilon _{l}|\Xi )  $
and $  \eta (\epsilon _{l}|P)  $, shown, respectively, in Figures \ref{model5/civetaXi}
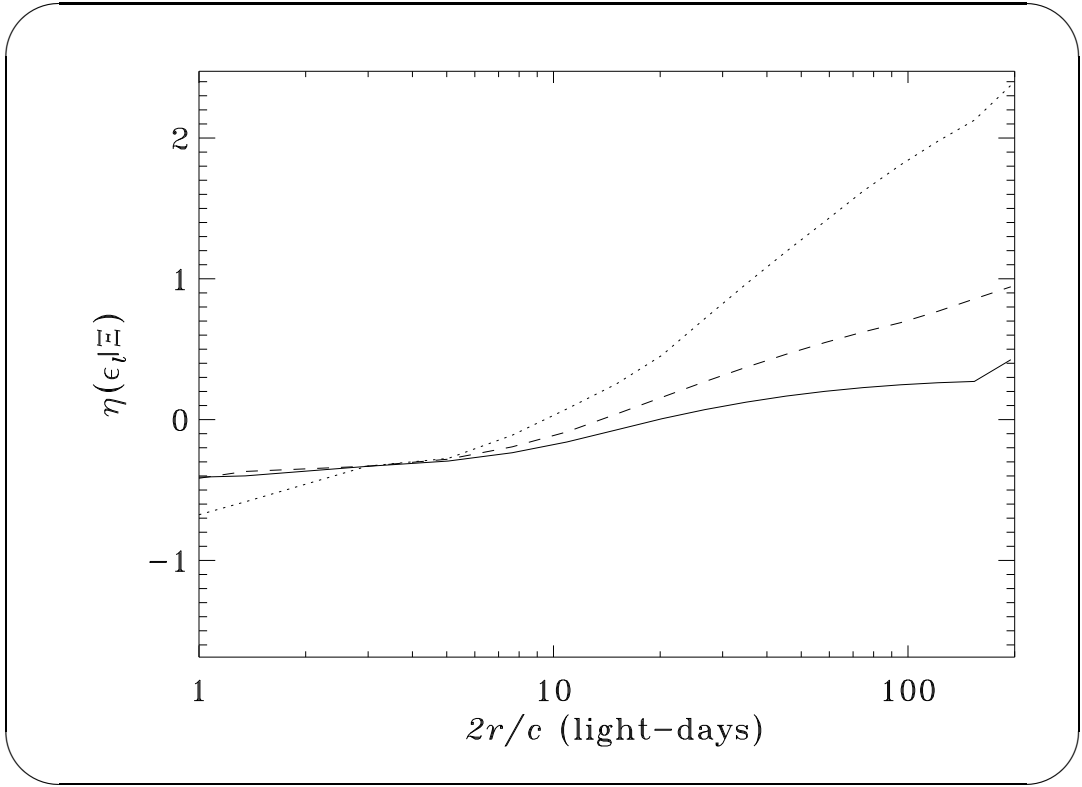
and \ref{model5/civetaP}.\fig{model5/civetaXi}{The asymptotic gain of the C
{\footnotesize IV} line emissivity due to variations of the ionization parameter.
Dotted line: model 5 ($  s=1.0  $). Dashed line: model 6 ($  s=1.5  $). Solid
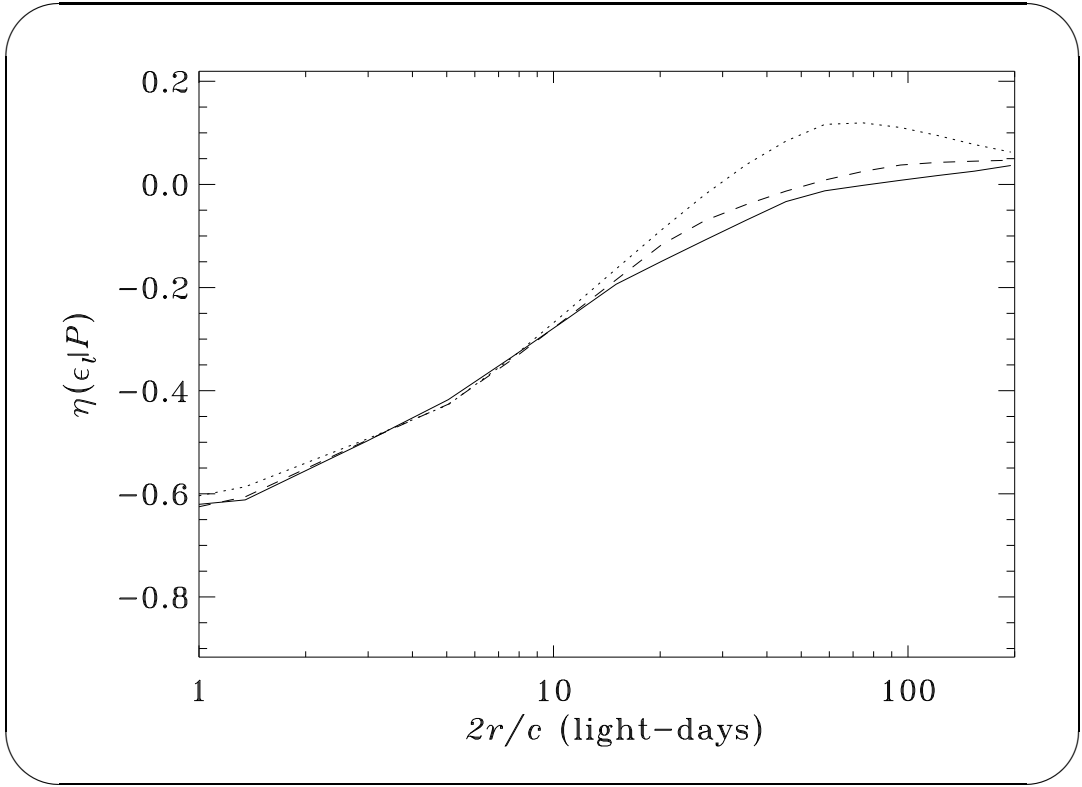
line: model 2 ($  s=1.9  $).}\fig{model5/civetaP}{The asymptotic gain of the
C {\footnotesize IV} line emissivity due to variations of the wind edge pressure.
Dotted line: model 5 ($  s=1.0  $). Dashed line: model 6 ($  s=1.5  $). Solid
line: model 2 ($  s=1.9  $).} Only the $  s=1.0  $ model has low enough ionization
for $  \eta (\epsilon _{l}|\Xi )\gg 1  $ to occur within 100 light-days from
the black hole.

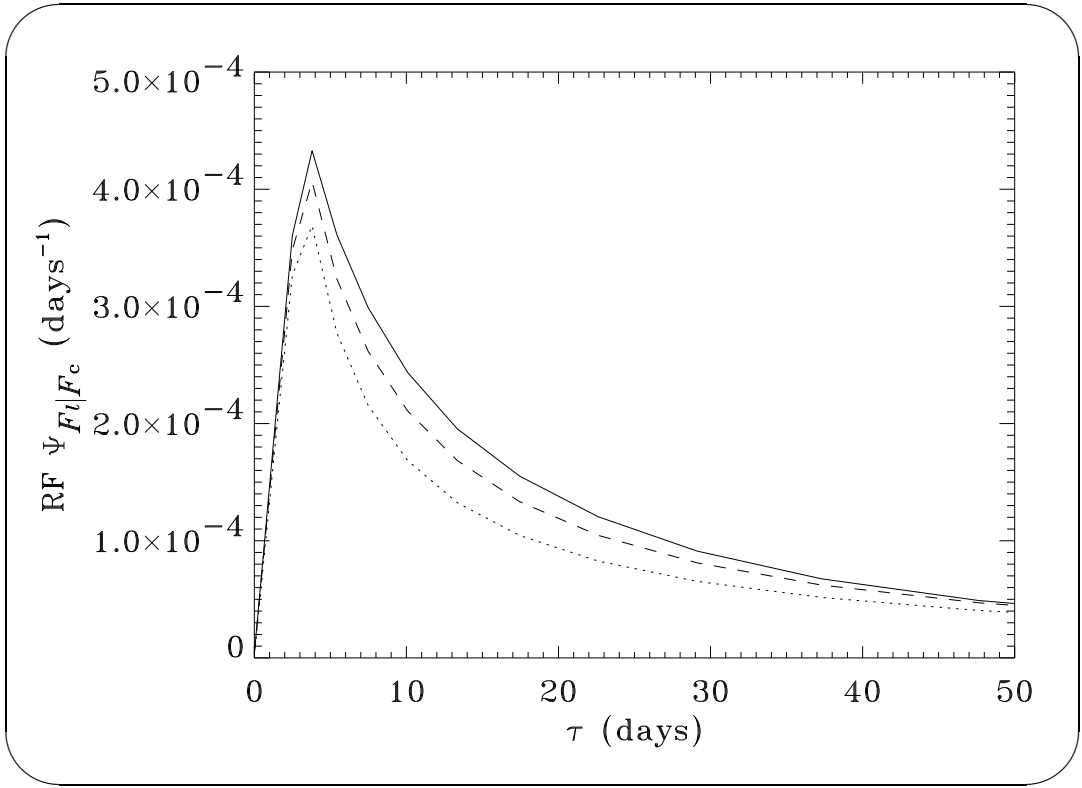
The Ly$  \alpha   $ response functions of models 5 and 6 are shown in \Fig{model5/Lyarf}{Ly$  \alpha   $
response functions. Dotted line: model 5 ($  s=1.0  $). Dashed line: model
6 ($  s=1.5  $). Solid line: model 2 ($  s=1.9  $). Though difficult to see
from this view of the data, the response function of model 5 falls the fastest
and has the weakest relative response at intermediate delays.}. As Figures \ref{model5/LyaetaP}, \ref{model5/LyaetaXi}, and
\ref{model5/Lyaeps} show, for these models Ly$  \alpha   $ is relatively more
sensitive to increases in pressure than ionization. This is due to high optical
depths and thermalization of Ly$  \alpha   $ at high pressures (as in Rees
et al. 1989). This causes model 2, which has the lowest average pressure (see
Fig. \ref{model5/area}), to have the strongest emissivity at intermediate distances
($  2r/c\sim 10  $ days). This difference gives rise to the relatively steep
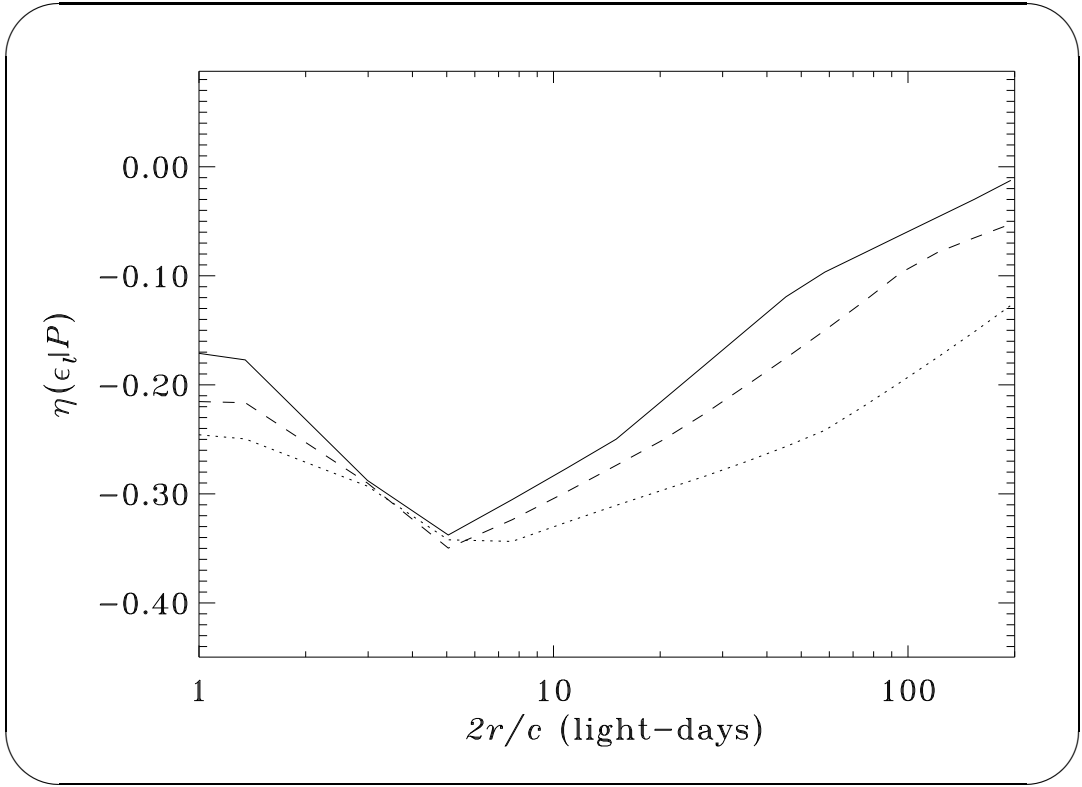
drop in the Ly$  \alpha   $ response function of model 5.\fig{model5/LyaetaP}{The
asymptotic gain of the Ly$  \alpha   $ line emissivity due to variations of
the wind edge pressure. Dotted line: model 5 ($  s=1.0  $). Dashed line: model
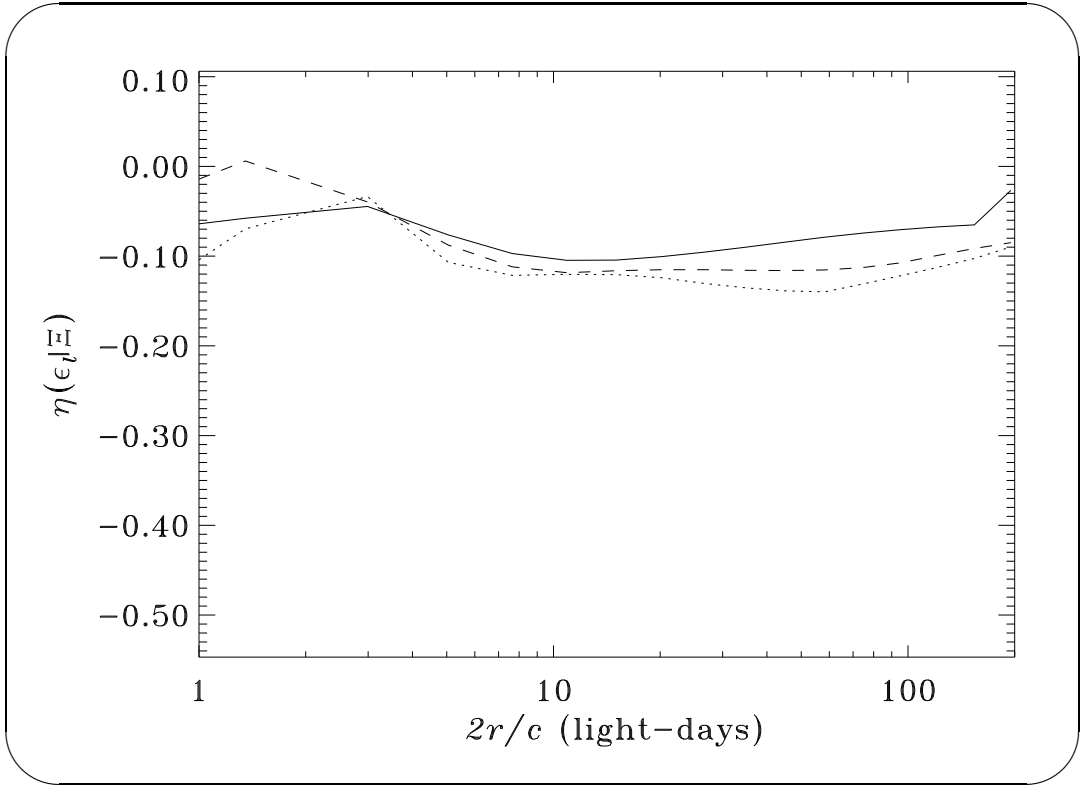
6 ($  s=1.5  $). Solid line: model 2 ($  s=1.9  $).}\fig{model5/LyaetaXi}{The
asymptotic gain of the Ly$  \alpha   $ line emissivity due to variations of
the wind edge ionization parameter. Dotted line: model 5 ($  s=1.0  $). Dashed
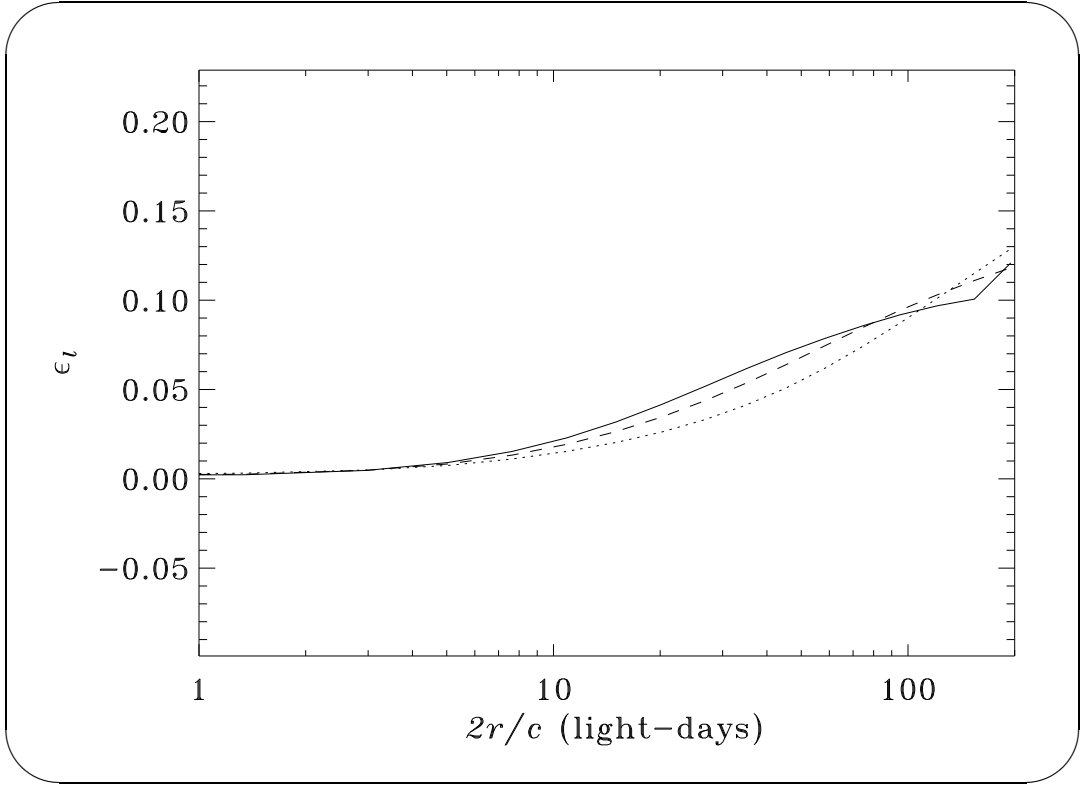
line: model 6 ($  s=1.5  $). Solid line: model 2 ($  s=1.9  $).}\fig{model5/Lyaeps}{The
emissivity of the Ly$  \alpha   $ line as a function of radius. Dotted line:
model 5 ($  s=1.0  $). Dashed line: model 6 ($  s=1.5  $). Solid line: model
2 ($  s=1.9  $). Note that the pressures are so high that this line becomes
thermalized near the black hole, especially for model 5.}

\section{Models 7, 8, and 9: Varying the Luminosity \protect$  L\protect  $ \label{luminosity}}

Models 3-6, discussed in the three previous sections, illustrate the effects
of changing parameters which affect the line ratios, but not the covering function.
In the remaining sections of this chapter, we present models which illustrate
the effects of varying not only the line ratios, but also the covering function,
which is in turn determined by the wind area and stellar distribution functions.
Unlike the results in \S \S~\ref{iontopratio}-\ref{s}, the results in the
remaining sections of this chapter probably have significance only if the stellar
wind line emission model is correct.

Model 7 is a variant of model 1. Its bolometric luminosity is $  10^{45}  $
ergs s$  ^{-1}  $ instead of the $  10^{44}  $ ergs s$  ^{-1}  $ of model
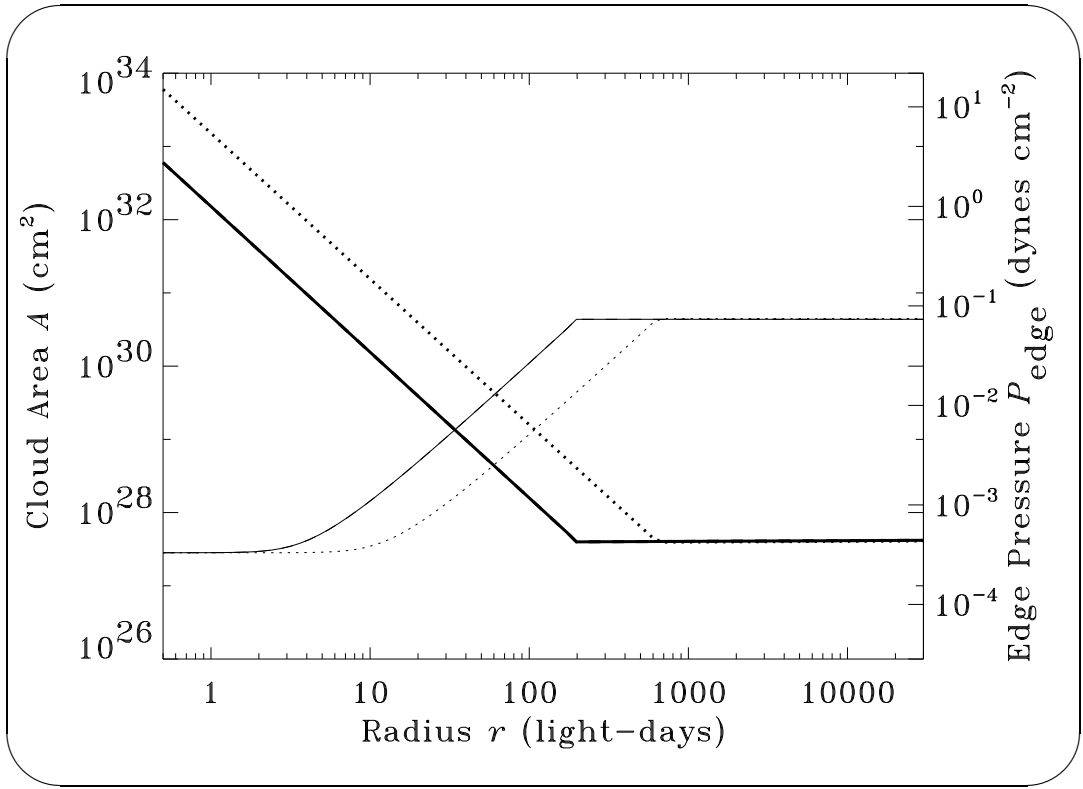
1. The wind area function of model 7 is shown in \Fig{model7/area}{Thin lines,
left axis: wind area functions. Thick lines, right axis: wind edge pressures.
Dotted lines: model 7. Solid lines: model 1.}. Note that the effective sizes
$  r_{\rm BLR}  $ of the line emission regions contributing towards the broader
profile components of models 1 and 7 are determined primarily by the wind area
functions rather than some other parameters of the covering functions, such
as the tidal radii. For models similar to models 1 and 7, the most important
effect of increasing the luminosity is that $  r_{12}  $ increases, with $  r_{12}\simeq r_{\rm BLR}\propto L^{1/2}  $
in particular, where $  r_{12}  $ again denotes the outermost position where
the winds begin to be fully ``stripped.'' 

This \emph{prediction} of model 1 is based upon the Kazanas (1989) paper. It
agrees with the first reverberation mapping campaign results for luminous quasars.
Kaspi et al. (1996), who published these results based on reverberation mapping
of 12 AGNs, conclude that they are indeed compatible with $  r_{\rm BLR}\propto L^{1/2}  $.
Any model which invokes evaporation of clouds that have internal pressure gradients
could yield this important result. However, with the notable exception of stellar
wind models, few such potential models could permit their clouds to draw upon
the required large reservoirs of replenishing gas. Therefore, the non-stellar
models are probably fundamentally incompatible with the evaporative mechanism
in the first place (\S~\ref{area}).

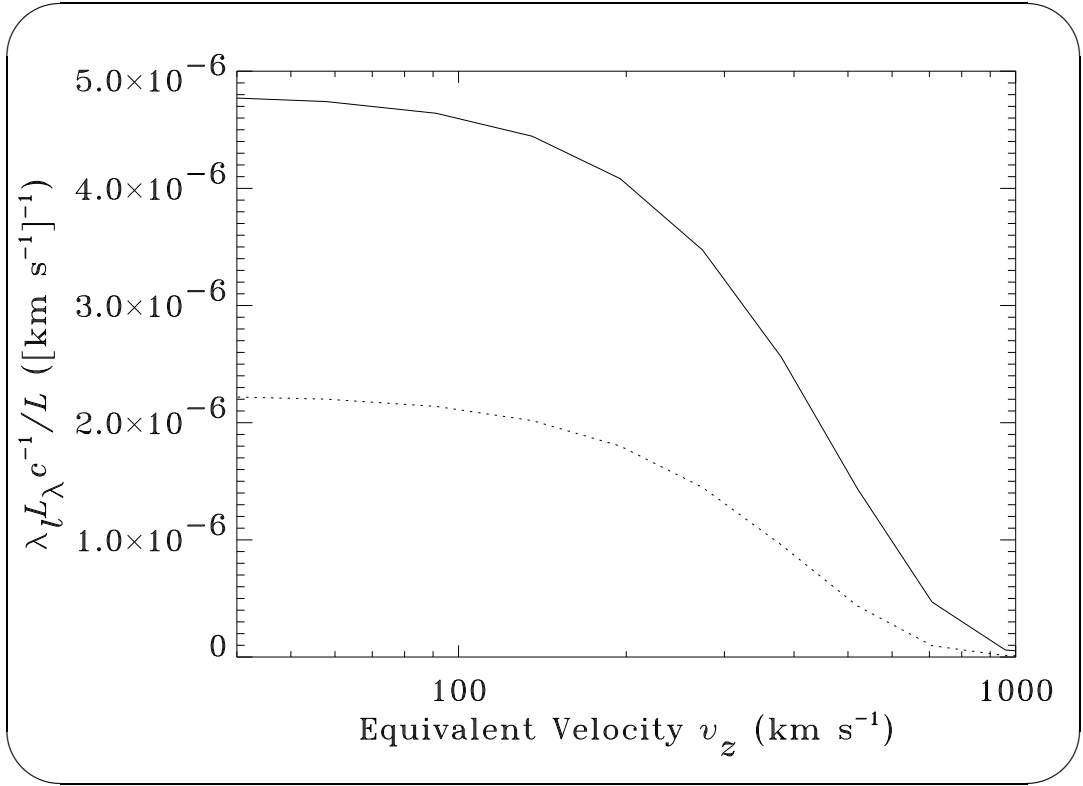
The line profile of Ly$  \alpha   $ for model 7 is shown in \Fig{model7/Lya}{Dotted
line: Half-profile of Ly$  \alpha   $ for model 7 ($  L=10^{45}  $ ergs s$  ^{-1}  $).
Solid line: half-profile of Ly$  \alpha   $ for model 2 ($  L=10^{44}  $ ergs
s$  ^{-1}  $).}. The FWHM for model 2 is 395 km s$  ^{-1}  $, while the FWHM
for model 7 is 330 km s$  ^{-1}  $. This narrowing is as expected, since the
higher fluxes strip/ablate the winds of the high-velocity stars nearest to the
black hole. As a result, the covering comes from farther out, and the broad
components that would contribute towards the profile are suppressed. Also, the
covering factor decreases. Incidentally, if one assumes that the black hole
dominates the stellar dynamics, one obtains for a hypothetical linear line the
relation $  v_{\rm FWHM}\propto L^{-1/4}  $ as the time-averaged luminosity
$  L  $ of an \emph{individual} AGN varies. This yields a much smaller FWHM
for model 7 of 222 km s$  ^{-1}  $. However, this relation is inapplicable
here because the dominant covering function for both of these models is from
narrow line clouds residing far enough away that the cluster is likely to dominate
the stellar dynamics.

Model 2 violates the theoretical $  r_{\rm BLR}\propto L^{1/2}  $ relationship
much more strongly than model 1. This is because the covering function of model
2 peaks at the tidal radius, regardless of the small variations in the wind
area function. This is illustrated by models 8 and 9, which are identical to
model 2 with the exception of their bolometric continuum luminosities of $  3.16\times 10^{44}  $
and $  3.16\times 10^{45}  $ ergs s$  ^{-1}  $ respectively. These luminosities
are a factor of 10 and 100 times higher than that of model 2. Unlike the models
shown in \S~\ref{L&P}, no other wind parameters were changed. As a result,
models 8 and 9 have different area and covering functions. The wind area and
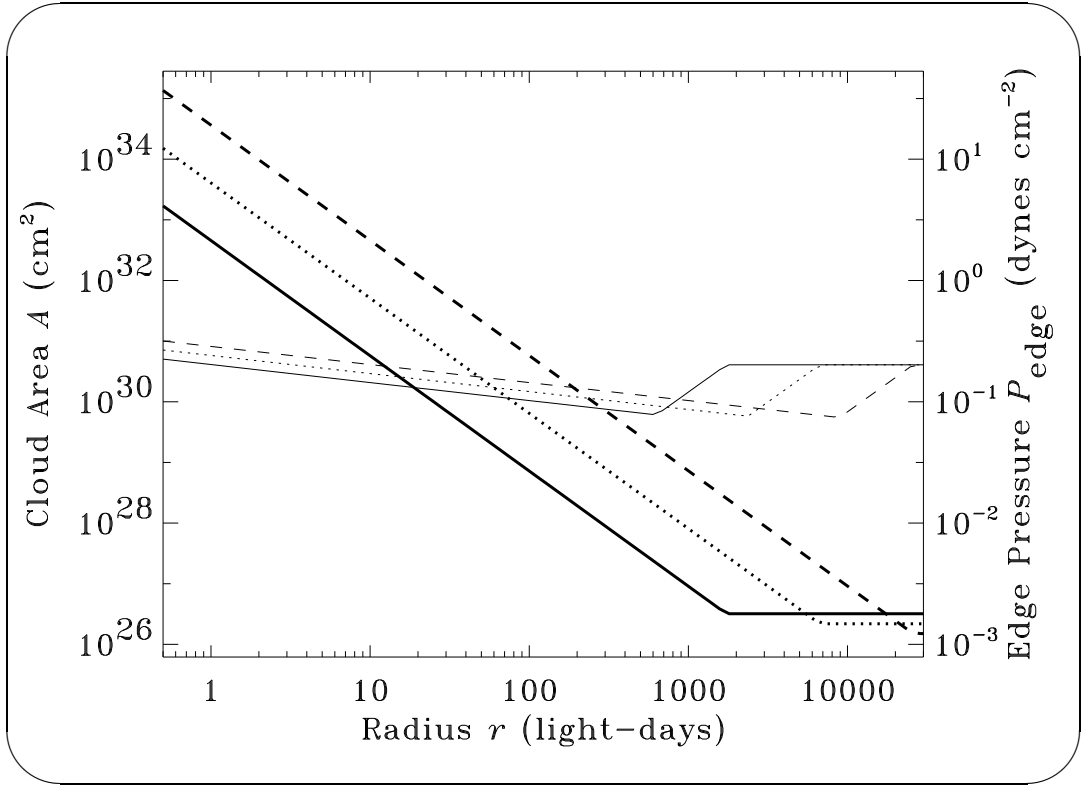
pressure functions of models 8 and 9 are shown in \Fig{model8/area}{Thin lines,
left axis: wind area functions. Thick lines, right axis: wind edge pressures.
Dotted lines: model 8 ($  L=3.16\times 10^{44}  $). Dashed lines: model 9 ($  L=3.16\times 10^{45}  $
ergs s$  ^{-1}  $). Solid lines: model 2 ($  L=3.16\times 10^{43}  $ ergs
s$  ^{-1}  $).}. The wind edge pressures of the models scale according to $  P\propto (L/r^{2})^{s/2}  $,
where $  s=1.9  $. As a result, nearly all of the additional flux is ``transferred''
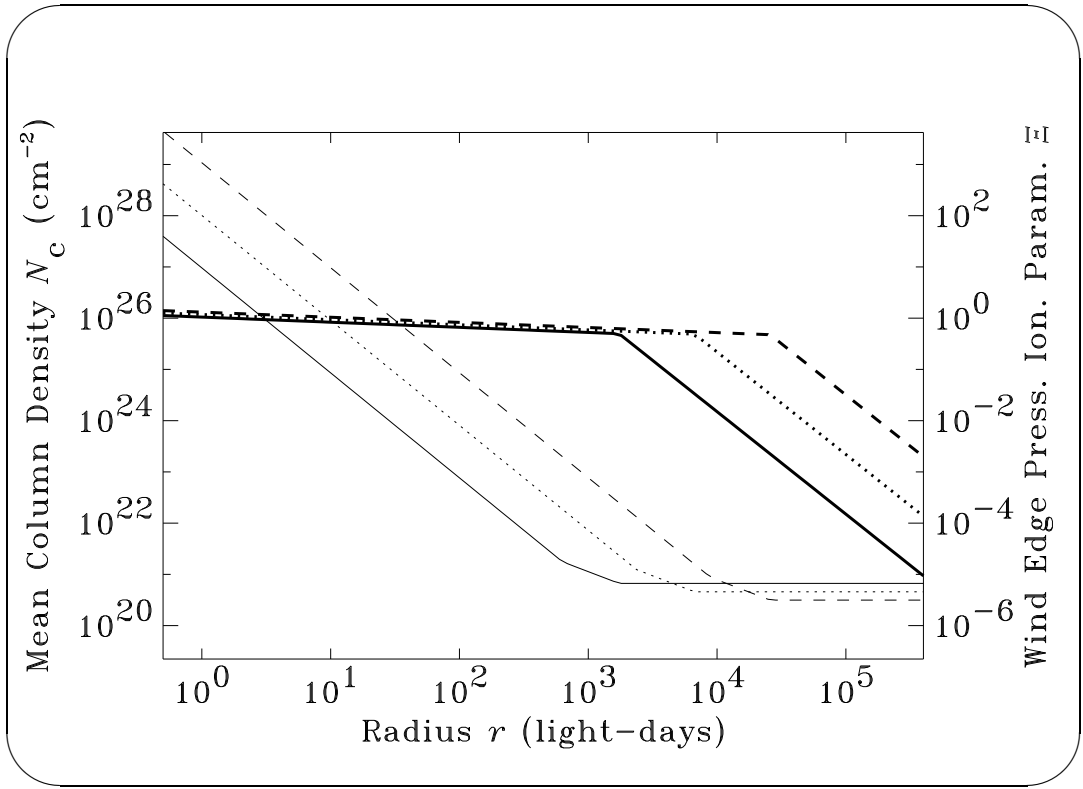
to the edge pressure rather than the edge ionization parameter shown in \Fig{model8/N_c+Xi}{Thin
lines, left axis: column density $  N_{\rm c}  $. Thick lines, right axis:
wind edge ionization parameter $  \Xi   $. Dotted lines: model 8. Dashed lines:
model 9. Solid lines: model 2.}. 

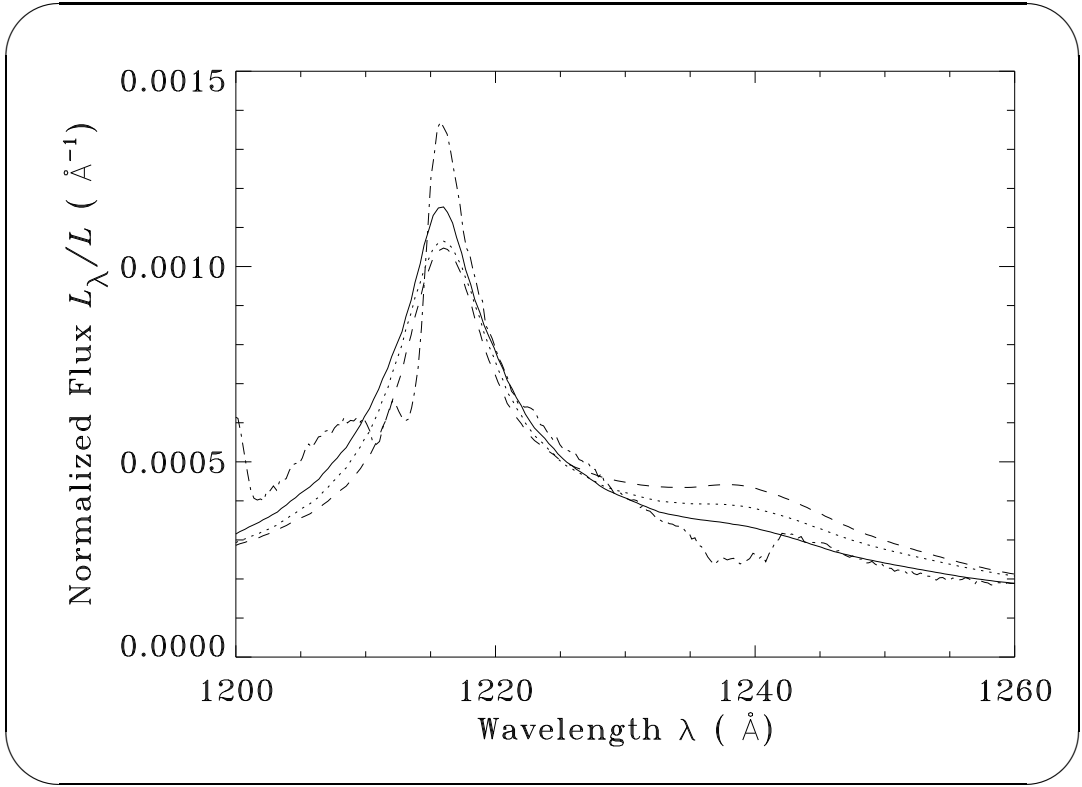
The Ly$  \alpha   $ profiles of models 8 and 9 are shown in \Fig{model8/Lya}{Ly$  \alpha   $
line profiles. Dotted line: model 8 ($  L=3.16\times 10^{44}  $). Dashed line:
model 9 ($  L=3.16\times 10^{45}  $ ergs s$  ^{-1}  $). Solid line: model
2 ($  L=3.16\times 10^{43}  $ ergs s$  ^{-1}  $). Dot-dashed line: NGC 5548.
Note that, for the reasons discussed in the text, results for models 9 are not
necessarily accurate.}. The profile strengths and shapes are similar to the
ones shown in \S~\ref{L&P} in that higher pressures result in less Ly$  \alpha   $
emission. However, the higher luminosity models have slightly enhanced emission
in many of the lines such as O {\footnotesize VI} and N {\footnotesize V}. This
is due to the fact these lines have much smaller optical depths and that $  \alpha =1.1>1  $
for these models; higher luminosities correspond to a larger summed line flux.

The electron density at the wind edges for clouds in model 9 at the tidal radius
is $  2.4\times 10^{12}  $ cm$  ^{-3}  $. This density is near the $  10^{12-13}  $
cm$  ^{-3}  $ limit beyond which XSTAR version 1.20 results may be invalid.\rfn{Fortunately,
recent versions of XSTAR account for certain processes that become important
at high pressures (e.g., three-body recombination), so it would be straightforward
to more accurately explore this important region of parameter space.}

The effect of varying $  L  $ upon the response function of the artificial
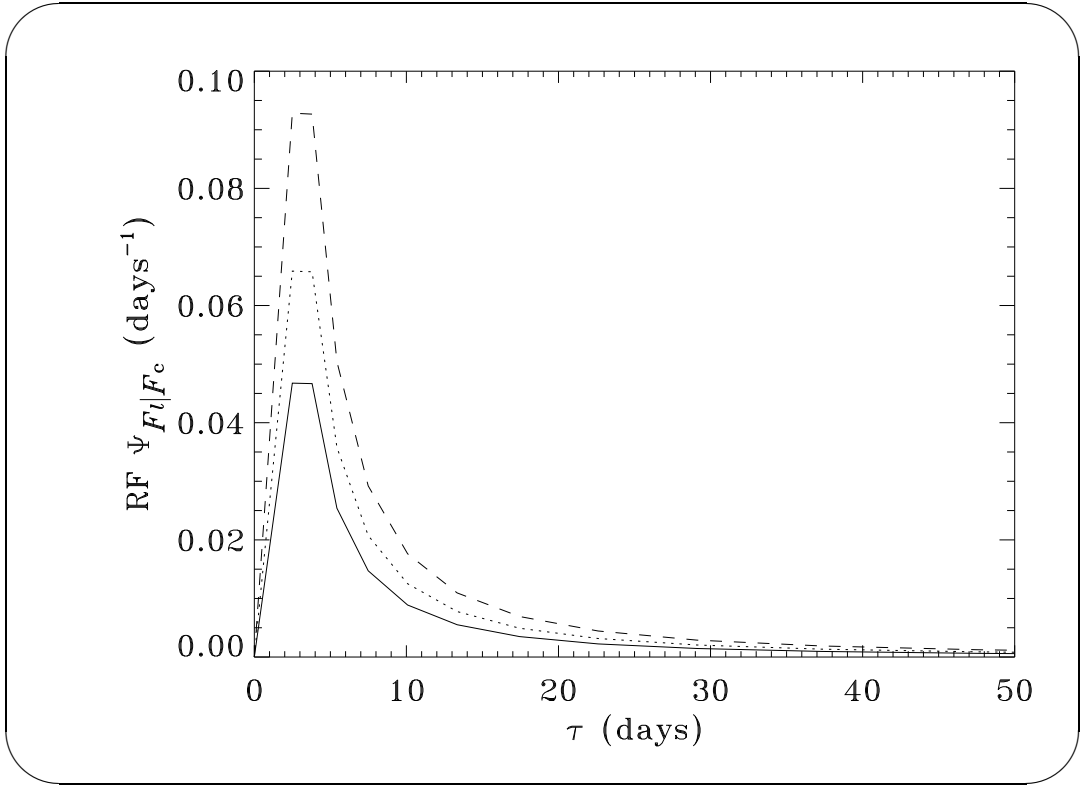
summed line is shown in \Fig{model8/summedlinerf}{The summed line response function.
Dotted line: model 8 ($  L=3.16\times 10^{44}  $). Dashed line: model 9 ($  L=3.16\times 10^{45}  $
ergs s$  ^{-1}  $). Solid line: model 2 ($  L=3.16\times 10^{43}  $ ergs s$  ^{-1}  $).}.
Because $  r_{12}\simeq r_{\rm t}\simeq r_{\rm BLR}  $, the luminosity alone
has little effect upon $  r_{\rm BLR}  $. However, since $  \dot{M}_{\rm h}  $
is likely to be a function of $  M_{\rm h}  $ (among other parameters), these
models do not necessarily indicate the measurable variation of the response
functions as a function of $  L  $ for models similar to model 2. Rather, they
merely indicate the intrinsic asymptotic changes for one particular AGN similar
to model 2 as its luminosity changes with time. They indicate, for example,
the linearized response functions that would be obtained from an AGN that switches
from a high state to a low state in a time that is much longer than $  r_{\rm BLR}/c  $.
(Incidentally, such a change from a high to low state could itself permit measurement
of the response functions on time scales much longer than $  r_{\rm BLR}/c  $.)
The most important feature of these response functions is that they all peak
at $  2r_{\rm t}/c  $ regardless of the luminosity. This implies, for profiles
of hypothetical lines that respond linearly, $  v_{\rm FWHM}\propto L^{0}  $
in AGNs with a fixed black hole mass.

The (nonlinear) effects of varying $  L  $ upon the Ly$  \alpha   $ response
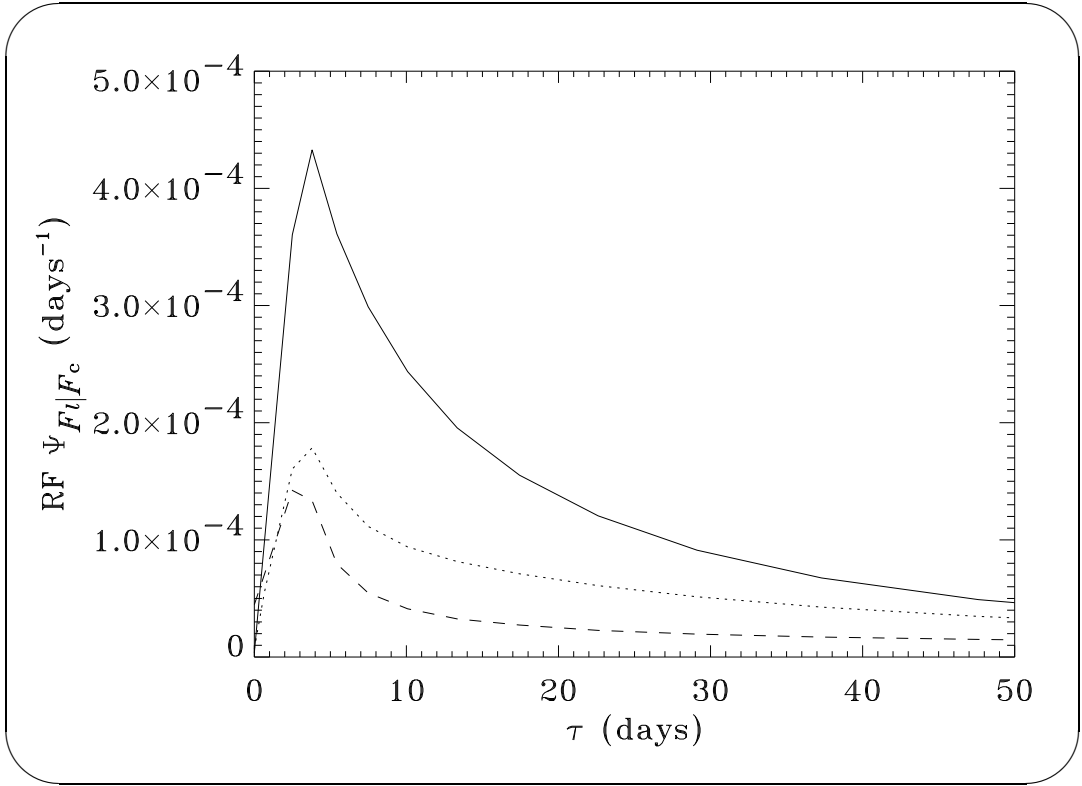
functions are shown in \Fig{model9/Lyarf}{Response functions of Ly$  \alpha   $.
Dotted line: model 8 ($  L=3.16\times 10^{44}  $). Dashed line: model 9 ($  L=3.16\times 10^{45}  $
ergs s$  ^{-1}  $). Solid line: model 2 ($  L=3.16\times 10^{43}  $ ergs s$  ^{-1}  $).
For the reasons discussed in the text, results for models 9 are not necessarily
accurate.}. Thermalization makes the line efficiency go down at higher luminosities.
This makes many of the strong lines, especially Ly$  \alpha   $, get weaker
even though the summed line gets stronger.

Another way of viewing the above results is through the behavior of $  -(2\alpha -s+3-\eta )  $,
the covering slope in the ``optically thick'' region. If this slope is steep
(negative) enough (e.g., $  \alpha \simeq 1  $ and $  3-\eta \simeq 2  $),
$  r_{\rm BLR}  $ is simply a function of the radius of the innermost BLR emission
of radius, which is $  r_{\rm t}  $ for these models. On the other hand, if
the slope is shallow (not as negative) enough, the outermost emission radius
becomes more important than the innermost emission radius and parameters such
as $  N_{\rm c}  $ and $  r_{\rm c}  $ would become important. Finally, if
the slope changes sharply with radius but is initially shallow because, e.g.,
$  \alpha \simeq 0  $ (as in model 1), then the BLR response depends upon the
wind parameters of which $  r_{12}  $ is a function and one can obtain $  r_{\rm BLR}\propto L^{1/2}  $.

\section{Models 10 and 11: Varying the Mass Loss Slope \protect$  \alpha \protect  $\label{alpha}}

Models 10 and 11 show the effects of varying mass loss slope parameter $  \alpha   $.
In model 10, $  \alpha =0.95  $. In model 11, $  \alpha =1.35  $. As equation
(\ref{coveringslope}) shows, the slope of the radially dependent covering function
$  An_{*}  $ includes a term of $  -2\alpha   $. This is because $  \alpha   $
affects the mass loss rates of the wind via $  \dot{M}\propto (L/r^{2})^{\alpha }  $.
For model 2, equation (\ref{coveringslope}) yields a covering slope in the
optically thick region of $  \simeq -1.8  $. Increasing $  \alpha   $ steepens
the covering functions, makes the profiles broader and less concave, and increases
the gain. Conversely, decreasing $  \alpha   $ narrows the profiles, makes
them more concave, and lowers the gain. 

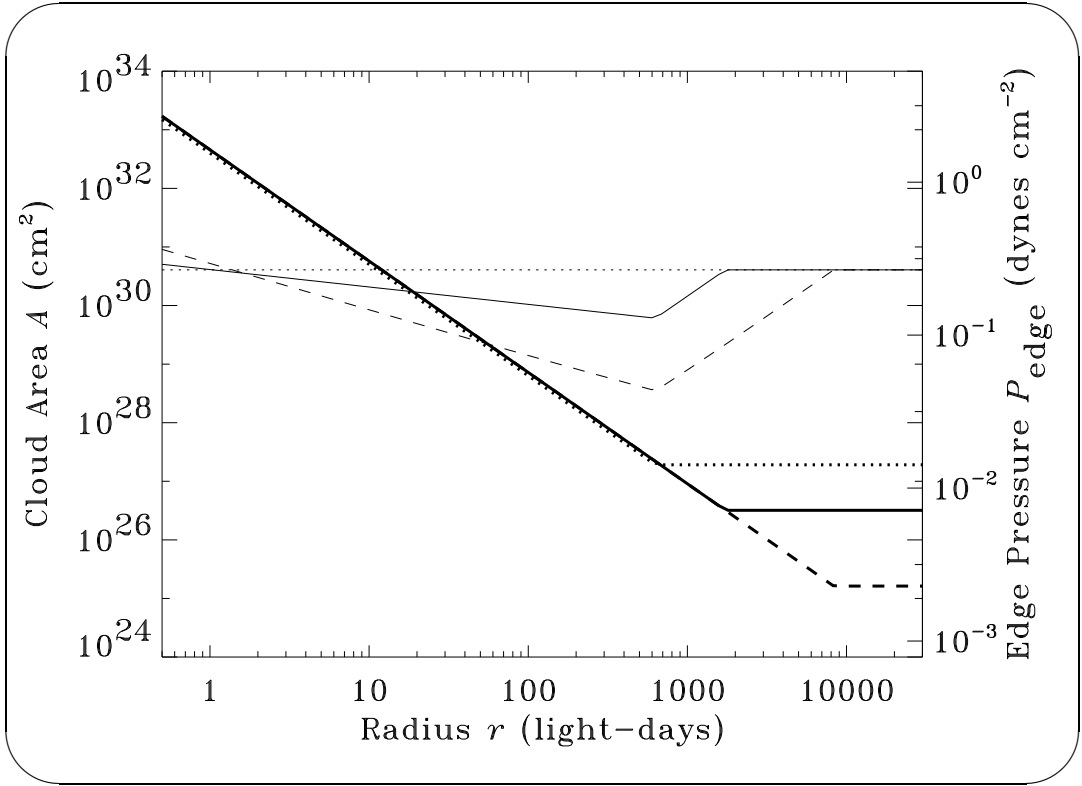
The wind area functions for models 10, 11, and 2 are shown in \Fig{model10/area}{Thin
lines, left axis: wind area functions for models 10, 11, and 2. Thick lines,
right axis: cloud edge pressures. Dotted lines: model 10 ($  \alpha =0.95  $).
Dashed lines: model 11 ($  \alpha =1.35  $). Solid lines: model 2 ($  \alpha =1.10  $).}. Scaled
profiles of the Ly$  \alpha   $ line for models 10, 11, and 2 are shown in
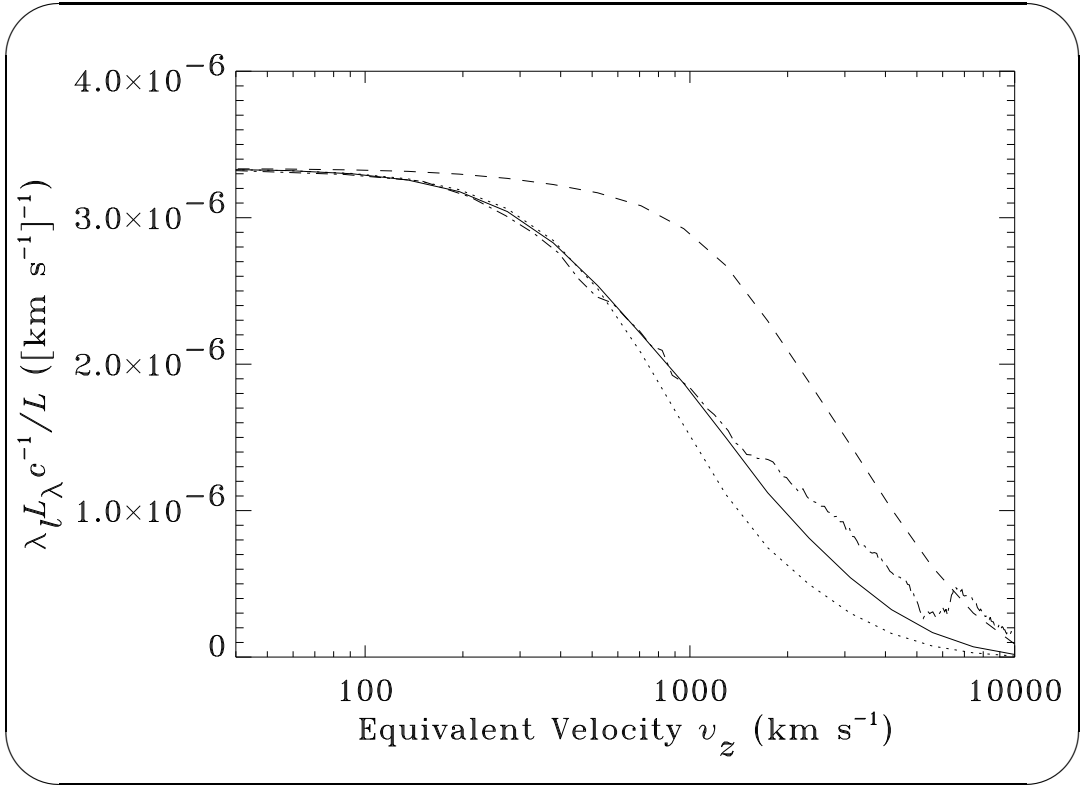
\Fig{model10/Lya}{Scaled Ly$  \alpha   $ line profiles. Dotted line: model
10 ($  \alpha =0.95  $), scaled to match the line peak of model 2. Dashed line:
model 11 ($  \alpha =1.35  $), scaled to match the line peak of model 2. Solid
line: model 2 ($  \alpha =1.10  $). Dot-dashed line: NGC 5548, scaled to match
line peak of model 2.}. As expected, the high-$  \alpha   $ model has the broadest
profile. Model 1 (in which $  \alpha =0  $) is an extreme case illustrating
how lowering $  \alpha   $ narrows the profiles (see Fig. \ref{model1/Lya}).
Unlike most of the cases discussed above, $  \alpha   $ has only a small effect
upon the BLR line ratios and, more precisely, the intensities as a function
of velocity dispersions. Thus, these changes in Ly$  \alpha   $ are representative
of the other lines. 

Nevertheless, because of the inherent radial dependences of the line emissivities
upon radius, $  \alpha   $ does indirectly affect several parameters including
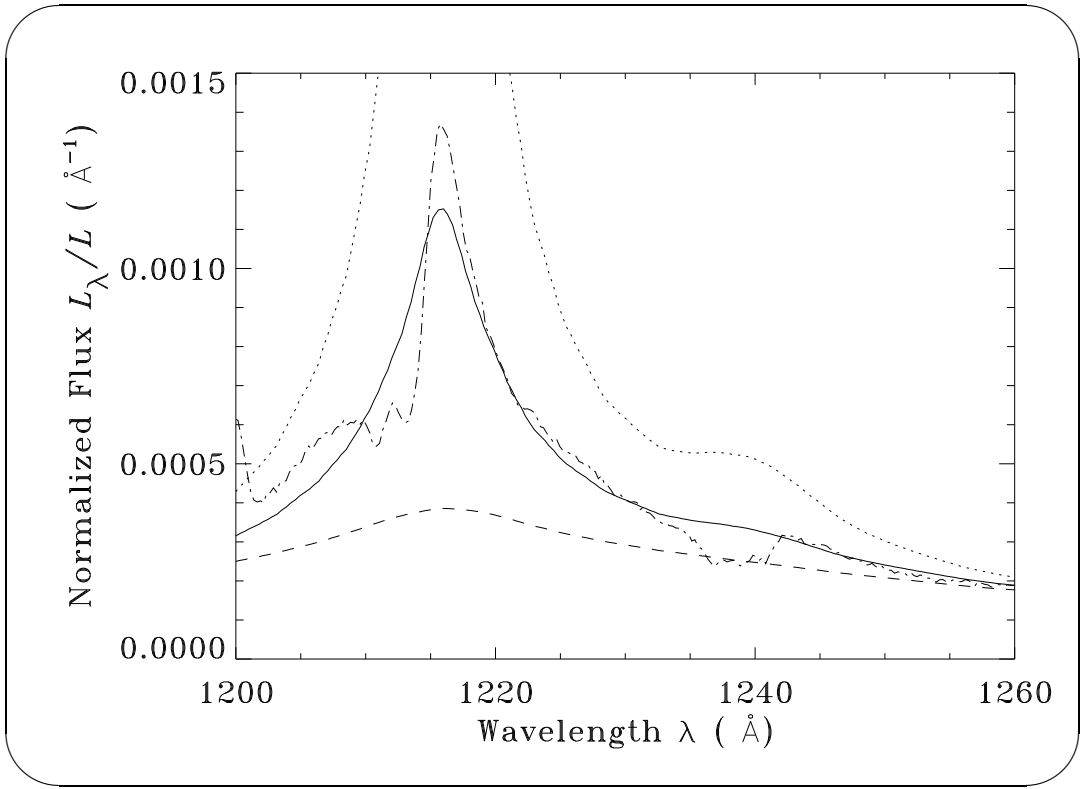
the line shapes and ratios. This is illustrated by \Fig{model10/Lyaspectrum}{Ly$  \alpha   $/N
{\footnotesize V} blend. Dotted line: model 10 ($  \alpha =0.95  $). Dashed
line: model 11 ($  \alpha =1.35  $). Solid line: model 2 ($  \alpha =1.10  $).
Dot-dashed line: NGC 5548. Note the relative narrowness of the N {\small V}
line of model 10.}. The profiles of the $  \alpha =0.95  $ models are so narrow
that the N {\footnotesize V} emission line appears similar in form to systems
with blueshifted absorption. No such absorption, however, was accounted for
in the model, and the effect is purely due to the narrowness of the $  \alpha =0.95  $
lines.

Figure \ref{model10/Lyaspectrum} also shows that raising $  \alpha   $ lowers
the covering factor. This is because the stellar wind mass loss functions of
models 2, 10, and 11 are normalized to be the same at $  r\simeq r_{\rm t}  $
(near the inner edge of the BLR), but model 11 has the steepest radial mass
loss gradient. Models 10 ($  \alpha =0.95  $) and 2 ($  \alpha =1.10  $) thus
have higher NLR coverings even though each of the models have similar very broad
line region coverings. 

These effects translate to the other observable quantities of the system, such
as the response functions. Response functions for the Ly$  \alpha   $ line
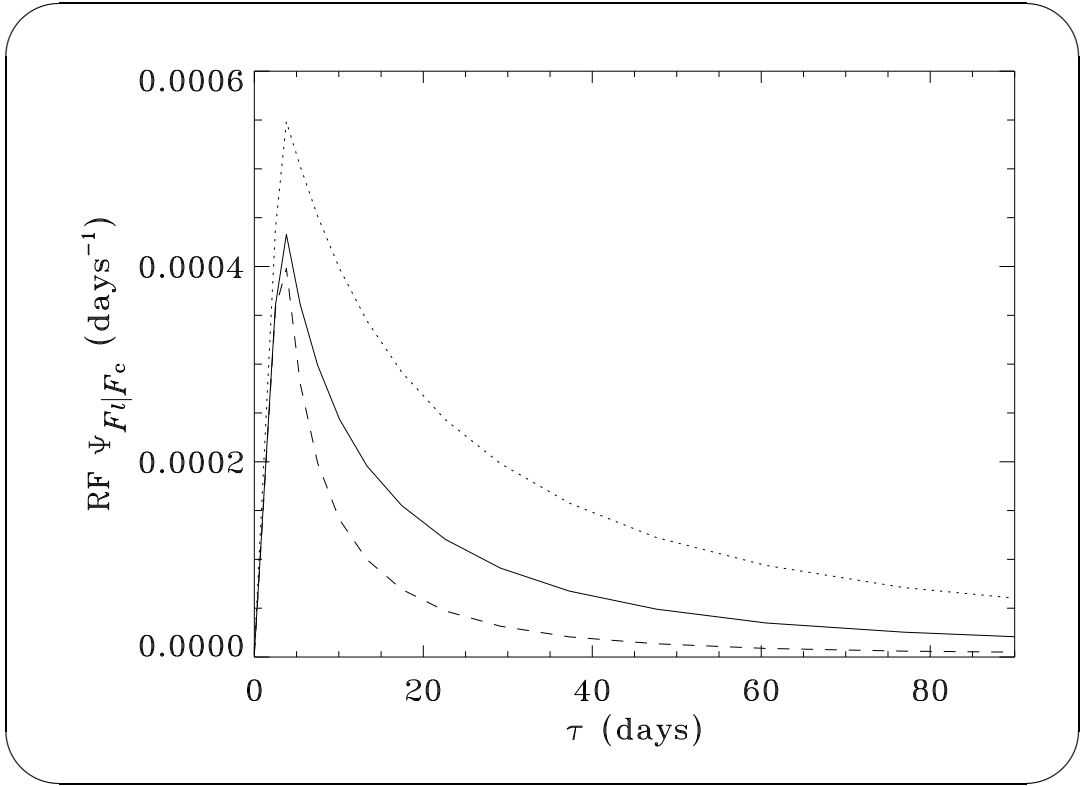
of models 10 and 11 are shown in \Fig{model10/Lyarf}{Response functions for
Ly$  \alpha   $. Dotted line: model 10 ($  \alpha =0.95  $). Dashed line:
model 11 ($  \alpha =1.35  $). Solid line: model 2 ($  \alpha =1.10  $).}.
Though the response function peak at $  \tau =2r_{\rm t}/c  $ for model 10
is only 1.4 times higher than that of model 11, it is 12 times higher at $  \tau =98  $
days. While the former increase is due primarily to the $  \alpha   $ term
in the expression of the gain (eq. {[}\ref{areagain}{]}) for clouds at $  r\sim r_{\rm t}  $,
the latter increase is due to the fact that $  \alpha   $ strongly affects
the ratio of the NLR covering to the BLR covering.

\section{Model 13: Varying the Terminal Wind Speed \protect$  v_{\infty }\protect  $
and the Mass Loss \protect$  \dot{M}\protect  $ \label{v_inf}}

The terminal wind speed $  v_{\infty }  $ is the maximum outward velocity attained
by the stellar winds that reprocess the continuum radiation. For models 1 and
2, we assume $  v_{\infty }=10  $ km s$  ^{-1}  $. As equation (\ref{windradius})
shows, changing $  v_{\infty }  $ would affect the cloud areas, with $  A\propto \dot{M}/v_{\infty }  $.
Thus, lowering $  v_{\infty }  $ or increasing $  \dot{M}  $ would increase
the covering factor, provided all other parameters remained the same. This is
discussed in more detail in \S~\ref{dmdt/vinf}. If both $  \dot{M}  $ and
$  v_{\infty }  $ were changed by the same factor, but $  N_{\rm {cmin}}  $
were unchanged, there would be only two major observable changes in the model: 

\begin{enumerate}
\item the cumulative mass lost by all of the stars, and 
\item the equivalent width of absorption features due to stellar winds. 
\end{enumerate}
In this section, we address the first of these two characteristics. Some aspects
of absorption are discussed in Appendix \ref{adpbal}.

In model 13, both $  v_{\infty }  $ and the mass loss function $  \dot{M}  $
were reduced to 6\% of the values assumed in model 2. The cumulative mass lost
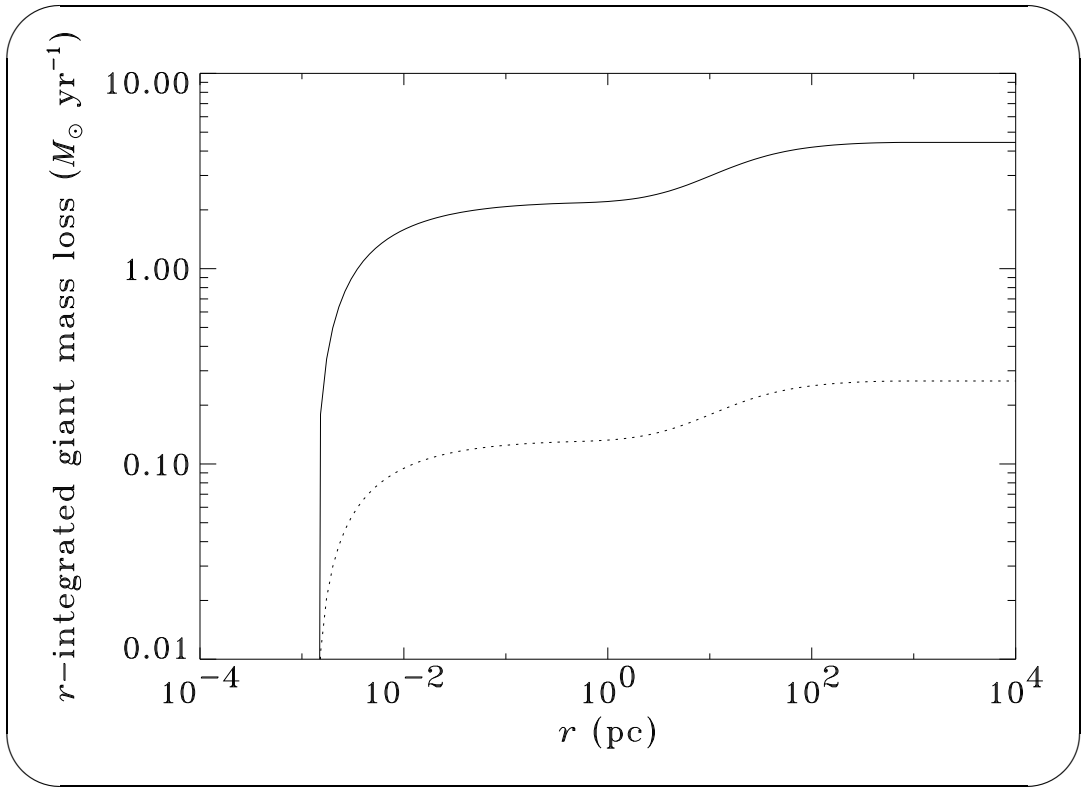
by all stars is shown in \Fig{model2/cummmass}{Total mass loss due to reprocessing
stellar winds inside distance from the black hole $  r  $. Dotted line: model
13 ($  v_{\infty }=0.6  $ km s$  ^{-1}  $). Solid line: model 2 ($  v_{\infty }=10  $
km s$  ^{-1}  $).}. If one assumes that all the gas lost by the stars is accreted,
lowering $  v_{\infty }  $ permits traditional accretion disk efficiencies
of $  \sim   $10\%.

It is possible, however, that only a small fraction of this mass would be accreted.
The hot plasma deposited into the intercloud medium by stellar winds could,
for instance, form a slow-moving outbound AGN superwind, accelerated in part
from the twisted magnetic fields of the accretion disk below.\qfn{check wording
here. do i want to say cooling flows here demos?} Alternatively, the hot plasma
deposited by winds could have a non-zero average angular momentum vector aligned
with the disk. In this case, the accretion disk must remove the angular momentum
by expanding before any additional material could be accreted onto the black
hole. The point is that the stellar wind mass loss rate is probably not equal
to the accretion rate onto the black hole (which is relatively easy to measure).

At any rate, $  v_{\infty }  $ is easily measurable in giants and supergiants.
It could therefore be argued that it is one of the least uncertain parameters
in the wind AGN line emission model. Common measurements yield $  v_{\infty }\simeq 10  $
km s$  ^{-1}  $. AN97 assumed $  \textrm{max}(v)\sim 0.1  $ km s$  ^{-1}  $
for some of their models. This maximal velocity is a factor of 100 less than
the observed value. It is a factor of $  10^{3}  $ less than the $  100  $
km s$  ^{-1}  $ gravitational escape velocities typically observed in giants.
Note that thermally driven winds with $  v_{\infty }\ll (2GM_{*}/R_{*})^{1/2}\sim 100  $
km s$  ^{-1}  $ should actually \emph{accrete} onto their stars. Though red
giant winds are not suspected as being thermally driven, we believe that very
slow winds are probably not plausible. 

Incidentally, very slow winds like those considered by AN97 would probably be
more susceptible to pressure from the intercloud medium. This is discussed in
detail in Appendix \ref{cloudshape}. In this context, it has been proposed that
the stars are not producing winds at all, but rather act as ``condensation
sites'' of a super-heated intercloud medium. Such models are not, however,
considered in this work.

\section{Models 16 and 17: Varying the Mass Loss to Terminal Wind Speed ratio \protect$  \dot{M}/v_{\infty }\protect  $
\label{dmdt/vinf}}

Models 16 and 17 are identical to model 2 except for their mass loss rates,
which are, respectively, 3.0 times lower and 3.0 higher than that of model 2.
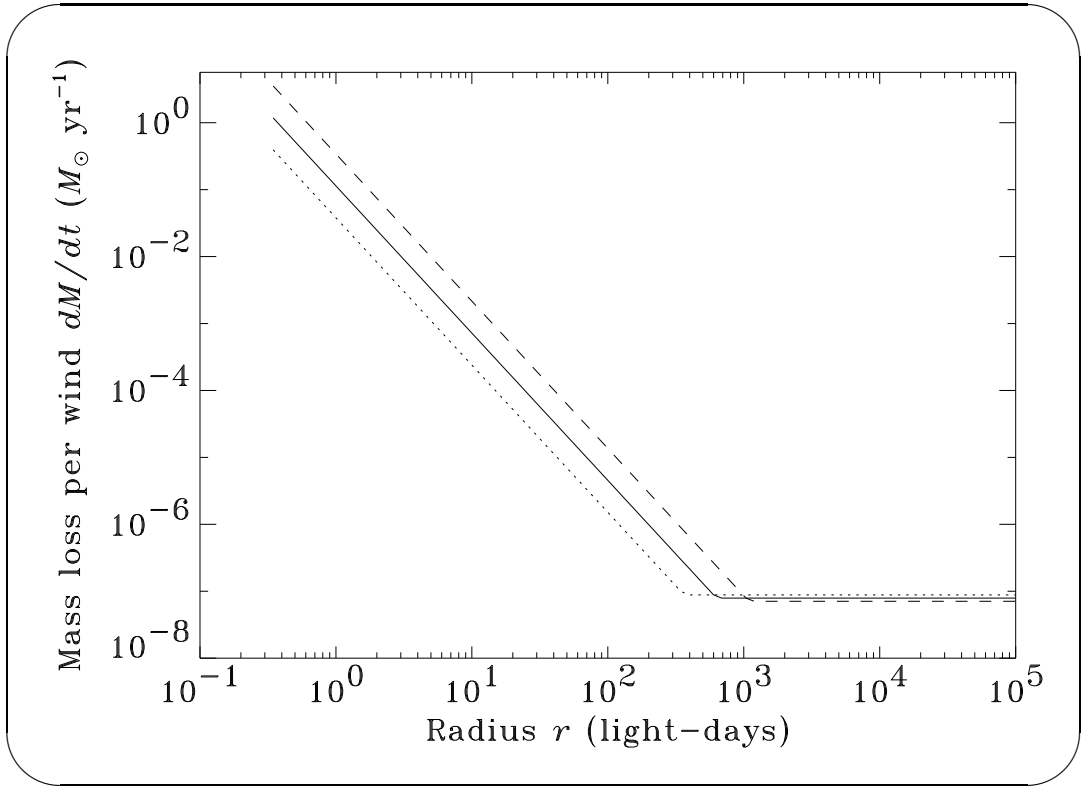
This is shown in \Fig{model16/massloss}{Mass loss rates per reprocessing wind.
Dotted line: model 16. Dashed line: model 17. Solid line: model 2.}. Since the
wind areas are proportional to $  \dot{M}/v_{\infty }  $, the lowest mass loss
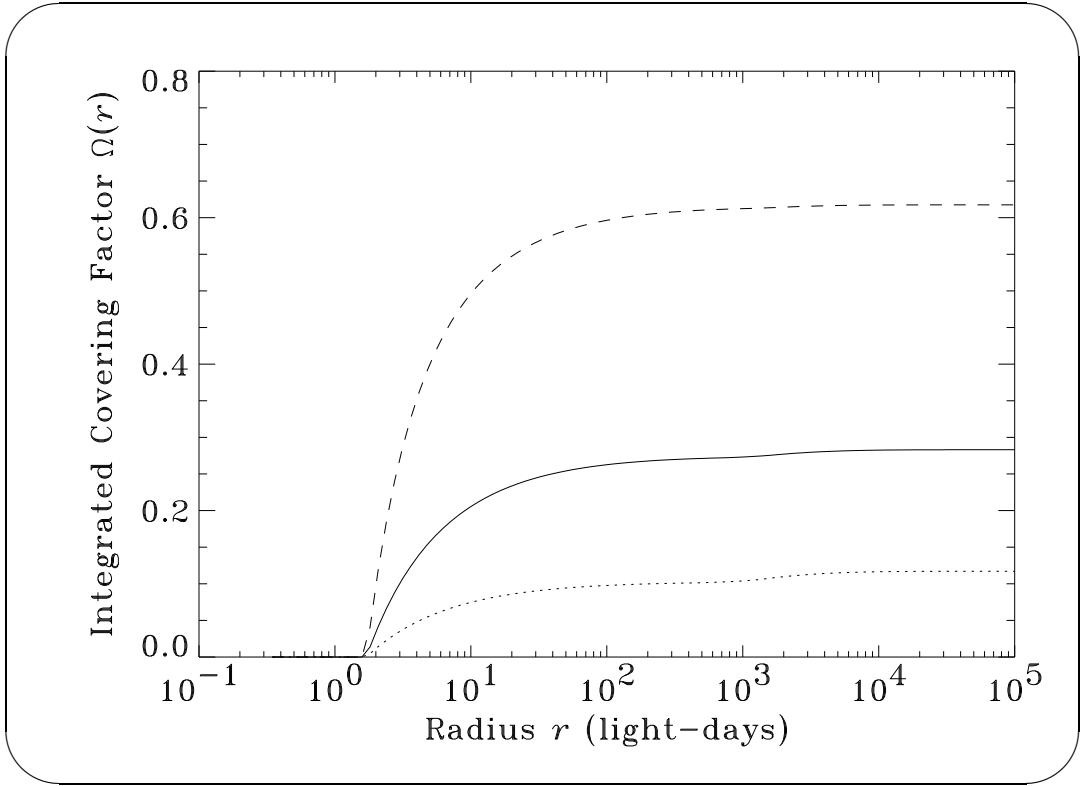
model (model 16) also has the least BLR covering. This is shown in \Fig{model16/covering}{Radial
covering functions of models 16, 17, and 2. Dotted line: model 16. Dashed line:
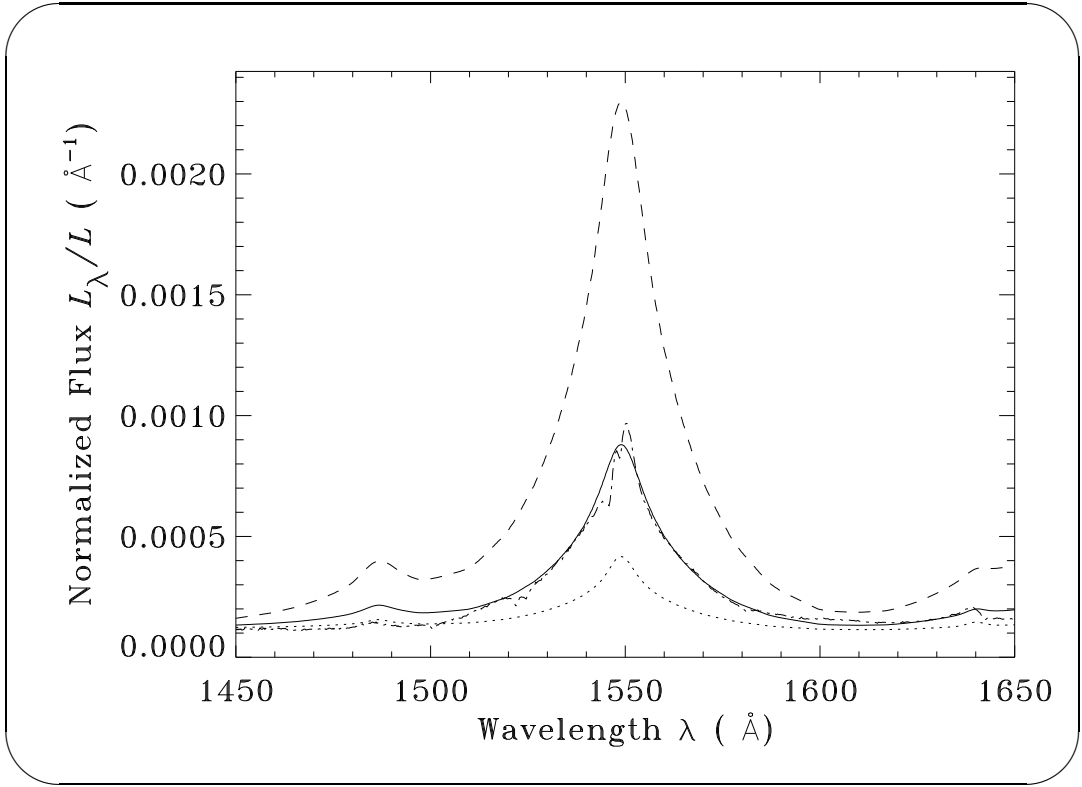
model 17. Solid line: model 2.}. \Fig{model16/civspectrum}{The spectral regions
near the C {\footnotesize IV} lines. Dotted line: model 16. Dashed line: model
15. Solid line: model 2. Dot-dashed line: NGC 5548.} shows the effect that varying
$  \dot{M}/v_{\infty }  $ has upon the line strengths. The response functions
scale with $  \dot{M}/v_{\infty }  $ in the same way. 

Since $  \dot{M}_{\rm min}  $ was not changed in these three models, the position
closest to the continuum source where $  \dot{M}=\dot{M}_{\rm min}  $ first
occurs is nearest to the black hole for model 16 and farthest from the black
hole in model 17. For this reason, the lowest mass loss model (model 16) has
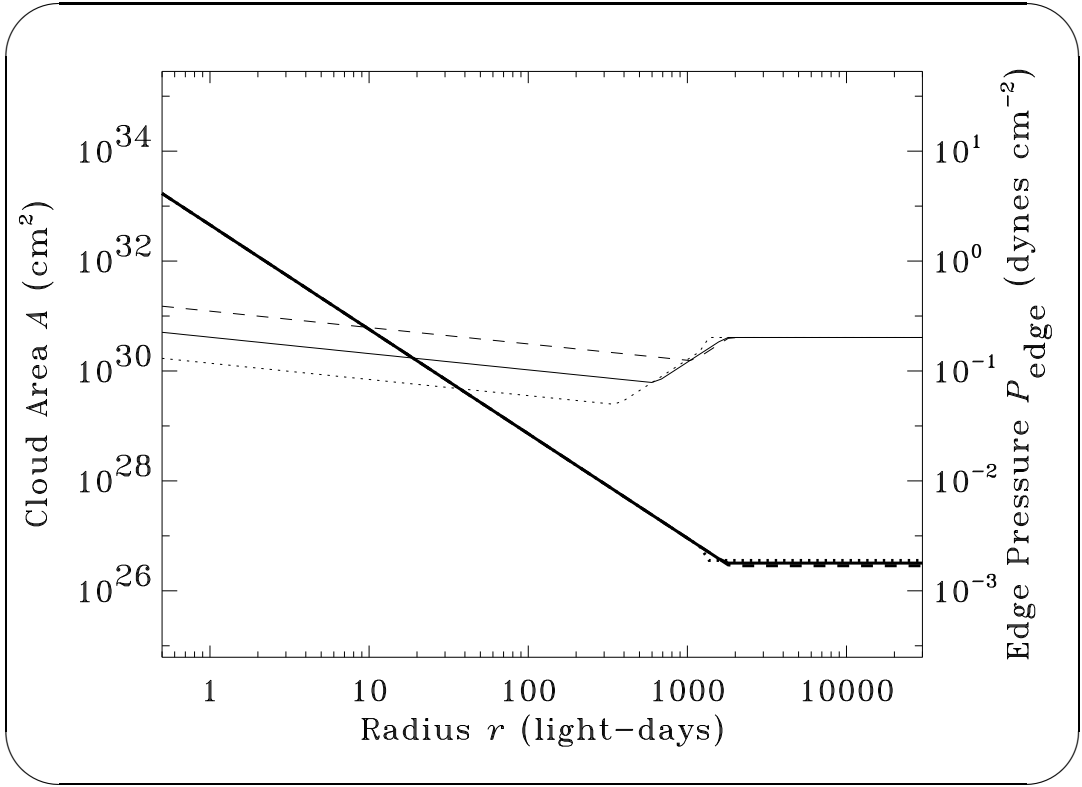
the highest NLR/BLR covering ratio, as Figures \ref{model16/covering} and \ref{model16/area}\fig{model16/area}{Thin
lines, left axis: wind area functions. Thick lines, right axis: wind edge pressures.
Dotted lines: model 16. Dashed lines: model 15. Solid lines: model 2.} show.
On the other hand, had $  \dot{M}_{\rm min}  $ also been changed by the same
factor that $  \dot{M}  $ had been changed, the scaled covering functions would
have been unaffected. The effect upon the Ly$  \alpha   $ profile shape of
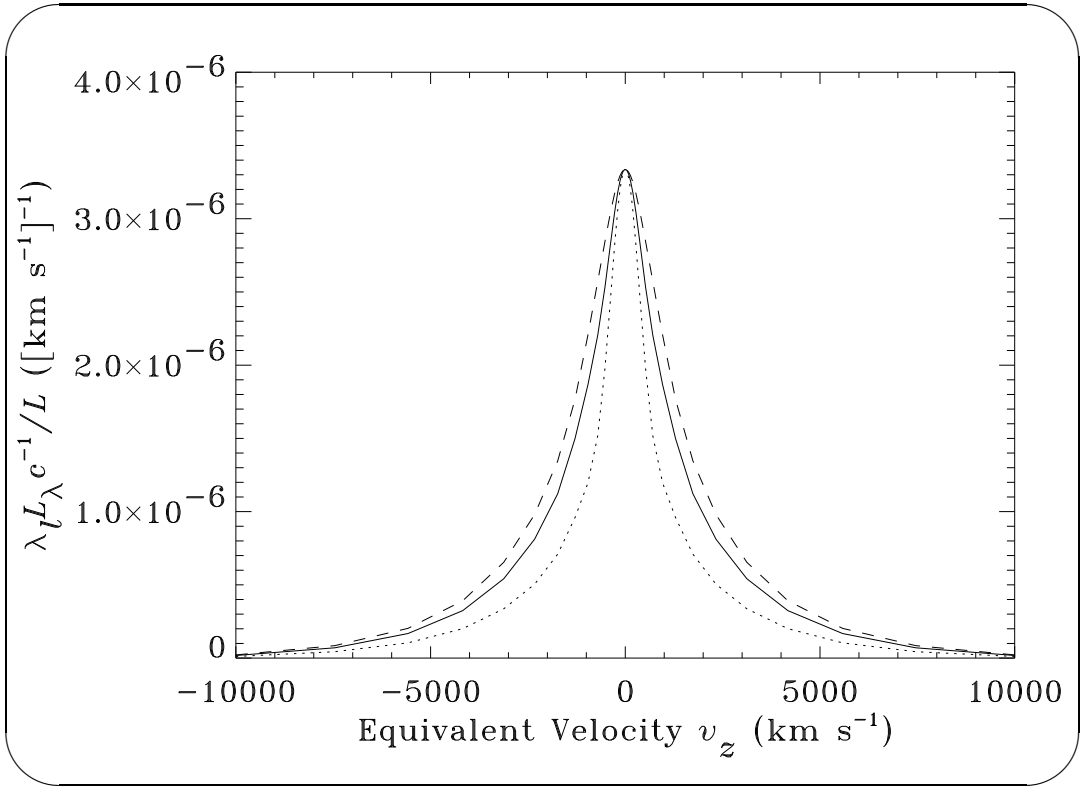
varying $  \dot{M}  $ is shown in \Fig{model16/Lya}{Ly$  \alpha   $ profile
shapes. Dotted line: model 16, scaled to the peak of the model 2 profile. Dashed
line: model 17, scaled to the peak of the model 2 profile. Solid line: model
2.}. As expected, model 16, which has the lowest $  \dot{M}  $, has the narrowest
Ly$  \alpha   $ profile.

Incidentally, the C {\footnotesize IV} profile shapes of these three models
are nearly identical. This is because the emissivity of C {\footnotesize IV}
is so low in the NLR.

\section{Models 14 and 15: Varying the Minimum Column Density \protect$  N_{\rm cmin}\protect  $\label{nc_min}}

\pfn{{[}need to say in other sections that these clouds are NOT optically thick
near r23!!! (done as far as I can tell) Also, must find out if emission of Ly$  \alpha   $
is caused by ionization parameter instead. ??? no idea what this refers to{]}
}As discussed previously, the minimum column densities of our models are denoted
by the parameter $  N_{\rm {cmin}}  $. This parameter serves as a reservoir
for storing our ignorance about possible disruption mechanisms that might affect
the more tenuous winds located far from the black hole.\pfn{{[}state in first
place where nc is talked about that we don't know if higher flux/vels would
make it worse far away though{]}- actually there is no need} Models 14 and 15
show the effect of varying $  N_{\rm {cmin}}  $. They are identical to model
2 except for their $  N_{\rm {cmin}}  $ parameters, which are $  N_{\rm {cmin}}=3\times 10^{21}  $
cm$  ^{-2}  $ for model 14 and $  N_{\rm {cmin}}=3\times 10^{22}  $ cm$  ^{-2}  $
for model 15. 

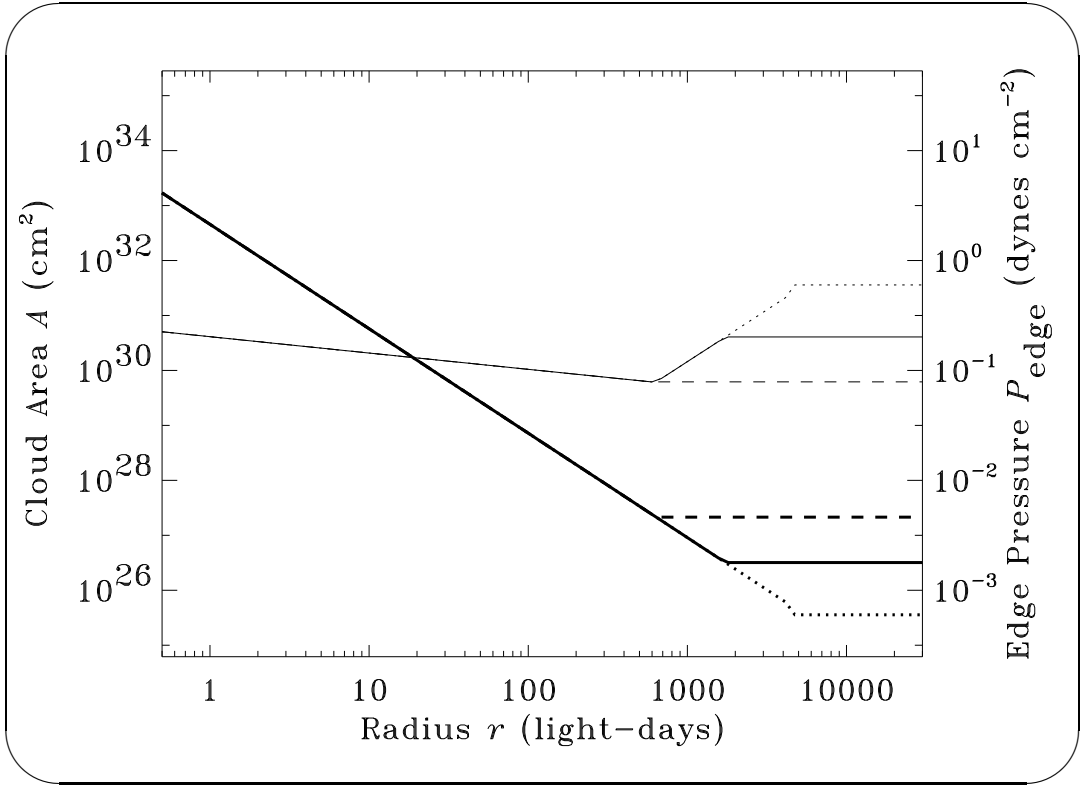
The area functions of these models are shown in \Fig{model14/area}{Thin lines,
left axis: wind area functions of models 14, 15, and 2. Thick lines, right axis:
wind edge pressures. Dotted lines: model 14 ($  N_{\rm {cmin}}=3\times 10^{21}  $
cm$  ^{-2}  $). Dashed lines: model 15 ($  N_{\rm {cmin}}=3\times 10^{22}  $
cm$  ^{-2}  $). Solid lines: model 2 ($  N_{\rm {cmin}}=8\times 10^{22}  $
cm$  ^{-2}  $). Because the edge pressures are fixed at the outermost emission
regions computed in the model (where $  N_{\rm {c}}=N_{\rm {cmin}}  $), the
edge pressures in the NLR are lowest for model 14. For the same reason, model
14 has the highest narrow line ionization parameter and NLR continuum optical
depths of these three models.}. Lowering $  N_{\rm {cmin}}  $ permits the winds
to follow their $  A\propto r^{2}  $ expansion phase out to regions farther
away from the black hole. (For the models considered here, $  N_{\rm {c}}=N_{\rm {cmin}}  $
occurs farther from the continuum source than does $  \dot{M}=\dot{M}_{\rm min}  $.) 

As a result, lowering $  N_{\rm {cmin}}  $ enhances the NLR covering. This
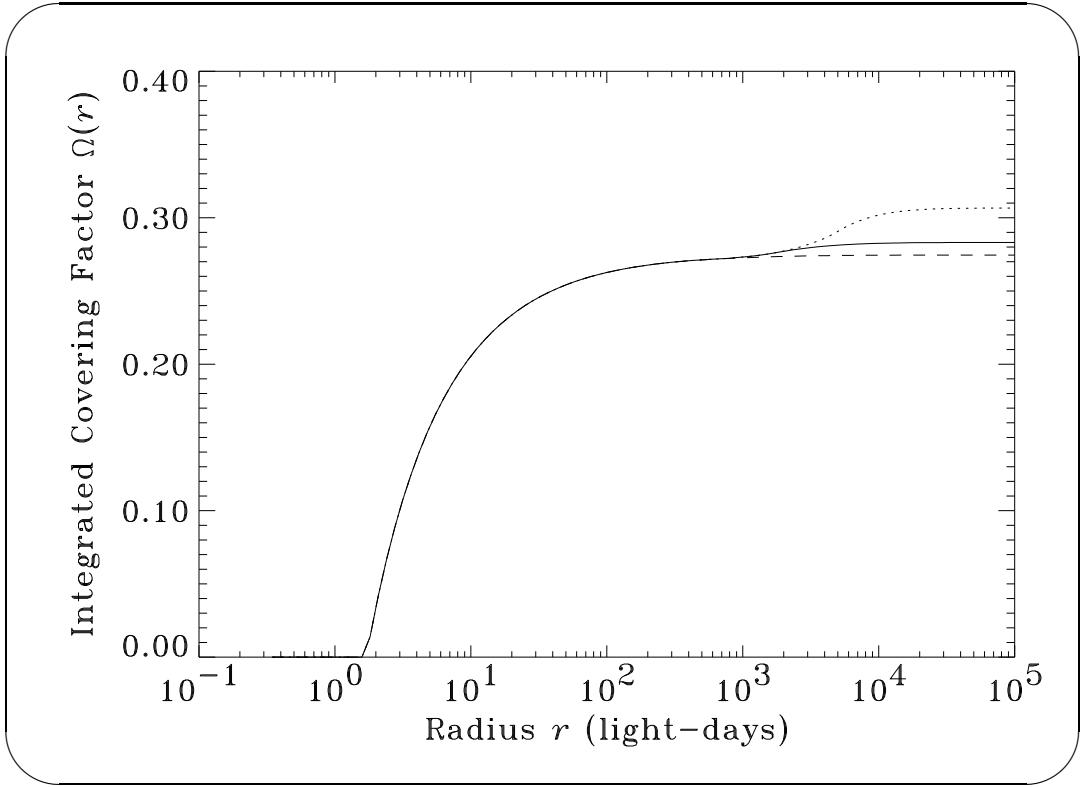
is shown in \Fig{model14/covering}{Integrated covering as a function of radius.
Dotted line: model 14 ($  N_{\rm {cmin}}=3\times 10^{21}  $ cm$  ^{-2}  $).
Dashed line: model 15 ($  N_{\rm {cmin}}=3\times 10^{22}  $ cm$  ^{-2}  $).
Solid line: model 2 ($  N_{\rm {cmin}}=8\times 10^{22}  $ cm$  ^{-2}  $).}.
The effects upon the profiles and other observables are as one would expect.
For instance, the Ly$  \alpha   $ profile for model 14 ($  N_{\rm {cmin}}=3\times 10^{21}  $
cm$  ^{-2}  $), which is shown in \Fig{model14/Lya}{Ly$  \alpha   $ profiles.
Dotted line: model 14 ($  N_{\rm {cmin}}=3\times 10^{21}  $ cm$  ^{-2}  $).
Dashed line: model 15 ($  N_{\rm {cmin}}=3\times 10^{22}  $ cm$  ^{-2}  $).
Solid line: model 2 ($  N_{\rm {cmin}}=8\times 10^{22}  $ cm$  ^{-2}  $).},
is narrower than that of model 15 ($  N_{\rm {cmin}}=3\times 10^{22}  $ cm$  ^{-2}  $).
Because the winds are optically thin near $  N_{\rm {c}}=N_{\rm {cmin}}  $,
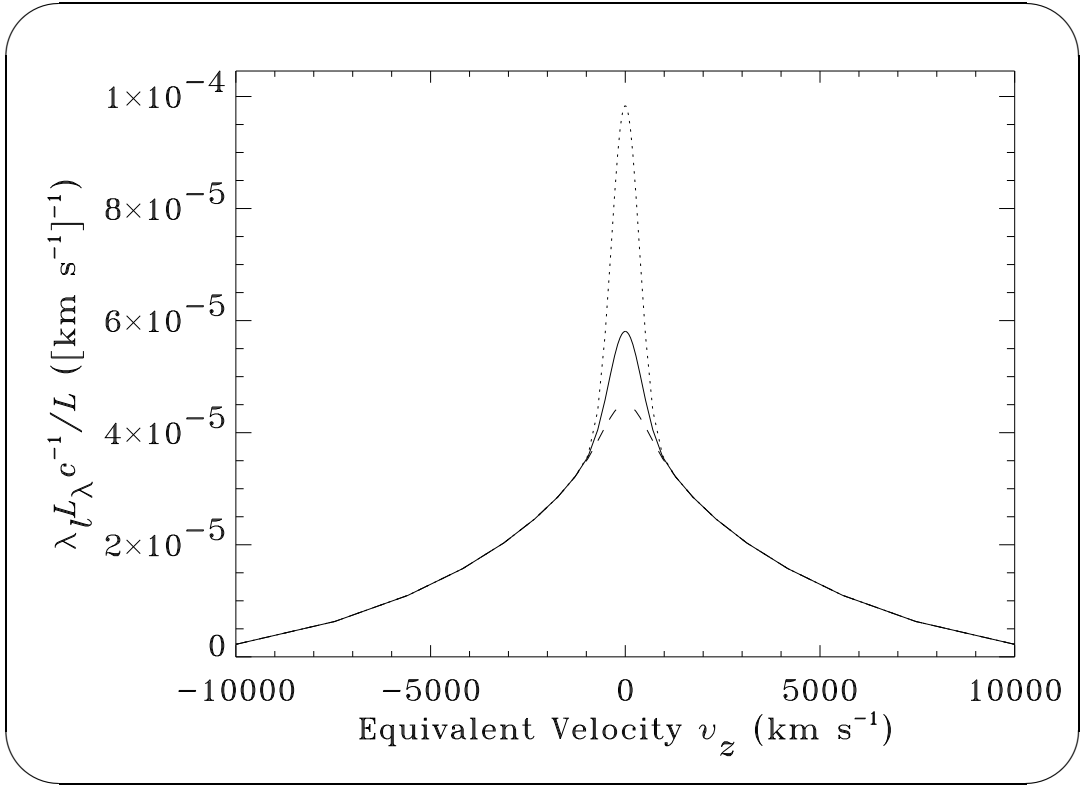
our approximation for the summed line profile (see \Fig{model14/Summedline}{Profiles
of the summed line using the assumption that all of the exposing continuum flux
is absorbed by the NLR winds, which is not the case. Dotted line: model 14 ($  N_{\rm {cmin}}=3\times 10^{21}  $
cm$  ^{-2}  $). Dashed line: model 15 ($  N_{\rm {cmin}}=3\times 10^{22}  $
cm$  ^{-2}  $). Solid line: model 2 ($  N_{\rm {cmin}}=8\times 10^{22}  $
cm$  ^{-2}  $). Note that the differences between these profiles are much greater
than those for the Ly$  \alpha   $ profiles (shown in Fig. \ref{model14/Lya}).})
is much more strongly affected by $  N_{\rm {cmin}}  $. 

For the volume of parameter space near model 2, the BLR covering is unaffected
by $  N_{\rm {cmin}}  $. Thus, the response functions at lags below $  \sim   $1000
days are not functions of $  N_{\rm {cmin}}  $. Lines which have negligible
emission in the NLR (e.g., C {\footnotesize IV}), are also unaffected by $  N_{\rm {cmin}}  $. 

However, this is not be the case for all possible models, especially those very
different from model 2. This is not only because $  N_{\rm {cmin}}  $ affects
the covering, but also because it affects the effective ionization of the outermost
clouds.

\section{Models 18 and 19: Varying the Minimum Mass Loss Rate \protect$  \dot{M}_{{\rm min}}\protect  $\label{mdot_min}}

\pfn{near perfect, not a perfect symmetry. This is because of relativistic effects.
Both general and special.}The mass loss functions of models 18 and 19 are shown
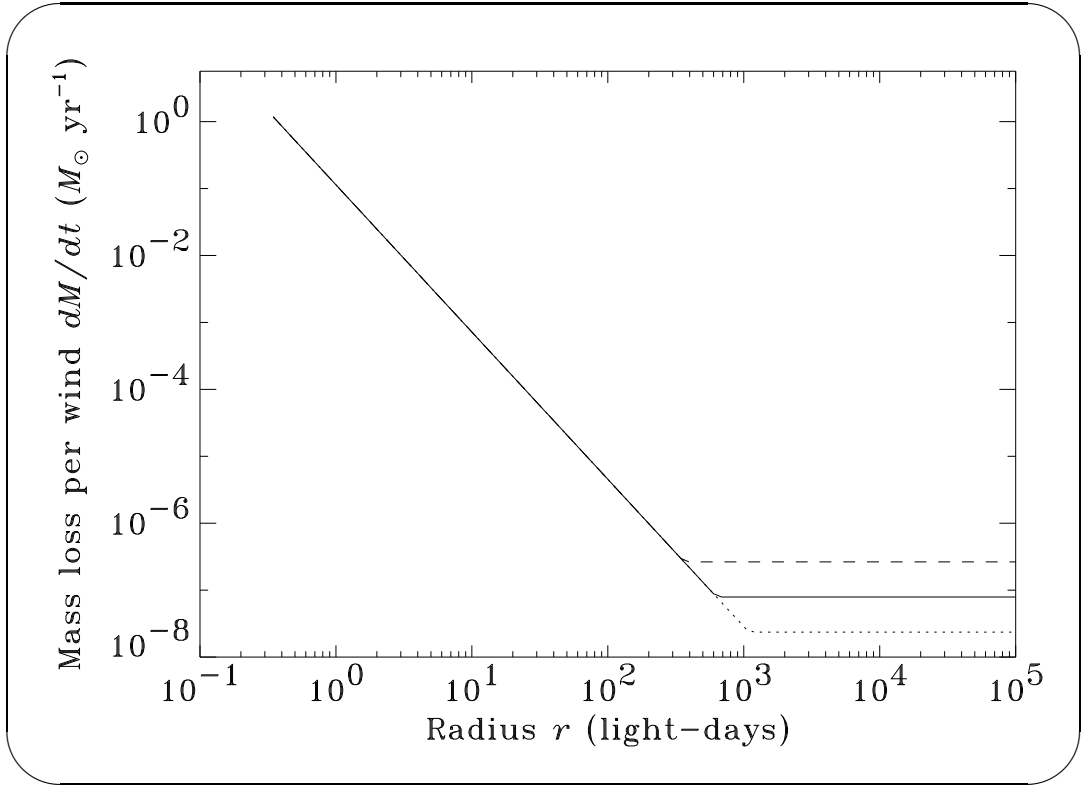
in \Fig{model18/massloss}{Mass loss rates for stellar winds assumed to be reprocessing
continuum radiation into line radiation. Dotted line: model 18 ($  \dot{M}_{{\rm min}}=2.4\times 10^{-8}\, \dot{M}_{\odot }  $
yr$  ^{-1}  $). Dashed line: model 19 ($  \dot{M}_{{\rm min}}=2.6\times 10^{-7}\, \dot{M}_{\odot }  $
yr$  ^{-1}  $). Solid line: model 2 ($  \dot{M}_{{\rm min}}=7.9\times 10^{-8}\, \dot{M}_{\odot }  $
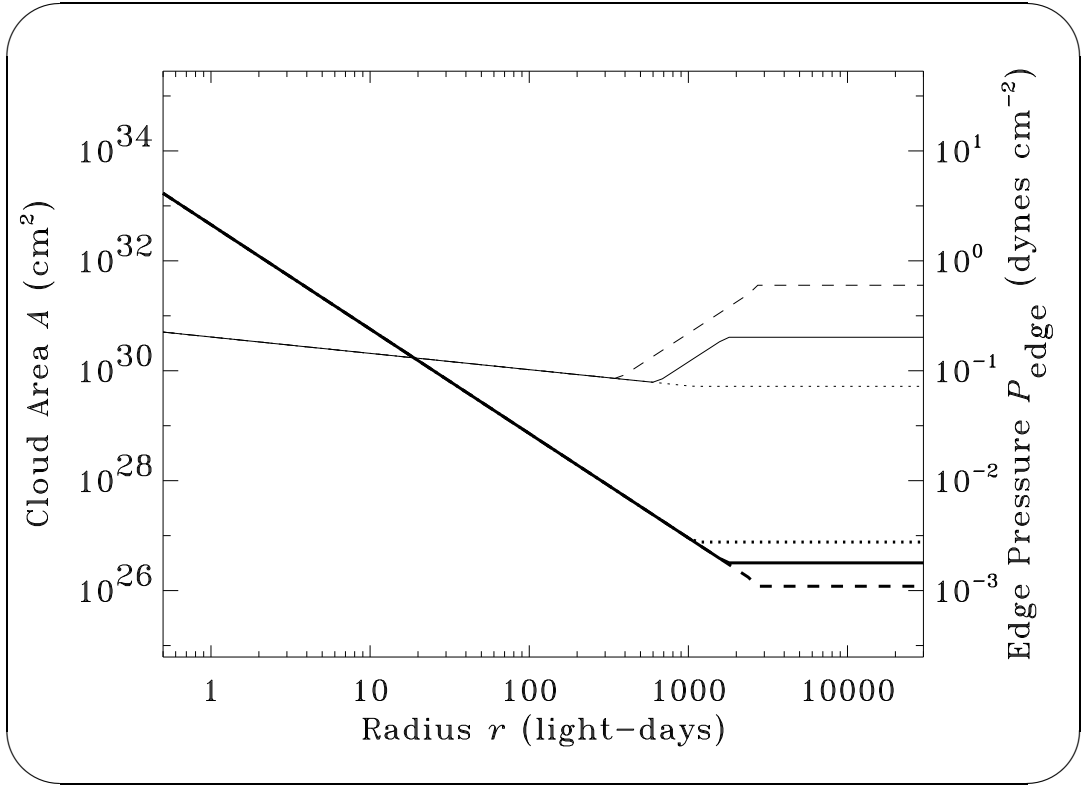
yr$  ^{-1}  $).}. \Fig{model18/area}{Thin lines, left axis: wind area functions.
Thick lines, right axis: wind edge pressures. Dotted lines: model 18 ($  \dot{M}_{{\rm min}}=2.4\times 10^{-8}\dot{M}_{\odot }  $
yr$  ^{-1}  $). Dashed lines: model 19 ($  \dot{M}_{{\rm min}}=2.6\times 10^{-7}\, \dot{M}_{\odot }  $
yr$  ^{-1}  $). Solid lines: model 2 ($  \dot{M}_{{\rm min}}=7.9\times 10^{-8}\, \dot{M}_{\odot }  $
yr$  ^{-1}  $).} shows the wind area functions of models 18 and 19, which have,
respectively, minimum mass loss rates 3.4 times less than and greater than that
of model 2. 

The minimum mass loss parameter $  \dot{M}_{{\rm min}}  $ is the mass loss
assumed for the giants far enough away from the AGN environment that they are
unaffected by it. The impact-parameter-averaged wind column densities obey $  N_{\rm c}\propto \sqrt{\dot{M}P/v_{\infty }}  $,
so increasing $  \dot{M}_{{\rm min}}  $ increases the column densities of the
NLR clouds. In $  \alpha >0  $ models such as the ones considered here, $  \dot{M}_{{\rm min}}  $
thus regulates where the condition $  N_{\rm {c}}\geq N_{\rm {cmin}}  $ forces
$  A  $ to be constant at its maximum value; increasing $  \dot{M}_{{\rm min}}  $
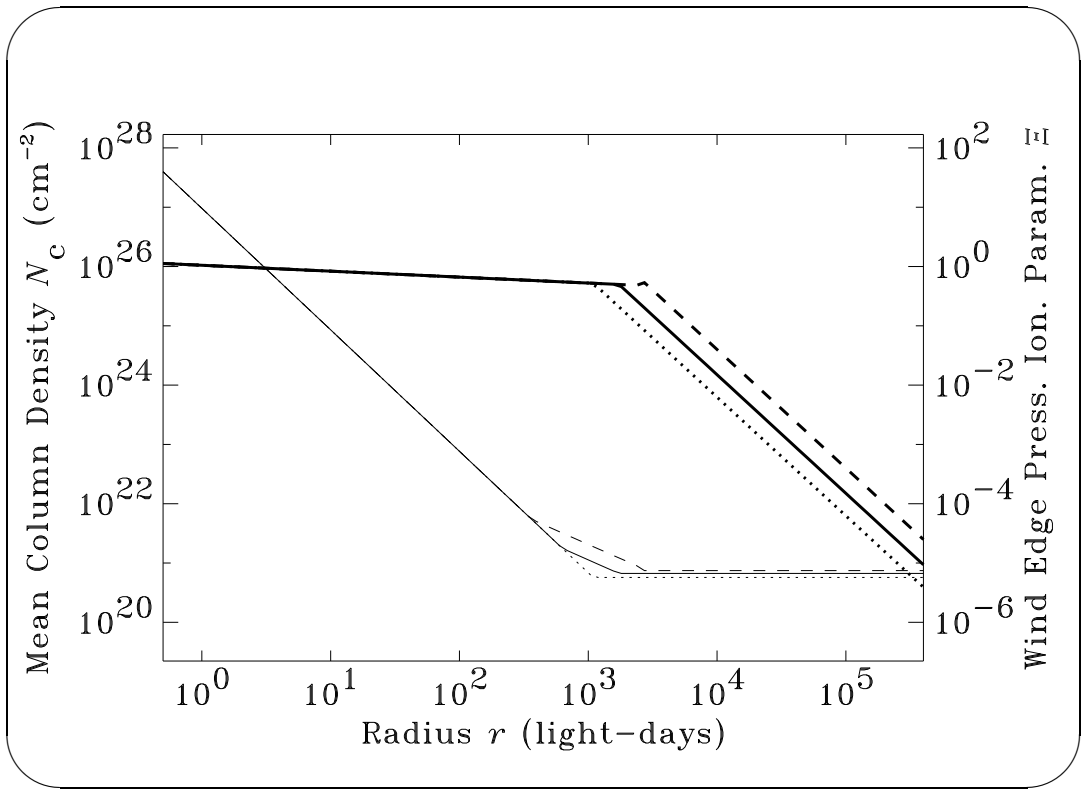
increases the (relatively small) region where $  A\propto r^{2}  $. As \Fig{model18/N_c+Xi}{Thin
lines, left axis: column density $  N_{\rm c}  $. Thick lines, right axis:
wind edge ionization parameter $  \Xi   $. Dotted lines: model 18 ($  \dot{M}_{{\rm min}}=2.4\times 10^{-8}\, \dot{M}_{\odot }  $
yr$  ^{-1}  $). Dashed lines: model 19 ($  \dot{M}_{{\rm min}}=2.6\times 10^{-7}\, \dot{M}_{\odot }  $
yr$  ^{-1}  $). Solid lines: model 2 ($  \dot{M}_{{\rm min}}=7.9\times 10^{-8}\, \dot{M}_{\odot }  $
yr$  ^{-1}  $).} shows, models with high $  \dot{M}_{{\rm min}}  $ have high
NLR mass losses and column densities. They attain $  \dot{M}=\dot{M}_{{\rm min}}  $
relatively near the continuum source and $  A(r)=\textrm{max}(A)  $ relatively
far from the continuum source. Their NLR geometrical wind areas are, therefore,
relatively large.

From a perspective of model fitting, $  \dot{M}_{{\rm min}}  $ does nearly
the opposite of the $  N_{\rm {cmin}}  $ parameter discussed in \S~\ref{nc_min}---increasing
$  \dot{M}_{{\rm min}}  $ or decreasing $  N_{\rm {cmin}}  $ will increase
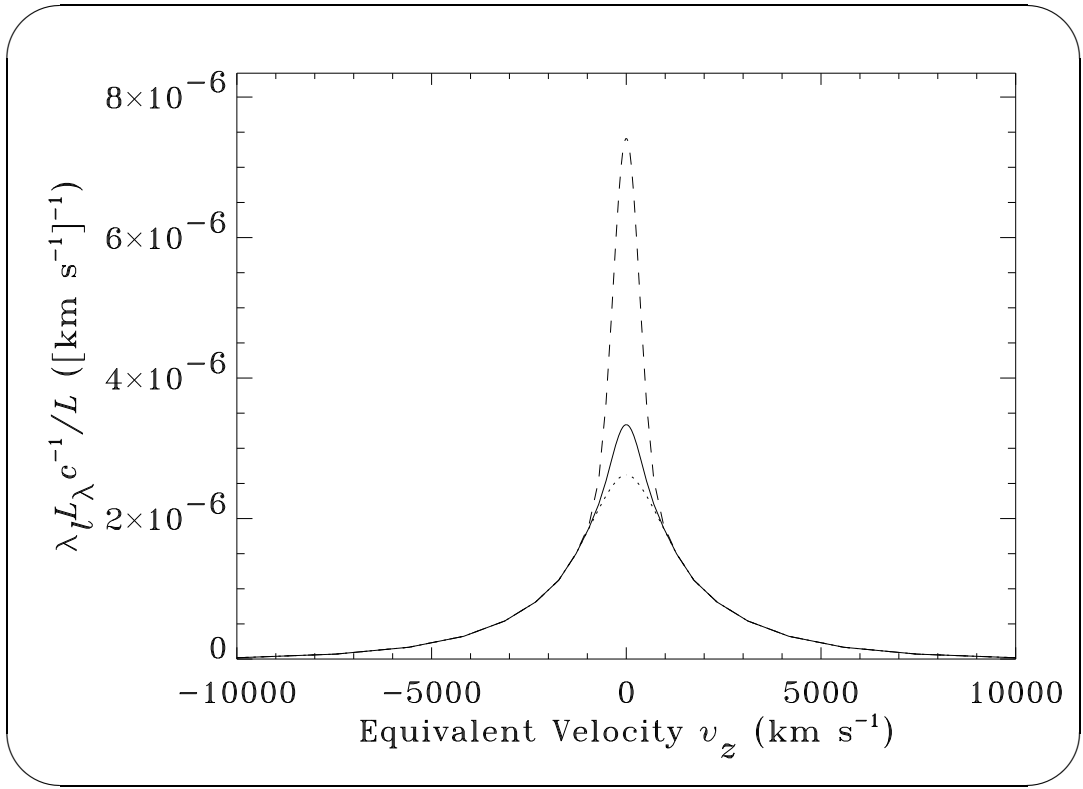
the NLR to BLR covering ratio, narrow the profiles (see \Fig{model18/Lya}{Ly$  \alpha   $
line profiles. Dotted line: model 18 ($  \dot{M}_{{\rm min}}=2.4\times 10^{-8}\, \dot{M}_{\odot }  $
yr$  ^{-1}  $). Dashed line: model 19 ($  \dot{M}_{{\rm min}}=2.6\times 10^{-7}\, \dot{M}_{\odot }  $
yr$  ^{-1}  $). Solid line: model 2 ($  \dot{M}_{{\rm min}}=7.9\times 10^{-8}\, \dot{M}_{\odot }  $
yr$  ^{-1}  $).}), and slightly flatten the response functions at extreme delays.
However, the optical depths of the NLR clouds are functions of both the $  \dot{M}_{{\rm min}}  $
and $  N_{\rm {cmin}}  $ parameters.

Because the results of varying $  \dot{M}_{{\rm min}}  $ are so nearly reversed
to those associated with varying $  N_{\rm {cmin}}  $, we do not show other
observable properties of models 18 and 19 here; readers interested in additional
observable effects of varying $  \dot{M}_{{\rm min}}  $ are directed to \S~\ref{nc_min}.

\section{Models 20 and 21: Varying the Density Slope \protect$  -(3-\eta )\protect  $
\label{eta}}

\pfn{Demos would like this section to be moved before the line emission section.
I think that it is best to start with the greased shoe. That is, start with
what people know and build on it. If I am going to learn a new computer language,
I would first like to be told how it is similar to, say, FORTRAN. So I am not
going to move it. Also, the earlier results are important to all AGN models.
So this ordering is more modest.}In the previous sections of this chapter, we
present models in which various wind area parameters are varied. Some of these
parameters affect the line ratios (see \S \S~\ref{iontopratio}-\ref{s}), some
affect the summed line (see \S \S~\ref{luminosity}-\ref{mdot_min}), and many
affect both. The stellar distribution functions $  f  $ of the models discussed
so far, however, have been identical. The distribution functions govern the
number densities $  n_{*}  $ and velocities of the stars. The remaining sections
of this chapter present models in which the distribution functions (which have
no direct effects upon the line emission characteristics of the clouds) have
been altered while the wind area parameters are fixed. 

Models 20 and 21 differ from model 2 by the values of their $  \eta   $ parameters,
which regulate the density slopes $  -(3-\eta )  $ via $  n_{*}\propto r^{-(3-\eta )}  $.
In model 20, $  -(3-\eta )=-1.0  $, while in model 21, $  -(3-\eta )=-1.8  $.
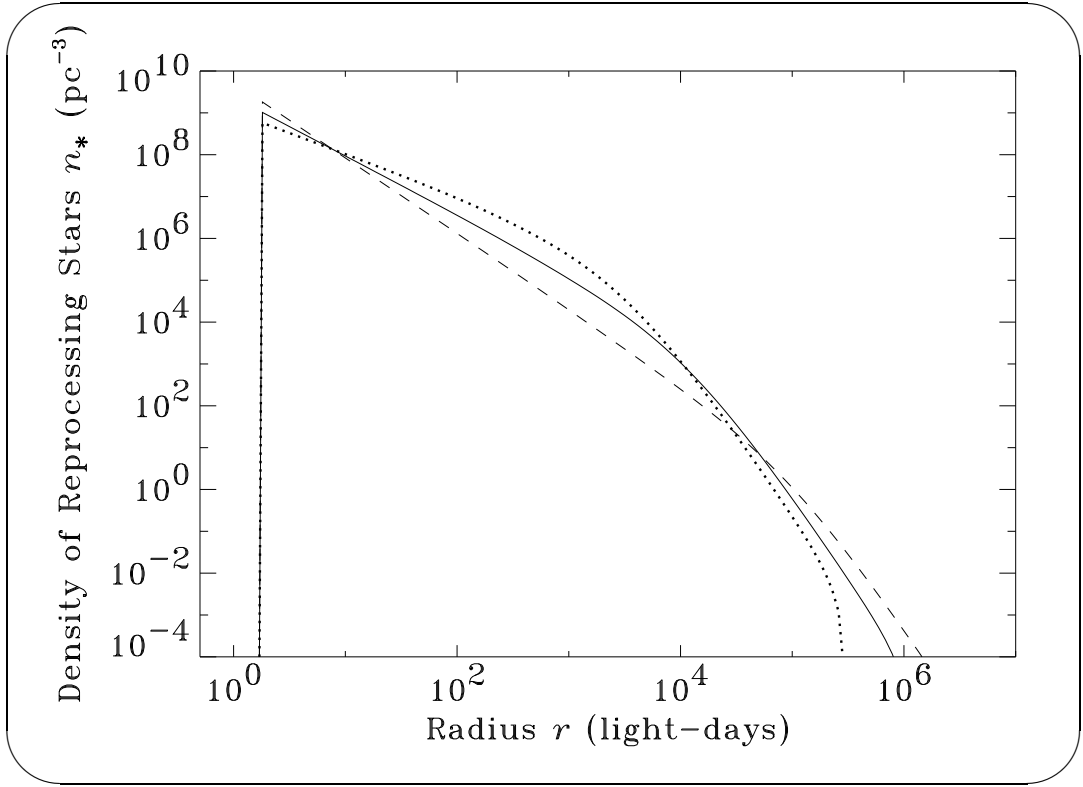
The stellar density functions of models 20 and 21 are shown in \Fig{model20/density}{Densities
of giant stars. Dotted line: model 20 (density slope of $  r^{-1.0}  $; $  An_{*}\propto r^{-1.4}  $
in the BLR). Dashed line: model 21 (density slope of $  r^{-1.8}  $; $  An_{*}\propto r^{-2.2}  $
in the BLR). Solid line: model 2 (density slope of $  r^{-1.4}  $; $  An_{*}\propto r^{-1.8}  $
in the BLR).}. For these spherically symmetric models, there is a one-to-one
correspondence between the stellar density functions and the distribution functions.
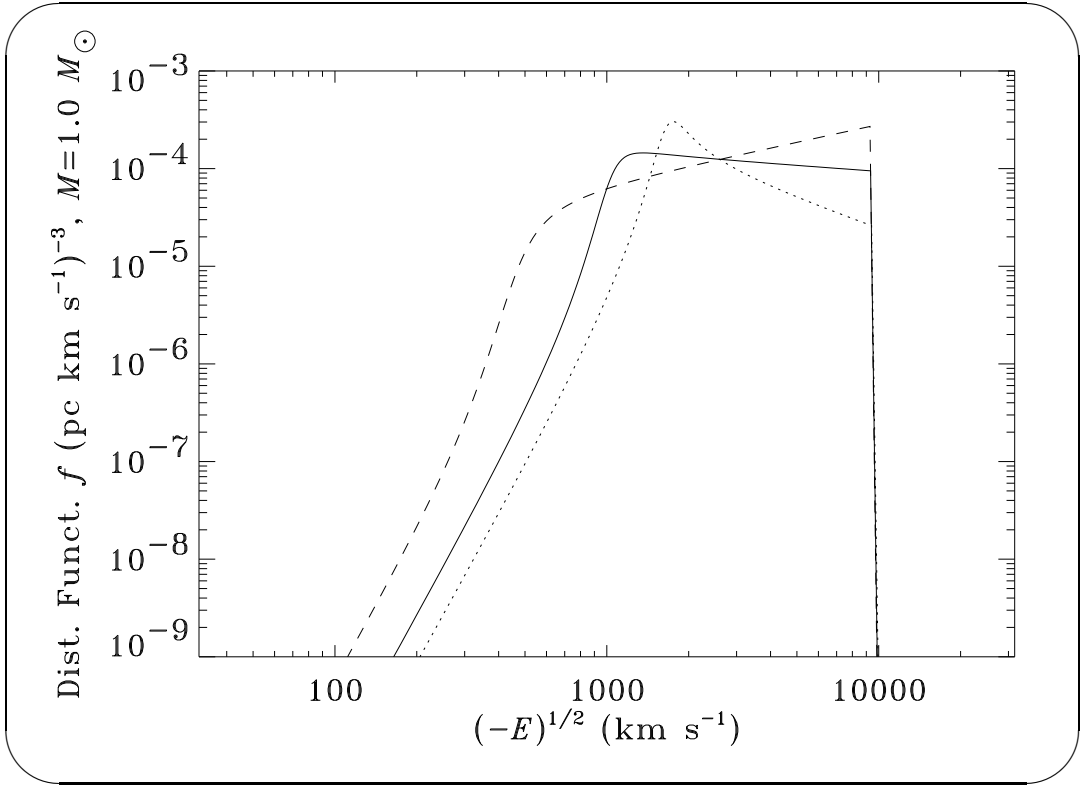
This is shown in \Fig{model20/f}{Distribution functions of giant stars $  f  $.
Dotted line: model 20 (density slope of $  r^{-1.0}  $). Dashed line: model
21 (density slope of $  r^{-1.8}  $). Solid line: model 2 (density slope of
$  r^{-1.4}  $). The slopes of the curves to the right of the ``bends'' (in
the high-velocity regions) regulate the slopes of the stellar densities at $  r\ll r_{\rm c}  $.
These slopes are different for each of the models shown. On the other hand,
the plots are similar to the left of the bends (in the low-velocity regions)
because $  n_{*}(r\gg r_{\rm c})\propto r^{-4}  $ for each model.}. 

From a model-fitting perspective, the $  \eta   $ parameter is similar to the
$  \alpha   $ parameter discussed in \S~\ref{alpha} in that it primarily affects
the slope of the covering function $  An_{*}  $. However, there are four main
differences between $  \eta   $ and $  \alpha   $:

\begin{lyxlist}{00.00.0000}
\item [1.]Though the covering $  An_{*}  $ depends upon both $  \eta   $ and $  \alpha   $,
the two parameters affect the covering with opposite signs. Specifically, the
relevant terms in the radial slope of the covering function in the ``optically
thick region'' are $  -2\alpha +\eta   $. 
\item [2.]A second difference between $  \eta   $ and $  \alpha   $ is that $  \eta   $
can change the velocity structure of the clouds, while $  \alpha   $ cannot.
This is because of the parameterization scheme we adopt. We assume that the
mass density function is proportional to the number density of reprocessing
winds (cf Appendix \ref{commentsonaandn}). The distribution function is computed
from the density function. For a given stellar cluster mass and BLR stellar
density, the core radius $  r_{\rm c}  $ is thus a function of $  \eta   $,
with steeper density functions resulting in larger values of $  r_{\rm c}  $.
Also, because the stellar velocities fall with increasing radius, relatively
steep density functions (yet less steep than $  r^{-2}  $) have relatively
slow NLR clouds.
\item [~]The velocity dispersions of the line-emitting stellar winds for models 20
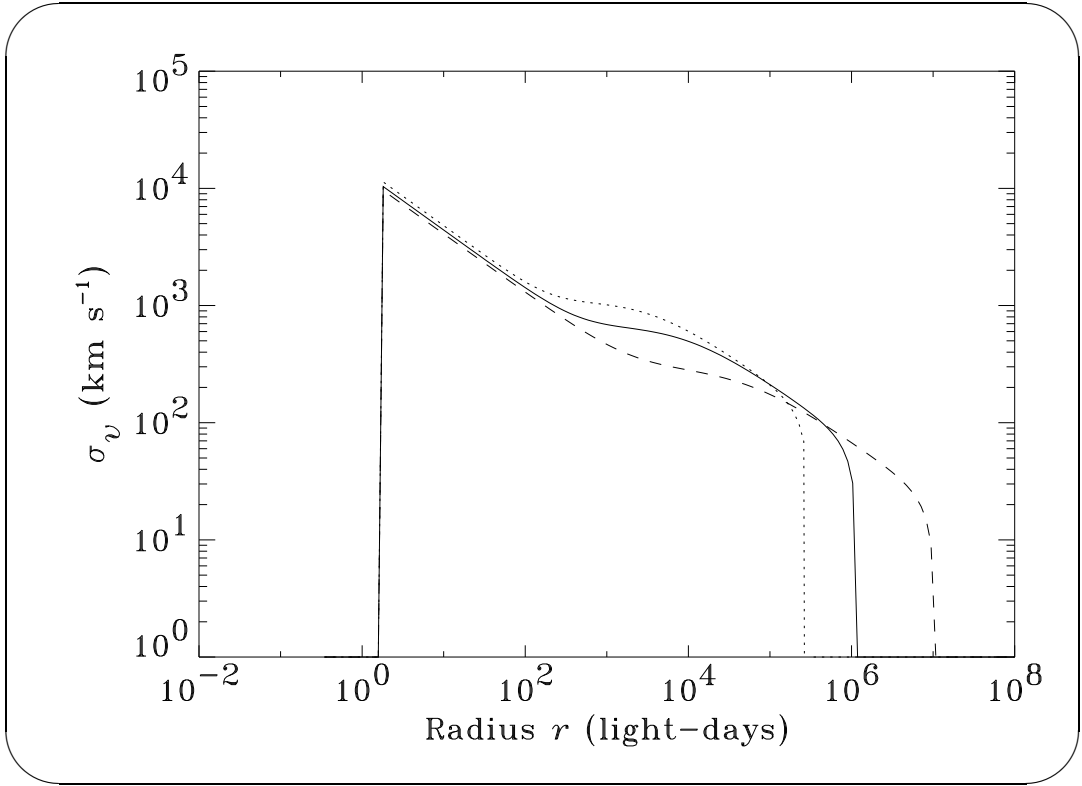
and 21 are shown in \Fig{model20/vel-rdays}{Velocity dispersions of stars assumed
to have line-emitting winds. Dotted line: model 20 (density slope of $  r^{-1.0}  $).
Dashed line: model 21 (density slope of $  r^{-1.8}  $). Solid line: model
2 (density slope of $  r^{-1.4}  $).}. As expected, the model with the steepest
slope (model 21) has the largest cluster core radius. Because model 21 has the
slowest NLR line-emitting clouds, its profiles (one of which is shown in Fig.
\ref{model20/Lya}) have the narrowest structural features. (That is, the velocities
at which the semi-logarithmic profiles become flat are lowest for this model.)
Thus $  \eta   $ is an important parameter for fitting profile cores. 
\item [~]Of course, it is not the only one. Because our parameterization scheme keeps
the inner stellar densities nearly constant when the cluster mass $  M_{\rm c}  $
alone is changed, $  M_{\rm c}  $ is another parameter which affects the core
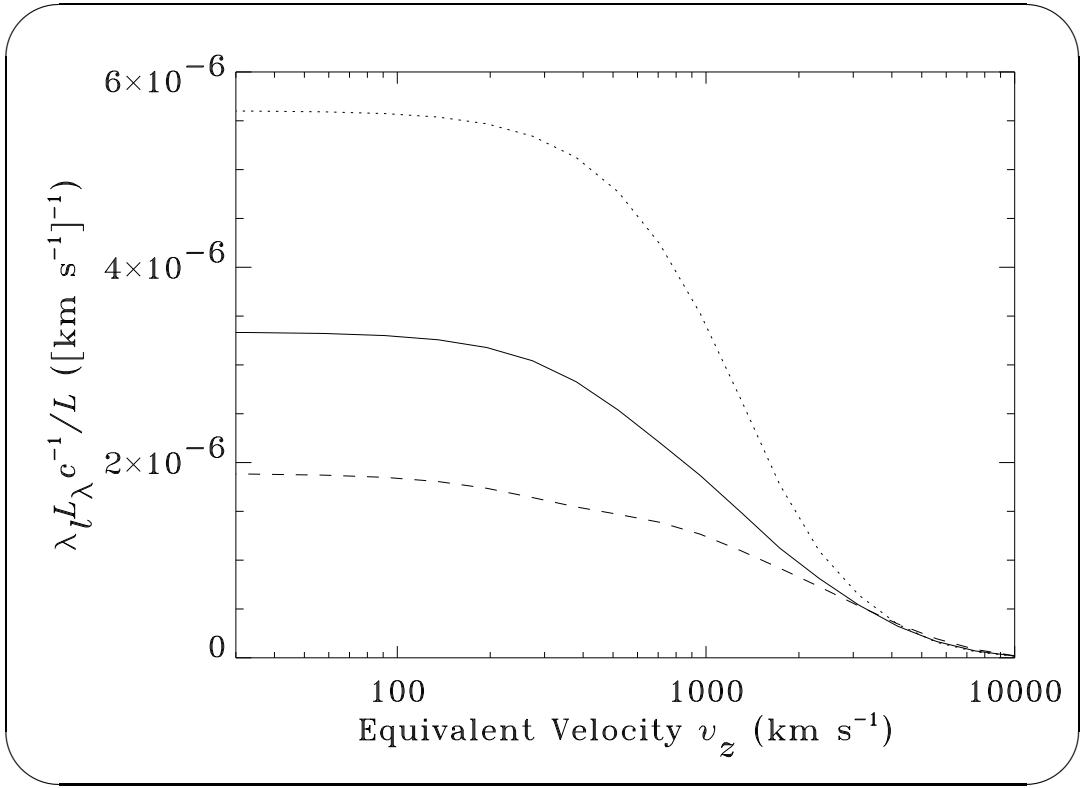
velocity (see also \S \S~\ref{m_h} \& \ref{m_core}).\fig{model20/Lya}{Line
profiles of Ly$  \alpha   $. Dotted line: model 20 (density slope of $  r^{-1.0}  $).
Dashed line: model 21 (density slope of $  r^{-1.8}  $). Solid line: model
2 (density slope of $  r^{-1.4}  $). Note that the velocity of the core (the
narrowest component/feature/structure apparent in the profiles) is lowest for
the model with the steepest density function and highest for the model with
the flattest density function.}
\item [~]Recall that the model shown in \S~\ref{alpha} with the steepest covering
function had the broadest profiles, while the model with the flattest covering
had the narrowest profiles and highest NLR/BLR ratio. The results shown in Figure
\ref{model20/Lya} do yield this behavior, but only at relatively large velocities.
As expected, opposite results occur for the velocities of the cores, with the
steepest density model (model 21) having the smallest core velocity. 
\item [3.]Two remaining differences between $  \alpha   $ and $  \eta   $ reveal
themselves by the response functions. The gains in the BLR, which are $  \simeq 1+\alpha -s/2+\eta (\epsilon _{l}|L)  $
for most of these models, depend upon $  \alpha   $ but not $  \eta   $. This
is because the area is a function of the local continuum flux but the stellar
number density is not. The response functions for the Ly$  \alpha   $ line
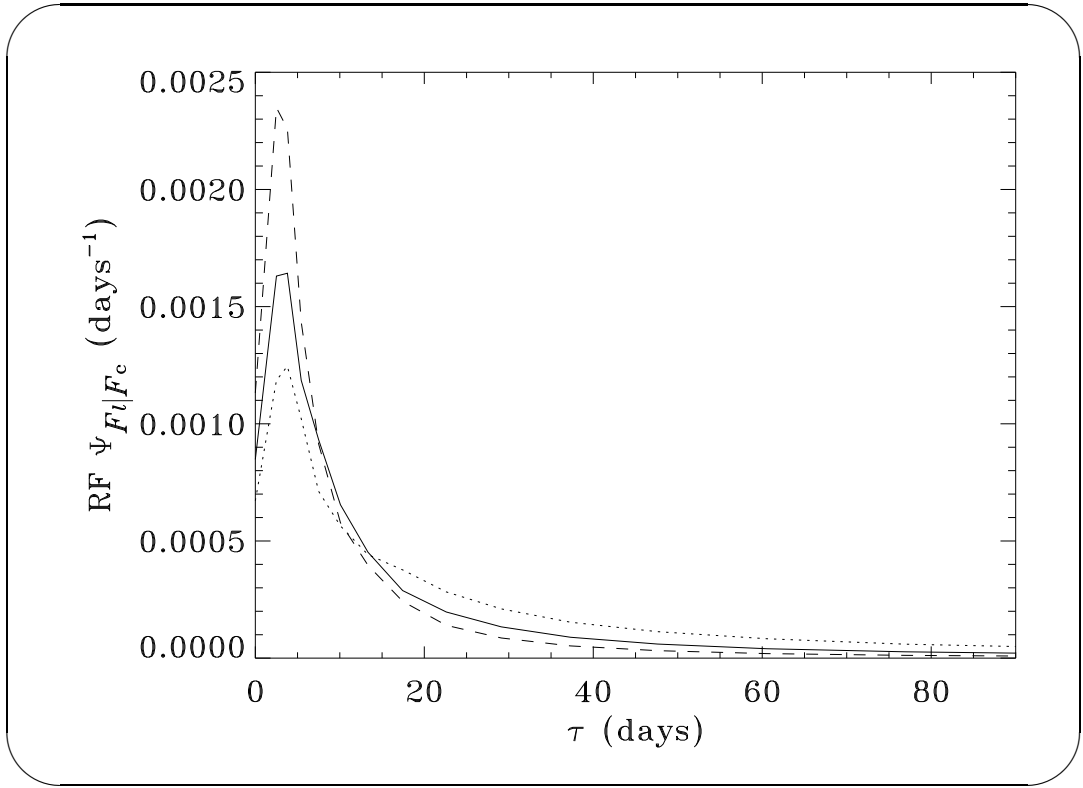
of models 20 and 21 are shown in \Fig{model20/civrf}{Response functions of the
Ly$  \alpha   $ line for models 20, 21, and 2. Dotted line: model 20 (density
slope of $  r^{-1.0}  $). Dashed line: model 21 (density slope of $  r^{-1.8}  $).
Solid line: model 2 (density slope of $  r^{-1.4}  $).}. As expected, the model
with the steepest density falloff (model 21) has the narrowest response function. 
\item [~]Note that for hypothetical models in which both the densities at $  r=r_{\rm t}  $
and the covering slopes $  -(2\alpha -s+3-\eta )  $ are the same, the BLR profile
components could be identical. However, the response function amplitudes of
the low-$  \alpha   $ models would be lower due to the factor of the gain.
Incidentally, perfect error-free measurements of the both the profiles and the
linearized response functions would only partially resolve this degeneracy.
This is because of the other parameters (such as $  s  $) in the expression
for the gain.
\item [4.]A fourth and relatively trivial difference between $  \alpha   $ and $  \eta   $
stems from the normalization conventions we adopted. For numerical reasons,
the models were normalized such that both the density at $  r=5r_{\rm t}  $
and the cluster to black hole mass ratios were identical. As a result, response
function peak heights are affected by $  \eta   $. In particular, they are
highest for the model with the steepest covering function, which is model 21.
\end{lyxlist}
These subtle differences between $  \alpha   $ and $  \eta   $ aside, models
20 and 21 show essentially the same things that models 10 and 11 (\S~\ref{alpha})
show: steepening the covering function makes the profiles broader, the profiles
more concave, and the response functions narrower. Flattening the covering function
does the opposite. Additional plots which illustrate these effects are provided
in \S~\ref{alpha}.\pfn{{[}redo so that these models have the same $  An_{*}  $
slopes as those in the alpha section of possible.{]} {[}also, make the dotted
one ??{]}That would be nice, but there is no way I have time to do that.}

\section{Models 22 and 23: Varying the Tidal Radius \protect$  r_{\rm t}\protect  $\label{tidal section}}

The tidal radius $  r_{\rm t}  $ is the innermost edge of the wind model BLR.
It affects the widths of the response functions and line profiles. If $  r_{\rm t}  $
is for some reason decreased, line emission can occur closer to the continuum
source. This narrows the response functions. Because stars move faster the closer
they are to the black hole, decreasing $  r_{\rm t}  $ also broadens the profiles
of lines that are strong at high pressures and local continuum fluxes. Increasing
$  r_{\rm t}  $ has the opposite effects.

The tidal radius of model 22 is 1.8 times smaller than that in model 2.\pfn{
(which was determined from equation \ref{tidalradiusequation} and the results
of Murphy et al. {[}1991{]})} The tidal radius of model 23, on the other hand,
is 1.8 times larger than that of model 2. The densities of red giant stars in
these models are shown in \Fig{model22/density}{Stellar densities. Dotted line:
model 22 ($  r_{\rm t}=0.90  $ light-days). Dashed line: model 23 ($  r_{\rm t}=2.9  $
light-days). Solid line: model 2 ($  r_{\rm t}=1.6  $ light-days).}. These
models are parameterized in order to have the same stellar density at the innermost
edge of their BLR.

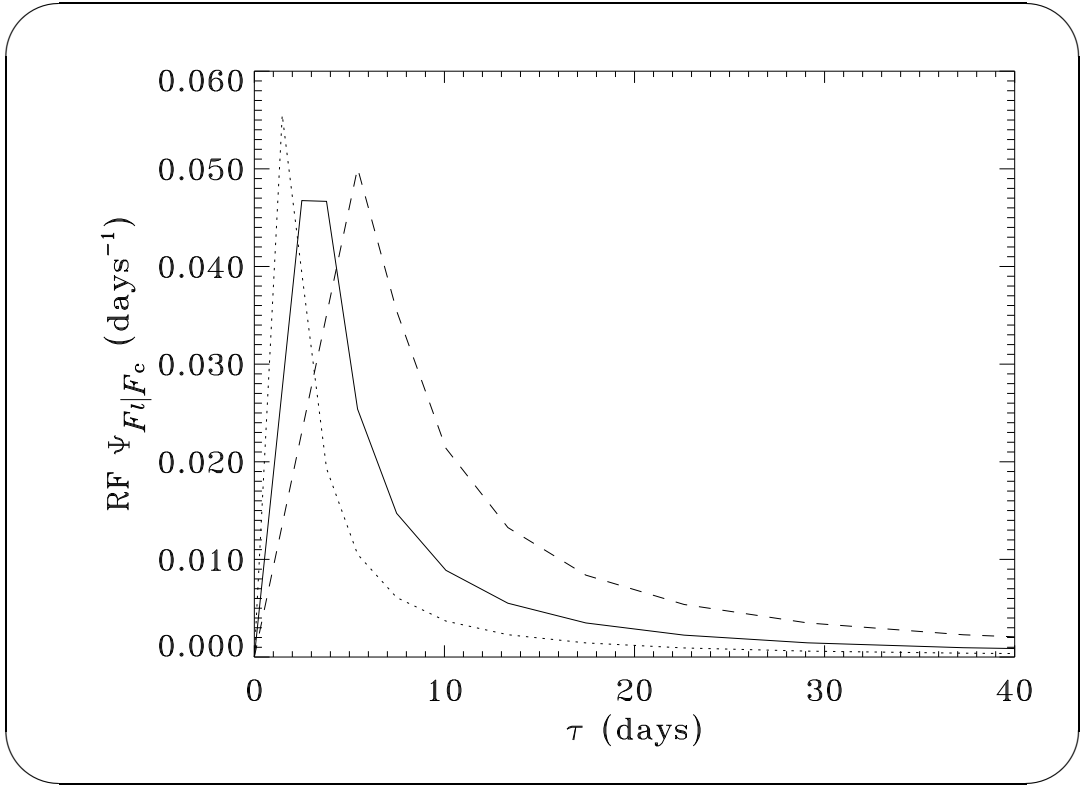
The Ly$  \alpha   $ response functions of models 22 and 23 are shown in \Fig{model22/summedlinerf}{Linearized
response functions for the synthetic summed line. Dotted line: model 22 ($  r_{\rm t}=0.90  $
light-days). Dashed line: model 23 ($  r_{\rm t}=2.9  $ light-days). Solid
line: model 2 ($  r_{\rm t}=1.6  $ light-days).}. As expected, the low-$  r_{\rm t}  $
model responds fastest, while the high-$  r_{\rm t}  $ model responds slowest.
Because the models are normalized to have the same densities at $  r=r_{\rm t}  $
and the radial slope of $  An_{*}  $ is less than $  -1.0  $, decreasing $  r_{\rm t}  $
lowers the stellar density at a given position in the AGN (see Figure \ref{model22/density}),
the covering, and the area under the response functions. These response functions
peak at $  \tau =2r_{\rm t}/c  $, as is expected of high-$  N_{\rm c}/\Xi   $,
spherically symmetric cloud models.

The velocity-integrated response functions of the C {\footnotesize IV} profile
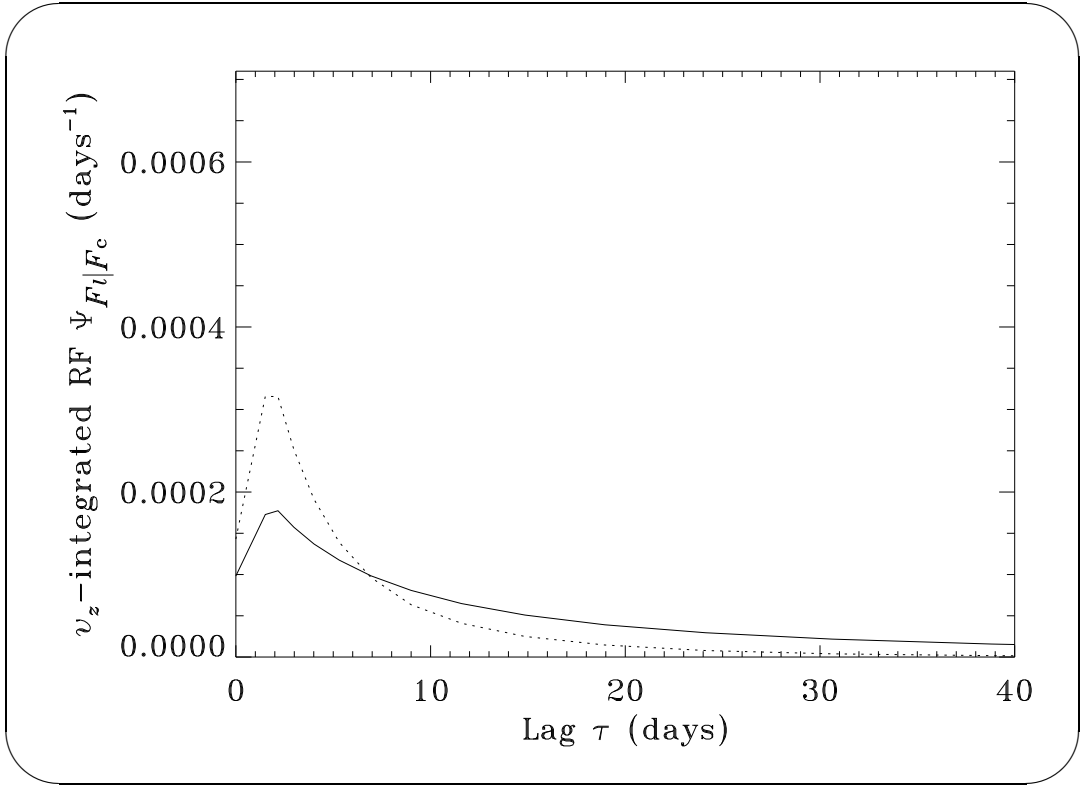
components for models 22 and 23 are shown in Figures \ref{model22/civvel-inta} and \ref{model22/civvel-intc}.\fig{model22/civvel-inta}{Response
functions of C {\scriptsize IV} profile components for model 22 ($  r_{\rm t}=0.90  $
light-days). Dotted line: wing component. Solid line: core component. Like the
response functions shown in \ref{model2/civvel-intb} for model 2, the wing/core
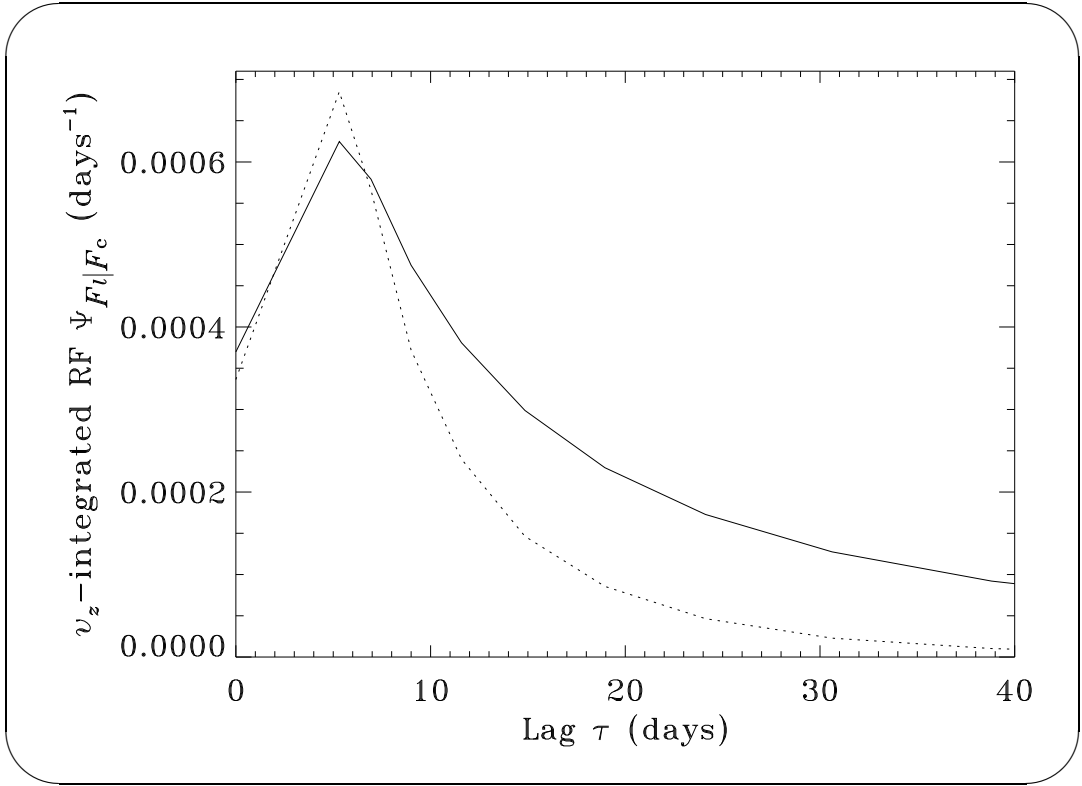
peak ratio is much greater than the observed value of $  \sim 1  $.}\fig{model22/civvel-intc}{Response
functions of C {\scriptsize IV} profile components for model 23 ($  r_{\rm t}=2.9  $
light-days). Dotted line: wing component. Solid line: core component. The wing/core
ratio of the response function peaks is 1.1 for this model, which is compatible
with the observed value of $  \sim 1  $.} As discussed in \S~4.2, $  r_{\rm t}  $
affects the wing/core response function peak ratio. In model 22, this ratio
is 1.8 for C {\small IV}. In model 23, which has a larger tidal radius, this
ratio lowers to 1.1. This value is consistent with the NGC 5548 ratio of $  \sim 1  $
(\S~4.2). The response functions of model 23, however, peak at 5.8 days, which
is $  \sim 2-\sigma   $ greater than the peaks observed for NGC 5548 by Wanders
et al. (1995) and Done \& Krolik (1996). Model 22 is less than $  \sim 1-\sigma   $
away from the observed peak, but has the poorest fitting wing/core ratio. Thus,
provided no other parameters are changed, no value of $  r_{\rm t}  $ will
result in models that fit the observations well. Varying the black hole mass
does affect the ratios of the FWHMs of the profiles and response functions,
but cannot fix this important problem with this model (\S~\ref{m_h}).

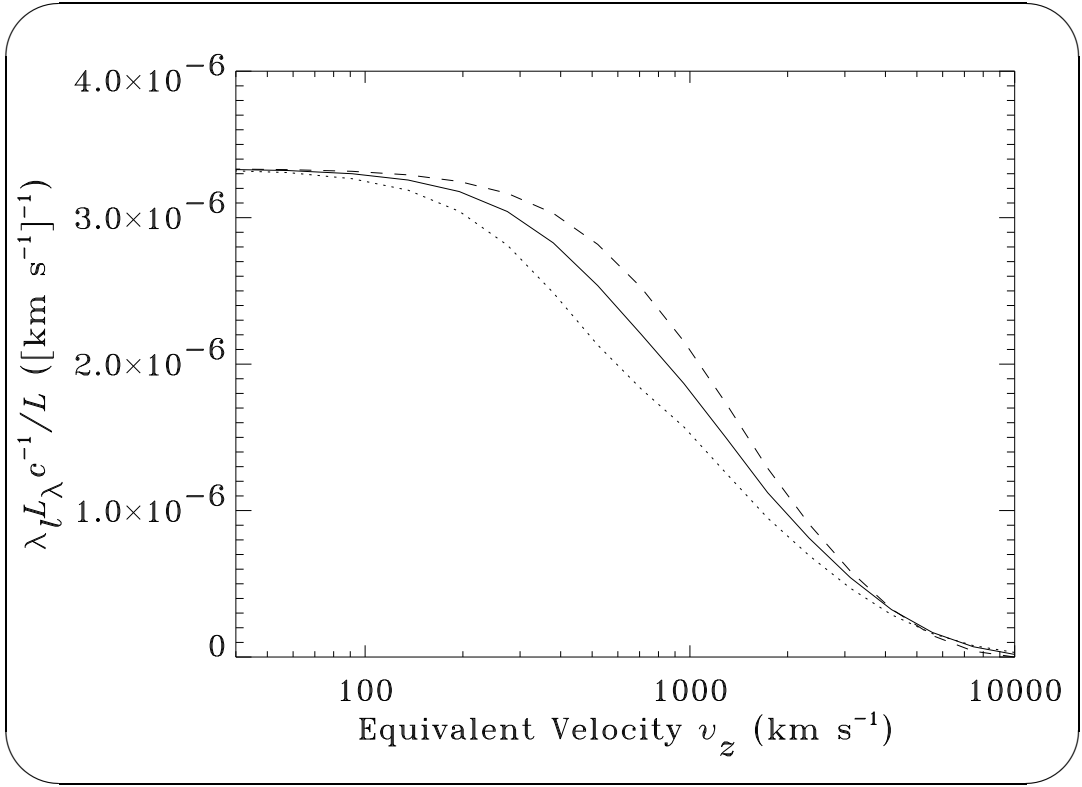
\Fig{model22/Lya}{Profiles of Ly$  \alpha   $. Dotted line: model 22 ($  r_{\rm t}=0.90  $
light-days), rescaled to profile peak of model 2. Dashed line: model 23 ($  r_{\rm t}=2.9  $
light-days), rescaled to profile peak of model 2. Solid line: model 2 ($  r_{\rm t}=1.6  $
light-days).} shows the Ly$  \alpha   $ profiles for models 22 and 23. In the
low-$  r_{\rm t}  $ model, the local continuum fluxes and pressures are so
high that there is not significant Ly$  \alpha   $ emission at $  r\gsim r_{\rm t}  $
(see \S \S~\ref{s} \& \ref{luminosity}). Most lines, on the other hand, are
less affected by thermalization. The profile base width of these lines follow
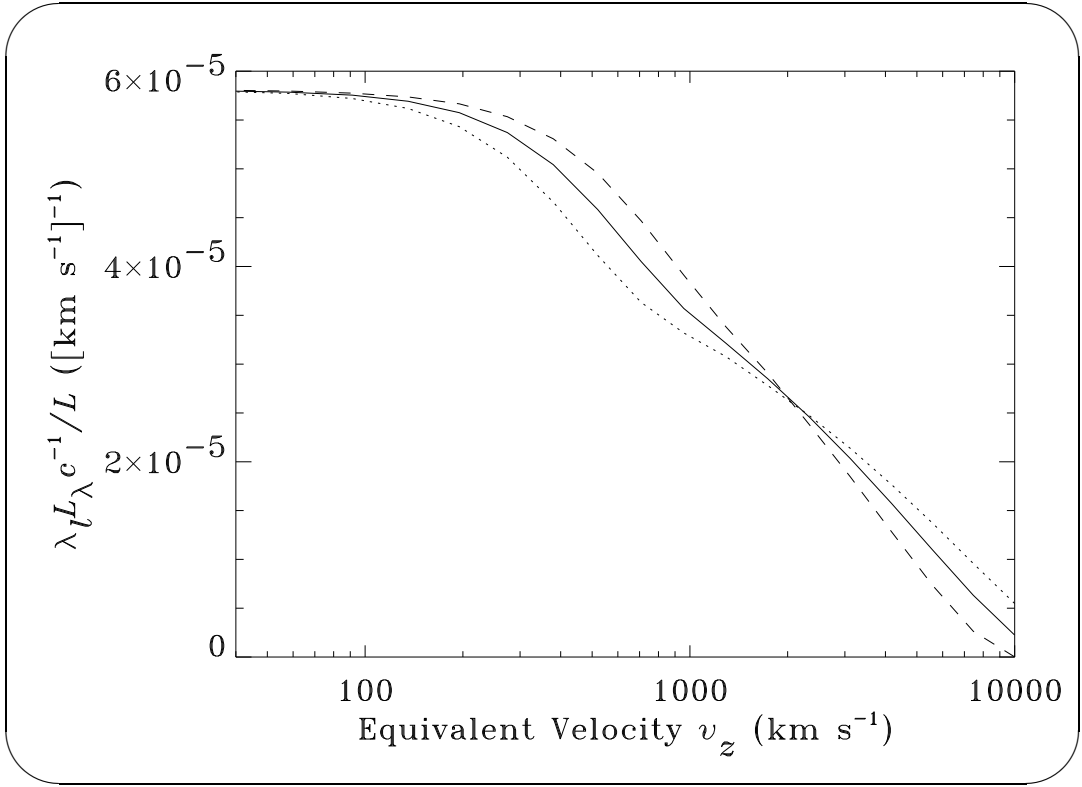
that of the summed line shown in \Fig{model22/summedline}{Profiles of the synthetic
summed line. Dotted line: model 22 ($  r_{\rm t}=0.90  $ light-days), rescaled
to match peak of model 2. Dashed line: model 23 ($  r_{\rm t}=2.9  $ light-days),
rescaled to match peak of model 2. Solid line: model 2 ($  r_{\rm t}=1.6  $
light-days).}. This line clearly broadens as $  r_{\rm t}  $ is reduced, as
is expected.

The covering function is steep enough in the models presented in this subsection
that the tidal radius is an important parameter for the wind models. However,
the tidal radius is not a free parameter of the stellar wind models. Rather,
it is a well-defined function (eq. {[}\ref{tidalradiusequation}{]}). The utility
of this subsection is thus primarily to test the viability of the stellar wind
model; if, in order to fit AGN data, the required values of the tidal radii
are much different from those given by equation (\ref{tidalradiusequation}),
the AGN line emission model, at least as described here, is probably wrong.
This section shows that equation (\ref{tidalradiusequation}) probably agrees
within a factor of 2 to the values one would obtain were $  r_{\rm t}  $ treated
as a free parameter and a $  \chi ^{2}  $ minimization were performed to the
NGC 5548 data. As the next subsection shows in more detail, this is a success
of the stellar wind model.\pfn{{[}check elsewhere that r\_tidal for model 2
is actually listed correctly since I had to change the numbers here{]} I checked
them and they seem OK. JAT 3/3/99}

\section{Models 24 and 25: Varying the Black Hole Mass \protect$  M_{\rm h}\protect  $\label{m_h}}

\pfn{{[}CONSIDER MOVING THIS SECTION EARLIER OR TAKING PARTS OF IT OUT AS IT
DUPLICATES STUFF SAID IN PREVIOUS SECTION.{]} }The black hole mass $  M_{\rm h}  $
of model 24 is $  9.5\times 10^{6}\, M_{\odot }  $. In model 25, $  M_{\rm h}=9.5\times 10^{7}\, M_{\odot }  $.
These values are, respectively, half a decade below and above the value of $  M_{\rm h}  $
assumed in model 2. The Ly$  \alpha   $ line profiles for models 24 and 25
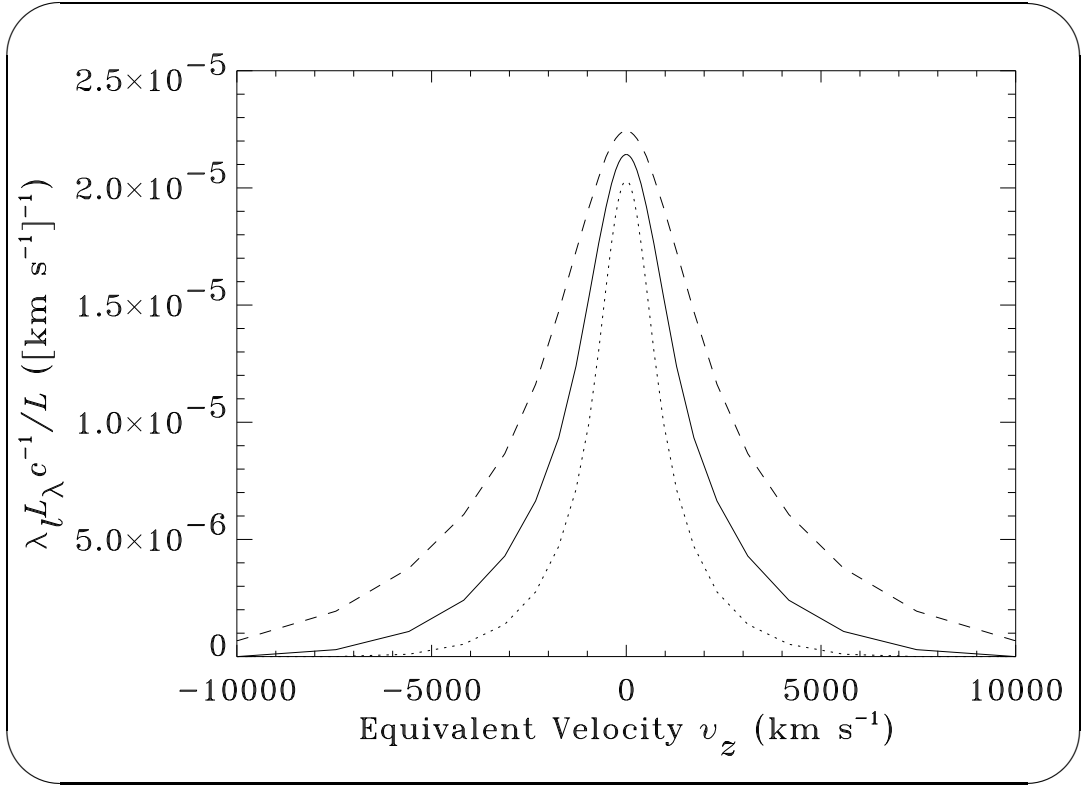
are shown in \Fig{model24/Lya}{Continuum-subtracted Ly$  \alpha   $ line profiles.
Dotted line: model 24 ($  M_{\rm h}=9.5\times 10^{6}\, M_{\odot }  $). Dashed
line: model 25 ($  M_{\rm h}=9.5\times 10^{7}\, M_{\odot }  $). Solid line:
model 2 ($  M_{\rm h}=3.0\times 10^{7}\, M_{\odot }  $).}. As one might naively
expect, the model with the heaviest black hole (model 25) has the broadest profiles.
Unlike $  r_{\rm t}  $, which can change some of the lines in different ways
(see \S~\ref{tidal section}), $  M_{\rm h}  $ affects each profile in the
same way. This is because $  M_{\rm h}  $ primarily changes just the speed
of the BLR clouds, not the local continuum flux to which they are exposed. 

The profiles shown in Figure \ref{model24/Lya} for models 2 and 24 not only
have narrower bases than the profile of models 25, but also have relatively
narrow peaks. The reason for this is that these models are parameterized to
have cluster to black hole mass ratios of 50, independent of $  M_{\rm h}  $.
Specifically, the cluster mass of model 24 is $  4.8\times 10^{8}\, M_{\odot }  $,
while the cluster mass of model 25 is $  4.8\times 10^{9}\, M_{\odot }  $.
Since the velocity fields of the NLR clouds in these models are regulated by
their stellar clusters rather than their black holes, the profile peaks of low-$  M_{\rm h}  $
models are narrower than high-$  M_{\rm h}  $ profile peaks.

The black hole mass also affects the tidal radius. Thus, changing $  M_{\rm h}  $
results in the changes discussed in the previous section. In particular, increasing
$  M_{\rm h}  $ increases $  r_{\rm t}  $ and the characteristic BLR delays
of the lines. Similarly, decreasing $  M_{\rm h}  $ decreases the characteristic
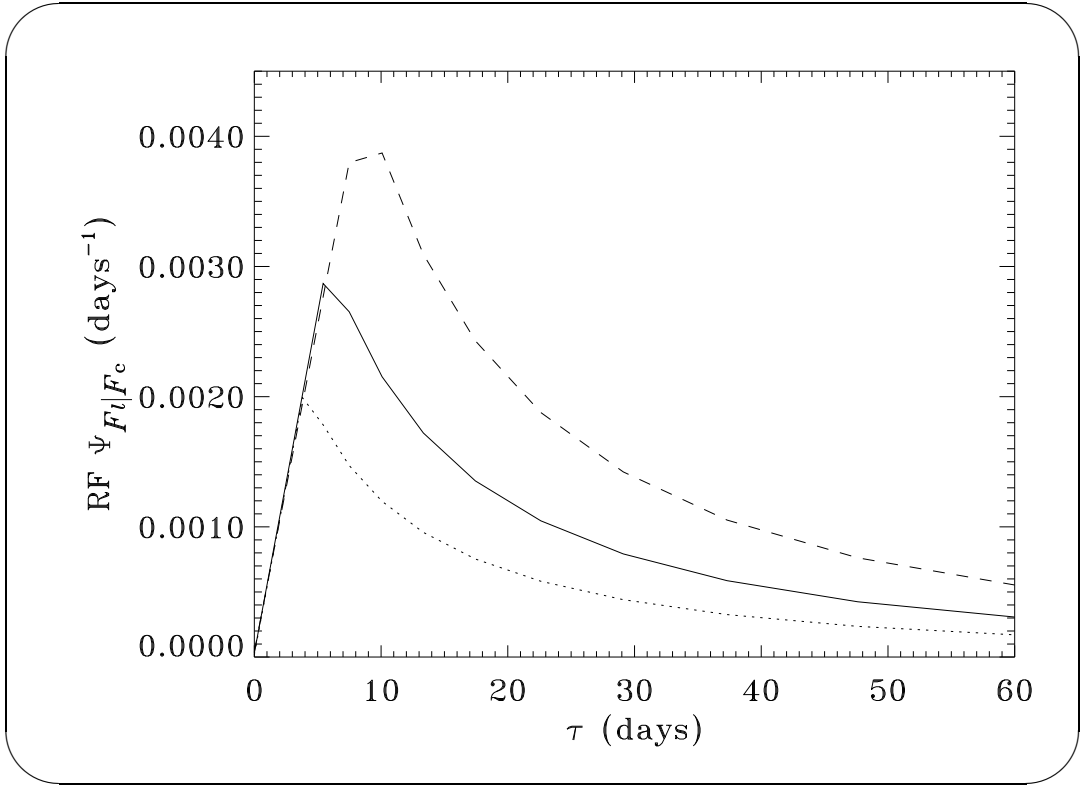
delays. This can be seen in \Fig{model24/Lyarf}{Linearized response functions
of Ly$  \alpha   $. Dotted line: model 24 ($  M_{\rm h}=9.5\times 10^{6}\, M_{\odot }  $).
Dashed line: model 25 ($  M_{\rm h}=9.5\times 10^{7}\, M_{\odot }  $). Solid
line: model 2 ($  M_{\rm h}=3.0\times 10^{7}\, M_{\odot }  $).}, which shows
the response functions of Ly$  \alpha   $ for models 24 and 25. 

An important difference between directly varying $  r_{\rm t}  $ (as is done
for the models discussed in the previous subsection) and indirectly varying
$  r_{\rm t}  $ via the $  M_{\rm h}  $ parameter (as is done for the models
discussed in this subsection) is that delays are linear to $  r_{\rm t}  $
but scale as $  M_{\rm h}^{1/3}  $ when just $  M_{\rm h}  $ is varied. The
relationship between $  M_{\rm h}  $ and the delays of the response function
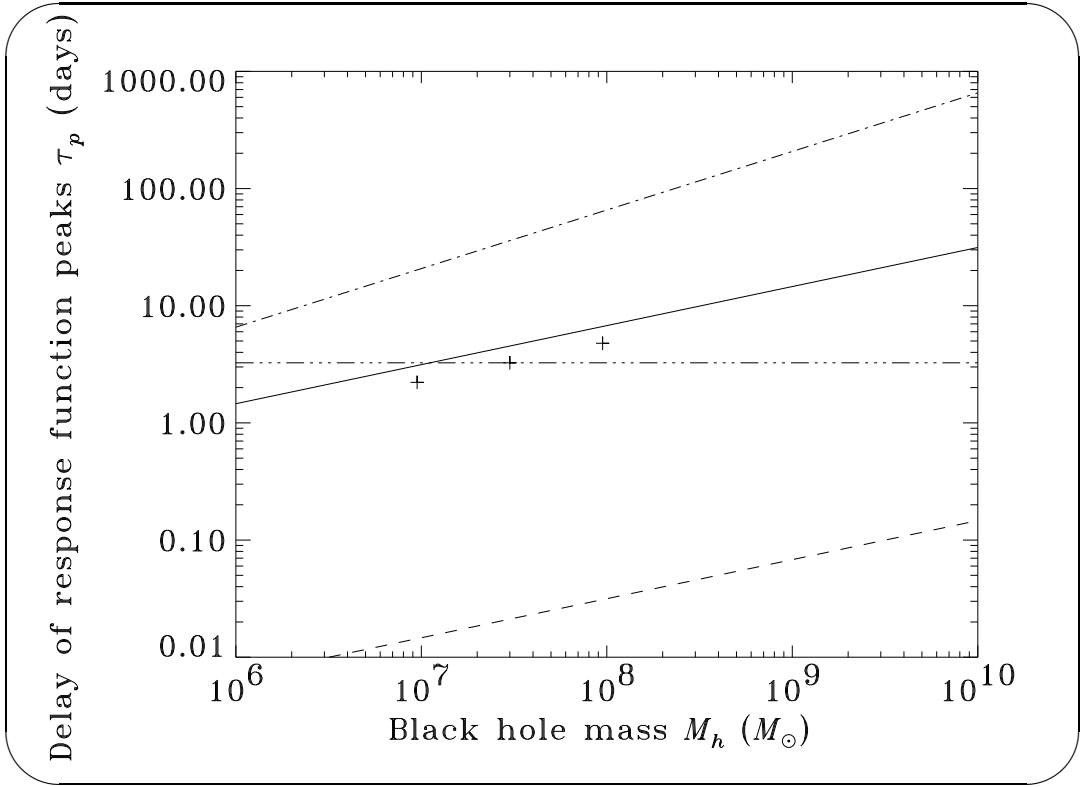
peaks for models 24 and 25 are shown in \Fig{model24/lagversesMh}{Plus-labelled
data points: delay of Ly$  \alpha   $ response function peaks for models 2,
24, and 25 plotted as a function of black hole mass. Solid line: delay of peaks
according to equation (\ref{delayatrtidal}) assuming $  R_{*}=1.0  $~A.U. and
$  M_{*}=1.0\, M_{\odot }  $ for each red giant (as in Kazanas 1989). Dot-dashed
line: delay of peaks if each AGN luminosity were its Eddington luminosity and
the innermost BLR ionization parameters and pressures of each AGN are identical
to model 2. Dash-triple-dotted line: delay of peaks if the innermost radius
is fixed. (This line is valid for models in which $  \eta (\Xi |L)+\eta (P|L)=1.0  $,
where the BLR ionization parameters and pressures are assumed to scale, respectively,
according to $  L^{\eta (\Xi |L)}  $ and $  L^{\eta (P|L)}  $, where $  L  $
is the time-averaged luminosity for various AGNs with different black hole masses.)
Dashed line: delay of peaks according to equation (\ref{delayatrtidal}) assuming
$  R_{*}=1.0\, R_{\odot }  $ and $  M_{*}=1.0\, M_{\odot }  $.}. Because $  d\ln (An_{*})/d\ln r=-1.8<-1.0  $
for these models, the dominant contribution towards their response covering
functions occurs near $  r=r_{\rm t}  $. Thus, the delay of the response function
peaks is only slightly less than the characteristic delay of the line emission
(defined as the integral of the delay-weighed response function). In other words,
because the covering functions are so steep in these models, the inner radius
(rather than the outer radius) regulates the covering.

This simplifies the analysis dramatically, and yields characteristic delays
that are similar to the delays of the response function peaks of $  \tau _{\rm p}  $.
These delays are 
\begin{equation}
\label{delayatrtidal}
\tau _{\rm p}=\frac{2R_{*}}{c}\left( \frac{2M_{\rm h}}{M_{*}}\right) ^{1/3}.
\end{equation}
 As the solid line in Figure \ref{model24/lagversesMh} shows, this equation
predicts that, if the various nonlinear effects are unimportant, the response
times are an increasing function of the black hole mass $  M_{\rm h}  $. 

An analysis analogous to the one performed above for the delays of the line
peaks can be performed for the line widths, and, in particular, the velocity
dispersions of the broadest possible profile components (corresponding to emission
from $  r\simeq r_{\rm t}  $). In order to do this, we make the crude approximation
that the local line profile and distribution function at the tidal radius are
Guassians proportional to $  e^{-v^{2}/(2\sigma ^{2}_{\rm t})}  $ and virialized
such that the magnitudes of their kinetic and potential energies are equal.
This approximation implies 
\begin{equation}
\label{sigmaatrtidal}
\sigma _{\rm t}=\sqrt{\frac{GM_{\rm h}}{3r_{\rm t}}}=\sqrt{\frac{GM_{\odot }}{3R_{\odot }}}\left( \frac{M_{\rm h}}{\sqrt{2}M_{\odot }}\right) ^{1/3}.
\end{equation}
 This equation yields an upper limit to the FWHMs of $  2.35\sigma _{\rm t}  $.
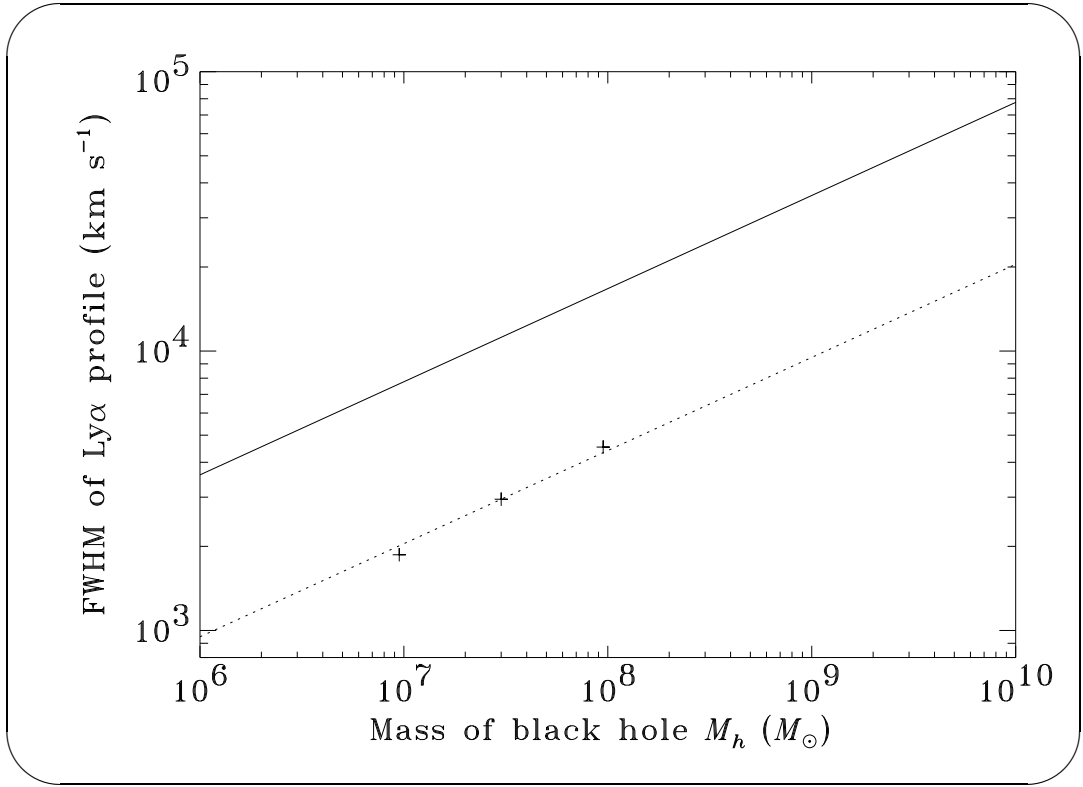
It is plotted in \Fig{model24/sigmaversesMh}{Plus-labelled data points: FWHMs
of Ly$  \alpha   $ line and black hole masses for models 2, 24, and 25. Solid
line: highest possible FWHMs of profile components according to equation (\ref{sigmaatrtidal}).
Dotted line: delay of a hypothetical line that scales according to equation
(\ref{sigmaatrtidal}) but is normalized to agree with the FWHM of the Ly$  \alpha   $
line of model 2.}. 

Equations (\ref{sigmaatrtidal}) and (\ref{delayatrtidal}) yield 
\begin{equation}
\label{tau-sigmarelation}
\tau _{\rm p}=\frac{2^{7/6}R_{*}}{c}\left( \frac{GM_{*}}{3R_{*}}\right) ^{-1/2}\sigma _{\rm t}.
\end{equation}
 This equation is an important prediction of the stellar wind AGN model. It
states that the response function peaks are proportional to the widths of the
broadest components of the lines. 

The beauty of equation (\ref{tau-sigmarelation}) is that it is straightforward
to falsify experimentally; \emph{if the stellar wind model described in the
main body of this dissertation is incorrect, we could know within a few years
simply by plotting $  \tau _{\rm p}  $ and} $  \sigma _{\rm t}  $ \emph{for
several different AGNs to see whether or not the relationship is linear}. Though
the proportionality constant between delay and velocity depends upon the precise
definitions assumed here and characteristics of the stars in AGNs, the fundamental
proportionality between delays and velocities as they vary in AGNs with different
black hole masses does not. This relationship between velocity and delay for
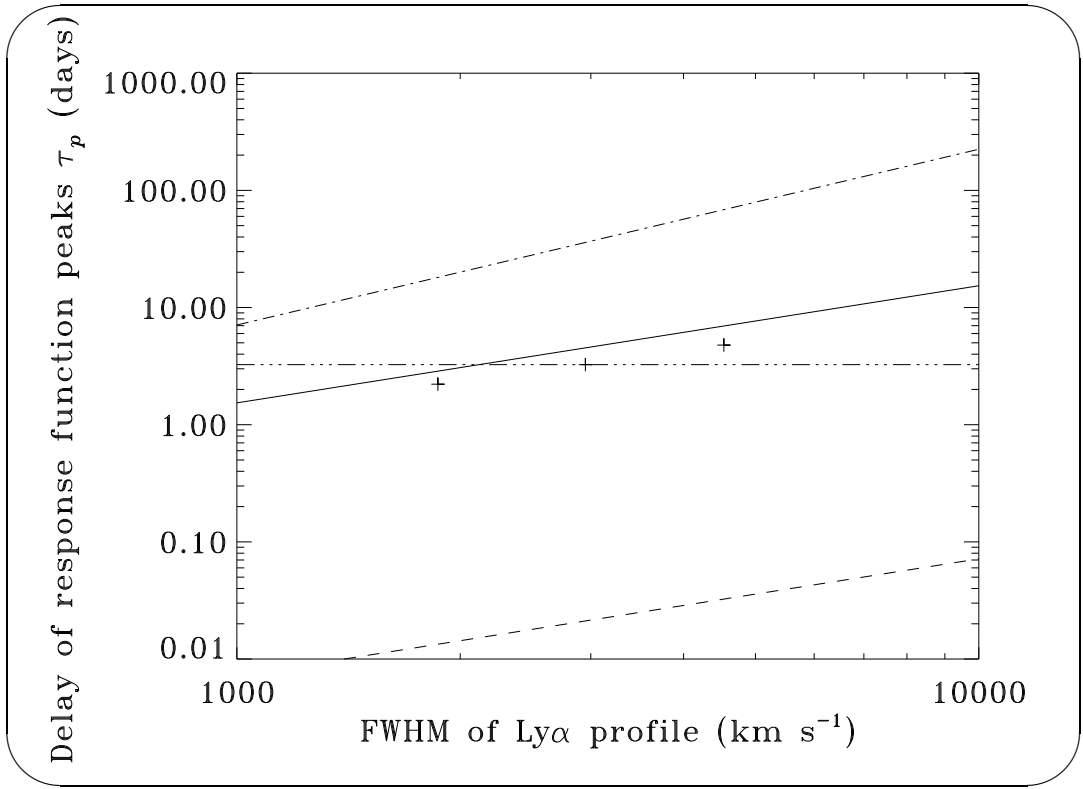
models 2, 24, and 25 is shown in \Fig{model24/lagversessigma}{Delays shown in
Figure \ref{model24/lagversesMh}, but with the $  x  $-axis (black hole mass)
coordinates transformed to velocity space according to the dotted line of Figure
\ref{model24/sigmaversesMh}.}. These velocity-delay correlations are testable
without making the questionable assumption that few AGNs are sub-Eddington (cf
Boller, Brandt, \& Fink 1996). They imply that NLS1s (narrow-line Seyfert 1s)
are simply AGNs with relatively small supermassive black holes. Incidentally,
scaling relations similar to the ones in this subsection can be used to show
that if the stellar wind models are correct, such AGNs with low black hole masses
should have relatively high edge pressures. Spectral features of NLS1s are also
consistent with this prediction of the stellar wind AGN line emission model.

\section{Models 26 and 27: Varying the Cluster to Black Hole Mass Ratio \protect$  M_{\rm c}/M_{\rm h}\protect  $
and the Core Radius \protect$  r_{\rm c}\protect  $\label{m_core}}

Two essential components of the stellar wind line emission model are the supermassive
black hole and the stellar cluster. This subsection concerns the ratio of the
masses of these two parameters, $  M_{\rm c}/M_{\rm h}  $. In the BLR, the
gravitational field is dominated by the black hole. On the other hand, far enough
away from the black hole, generally near the ``inner edge'' of the NLR, the
gravitational field is dominated by the stellar cluster. For this reason, the
$  M_{\rm c}/M_{\rm h}  $ parameter primarily affects only the narrow components
of the line profiles. 

Models 26 and 27 illustrate this. These models are identical to model 2 except
for their stellar cluster to black hole mass ratios, which are $  5.0  $ for
model 26 and 0.10 for model 27. These ratios correspond to cluster masses of
$  3.1\times 10^{6}\, M_{\odot }  $ for model 26 and $  1.5\times 10^{8}\, M_{\odot }  $
for model 27. For comparison, the $  M_{\rm c}/M_{\rm h}  $ ratio for model
2 is $  50  $. Because numerical models and observations of zero-redshift galactic
nuclei suggest cluster to hole mass ratios greater than or approximately equal
to unity, model 26 ($  M_{\rm c}/M_{\rm h}=0.10  $) is probably only hypothetical;
it was computed merely to illustrate what would happen without the expected
stellar cluster. 

In each model, the stellar densities at $  r=r_{\rm t}  $ were normalized to
be identical. With this parameterization, $  M_{\rm c}/M_{\rm h}  $ simply
regulates the core radius $  r_{\rm c}  $ (see eq. {[}\ref{Inputdensity}{]}).
Thus, increasing $  M_{\rm c}/M_{\rm h}  $ extends the cluster outwards, while
decreasing $  M_{\rm c}/M_{\rm h}  $ makes the cluster smaller (models 26 and
27 have density functions flatter than $  r^{-2}  $). These features are shown
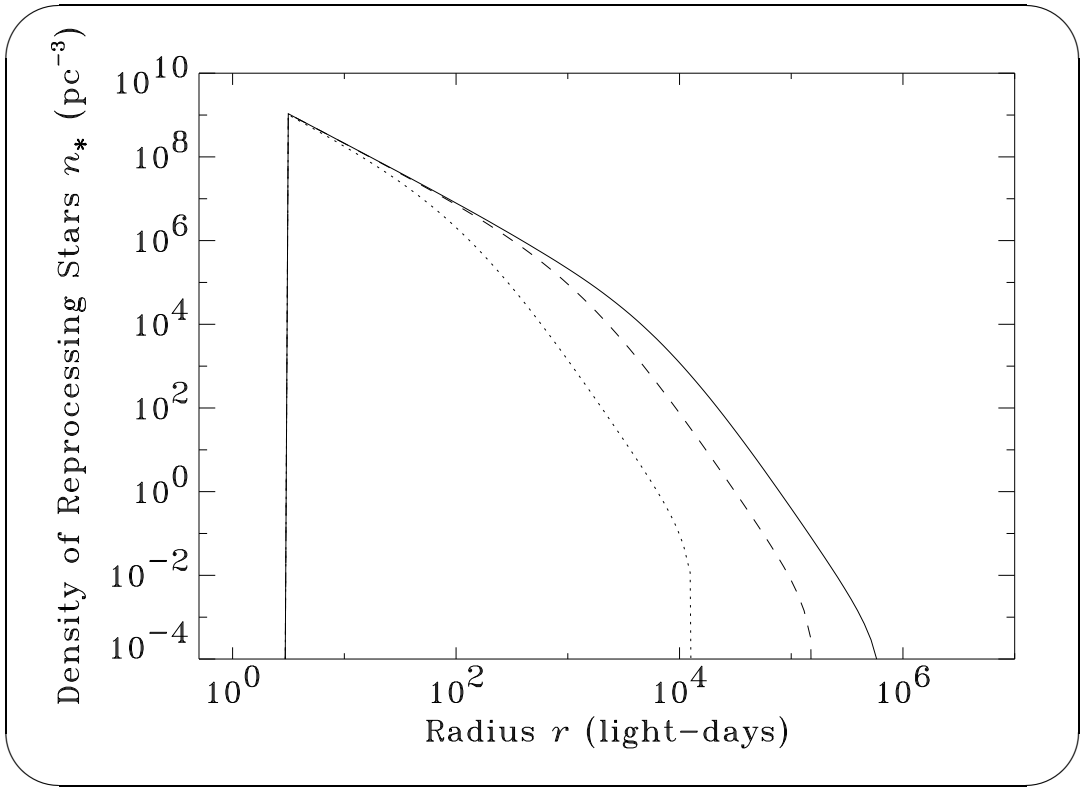
in \Fig{model26/density}{Dotted line: density of stars with reprocessing stellar
winds for model 26 ($  M_{\rm c}/M_{\rm h}=0.10  $). Dashed line: density for
model 27 ($  M_{\rm c}/M_{\rm h}=5.0  $). Solid line: density for model 2 ($  M_{\rm c}/M_{\rm h}=50  $).}. 

The Ly$  \alpha   $ profiles of models 26 and 27 are shown in \Fig{model26/Lya}{Ly$  \alpha   $
line profiles for different cluster masses. Dotted line: model 26 ($  M_{\rm c}/M_{\rm h}=0.10  $).
Dashed line: model 27 ($  M_{\rm c}/M_{\rm h}=5.0  $). Solid line: model 2
($  M_{\rm c}/M_{\rm h}=50  $).}. The Ly$  \alpha   $ narrow component (already
weak in model 2) is not apparent in the low-$  M_{\rm c}/M_{\rm h}  $ model.
This is because this model has little covering in its NLR. Ironically, at a
given position, the stars in the low-$  M_{\rm c}/M_{\rm h}  $ model move slower
than in the other models. This is because the distribution function of this
model is not influenced by the potential well of its cluster. This is shown
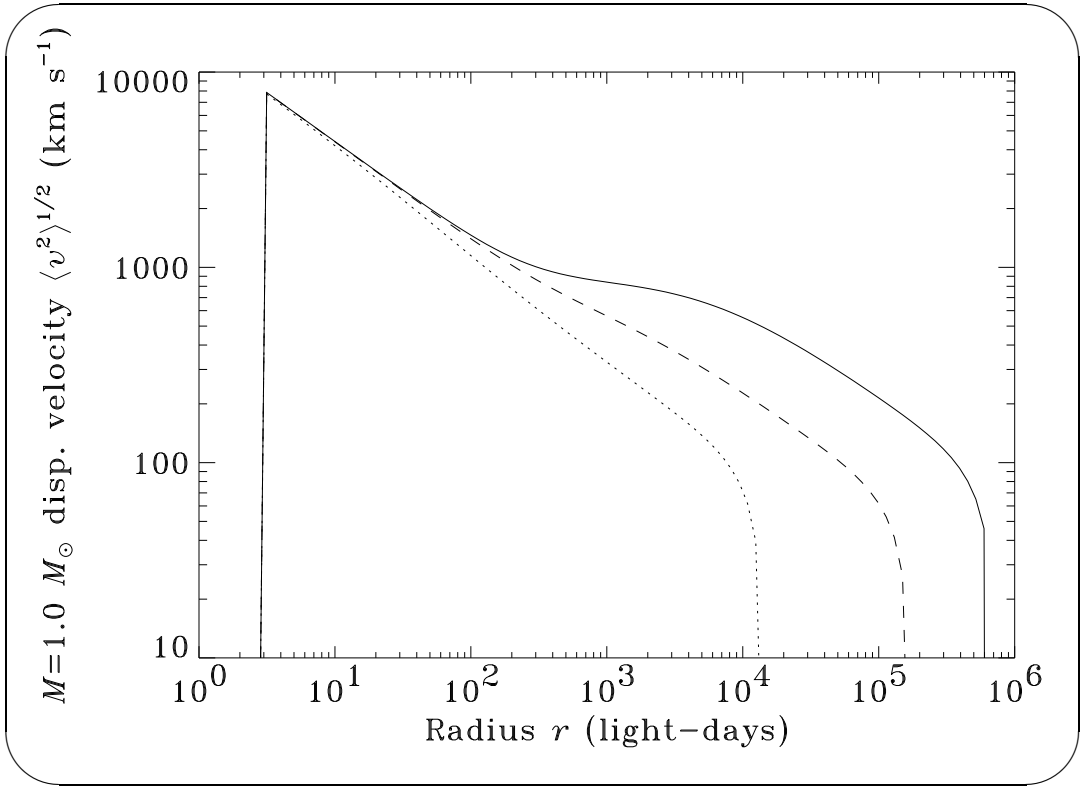
in \Fig{model26/vel-rdays}{Dotted line: dispersion velocity of solar-mass stars
for model 26 ($  M_{\rm c}/M_{\rm h}=0.10  $). Dashed line: dispersion velocity
for model 27 ($  M_{\rm c}/M_{\rm h}=5.0  $). Solid line: dispersion velocity
for model 2 ($  M_{\rm c}/M_{\rm h}=50  $). Note that the model 26 stellar
cluster is light enough that the velocity dispersion falls as $  r^{-1/2}  $
everywhere.}. For models with substantial narrow line covering, small variations
in $  M_{\rm c}  $ or $  r_{\rm c}  $ regulate the width of the narrowest
profile component. This width is simply the velocity where the log-linear profile
begins to flatten. Increasing $  M_{\rm c}/M_{\rm h}  $ increases both this
width and the overall covering in the NLR.

Most of the UV lines are less affected by the $  M_{\rm c}/M_{\rm h}  $ ratio
than is Ly$  \alpha   $. This is the case, for instance, with the C {\small IV}
line. As \Fig{model26/civ}{C {\small IV} profiles for different cluster masses.
Dotted line: model 26 ($  M_{\rm c}/M_{\rm h}=0.10  $). Dashed line: model
27 ($  M_{\rm c}/M_{\rm h}=5.0  $). Solid line: model 2 ($  M_{\rm c}/M_{\rm h}=50  $).}
shows, the profile of this line is nearly the same for both the $  M_{\rm c}/M_{\rm h}=5.0  $
and $  M_{\rm c}/M_{\rm h}=50  $ models. The reason for this is simply that
the line is not strong in the NLR. 

As one might naively expect, the response functions at delays less than $  \sim   $
100 days are even less affected by the $  M_{\rm c}/M_{\rm h}  $ ratio. This
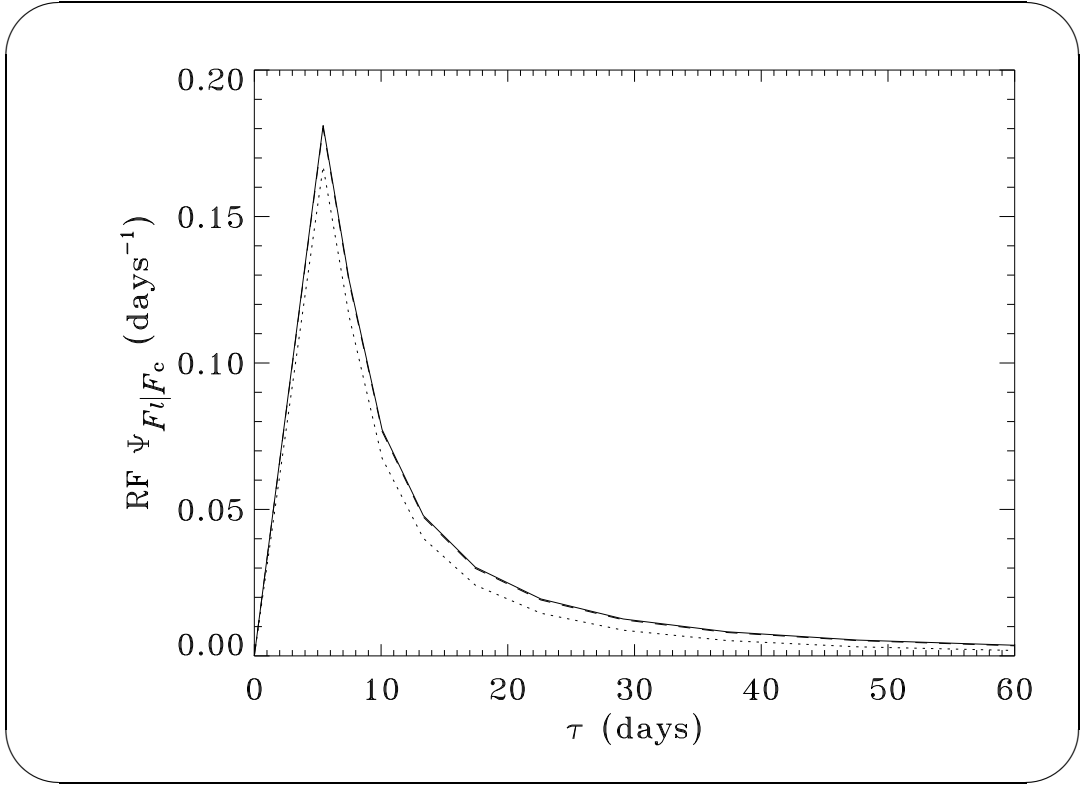
is shown in \Fig{model26/summedlinerf}{Response functions of the summed line
for different cluster masses. Dotted line: model 26 ($  M_{\rm c}/M_{\rm h}=0.10  $).
Solid line: model 27 ($  M_{\rm c}/M_{\rm h}=5.0  $) and model 2 ($  M_{\rm c}/M_{\rm h}=50  $).
The results are similar for each model because the $  M_{\rm c}/M_{\rm h}  $ parameter
primarily affects only the NLR, not the BLR.} for the summed line .\pfn{Model
28: a Model With a Steep Stellar Slope \label{modelwithsteepslope} 

The stellar density function of model 28 is shown in \textbackslash{}Fig\{\}\{\}.
THIS WOULD BE NICE BUT DO I REALLY HAVE THE TIME FOR IT???? NO.}

\chapter{Summary and Discussion\label{summaryanddiscussion} }

We have calculated the observational features of the stellar wind model proposed
by Norman \& Scoville (1988) \& Kazanas (1989). The main results of our study
can be summarized as follows:

\begin{enumerate}
\item The covering function $  An_{*}  $ in the Kazanas (1989) model is too low and
does not drop fast enough with increasing radial distance to match typical AGN
line profile strengths and shapes. Using a stellar distribution function designed
to be compatible with what is expected for Seyfert-class galactic nuclei, we
obtain a covering factor of only 0.04. This is one-tenth the covering typically
observed in Seyfert 1s. Moreover, the covering slope is very high ($  d\textrm{ln}[An_{*}]/d\textrm{ln}r\simeq +0.6  $),
so this covering corresponds to emission from the NLR rather than the BLR (\S~\ref{model1}). 
\item If the conditions in AGNs are, for poorly understood reasons, severe enough
that wind mass loss rates of stars nearest to the black hole are substantially
enhanced, then both the covering and its integral change. Assuming $  \dot{M}\propto (L/r^{2})^{\alpha }  $,
where $  \alpha =1.15  $, we obtain a covering factor of 0.28 and a covering
slope of $  d\textrm{ln}(An_{*})/d\textrm{ln}r=-1.8  $. The resulting optically
thick area functions are approximately constant. This change results in line
profile strengths and shapes similar to those in NGC 5548 (\S~\ref{model2}). 
\item Though a $  \chi ^{2}  $ fit of the various stellar wind AGN model parameters
to the NGC 5548 spectral data was not performed in this study, the line ratios
of model 2 are similar to those of NGC 5548. The primary reason for this is
that model 2 has $  \Xi =1.0  $ instead of $  \Xi =10  $ as was originally
postulated in Kazanas (1989) (\S~\ref{model2}). Even with this change, however,
there were important differences between model 2 and the NGC 5548 data. For
instance, the N {\footnotesize IV} $  \lambda   $1486 line peak is a factor
of 3.7 times stronger than that in NGC 5548, while the O {\footnotesize III}{]}
$  \lambda   $1664 line flux peak is a factor of 2.5 times stronger. Some of
these discrepancies are probably due to the high metallicity we assumed. However,
no adjustment of the abundances can resolve all of the line ratio discrepancies.
In particular, Ly$  \alpha   $/H$  \beta   $, which is 20.7 in model 2, is
significantly higher than the observed ratio of ratio 12.6$  \pm 2.0  $. This
is despite the fact that the Ly$  \alpha   $ line in model 2 is highly suppressed
by thermalization. Also, the C {\footnotesize III}{]}/C {\footnotesize IV} line
ratios of the broad components of the profiles of our models are lower than
those observed in NGC 5548. This problem arises because the continuum flux in
the broad line region of NGC 5548 is so high for the distance we assumed (Krolik
et al.~1991). It occurs despite our accounting for cloud/wind regions that are
both angled and optically thick to continuum radiation (\S \S~\ref{iontopratio} \&
\ref{L&P}).
\item The wind response functions and profiles of the stellar wind models we analyzed
have a near-perfect red to blue symmetry and are not significantly shifted.
Though self-absorption features were not accounted for, under the assumptions
of the models they would be very weak and also not shifted (\S~\ref{lineprofilessection}, Appendix
\ref{adpbal}; cf Appendix \ref{shiftsap}). These models, therefore, are unable
to match the shifts and asymmetric response features in NGC 5548 (and other
well-studied AGNs), such as the blueshift of the C {\small III}{]} profile peak
and the difference between the C {\small IV} red and blue wing delays (\S~\ref{model2}). 
\item Due to the highly asymmetric cloud line emission and an assumed optically thin
intercloud BLR, our models, which for simplicity ignore local delays (cf Appendix
\ref{localdelays}), yield response functions that are approximately zero at
$  \tau =0  $ (\S~\ref{model2}). This result is different from the published
response functions of NGC 5548 that probed relatively high frequencies and which
yielded C{\footnotesize ~IV} response functions that peaked at $  0\pm 2  $
days (Wanders et al. 1995; Done \& Krolik 1996). This difference is less severe
in models in which both $  r_{\rm t}  $ and $  M_{\rm h}  $ are artificially
lowered (\S \S~\ref{tidal section} \& \ref{m_h}). However, such models have
three problems. First, they have broad to narrow intensity C {\small IV} response
function peak ratios of $  \gsim 1.8  $ (\S~\ref{model2}). The C {\small IV}
response functions of both Wanders et al. (1995) and Done \& Krolik (1996) have
ratios of $  \sim 1.0  $. Second, such models have C {\footnotesize IV} response
function FWHMs of $  \lsim 8  $ days, rather than the $  \sim 15  $ days suggested
by observations. Third, $  r_{\rm t}  $ and $  M_{\rm h}  $ must be lowered
by equal fractions to prevent the profiles from narrowing. Wind models predict
a much different relation of $  r_{\rm t}\propto M_{\rm h}^{1/3}  $ (\S~\ref{m_h}).
However, since the proportionality constant in this equation is not tightly
constrained in the first place, this last problem is probably not significant.
\item If $  s  $ is treated as a free parameter, the profile shapes suggest $  s\simeq 1.7  $,
where $  P\propto r^{-s}  $. While this agrees with the Kazanas (1989) prediction
of $  s\simeq 2.0  $, it should be pointed out that some of the line flux ratios
in the $  s\sim 1.0  $ models we computed came closer to matching those in
NGC 5548 (\S~\ref{s}).
\item It is possible that the stellar mass functions in AGNs are ``top heavy.''
In this case, our calculations, which ignore winds from hot stars, may be incorrect,
especially for application towards Seyfert 2s and LINERs (Appendix \ref{Ostararea}).
The implied IMF gradients are probably too weak to affect the covering functions
significantly (Appendix \ref{commentsonaandn}). They are, however, probably
strong enough to produce severe radial dependences in the mass to light ratios
(Appendix \ref{darkma}) and high QSO metallicities (Appendix \ref{agnabundances}).
\item Each of the various wind models yield specific and falsifiable predictions regarding
the scaling of the response functions with the various AGN parameters. For wind
models similar to model 1 (which has $  \alpha -s/2\simeq -1  $), we obtain
$  \tau _{\rm p}\propto L^{1/2}  $, where $  \tau _{\rm p}  $ is the delay
of the response function peaks (\S~\ref{luminosity}). Since this delay is approximately
proportional to the characteristic delay of the response function, it agrees
with the empirical results of Kaspi et al. (1996). Models similar to our preferred
model (model 2, which has $  \alpha -s/2\sim 0  $) yield $  \sigma _{\rm t}\propto M_{\rm h}^{1/3}  $
and $  \tau _{\rm p}\propto \sigma _{\rm t}  $ (eq. {[}\ref{tau-sigmarelation}{]}).
This implies that the black holes masses of NLS1s are relatively low. In model
2, the tidal radius was increased by a factor of $  1.8  $ above the value
in the CMD runs. This supports the stellar wind line emission model because
this increase is much smaller than $  r_{\rm t}/(3r_{\rm S})=270  $, where
$  3r_{\rm S}  $ is the only length scale common to most hypothesized AGN models
with the possible exception of the pressure- and luminosity-dependent $  \Xi >\Xi ^{*}_{\rm c}  $
one. These results suggest that measuring the correlations between $  <\tau >  $
and $  \sigma   $ in various AGNs should prove useful in testing the viabilities
of the various wind models (\S \S~\ref{tidal section} \& \ref{m_h}); reverberation
studies exploring both of the $  <L>  $ and $  M_{\rm h}  $ dimensions (which,
despite the obvious statistical correlation, are actually separate epoch-dependent
parameters for most Seyferts) should be performed. The recent interests in NLS1s
and QSO variability are good starting points in this endeavor. Future efforts
to perform velocity deconvolution (Appendix \ref{veldtempa}) and absorption
reverberation mapping (Appendix \ref{adpbal}) will also help.
\end{enumerate}
~

Some of the results for model 1 are similar to those found in AN94 and AN97,
which also presents models of AGN line emission from stellar winds. For instance,
model 1 has narrow profiles that are deficient in low-ionization broad line
emission. On the other hand, our results for model 2 are different from their
results. These differences arise primarily because of differences in our assumptions.
We invoke a position-dependent ionization parameter in an attempt to let the
stellar wind evaporation rates balance the mass loss rates (\S~\ref{area}).
Also, instead of permitting the reprocessing stars to have a special density
function or terminal mass loss velocity, we assume that the BLR conditions are
extreme enough to cause the mass loss rates to vary with radius (\S~\ref{area},
Appendix \ref{commentsonaandn}). As a result, the winds in the inner part of
the BLR of our model 2 have more covering per stellar wind than those in AN94
and AN97.

From some perspectives, the stellar wind model is a failure. A basic assumption
of the model, for instance, is that the wind areas increase with distance from
the black hole. Such models do not appear to match observations. This failure,
however, is less severe than the failures of most of the BLR models discussed
in Chapter 1, which appear to have even more fundamental problems, such as violating
well-established principles of physics. Moreover, the resolution of the main
problems of model 1, which involved assumptions about the possible complicated
interactions between the stellar winds and the hostile environment of AGNs,
can be viewed more as an engineering problem than one of basic physics. From
this perspective, the wind AGN model must be regarded as a viable, though, unfortunately,
not simplistic solution to the AGN line emission problem. Ideally, a more complete
analysis of the structure of heated winds from stars moving supersonically through
the intercloud medium would be simulated using a rigorous MHD/radiative transfer
code. Such an approach might answer some of the basic questions that have been
raised by this study. As long as this engineering problem remains unsolved,
it will probably be difficult to make firm assessments of the viability of this
promising AGN model.

\begin{appendix}

\chapter{UV Cross Sections of Stellar Winds for Various Stars\label{Ostararea} 
}

The first step in modeling the AGN emission line region with heated stellar
winds is to determine which types of stellar winds are expected to have the
largest contribution towards line emission. In this appendix, we describe why
only red giant and red supergiant stellar winds were taken into account in our
calculations of line emission from stellar winds in AGN.  

There are two distinct possible types of line emission from stellar winds in
an AGN: 

\begin{enumerate}
\item the ``reprocessed'' line emission that occurs due to line cooling after heating
from the UV continuum radiation and 
\item the ``intrinsic'' line emission that would occur even without external heating
from the AGN continuum source. 
\end{enumerate}
Reprocessed emission should be important for stellar winds which have a substantial
UV cross section and which would be heated by the AGN continuum. Provided the
heated wind temperature remains below $  \sim 10^{5}  $ K, this additional
heating should be balanced by additional line emission cooling. For completeness,
both reprocessed and intrinsic line emission should be included in any fully
self-consistent model. However, if the stellar wind AGN model is successful
at describing the BLR, the reprocessed line emission must dominate the intrinsic
emission, as broad line emission in quasars and Seyfert 1s is observed to vary
with the local continuum intensity. Narrow line emission, which has not yet
been observed to vary with the long-term, time-averaged continuum level, does
not currently have this additional constraint.  

The strongest stellar winds are O-star winds. O-star winds are frequently approximated
as being isothermal. The Eddington approximation for an LTE grey atmosphere
yields a wind temperature of $  T_{\rm {wind}}(\tau =0)=(1/2)^{1/4}T_{\rm {eff}}\sim 0.8T_{\rm {eff}}  $.
More detailed models indicate that temperatures in radiatively driven winds
should decrease slightly with increasing radius. They also predict an ionization state of the gas that
increases with radius until the effects from the interstellar medium become
important. O-stars are too hot to contain neutral hydrogen and are, therefore, optically
thin to UV continuum radiation near 1000 \AA. Because this region of the spectrum
is near the apparent AGN ``big blue bump'' $  \nu F_{\nu }  $ peak, most
stellar winds should have an effective absorptive UV cross sectional area only
as large as their associated radius. Thus, the situation for UV-heated O-star
winds is surprisingly similar to that for UV-heated coronal winds like that
of the sun.  

At wavelengths below 504 and 228 \AA, however, the He I and He II ionization
edges give significant opacity, which can cause a hot stellar wind to be optically
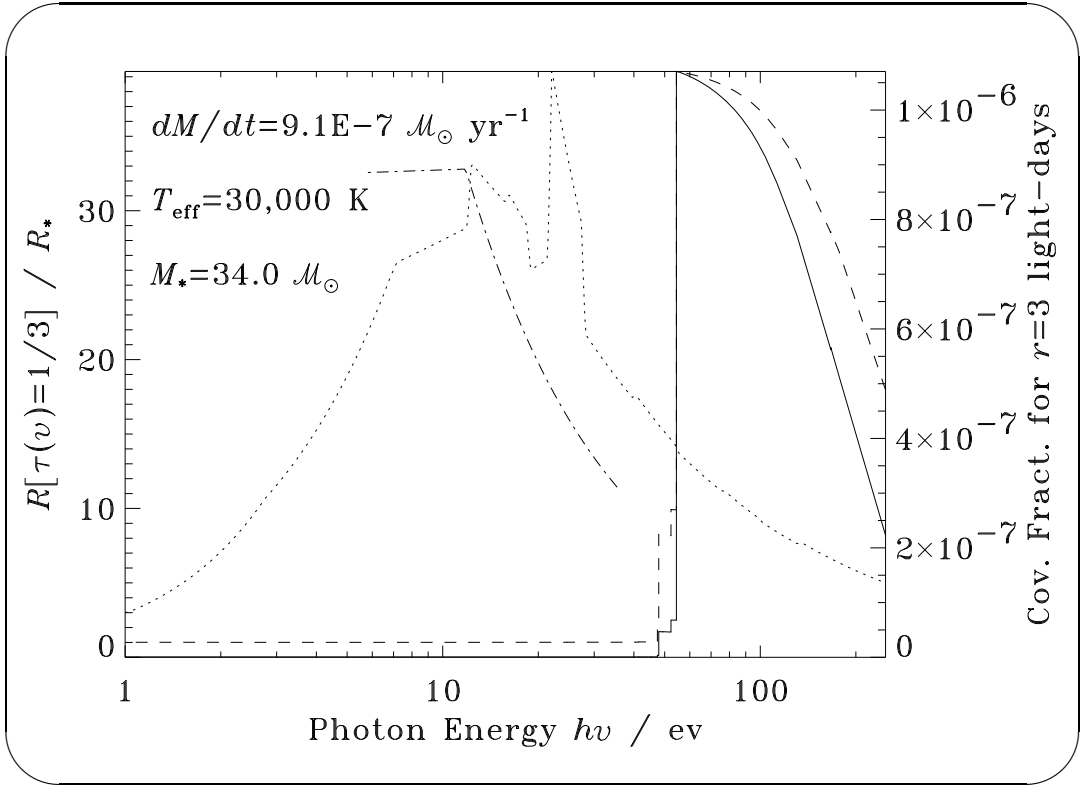
thick in this region. This is shown in \Fig{../stellarmodels/O-starcovering}{Radius
of the photosphere of a $  M_{*}=34.0\, M_{\odot }  $, $  T_{\rm {eff}}=30,000  $
K, $  \dot{M}=9.6\times 10^{-7}\, M_{\odot }  $ yr$  ^{-1}  $ stellar atmosphere
(solid line, left axis) as a function of photon frequency $  \nu   $. The dotted
line is the AGN continuum assumed for the models shown in this dissertation.
The dot-dashed line shows the composite AGN continuum from Zheng et al. (1997).
The dashed line and right axis give the covering fraction due to this specific
stellar wind. Note that, provided these models apply to O-star winds in AGN,
even with 10$  ^{6}  $ stars of this type at a distance of 3.0 light days from
the continuum source, only $  \sim 1/3  $ of the radiation from these continua
could be absorbed. This data was kindly supplied by A. de Koter using version
3.27 of the ISA stellar wind code (e.g., Schaerer \& de Koter 1997).}. In AGNs
with continua in the BLR that are similar to the ones shown in Figure \ref{../stellarmodels/O-starcovering},
which do not have substantial EUV/soft X-ray flux, O-star winds should be unaffected
by the additional AGN radiation. For this reason, and also to simplify calculations,
line emission from O-star winds was neglected in this work. However, determining
the actual continua in the BLR depends upon estimates of the intrinsic absorption
along the line of sight. For instance, if the He I and He II column from the
broad line region is high enough that the AGN spectra we observe have already
been significantly absorbed below the Lyman limit, line emission from heated
O-star winds would be more important than is assumed in this work. Incidentally,
recent NTLE theoretical estimates of AGN accretion disk spectra do in fact predict
$  10^{1}-10^{2}  $ times more He II ionizing photons than the older LTE accretion
disk models (Hubeny \& Hubeny 1997).

In contrast to stellar winds from O-stars, winds from cooler late-type stars
can have a significant fractional abundance of neutral hydrogen. This permits
such stars to be much more efficient at absorbing AGN continuum radiation. Therefore,
in an AGN, the late-type stars should be important sources of UV line emission.
Moreover, over the course of the $  M\lsim 20\, M_{\odot }  $ lifetime of a
star, mass loss is greatest while in the giant phase of its evolution. This
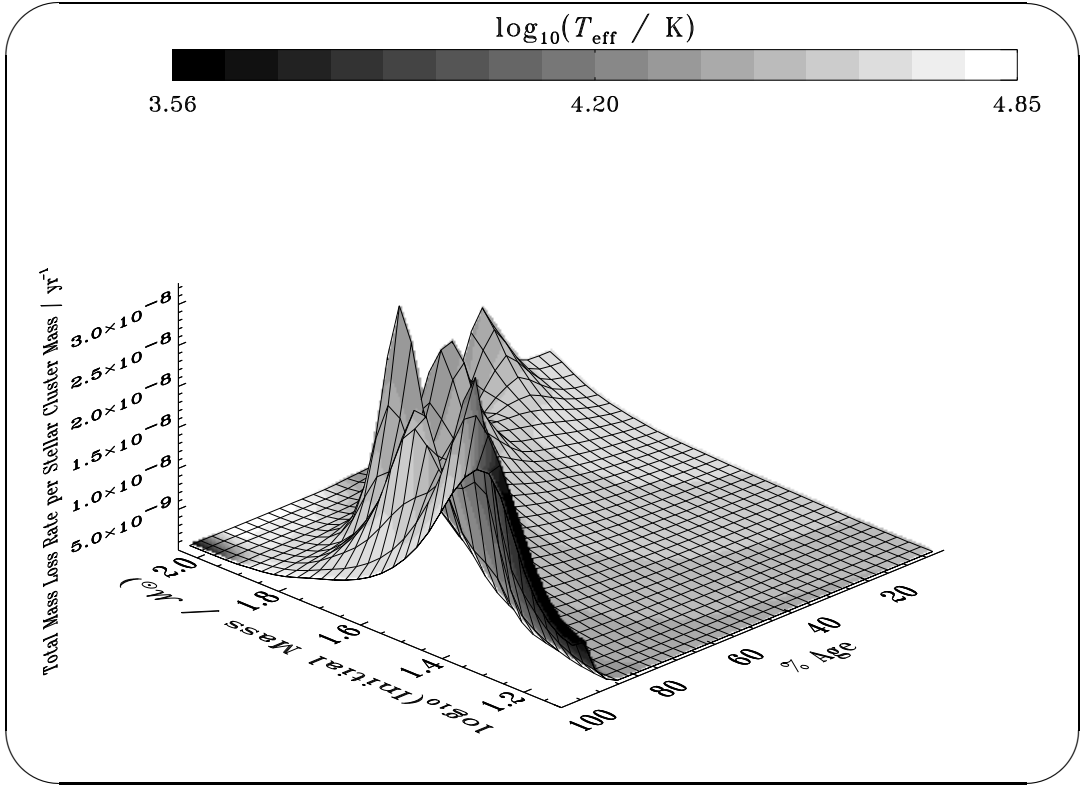
is shown in \Fig{../stellarmodels/saltpetermasslossandtemp}{Integrated stellar
wind mass loss rates for massive stars as a function of scaled age and logarithm
of the initial (post-Hayashi track) mass. The integrated volume under the surface
is the mass loss of the stellar winds per unit stellar cluster mass.  This plot
is particularly useful in determining the integrated mass loss when the stellar
mass density is well constrained but the stellar present day mass function is
not. Under the assumptions made, injection into the interstellar medium from
stellar winds is produced from stars with a wide range of initial masses, but
with the heaviest stars being least important due to their sparseness. The darkness
of the surface indicates the effective temperatures of the stellar photospheres.
 With the exception of $  M>70\, M_{\odot }  $ stars near the ends of their
fusion-burning lifetimes, the mass loss for a given star is highest when the
effective temperature is at its lowest. This is interesting because it is precisely
the low-temperature winds that would have the largest UV cross sections. The
effective temperature of the stars with winds that dominate the mass loss is $  T_{\rm {eff}}\la 10^{4.3}  $
K. This data is from Meynet et al. (1994) for mass loss rates twice the value
predicted from simple photoionization codes and with $  z=0.04  $. The mass
function index was assumed to be be $  dN/dM\propto M^{-2.35}  $ ($  \Gamma =-1.35  $)
with $  M_{\rm {lower}}=0.1\, M_{\odot }  $, in crude accordance with Salpeter
(1955).}. This Figure also shows that the stellar winds from O-stars are too
hot to be optically thick at the $  \nu F_{\nu }  $ peak and the cooler winds
from red giants contribute at least as much wind mass loss. For these and other
reasons it was decided that our calculations would not account for line emission
from O-star winds.

The mass loss rates shown in Figure \ref{../stellarmodels/saltpetermasslossandtemp}
are actually only valid if the mass function is similar to that of Salpeter
(1955). For reasons discussed in Appendices \ref{commentsonaandn} \& \ref{darkma},
the mass function in AGN may, however, be ``top heavy,'' with a higher (less
negative) exponent in the stellar density distribution function. The consequences
of a top heavy present day mass function upon the wind mass loss rates are shown
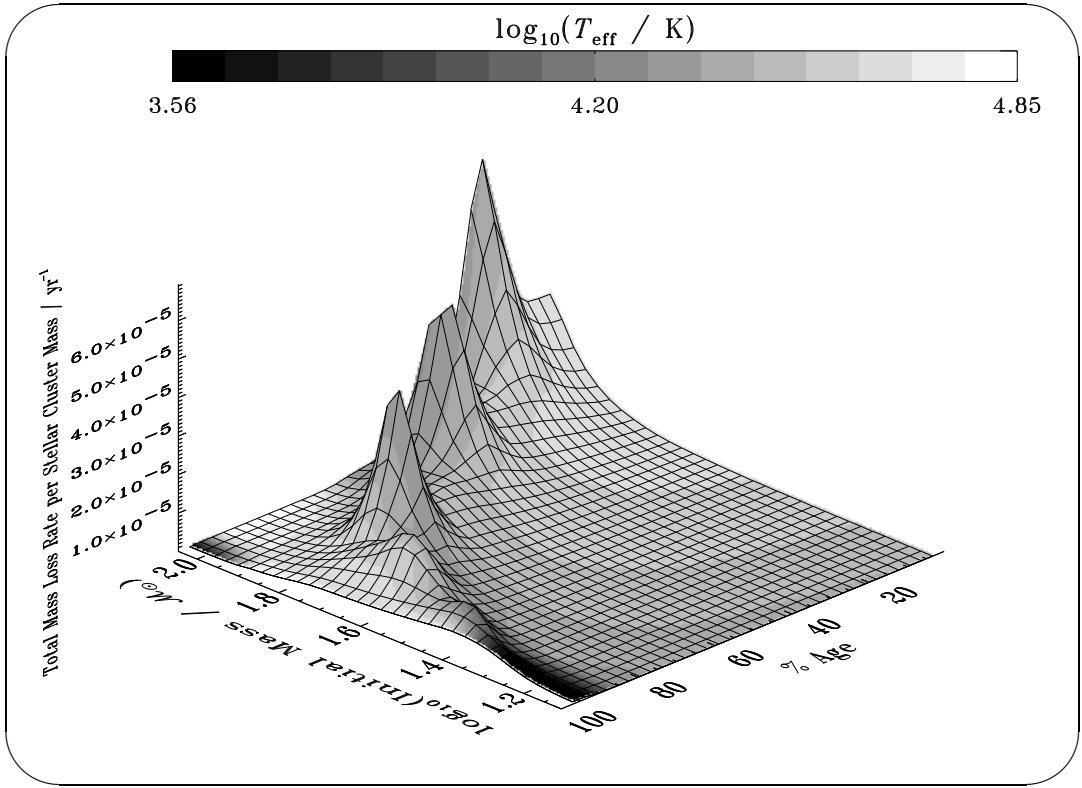
in \Fig{../stellarmodels/agnmasslossandtemp}{The integrated stellar mass loss
rates as in \ref{../stellarmodels/saltpetermasslossandtemp}, but under the
assumptions of a very ``top heavy'' mass function index of -0.5. For this
mass function, mass loss from $  T_{\rm {eff}}\sim 10^{3.6}  $ K red giants
is much less than the mass loss from the warmer $  T_{\rm {eff}}\sim 10^{4.3}  $
supergiants. Note also that the normalization is a factor of $  \sim   $2000
times higher than that for the mass function assumed in Figure \ref{../stellarmodels/saltpetermasslossandtemp}.
This is primarily due to the existence of non-plotted winds from $  M<12\, M_{\odot }  $
stars and $  M_{\rm {lower}}=0.1\, M_{\odot }  $. However, because massive
stars are so much more efficient at recycling mass into the interstellar medium
and their lifetimes are so short compared to solar-mass stars (see Figure \ref{../stellarmodels/efficiency}),
a top heavy mass function still results in enhanced $  M\lsim 10\, M_{\odot }  $
mass loss rates.}. Not only would a top heavy mass function increase the contribution
that winds of $  T_{\rm {eff}}>10^{3.8}  $ K stars might have upon the reprocessed
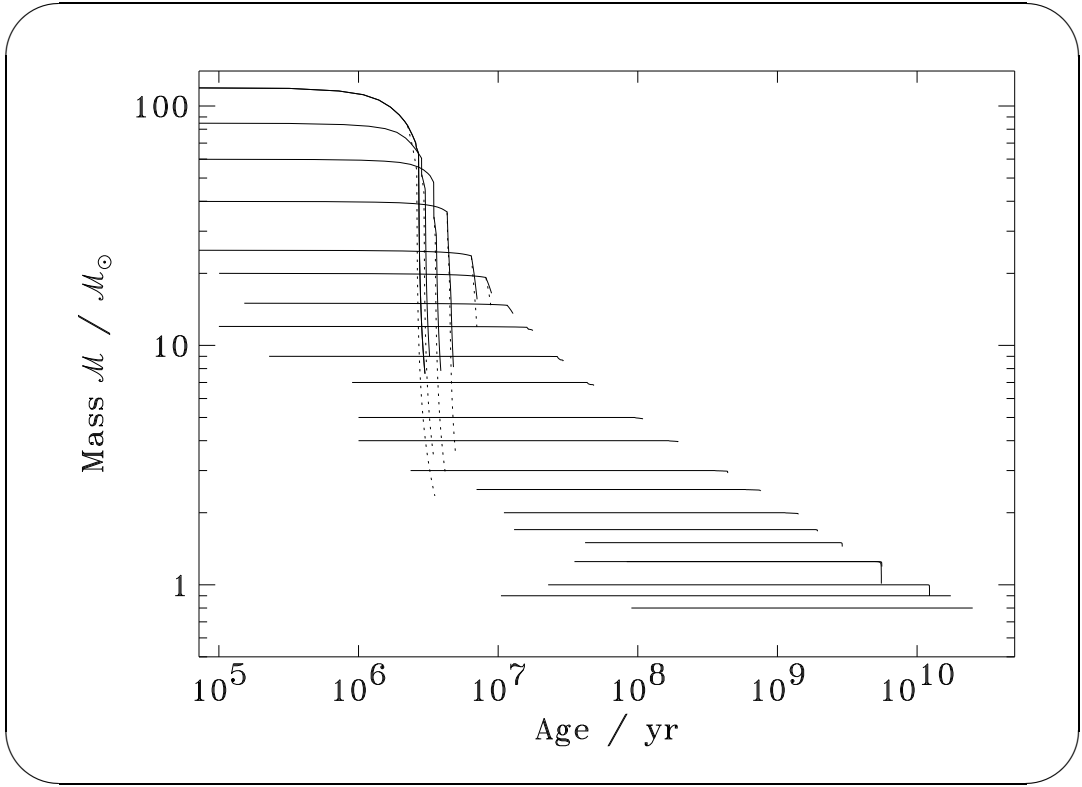
line emission, but, as is shown in \Fig{../stellarmodels/efficiency}{Integrated
mass loss as a function of stellar age with overshooting for $  z=0.02  $.
Models with enhanced post main sequence mass loss rates are shown as dotted
lines. The post main sequence enhanced mass loss stars with initial masses of
$  M=120\, M_{\odot }  $, $  M=85\, M_{\odot }  $, $  M=60\, M_{\odot }  $,
and $  M=40\, M_{\odot }  $ each have ``wind conversion efficiencies'' above
90\%. Data from Schaller et al. (1992).}, it would also increase the overall
mass lost from stellar winds. In fact, one of the issues involved with the possibility
of having an AGN mass function as ``top heavy'' as that shown in Figure \ref{../stellarmodels/agnmasslossandtemp}
is that the total mass injection rate from the winds of the cluster is several
decades higher than the Eddington luminosity accretion rate. Incidentally, this
is a problem only for models in which all of the mass lost from stellar winds
is assumed to be accreted by the supermassive black hole---models in which,
e.g., the hot, inter-cloud/stellar medium (HIM) is blown outwards into a superwind
and recycled into massive stars would not suffer from this problem. Note that
before massive stars become red supergiants, they would emit significant intrinsic
line emission. Thus, a top heavy mass function would increase both the reprocessed
and intrinsic types of stellar wind line emission. Therefore, LINERs and Seyfert
2s could simply be the result of line emission from hot stars provided $  \sim   $50\%
of galactic nuclei have very top-heavy mass functions.

In conclusion, winds from O-stars might be important contributors to AGN line
emission if AGN mass functions are top heavy and the BLR soft X-ray continuum
flux is high. Nevertheless, due to the current uncertainties in the present
day mass function in AGN and the additional complexities/parameters involved
with properly modeling warmer winds exposed to continuum radiation, we have
not attempted to account for such effects---only winds from red giant stars
were accounted for in our models. Future research in these areas is, however,
certainly warranted.

\chapter{Comments on the Spatial Distribution of ``Bloated Stars''\label{commentsonaandn}
}

As discussed in \S\S~\ref{f} and \ref{eta}, all the observables of the stellar
wind model except for the line ratios are strong functions of the particular
choice of stellar density functions. In Alexander \& Netzer (1994, 1997; hereafter,
``AN94'' \& ``AN97''), the winds emanating from the BLR clouds were assumed
to be caused by stars ``bloated'' by the AGN central source. The fundamental
physical properties (such as the mass loss rates) of the bloated stars were
assumed to be independent of radius. The inner mass density function assumed
in AN94 and AN97 is based upon the stellar densities of the models of Murphy,
Cohn, \& Durisen (1991; hereafter ``MCD'') for $  M=0.8M      _{\odot }  $
stars. It is $  \propto r^{-1/2}  $ for the first decade in radius, $  \propto r^{-1}  $
for the next 1.5 decades (farther out in radius), and $  \propto r^{-4}  $
at larger distances. In AN97, the number density of bloated stars was assumed
to be two powers in $  r  $ steeper than the mass density function. From an
observational perspective, this extra steepness affects the models in a fashion
similar to the way in which a radius-dependent mass loss rate (as was essentially
done in this dissertation) affects the models. It permits line profiles and
response functions that agree with observational data. In this appendix, we
attempt to estimate the viability of the AN97 bloated star density function.
To do this, we first discuss the likely masses of bloated stars. We then discuss
the theoretical and experimental indications of what the bloated star spatial
densities might be in AGNs. Finally, we estimate the approximate likeliness
that the relative stellar density gradients could have the magnitude and sign
that were assumed in AN97. We find that this likelihood is small.

\section{Fundamental Issues}

Before addressing the question of the density function that bloated stars might
have, some of their fundamental physical properties, such as their masses, must
first be estimated. AN94 suggest that their bloated stars may be red giants
or supergiants that have been bloated by continuum heating due to the AGN. A
detailed discussion of stellar bloating is provided in \S~\ref{Heated stars?}.
Bloating\rfn{Where, to be precise, the word ``bloating'' is used here to denote
an increase in photosphere radius.} can occur in the convective region of stars
when the external continuum heating is sufficiently high. Kaspi et al. (1996)
shows that most AGNs have luminosity-independent BLR fluxes in the range of
10$  ^{8}  $-10$  ^{12}  $ ergs s$  ^{-1}  $ cm$  ^{-2}  $. These numbers
are compatible with an AGN that has, e.g., a luminosity of 10$  ^{44}  $ ergs
s$  ^{-1}  $, a BLR inner radius of 1 light day, and a BLR outer radius of
30 light days. According to the results discussed in Chapter 1 regarding stellar
bloating, this implies that, for stars in hypothetical circular orbits in the
BLR, stellar bloating could only be substantial for main sequence stars less
massive than $  \sim 1\, M_{\odot }  $. Bloating increases the radii of the
low mass stars by only $  \sim   $20\%. The relevant time scale for this bloating
to occur is the Kelvin-Helmholtz time scale of the convective envelope, which
is $  \sim 10^{7}  $ years. As AN97 point out, the BLR stars are likely to
be on highly eccentric orbits. The BLR crossing times are of the order of a
year, which is much smaller than this bloating time. Therefore, the continuum
flux seen by the star (in regards to any possible bloating) is the orbit-averaged
flux, which is likely to be much lower than the $  \gsim   $10$  ^{11}  $
ergs s$  ^{-1}  $ cm$  ^{-2}  $ required to affect $  M>1.0\, M_{\odot }  $
main sequence stars. Note that low mass stars normally have such low mass loss
rates and resultant theoretical line-emitting cross sections that even with
extreme bloating they should still be insignificant contributors to AGN line
emission. Thus, bloated main sequence stars are probably not good AGN line emission
candidates if the heating is due to the continuum radiation. 

AN94 mention that MacDonald, Stanev, \& Biermann (1991) have suggested that
neutrino heating might actually be the primary cause of stellar bloating. There
are at least three arguments that can be made against this proposition. First,
since the expected reaction pathways for high-energy neutrino production may
result in a neutrino to bolometric luminosity ratio that is much smaller than
unity, the effects of possible neutrino heating may intrinsically be much less
important than those due to any continuum heating. Second, at least in some
models, the neutrinos would not be radiated isotropically, but rather would
be emitted along the jets. In this case, only a small fraction of stars could
be affected. A third and more model-independent argument that could be made
against neutrino-bloated stars is that for such models the covering factors
should be a strong function of the neutrino luminosity. The neutrino to bolometric
luminosity ratio is expected be positively correlated with the $  \gamma   $-ray
to bolometric luminosity ratio. Approximately $  \sim   $10\% of AGNs (only
the radio-loud ones) appear to have substantial high-energy nonthermal radiation.
Yet the covering factors in these objects, which presumably should have relatively
high neutrino fluxes, does not appear to be significantly higher than those
of the radio-quiet objects. 

For the above reasons, we can probably rule out main sequence stars as bloated
star candidates. This implies that bloated stars, if they exist, are (as AN94
\& AN97 indeed proposed) probably late-type stars. This in turn implies $  M\gsim 1\, M_{\odot }  $,
because $  M\lsim 1\, M_{\odot }  $ stars have main sequence lifetimes greater
than $  \sim   $10$  ^{10}  $ yr. Moreover, if the bloated stars are similar
to red supergiants in particular, they are ``intermediate-mass'' stars, with
$  M\ga 8\, M_{\odot }  $. For simplicity, we will henceforth assume this to
be the case.

Because this assumed mass range is so high, bloated stars are probably very
scarce. The precise shapes of the bloated star density functions are functions
of the position-dependent present day mass functions in AGNs. In the next two
sections we discuss some of the theoretical and empirical constraints that can
be placed upon these functions and the density functions of bloated stars in
particular.

\section{Theoretical Considerations
}

Though the mass functions of stars in AGNs are not well known, several theoretical
constraints can be made. The present day bloated star mass functions should
(according to the above assumption) be proportional to the main sequence lifetime
of a $  M\simeq 8\, M_{\odot }  $ star, which is approximately $  3\times 10^{7}  $
years.\rfn{Incidentally, $  \sim 10\%  $ of galaxies are Seyfert 1s and AGNs
are estimated to be active for a much longer duration of $  \sim 10^{10-1}=10^{9}\, \textrm{years}  $.
Therefore, the observation of these intermediate-mass stars in galactic nuclei
implies that star formation in AGNs, unlike that which has been proposed for
globular clusters, must be an ongoing process.}\pfn{Therefore, in AGNs the red
supergiant density function is probably not directly a function of galactic
age.}

The density functions are also dependent upon the particular AGN conditions.
Theoretical studies by MCD indicate that there are two extreme cases to consider
in AGNs regarding stellar density functions. For stellar densities significantly
below $  \sim 10^{7}\, M_{\odot }  $ pc$  ^{-3}  $, physical collision (also
called coalescence) rates are small enough to be unimportant ($  df/dt\sim 0  $)
compared to the loss rates due to tidal disruption. In this case, numerical
work by both Bahcall \& Wolf (1977) and MCD indicates that, if evolution is
ignored, the intermediate-mass stars (the supergiant progenitors) follow a $  n\propto r^{-7/4}  $
density distribution, while the density of the lighter stars scales as $  n\propto r^{-6/4}  $.
Incidentally, neither of these functions approaches the steepness of the $  n\propto \sim r^{-10/4}  $
and $  n\propto \sim r^{-12/4}  $ functions assumed by AN97. The reason that
the massive stars have a steeper falloff is because of dynamical friction, which
tends to virialize energies rather than velocities. However, the late-type red
stars, which have the largest size to mass ratios, are particularly susceptible
to tidal disruption. MCD show that, unlike dynamical friction, this process
``flattens'' the density function of the giants. Presumably, any possible
bloating would enhance this process. 

The other extreme case to consider is that of a density significantly above
$  \sim 10^{7}\, M_{\odot }  $ pc$  ^{-3}  $. For this case, MCD show that
physical collisions are the leading contributor to mass loss. Because such collisions
scale with the inverse power of the density squared, they also ``flatten''
the density distribution function of the largest stars. Since the supergiant
stars have the largest cross sections, they again are the ones most subject
to this process and therefore have the \textit{least} steep density function
in very dense stellar systems. Thus, whether or not the densities are high enough
for collisions to be important in a system, it would seem unlikely that the
giant stars would have a density function that falls off significantly faster
than the normal populations.  

One might argue that in ``active'' wind models, where the evolved stars are
assumed to be affected by the intense continuum radiation (the ``bloated stars''
scenario), the situation might be different. In active stellar wind models the
innermost late-type stars lose mass at an extreme rate. However, this reduces
the time spent in the giant phase and hence again acts toward flattening their
density function (Tout et al. 1989).  

MCD included physical processes in the Fokker-Plank equation with $  df/dt\neq 0  $.
However, they ignored several other potentially important processes that could
affect the stellar densities. In particular, they ignored any possible position
dependence in the initial mass function. There are several reasons that such
a position dependence might exist. For instance, the Jean's mass is
\begin{equation}
\label{B1}
M_{\rm {J}}={\pi ^{5/2}\rho _{0}^{-1/2}\over 6}\left( \gamma kT\over G\mu m_{\rm {p}}\right) ^{3/2},
\end{equation}
where $  \rho _{0}  $ is the mass density, $  k  $ is the Boltzmann constant,
$  T  $ is the temperature, $  m_{\rm {p}}  $ is the mass of the proton, $  G  $
is the gravitational constant, and $  \mu   $ is the mass of the heaviest dust
grains\rfn{In most derivations of the Jean's mass, the gas is assumed for simplicity
to be pure molecular hydrogen. Here I have attempted to generalize the result
for the case in which the gas is composed of several different particle species
of various masses in thermal equilibrium with one another. As a result of diffusive
drift of the heavier particles through the gas, the pressure gradient for a
given particle that is required to resist self-gravitational forces in a turbulent-free
medium is dependent upon its mass. In this case, the Jean's length of a statically
inhomogeneous medium is also dependent upon the particle mass, with dust particles
being able to self-gravitate in a body that is predominantly supported by pressure
gradients of lighter particles. Incidentally, in the turbulent case, the diffusion
could be halted at any time. Here equation (\ref{B1}) yields (assuming a minimum
possible temperature of 5K, a maximum possible dust particle mass of $  7\times 10^{-14}  $~g,
and a maximum possible ``post-condensation density'' of 1g cm$  ^{-3}  $)
a minimum possible ``comet'' diameter of only $  \sim 10^{3}  $ cm. Conversely,
equation (\ref{B1}) yields a minimum possible diameter for hydrogen-containing
objects of $  \sim       10^{8}  $ cm.} in the gas in units of $  m_{\rm {p}}  $.\pfn{While
this equation is accurate at predicting the masses of galaxies and globular
clusters, it can fail to predict stellar masses. This may be because incorrect
values of $  \mu   $ are often assumed when equation (\ref{B1}) is used. It
may also be because equation (\ref{B1}) ignores the time dependence of $  \rho   $,
the time dependence of $  \mu   $, diffusion, magnetic pressure, and Hayashi
track wind-induced loss. At any rate, } Temperatures near the AGN environment
are probably higher than those in quiescent galactic nuclei. Therefore, equation
(\ref{B1}) implies that the mean self-gravitating mass would be relatively
large near AGNs. It also suggests the IMF of the massive stars has a radial
dependence. 

A more detailed analysis by Padoan, Nordlund, \& Jones (1997) results in an
upper limit to the star formation mass of 
\begin{equation}
\label{maxmass}
M_{\rm {max}}=0.2M_{\odot }\left( n\over 1\times 10^{3}\, \rm {cm}^{-3}\right) ^{-1/2}\left( T\over 10\, {\rm K}\right) ^{2}\left( \sigma _{v}\over 2.5\, {\rm {km}\, \rm {s}^{-1}}\right) ^{-1},
\end{equation}
 where $  \sigma _{v}  $ is the velocity dispersion of the molecular cloud.
Padoan et al. (1997) show that this expression agrees well with a Salpeter IMF
for $  T\sim 5-40\, \textrm{K}  $. This relation has an even stronger dependence
upon the temperature than that of the Jean's mass. Padoan et al. (1997) show
that a temperature of $  T\ga 60\, \textrm{K}  $ results in ``top heavy''
IMFs similar to the one observed by Malumuth \& Heap (1994) for R136 in 30 Doradus
and starburst galaxies. Note that even with the increased temperatures associated
with AGNs, extending this result to AGNs might require taking into account the
higher velocity dispersions in the vicinity of a supermassive black hole. However,
for the models shown in this dissertation, the gravitational influence of the
black hole is overshadowed by the rest of the galaxy near and outside the NLR.
Therefore, for star formation that occurs outside the BLR, the role of velocity
dispersion might not be as important.  

Adams \& Fatuzzo (1996) derive a semi-empirical formula of 
\begin{equation}
M_{\rm {max}}\propto {T\over \mu }^{a}
\end{equation}
 where $  1\le a\le 3/2  $. Although the assumptions Adams \& Fatuzzo make
are completely different than those made by Padoan et al. (1997), the final
result is quite similar in that both formulas appear to predict more heavy stars
near the heated AGN environment. 

The innermost radius at which stars are able to form in an AGN and the associated
local physical conditions such as temperature, however, are not at all known.
Thus, neither the theory of Adams \& Fatuzzo (1996) nor the theory of Padoan
et al. (1997) appear to be robust enough to make quantitative predictions about
the IMF or supergiant density function in AGNs at this time.

\section{Empirical Considerations}

One approach which bypasses some of the theoretical uncertainties discussed
in the previous sections is to appeal to empirical observations. Because stars
in AGNs cannot be resolved, direct measurement of the AGN mass function is not
feasible. However, we can look at nearby globular clusters and galaxies in the
hope that they may possess some of the same features of AGNs  albeit on a much
more compact scale. 

If high stellar densities generally result in top-heavy IMFs, mass function
gradients should exist even within the Milky Way. Massey, Johnson, \& DeGioia-Eastwood
(1995) measured\rfn{Scalo (1986) defines $  \Gamma   $ as $  d\textrm{log}\xi (\textrm{log}M)/d\textrm{log}M  $,
where $  \xi (\textrm{log}M)  $ is the number of stars of mass $  M  $ per
logarithmic mass bin per kpc$  ^{-2}  $. With this definition, the number of
stars per unit mass is proportional to $  M^{\Gamma -1}  $.} $  \Gamma   $
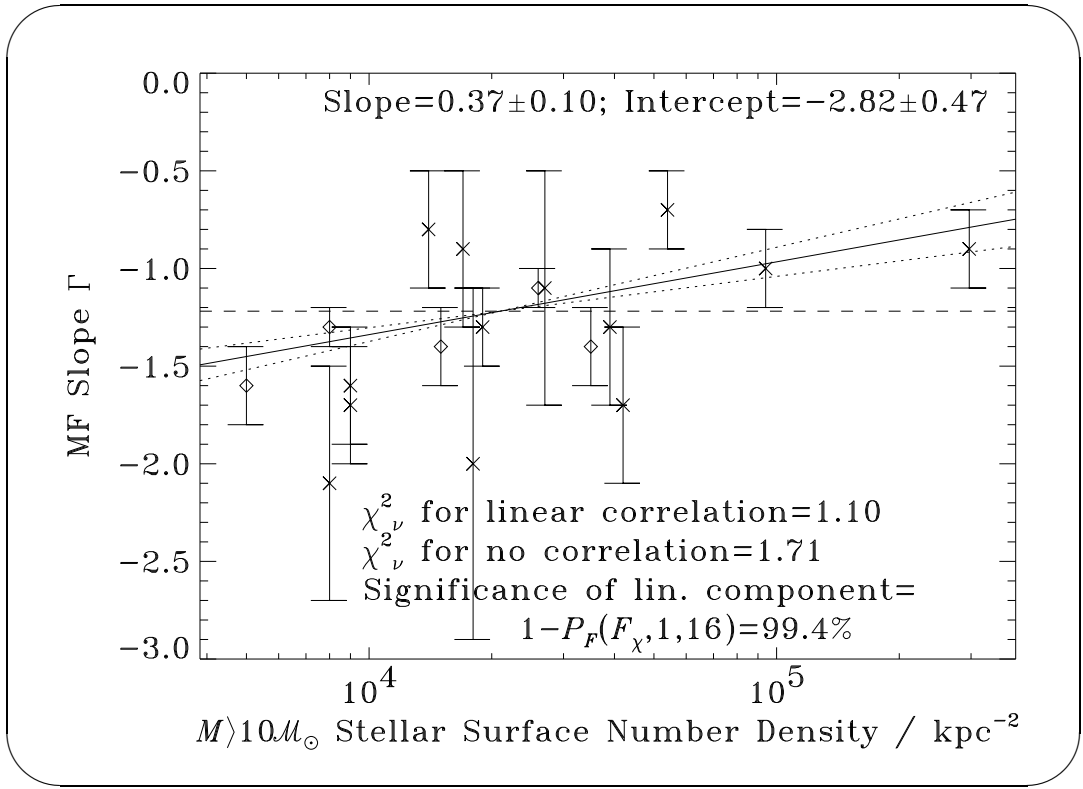
for 18 OB associations. Their results are shown in \Fig{../R136/massey}{The
present-day mass function slope as a function of OB association surface density.
The dashed line shows the best density-independent fit to the data, the solid
line shows the best linear fit, and the dotted lines show the 1-$  \sigma   $
deviations in slope from the best fit.}. Although Massey et al. (1995) stated
that there was no apparent correlation between $  \Gamma   $ and the local
surface density, \emph{application of an F-test to their data indicates a 99.5\%
significance of a linear dependence}. There are two differences between the
analysis presented in \ref{../R136/massey} and that shown in Figure 7 of Massey
et al. (1995). First, the independent variable was assumed to be the logarithm
of the surface density, rather than the surface density itself. Second, NGC
7235 was not completely removed in this analysis; the error of $  \Gamma   $
for this object was obtained from graphical analysis of Figure 5 of Massey et
al. (1995) to be 0.9. While the results shown in Figure \ref{../R136/massey} are
suggestive of a top heavy mass function in dense regions, the relationship between
surface density and position in AGNs itself is not known, so these data probably
do not provide a firm means of estimating the mass function in AGNs.  

Next to the galactic center, the nearest region with extremely high stellar
densities is R136 in NGC 2070, which itself is in 30 Doradus. Analysis of measurements
by Hunter et al. (1996) using refurbished $  HST  $ WFPC2 data appears to indicate
a spatially dependent intermediate-mass mass function. However, the errors obtained
from their methods were very large, and Hunter et al. (1996) concluded that
a spatially independent intermediate-mass mass function is not ruled out. Hunter
et al.'s data are shown in \Fig{../R136/hunter}{The mass function parameter
$  \Gamma   $ as a function of distance from the R136 center. The dashed line
is the best fit for a position-independent IMF. The solid line is the best linear
fit. The dotted lines show the 1-$  \sigma   $ uncertainties in slope. This
data is from Hunter et al. (1996).}. An F-test, which is insensitive to global
scale factors in errors, indicates that a linear relationship \emph{is} actually
warranted at the 97.2\% confidence level.  

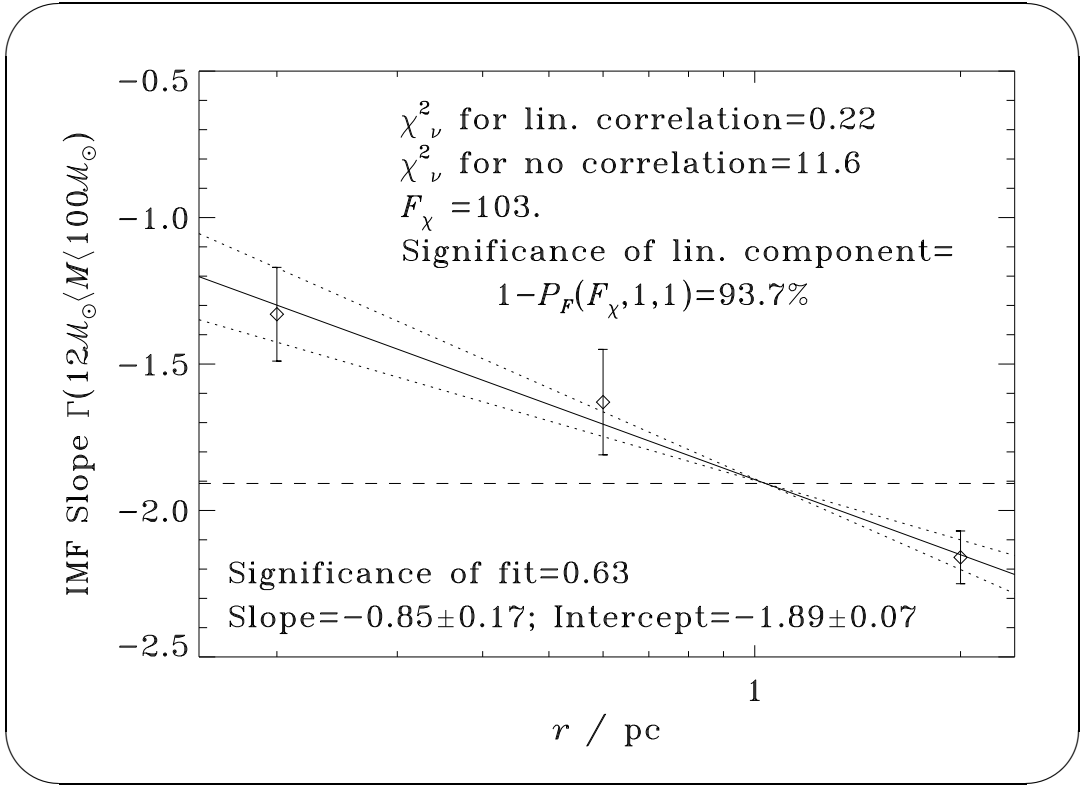
\Fig{../R136/brandl}{The mass function parameter $  \Gamma   $ for massive
stars as a function of distance from the R136 center. The dashed line is the
best fit for a position-independent IMF. The solid line is the best linear fit.
The dotted lines show the 1-$  \sigma   $ uncertainties in slope. Data from
Brandl et al. (1996).} shows the mass function parameter $  \Gamma   $ for
massive stars as a function of distance from the R136 center. In this case,
the slope is much higher than for the intermediate-mass stars of Figure \ref{../R136/hunter},
but the significance of a linear component is not.  

Presumably, the relationship between $  \Gamma   $ and $  r  $ also exists
for other parameters which might more accurately reflect the local physical
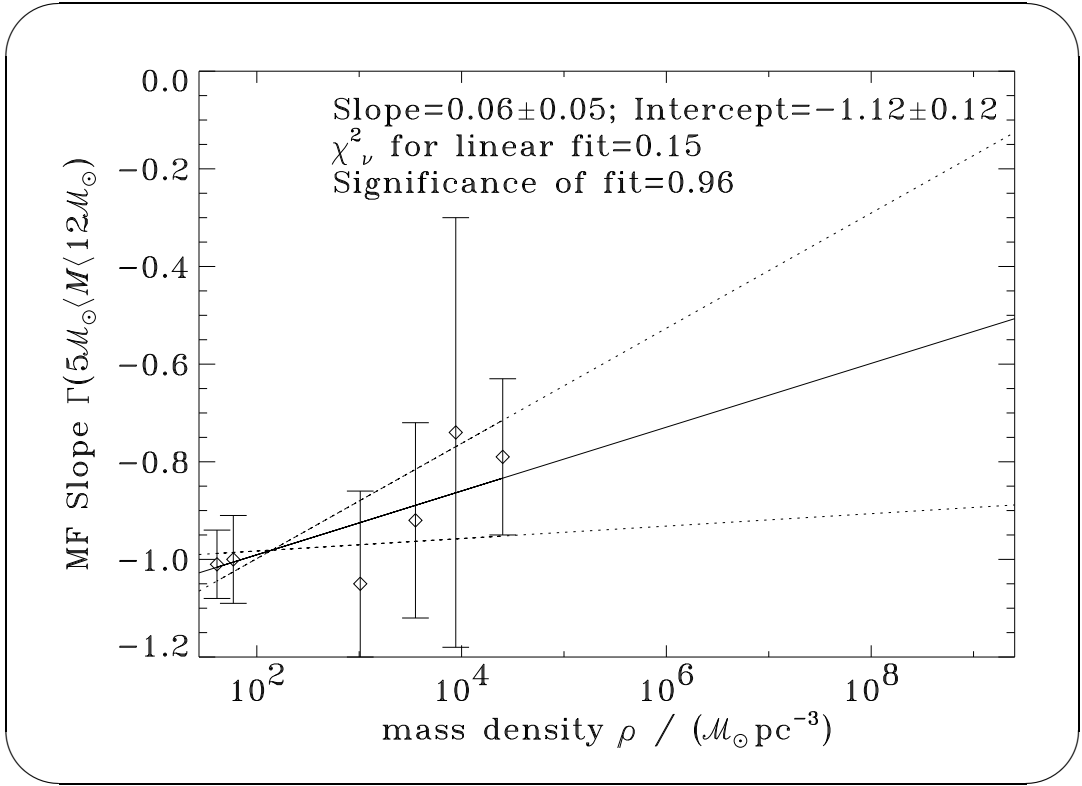
conditions of R136 that are responsible for the mass function gradient. \Fig{../R136/lights}{As
in Figure \ref{../R136/hunter}, the mass function parameter $  \Gamma   $
for intermediate-mass stars, but as a function of mass density rather than radius.}
shows the result of replacing the independent variable of radius with the local
mass density as inferred from the fit of Brandl et al. (1996) of the mass surface
density to a standard King profile. The significance of fit is the same as the
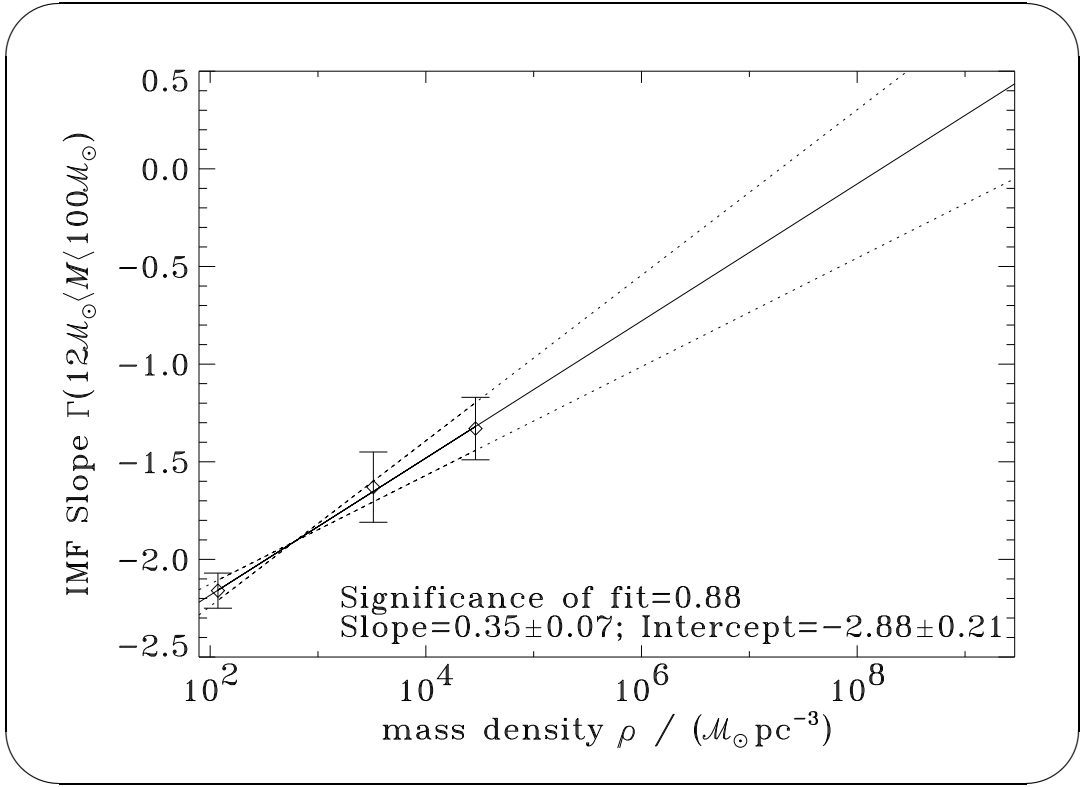
one shown in Figure \ref{../R136/hunter}. \Fig{../R136/heavies}{As in Figure
\ref{../R136/brandl}, the mass function parameter $  \Gamma   $ for massive
stars, but as a function of the local mass density.} shows that such a variable
transformation increases the significance of fit from 63\% to 88\%, which implies
that for massive stars the logarithm of the local mass density has a more linear
relationship with $  \Gamma   $ than does the logarithm of the distance from
the cluster center.

\section{A Crude Yet Quantitative Estimate of the Viability of the Bloated Star Density
Assumption}

We can use the results of the previous section to examine quantitatively the
viability of the bloated star density function employed by AN97. If we represent
the initial mass function by a simple power law, the number density per unit
mass of stars more massive than $  M_{\rm BS}  $ is proportional to $  \sim -M_{\rm BS}^{\Gamma }/\Gamma   $.
Combining this result with numerical differentiation of the results shown in
Figure \ref{../R136/hunter} for $  5M_{\odot }<M<12M_{\odot }  $ stars in R136
yields an approximate bloated star logarithmic density slope enhancement of
$  -0.22_{-0.18}^{+0.20}  $ (assuming R136 is similar to an AGN, all flattening
can be ignored, etc.). This implies that a reasonable enhancement factor of
the density function of bloated stars might be $  r^{-0.22}  $. Similarly,
an enhancement of $  r^{-0.42}  $ would be an acceptable $  -1\sigma   $ deviation
from this. The $  r^{-2}  $ enhancement factor assumed by AN97 is $  -9      \sigma   $
below the limits implied by the R136 results. Based upon the results of the
previous sections, if the bloated stars employed by AN97 were somehow less massive
than 5 $  M_{\odot }  $, the slope enhancement probably would be smaller in
magnitude. Also, relative flattening of the bloated stellar density function
due to collisions and tidal disruptions was ignored in this calculation. In
dense stellar systems, the flattening may be more important than gravitational
friction and intrinsic IMF gradients, in which case the bloated stars may actually
have the \emph{flattest} density function. Thus, the $  9\sigma   $ deviation
in amplitude result could be considered as a lower limit to the actual magnitude
by which the AN97 assumption deviates from what is expected in AGNs. On the
other hand, it should be pointed out that several systematic errors were not
included in this calculation. Inclusion of such errors might decrease the significance
of the deviation.

Incidentally, if the mass gradients implicitly assumed in AN97 were extended
to the rest of the galaxy, unusual \emph{mass} density functions would probably
result. In particular, since the surface brightness profiles are well known,
the implied stellar mass density functions could \emph{increase} with galactocentric
distance. Since this is highly unlikely, it is another possible problem with
the AN97 models. More discussion of the relationship between stellar mass function
gradient enhancements and the stellar mass functions is provided in Appendix
C.

\section{Conclusions and General Discussion}

If we assume that the bloated stars in AN97 are intermediate-mass supergiants,
dynamical friction and intrinsic IMF gradients may contribute to an increase
in the steepness of their density functions. If the observations of R136 mass
segregation are extrapolated to AGNs, the logarithmic density function slope
of stars bloated by the AGN continuum flux is approximately $  0.22_{-0.18}^{+0.20}  $
steeper than the underlying mass function. The $  r^{-2}  $ density enhancement
assumed by AN97 is 9$  \sigma   $ steeper than this. Moreover, it is likely
that flattening of the density function of bloated stars due to enhanced collision
and tidal disruption rates is more important than mass segregation. If this
is true, the bloated star density function assumed by AN97 is even more questionable.

What is interesting about the assumption this appendix analyzes in detail is
that, on the surface, it could appear to be relatively innocuous and even unimportant.
This may be because the AGN system, like many others in astrophysics, is inherently
complex. Thus, the theories are not well constrained. In these cases, it is
actually common to introduce parameters which, when appropriately adjusted,
result in reasonable fits with data. But when analyzed in more detail, the physical
justification for the introduction of these parameters is often weak. The results
of this appendix suggest that this is the case with the assumption regarding
the density of bloated stars made in AN97.\pfn{It is frequently assumed in galactic
evolution models that the initial mass function of stars is homogeneous and
that deviations from Salpeter (1955) mass functions are the result of sudden
starbursts and age segregation. However, recent observations (e.g., xxx et al.
199?) indicate that the ?80tar bursts less than ?10$  ^{6}  $ years ago. The
associated frequency of these bursts is presumably only once every ?10$  ^{8}  $?
years. The probability that it is a coincidence that we happen to be observing
the 100 or so starbursting galaxies is only 10$  ^{-4}  $\%. In other words,
at a confidence level of 1-10$  ^{-5}  $\%, the traditional galactic evolution
models appear to violate the Copernican Principle.  

For this reason, alternative galactic evolution models have been constructed
in which this and related problems are not as severe. {[}more review crap here.{]}
Thus, at least for dense stellar systems, current models have important shortcomings.

rewrite this to indicate that we really are not sure which stars are giving
out the wind. Demo's model assumes one thing, AN97 assume another thing. This
is a general mini paper below. I should really start with an introduction. This
is also relevant toward dark matter, etc.  

In conclusion, AN97 wrong. Me good and crafty ... However, there is an important
point to be made here. Each of the above arguments made against extremely steep
broad line region supergiant density functions rests on one assumption: that
because these stars are more suscsusceptableeptable to enhanced mass loss in
the BLR they would not be able to exist as long as otherwise. Thus, the same
mechanisms that produce candidate material for broad line emission are the ones
that imply it should not be there for long periods. The way this dissertation
addresses this point is by enhancing the mass loss rates of the red supergiants
in AGN instead of their density function. 

{[}al. (1992) show that the general trend for three of the four galaxies they
observed was to redden with increasing angular distance from the galactic center.
While some of the galaxies in their sample indicate reddened central pixels,
these observations were not corrected for dust yet are particularly sensitive
to it. As pointed out, e.g., in Netzer et al. (1995), such reddening with $  E_{B-V}\sim 0.3  $
is quite plausible. With the questionable assumption of spherical symmetry,
reddening is a monotonically decreasing function of angular distance from the
galactic center, especially in the inactive nuclei (where evaporation of dust
is less important) observed here. For this and the other reason mentioned above,
models in which the giant density falls significantly faster than $  n\propto r^{-2}  $
are probably not viable. {]}

Finally, we note that there is little experimental evidence to support nuclei
with late-type stars that have a significantly steeper density function than
the normal stars. There \textit{is} evidence for blueward color gradients in
the UV For instance, observations by Balcells et al. (1994) indicate a color
gradient of $  d\textrm{ln}(U-R)/d\textrm{ln}(r)\sim -0.2  $ and $  d\textrm{ln}(B-R)/d\textrm{ln}(r)\sim -0.1  $.
This translates into three dimensional color distribution function that falls
with powers of xxxx. If one assumes that the supergiant density scales with
xx power of the . Finally, if we de-project this into three dimensions we find
a APGB supergiant density that scales with $  r^{-2.xx}  $ and a main sequence
density that falls off with radius as $  r^{-2.xx}  $. Thus a galaxy with a
supergiant density that fell off with radius as fast as $  r^{2.3}  $ would
be ?10? standard deviations away from the mean of the galaxies measured by Belcells
et al. (1994).

In the spirit of Occam's razor, one is tempted to assume that nearly all of
the dark matter associated with galaxies similar to our own is either spherically
shaped or disk-shaped, but not both. Because disk-shaped dark matter is required
to explain the stellar velocity distribution in solar neighborhood (ref is in
Binney \& trmaine), which, due to its proximity, is probably the best understand
portion of the galaxy, one is pulled toward the conclusion that nearly all dark
matter associated with spiral galaxies resides along an extention of their disks.
... }

\chapter{Extent of Mass Function Gradients and Their Implications\label{darkma} 
}

\infootbegplain
\noindent 
\affil{Department of Physics, United States Naval Academy, Annapolis, MD
21402-5026, USA; E-mail (Internet): taylor@milkyway.gsfc.nasa.gov}
\par
\noindent 
\par
\noindent As discussed in Appendix \ref{commentsonaandn}, there is some evidence that the present day mass
function may vary with position (though not as strongly as implied in
Alexander \& Netzer [1997]).  In this appendix, I discuss the possible
implications of this towards our understanding of the dark matter problem.
\noindent 
\par
\noindent \keywords{galaxies: luminosity function, mass function --- galaxies:
kinematics and dynamics --- dark matter --- galaxies: halos --- Magellanic
Clouds --- galaxies: evolution}
\par
\noindent {}\setcounter{section}{1}{}\addtocounter{section}{-1}\section{Brief Discussion of Evidence for and Against IMF Gradients}
\def\figdirprefix{\home/stellarmodels/}
\medskip
In a
recent paper, Padoan, Nordlund, \& Jones (1997) claimed on theoretical grounds
that the initial mass function (IMF) should be a function of the local
temperature ${\it T}$ of the original molecular clouds.  Padoan et al. (1997)
argued that dense star forming regions, such as those in starburst galaxies,
should be warmer than sparser star forming regions.  In fact, if the
temperature dependence of the clouds is not drastically different from that of
a blackbody, then $T\propto \rho ^{1/4}_{{\scriptsize\rm l}}$, where $\rho _{{\scriptsize\rm l}}$ is the local mean luminosity once star
formation has already started.  Padoan et al. claimed that starburst regions
should therefore have a flatter IMF and be more ``top heavy."  Similar
reasoning would imply that the IMF in cooler regions of galaxies should favor
low mass star formation and be steeper.
\par
In support of their star formation model, Padoan et al. (1997) noted that
for $T\gtrsim 60$ K, their models predict a top heavy IMF similar to that found in the
center of R136~(Malumuth \& Heap 1994, Brandl et al. 1996), the bright stellar
cluster in 30 Doradus.  Due to its proximity
and the fact that it is the most massive H {\scriptsize\rm II} region in the Local Group, 30
Doradus is perhaps the best star formation ``laboratory" accessible to us.
However, the relaxation time in R136 may be
less than its age (Campbell et al. 1992), so dynamic friction may also
contribute toward the R136 present-day mass function gradient.
\par
Fortunately, many other avenues of testing Padoan et al.'s model exist.
The O-star catalog of Garmany, Conti, \& Chiosi (1982) shows a flattening of
the IMF slope toward the Galactic center (cf Humphreys \& McElroy 1984). This
catalog also reveals a strong Wolf-Rayet/O-star ratio gradient. This supports
Padoan et al.'s model since higher surface brightness
regions would, on the average, yield higher temperatures and flatter IMFs.
The blue-to-red O-star supergiant ratio is also a strongly decreasing
function of Galactocentric distance within galaxies including the Milky Way
(e.g., {Hartwick 1970; }Humphreys 1978).  Initially, this gradient was
ascribed to metallicity gradients (e.g., McClure \& van den Bergh 1968).  But
stellar models actually predict a decrease in the blue-to-red ratio
with an increase in the metallicity. (For a review of the blue-to-red gradient
problem, see Langer \& Maeder [1995].) As Figure
\ref{../stellarmodels/saltpetermasslossandtemp} shows,
they also demonstrate a smooth increase in effective temperature with
supergiant progenitor mass. In particular, progenitors
of blue supergiants are more massive $(m\sim 40)$ than those of red supergiants
$(m\sim 10)$ for a given metallicity (e.g., Schaerer et al. 1993).  Therefore, the
solution to this problem may simply be the existence of an IMF gradient,
in which case metallicity is only a passive yet correlated variable.
\par
\medskip
Direct measurements of the IMF outside of the Milky Way are impossible
for all but the nearest galaxies and indirect indicators are plagued by
uncertainties in dust, metallicity, and the star formation rate.  Despite
these complications, several of the models which attempt to explain
correlations between local surface brightness, color, line ratios,
metallicity, and the star formation rate assume luminosity-dependent IMFs
(e.g., Edmunds \& Phillipps 1989; Phillipps, Edmunds, \& Davies 1990).  On the
other hand, the detailed models of de Jong (1996) permit dust, metallicity,
star formation rate, but {\it not} the IMF to be spatially dependent parameters.
de Jong could not, however, get his models to agree with observed color
gradients without also requiring them to have a very large range of
metallicities in the central disk regions.  Several evolutionary models of
inner regions of starburst galaxies assume low mass cutoffs or top heavy IMFs
(e.g., Rieke et al. 1980; Augarde \& Lequeux 1985; Doane \& Mathews 1993; Doyon,
Joseph, \& Wright 1994).  Finally, independent theoretical arguments supporting
IMF gradients range from models which are consistent with the simple form of
the Jeans expression for the typical stellar mass in solar units of
$<m>\propto T^{3/2}~($e.g.; Larson 1982; Bodenheimer, Tohline, \& Black 1980) to much more
complicated models, such as the outflow-regulated model of Adams \& Fatuzzo
(1996), which predicts $<m>\propto T^{a}$, where $1\le a\le 3/2.$ 
\par
If IMFs are actually a function of $\rho _{{\scriptsize\rm l}}$ or $T$, there would be several
important astrophysical consequences.  For instance, there would be a
position-dependence in the mean mass to light ratio.  This is due to the
strong dependence of the mass to light ratio upon the IMF.  In R136, this
makes the mass density function $\rho _{{\scriptsize\rm m}}$ much different from $\rho _{{\scriptsize\rm l}}~($Malumuth \&
Heap 1994, Brandl et al. 1996) and complicates
estimates of the total mass.  Padoan et al.'s
results indicate that similar effects might occur in spiral galaxies.  If the
luminosity of a star is taken as $L\simeq L_{\odot }m^{y}$, where $y\simeq 3.5,$ the Jeans expression
above would suggest the crude relation $<m>\propto \rho ^{3/8}_{{\scriptsize\rm l}}$ and yield
$\rho _{{\scriptsize\rm m}}\propto \rho ^{1+3(1-y)/8}_{{\scriptsize\rm l}}\simeq \rho ^{0.06}_{{\scriptsize\rm l}}$.  Unfortunately, previous
works
have assumed that IMFs are independent of time and position with,
specifically, $\rho _{{\scriptsize\rm m}}\propto \rho ^{1.0}_{{\scriptsize\rm l}}$ throughout a given spiral galaxy (e.g., van Albada et
al. 1985).  In this appendix, surface mass densities of spiral
galaxies are
computed, for the first time, by explicitly accounting for the possible types
of IMF gradients that might exist if theories like those of Padoan et al. are
correct.
\par
\noindent {}\setcounter{section}{2}{}\addtocounter{section}{-1}\section{An Empirical Estimate of the R136 IMF Gradient}
\par
Since position-dependent measurements in R136
of both $\rho _{{\scriptsize\rm l}}$ and the IMF slope $\Gamma $ (where ${\it dN}/{\it dM\propto m}^{\Gamma -1}$ is the number of stars per
unit mass in solar units) have already been made, computing the dependence of
the R136 IMF upon the local luminosity is straightforward.  Doing this will provide a useful
starting point in obtaining a crude yet quantitative estimate of the possible
types of IMF gradients that might generally exist in all galaxies including
the Milky Way.
\begin{table}
\begin{center}
\caption{IMFs in R136} \label{IMFSinR136}
\begin{tabular}{c c c c c} \tableline \tableline
$R$/pc & $\Gamma(R)$     & $m_{\rm l}$\qquad & $m_{\rm u}$ &
$\rho_{\rm l}/(L_\odot{\rm pc}^{-3})$ \\ \tableline
0.20 & $-1.29\pm 0.20$ & 5.6      & 120   & $1.5\times10^6$ \\
0.60 & $-1.46\pm 0.23$ & 3.6      & 76    & $1.5\times10^5$ \\
2.0 & $-2.12\pm 0.09$ & $\le2.0$ & 48 & $1.5\times10^3$ \\
\tableline
\end{tabular}
\end{center}
\tablecomments{Data adapted from Brandl et al. (1996) for (age-spread
restricted) stars 2.5---3.5 Myr old.}
\end{table}
\par
Table \ref{IMFSinR136} summarizes Brandl et al.'s (1996) results for the
IMF based
upon high resolution 5-color photometry of the stars in R136 estimated to be
between 2.5 and 3.5 Myrs years old.  The right-most entry
of Table \ref{IMFSinR136} shows the results of
performing the coordinate transformation between ${\it R}$ and $\rho _{{\scriptsize\rm l}}$ using Figure 15 of
Hunter et al. (1996).  Though Brandl et al. (1996) did not make explicit
measurements of the upper and lower stellar mass cutoffs $m_{{\scriptsize\rm l}}$ and $m_{{\scriptsize\rm u}}$ to the
power-law approximation of the IMF, Table \ref{IMFSinR136} includes estimates
of their
dependences upon the local surface brightness.  The lower mass limits were
obtained from the peaks of Brandl et al.'s mass functions, while the upper
limits were taken from the highest masses observed per radius bin. Both
$\log _{10}(m_{{\scriptsize\rm l}})$ and $\log _{10}(m_{{\scriptsize\rm u}})$ are found to decrease by $\simeq 0.2$ with each successive
increase in radius.  Brandl et al. (1996) performed completeness corrections,
so the depletion of low mass stars in all but the outer regions of R136 is
presumably real.  The results of performing a linear fit of the IMF parameters
of R136 to $\log _{10}(\rho _{{\scriptsize\rm l}})$ are shown in
columns 2-7 of Table \ref{ModelParameters} as Model A. Uncertainties of
parameters
calculated from more than two radius bins are shown in parenthesis.
\par
\noindent \begin{table}
\begin{center}
\caption{Model Parameters} \label{ModelParameters}
{\scriptsize \begin{tabular}{c c c c c c c c c c c} \tableline \tableline
Model & $\Gamma_0$ & $\Gamma_1$ & $m_{\rm l0}$ & $m_{\rm l1}$ & $m_{\rm u0}$ &
$m_{\rm u1}$ & $\rho_{\rm m\odot}/ $  & $\Sigma_{\rm
m\odot}/ $ & $<\Sigma_{\rm m\odot}>/
$ \\
&   &   &  &  &   &     & $M_\odot{\rm pc}^{-3} $  & $
M_\odot{\rm pc}^{-2} $ & $ M_\odot{\rm pc}^{-2} $ \\
\tableline
A & $-3.03 $ & $0.28 $ & $-0.08$ & $0.13$ & $1.25 $
& $0.12 $ & 0.12 & 67 & 67 \\
& $(\pm0.26)$ & $(\pm0.06)$ & $\ldots 
$ & $\ldots 
$ & $(\pm0.23)$ &
$(\pm0.04)$ & ... & ... & ... \\  B & -1.11 & 0.42 & -1.52 & 0.00 & 1.30 & 0.00 & 0.31 & 179 & 179 \\
C & -0.55 & 0.40 & -1.52 & 0.00 & 1.60 & 0.00 & 0.10 & 60  &195
\\
\tableline
\end{tabular}}
\end{center}
\tablecomments{This assumes $f=f_0+ f_1{\rm log}_{10}[\rho_{\rm
l}/(L_\odot{\rm
pc}^{-3})]$, for $f=\Gamma$, ${\rm log}_{10}(m_{\rm l})$, or ${\rm
log}_{10}(m_{\rm u})$.}
\end{table}
\par
\noindent {}\setcounter{section}{3}{}\addtocounter{section}{-1}\section{Dynamical Properties of Spiral Galaxies with IMF Gradients}
\par
The IMF gradient of Model A implies a surface mass density that is
different from what would be obtained were the mass to light ratio
constant.  The surface mass density for Model A, if scaled according to the
surface luminosity function suspected for the Galaxy, is shown in the top
panel of Figure {}\ref{v_circstriplet}.  The disk scale length $R_{0}=4.5$
kpc and solar Galactocentric radius $R_{\odot }=7.8$ kpc were taken from Kuijken \&
Gilmore's (1989a) model of the Galaxy.  For simplicity, $\rho _{{\scriptsize\rm l}}$ at a given radius
was assumed to be constant throughout a disk thickness of
575 pc.  The surface brightness was normalized to be $22.5~L_{\odot }$pc$^{-2}$ at $R=R_{\odot }$,
which results in $\rho _{{\scriptsize\rm l\odot }}\equiv \rho _{{\scriptsize\rm l}}(R_{\odot })=0.037~L_{\odot }$pc$^{-3}$.  For each
radius bin, the IMF was obtained from $\rho _{{\scriptsize\rm l}}$ and the coefficients shown in Table
\ref{ModelParameters} for Model A.  This IMF was converted to present
day mass and luminosity functions by assuming (purely for simplicity) a
constant star formation rate for the past $1.0\times 10^{10}$ yrs.  The main sequence
lifetime-luminosity-mass relationships used to obtain the mass to light ratio
as a function of the IMF were obtained from logarithmic-linear interpolation
of $m\ge 0.8$ models published by Schaller et al. (1992) for $z=0.02,$ overshooting
of the $m\ge 1.5$ stars, and standard mass loss rates.  For $m<0.8,
L|_{m=0.25}=7.8\times 10^{-4}L_{\odot },~L|_{m=0.08}=6.55\times 10^{-9}L_{\odot }$,~and $L|_{m\le 0.07}=5.0\times 10^{-12}L_{\odot }$ were
assumed.
\par
\medskip
The surface densities both of Model A and of the constant mass to light
ratio model fall off exponentially with increasing radius.  The effective
scale length of Model A is $\simeq 7.5$ kpc, which is $\simeq 1.7$ times larger than that of
the surface brightness function.  This increase in the scale length is a
result of the fraction of low mass stars (and the mass to light ratio)
increasing with radius.
\par
\medskip
From the surface density, other dynamical properties of the galaxy can
also be calculated.  The circular velocity (i.e., the rotation curve)
corresponding to the surface density of Model A is shown in the upper middle
panel
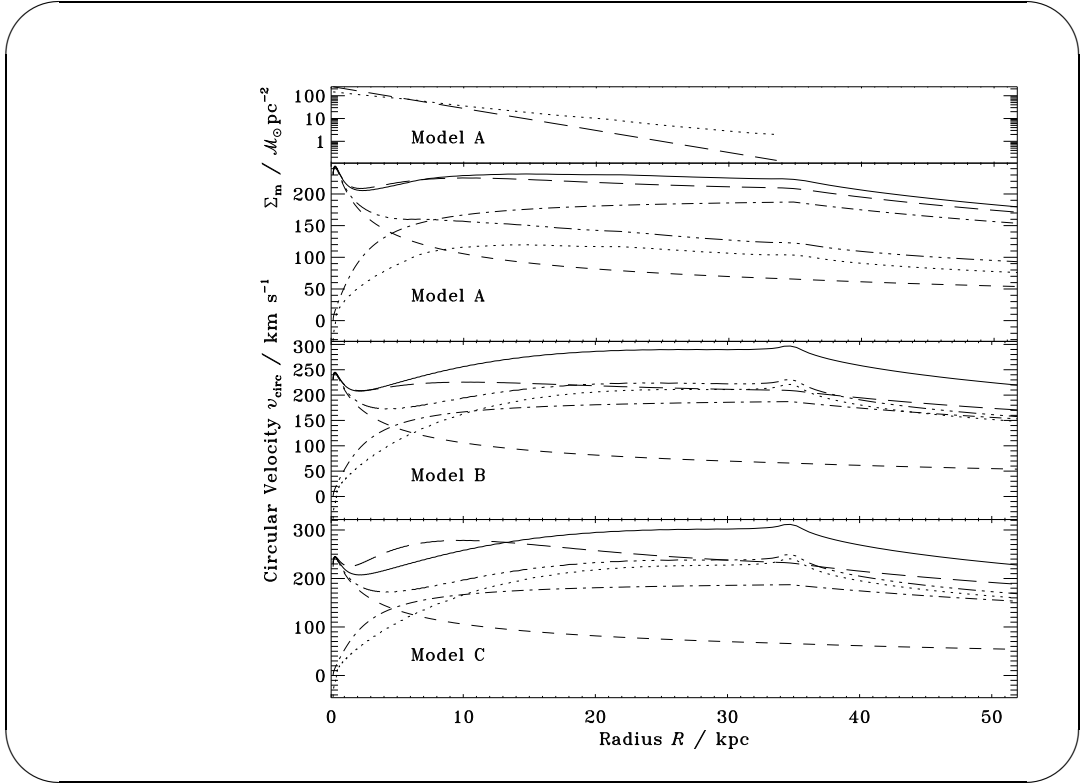
of \Fig{v_circstriplet}{{\it Top panel}: the surface density of a spiral galaxy
similar to the Milky Way but with the IMF of Model A.  The dotted line is the
surface density assuming that all stars lie on the main-sequence.  The dashed
line is the surface density if the {\it V} band mass to light ratio were constant
at $\gamma _{{\scriptsize\rm V}}=2.0M_{\odot }/L_{\odot }$.  {\it Lower three panels}: circular velocities of Models A (upper
middle), B (lower middle), and C (bottom). The circular velocities of Model A
correspond to the surface density function shown in the top panel.  For each
model, the solid curve accounts for all components of mass, the dot-dashed
curve accounts for just the halo, the dotted curve accounts for just the disk,
the short-dashed curve accounts for just the bulge and spheroid stars, the
dash-triple-dotted curve accounts for everything except the halo, and the
long-dashed curve represents a model with a constant mass to light ratio of
$\gamma _{{\scriptsize\rm V}}=2.0M_{\odot }/L_{\odot }$. The thicknesses of the disks were ignored in these computations.
At ${\it R<}0.30$ kpc, the gravitational force in Model A due to the disk-bound mass
is outward. The corresponding value of the circular velocity is technically
imaginary, but is plotted here as negative. A similar but opposite effect
occurs at the outer edge of the disk, which is at 35.0 kpc in these models,
where the circular velocities level off or even rise just before following
nearly Keplerian motion.}.  The parameters for the bulge, spheroid, and halo
were taken from Table \ref{IMFSinR136} and Figure 5 of Kuijken \& Gilmore's
(1989a) model of the Galaxy.  To avoid a divergent and unphysical total mass,
the additional assumption that all components of the Galaxy terminate at an
arbitrarily selected maximum radius of 35.0 kpc was also made.  For Model A,
this results in a total mass of the halo, bulge/spheroid, and disk of, respectively, $2.8\times 10^{11} M_{\odot }, 3.5\times 10^{10}
M_{\odot }$, and $6.3\times 10^{10} M_{\odot }$.  In comparison, the integrated disk mass of the
$\gamma _{{\scriptsize\rm V}}=2.0M_{\odot }/L_{\odot }$ model is only $3.2\times 10^{10} M_{\odot }$ and increases much faster with radius.
Similar results were obtained using the Bahcall \& Soniera (1984) Galaxy model,
though their smaller disk scale length of 3.5kpc causes the intermediate
regions of the disk to have a higher density.  For simplicity, the surface
mass density of stellar remnants and gas was assumed throughout the disk to be
1/3 that of the stars.  Because the halo dominates the mass distribution, the
circular velocity curve (solid line) is nearly flat.  Without the halo, the
circular velocity curve falls from 185 km s$^{-1}$ at $R=2.0$ kpc to
124 km s$^{-1}$ at $R=34$ kpc. Though the surface density of Model A
corrected for IMF gradients is different from that previously obtained for
spiral galaxies, Figure {}\ref{v_circstriplet} shows that the change is not enough
to dramatically affect the dynamical properties of the disk, such as the
circular velocity curve.  \clearpage
\par
\medskip
Another, more direct, effect of Model A's IMF gradient is shown in
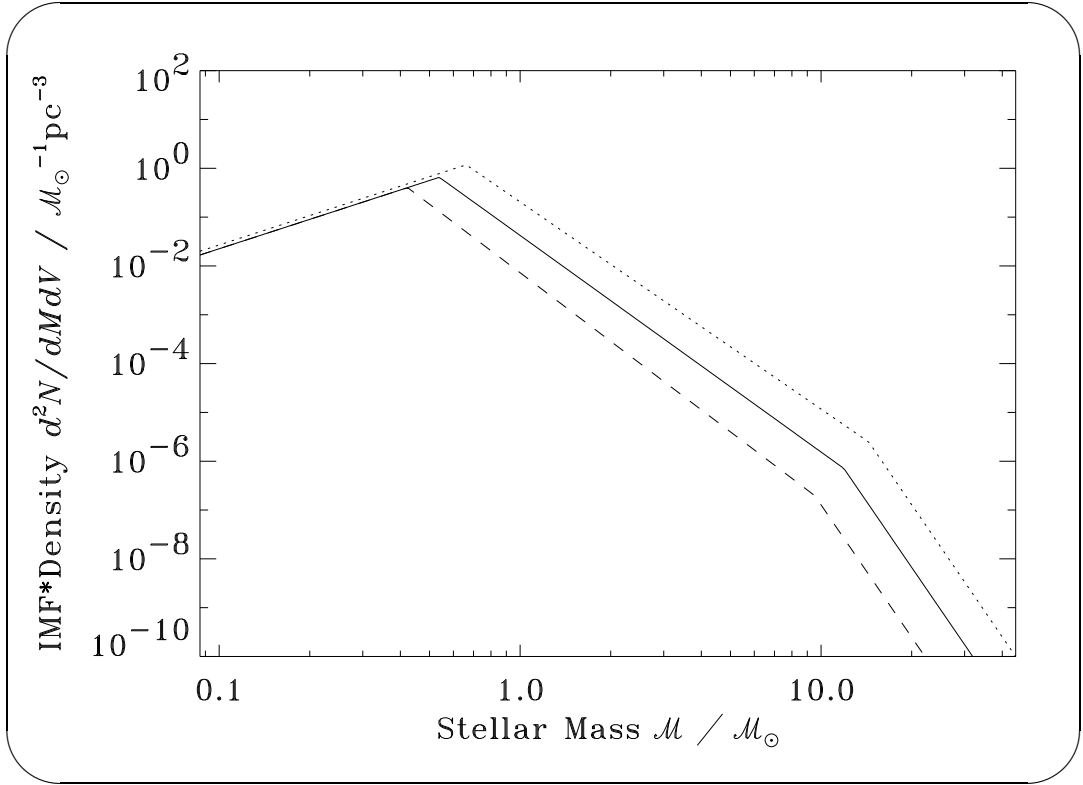
\Fig{IMFs}{The IMF of Model A at a Galactocentric radius of 1.0
kpc (dotted
line), 7.8 kpc (solid line), and 15 kpc (dashed line).  Though the
differences between these three IMFs are small in terms of parameters
traditionally computed from IMFs, they correspond to large differences in
the mass to light ratios.}. For numerical
reasons, the slopes at $m<m_{{\scriptsize\rm l}}$ and $m>m_{{\scriptsize\rm u}}$ were set, respectively, to be $\Gamma =+3$ and
$\Gamma =-8$ rather than positive and negative infinity.  The IMF of Model A is very
negative at all radii, with $\Gamma |_{R=1.0\hbox{~{\scriptsize\rm kpc}}}=-3.2, \Gamma |_{R=R_{\odot }}=-3.4,$ and
$\Gamma |_{R=15.0\hbox{~{\scriptsize\rm kpc}}}=-3.6.$  This occurs even though R136's spatially averaged
IMF is typical and its IMF gradient is small only because it has a luminosity
density that is $\sim 10^{4}-10^{8}$ times higher than typical regions of spiral galaxies.
For
comparison, Salpeter (1955) found $\Gamma =-1.35$ for $0.4\le m\le 10,$ while Miller \& Scalo
(1979) obtained much higher values of $\Gamma =-0.4, -1.5,$ and -2.3 for,
respectively, $0.1<m<1.0, 1.0<m<10,$ and $m>10.$
\par
These differences suggest that the IMFs in R136 and the Milky Way do not
directly scale to one another via the $\rho _{{\scriptsize\rm l}}$ variable.  This may be because R136
has recently experienced its first starburst.  In this case, a better
correlation variable to employ might be $<\rho _{l}>_{t}$, the {\it time-averaged} luminosity
density.  It may also be because of other inherent differences
between R136 and the Milky Way.  For instance, the bright, early-type stars in
spiral galaxies are generally confined to relatively narrow Galactocentric
radii near that of their initial birth sites.
In contrast, stars in elliptical galaxies similar to R136 undergo substantial
mixing due to their highly eccentric orbits.  At any rate, the IMFs in spirals
like the Milky Way may scale differently than the scaling in R136.
\par
\medskip
For these reasons, other models were also considered.  Model B was
constructed in order to help answer the question of just how necessary the
dark halo is for circular velocity curves to be flat.  The IMF gradient $\Gamma _{1}$
was adjusted to minimize the curvature of the outer circular velocity curve,
while $\Gamma _{0}$ was adjusted such that the rotation velocity was $\simeq 220$ km s$^{-1}$.
For simplicity, $m_{{\scriptsize\rm l}}$ and $m_{{\scriptsize\rm u}}$ were fixed.  The lower middle panel of Figure {}\ref{v_circstriplet}
shows that the
circular velocity curve of Model B is surprisingly flat throughout most of the
outer regions of the disk before the halo component is included.  The total
disk mass for Model B is $2.4\times 10^{11}M_{\odot }$, with $<\gamma _{{\scriptsize\rm v}}>_{\hbox{{\scriptsize\rm disk}}}{\scriptsize\rm =}14.5M_{\odot }/L_{\odot }$, which is 7.3
times
larger than the $\gamma _{{\scriptsize\rm V}}=2.0M_{\odot }/L_{\odot }$ model.  The value of $\Gamma _{1}$ for Model B is 0.42.  This
is
$50\%$ higher than the IMF gradient in R136.  The change within the Milky Way of
$\Gamma $ measured by Garmany et al. (1982) between the inner and outer semicircular
regions of radius 2.5 kpc surrounding the Sun was -0.8, which for a disk scale
length of $R_{0}=4.5$ kpc corresponds to $\Gamma _{1}=0.8\times 3\pi R_{0}/(8$log{\it e}$\times 2.5$~kpc)=3.9.  This is
much higher than the value in Model B.  Thus, the IMF gradient of Model B is
well below empirical constraints.
\par
\medskip
However, there are at least five potential problems with the
halo-less
form of Model B:
\par
\noindent \begin{enumerate}\item In the adopted solar vicinity $(R_{\odot }=7.8$ kpc), the surface
density is $179M_{\odot }$pc$^{-2}$.  This is an unacceptable 15
standard deviations higher than the local value of $46\pm 9~M_{\odot }$pc$^{-2}$ measured by
Kuijken \& Gilmore (1989b).  The corresponding mass to light ratio is $7.9M_{\odot }/L_{\odot }$.
This is $60\%$ higher than the local value adopted in standard texts such as
Binney \& Tremaine (1987).  Similarly, the IMF slope at this radius is $\Gamma =-1.7.$
For the low mass $(m\lesssim 0.5)$ stars, this value is incompatible with Miller \&
Scalo's (1979) result of $\Gamma =-0.4.$\item Mestelian disks, which are similar to
Model B, are
commonly thought to be unstable to bar formation.  The Toomre instability
parameter $Q$ is $\sigma _{R}\kappa /(2.9G\Sigma _{{\scriptsize\rm m\odot }})$, where $\kappa \simeq 36$ km s$^{-1}$ kpc$^{-1}$ is the epicycle
frequency and $\sigma _{R}$ is the mass-weighted stellar velocity dispersion (Toomre
1974).  Published estimates are $Q\simeq 1-3$ in the solar vicinity.  Because stellar
velocity dispersions are empirically observed to decrease with mass even for
stars with lifetimes greater than the age of the Galaxy, estimates of $\sigma _{R}$ are
sensitive to $m_{{\scriptsize\rm l}}$.  Wielen (1977) obtained $\sigma _{R}=62\pm 12$ km s$^{-1}$  for $0.1\lesssim m\lesssim 0.8$~K and
M dwarfs, which implies $Q\gtrsim 1.0\pm 0.2$ for Model B.  This lower limit is low enough
to sustain spiral arm structure which numerical simulations show would rapidly
dissipate otherwise.  However, it is too near unity to prevent the growth of substantial
arm/interarm stellar mass density contrasts.  Though such mass contrasts are
now known to exist in normal spirals (e.g., Rix \& Zaritsky 1995, Gonz\'alez \&
Graham 1996), they are not accounted for in Model B. Incidentally, the halo
component does not necessarily affect this instability (Sellwood 1985).\item
The circular velocity curve at $R\lesssim 35$ kpc is not precisely flat, but actually
rises before attaining a nearly Keplerian fall off.  This is the result of the non-spherical potential. \item The circular
velocity drops below 200 km s$^{-1}$ in the inner regions of the disk.  This result
is expected.  For Model B, the mass to light ratio is
\par
\s
\simeq m^{1-y}_{{\scriptsize\rm u}}(m_{{\scriptsize\rm u}}/m_{{\scriptsize\rm l}})^{-\Gamma -1}(y+\Gamma )(M_{\odot }/L_{\odot })/(-\Gamma -1),
\labep{14.1}\e
\noindent where $L\propto m^{y}$ is assumed for $m\lesssim m_{{\scriptsize\rm u}}$.  If
\par
\s
\Gamma _{1}=[\ln (m_{{\scriptsize\rm u}}/m_{{\scriptsize\rm l}})\log _{10}{\it e}]^{-1},
\labep{14.2}\e
\noindent which Model B obeys to within $15\%$, the mass to light ratio would scale as
$\simeq e^{R/R_{0}}$.  This in turn would imply a disk surface density that is relatively
constant.  The circular velocities of such disks increase monotonically with $R$
and are zero at $R=0.$  This problem with low inner disk velocities is probably
not serious because circular velocity curves are frequently
compatible even with constant mass to light ratio, halo-less models throughout
their entire optically bright regions (e.g., Kent 1986).  Furthermore,
flatter, halo-less velocity curves could probably be attained by including the
following: galaxy
parameters slightly different than those of Kuijken \& Gilmore (1989a), a (more
realistic$)
\log $-normal IMF (Miller \& Scalo 1979), expected spatial dependence in remnant
and gas mass fractions, and variations of $m_{{\scriptsize\rm l}}$ or $m_{{\scriptsize\rm u}}$ with $\rho _{{\scriptsize\rm l}}$.  For instance, the velocity dip is better masked by the bulge
if Bahcall \& Soniera's (1984) smaller disk scale length of 3.5 kpc is assumed.
\item The
IMF gradient of Model B appears to be too {\it small} to be compatible with the
measurement of Garmany et al. (1982).  Equation ({}\ref{14.2}) suggests
that
this discrepancy would be less if a smaller $m_{{\scriptsize\rm u}}/m_{{\scriptsize\rm l}}$ ratio had been employed.
\end{enumerate}
\par
Of the above potential problems, only the first two appear to be
significant at this time.  Both can be overcome by taking into account the
arm/interarm density contrasts observed in spiral galaxies; Model C was
constructed to be similar to Model B, but has an azimuthally averaged
light and mass density that is 3.25 times greater than the interarm values in
which the Sun presumably resides.  The circular velocity curve of Model C,
shown in the lower panel of Figure {}\ref{v_circstriplet}, is
slightly higher, but
otherwise similar to that of Model B. However, the solar-vicinity disk surface
density is only $60 M_{\odot }$pc$^{-2}$.  This is a much more reasonable 1.6 standard
deviations above the
value determined by Kuijken \& Gilmore (1989b) and is actually lower than
Bahcall \& Soniera's (1984) value of $\simeq 85 M_{\odot }$pc$^{-2}$.
\par
\noindent {}\setcounter{section}{4}{}\addtocounter{section}{-1}\section{Discussion}
\par
A direct scaling of R136's IMF to the Galaxy does not dramatically alter
the circular velocity curve.  However, Models B and C, with their higher, yet
modest IMF gradients, have nearly flat $v_{\hbox{{\scriptsize\rm circ}}}\lesssim 220$ km s$^{-1}$ circular velocity
curves only before the traditional dark halo component is included.  Note that if one assumes that these types of
models and their $\sim 10^{1}$-fold mass enhancements are representative of most
galaxies, that the fiducial stellar contribution towards the closure
density is $\Omega _{*}\simeq 0.004~($e.g., Peebles 1993) before accounting for IMF gradients,
that the cosmological constant is zero, and that there is no hot dark matter,
one would obtain
\par
\s
\Omega \simeq \Omega _{\hbox{{\scriptsize\rm baryon}}}\simeq 0.04+\Omega _{\hbox{{\scriptsize\rm gas}}}.
\e
\noindent In this equation, $0.007\lesssim \Omega _{\hbox{{\scriptsize\rm gas}}}$$\lesssim 0.08~($Mulchaey et al.
1996), where $\Omega _{\hbox{{\scriptsize\rm gas}}}$ is the closure fraction due to all gas including hot plasma
in galactic clusters.
\par
\medskip
Current models of galactic evolution (e.g., Dwek 1998, Worthey $1994, \&$ de
Jong 1996) do not account for IMFs that might vary with time and position via
the
temperature.  This is true despite prior warnings that the IMF
probably has important dependences upon time and position (e.g., Mihalas \&
Binney 1978).  In light of the above results, accounting for IMFs with such
dependences may be necessary even to obtain results that are only accurate to
first order.  Accounting for these dependences may, for relatively obvious
reasons, clarify our understanding of several astrophysical phenomena
including the G-dwarf problem, intrinsic (as a function of radius) and
extrinsic (as a function of galactic morphology) metallicity and color
gradients, and the Tully-Fisher relation.
\par
\medskip
However, some problems remain
with assuming that the above models are even representative of spiral
galaxies.  For instance, why would the disk-edge peaks of Models B and C not
have been observed?  Also, why would spiral galaxies conspire to obey a
relation similar to equation ({}\ref{14.2})?  Questions similar to these
will be addressed in future work.
\par
\noindent 
\par
\noindent 
\par
\noindent 
\par
\infootendplain

\chapter{Obtaining AGN Abundances from Galactic Abundance Gradients and the {[}O/Fe{]}
Ratio\label{agnabundances} 
}

\section{Introduction}

Abundances in AGNs are generally estimated by analyzing spectral features like
line ratios. Some of these estimates have yielded extremely unusual abundances.
For instance, Marshall et al. (1993) obtained results that yield {[}O/Fe{]}$  \lsim -0.9  $
in active galaxy NGC 1068, where $  [\textrm{X}/\textrm{H}]  $ is defined to
be $  \log \left( A\left( \textrm{X}\right) /A_{\odot }\left( \textrm{X}\right) \right)   $,
$  A(\textrm{X})  $ is the abundance of element X, and $  A_{\odot }(\textrm{X})  $
is the solar abundance of X.

As universal as the general procedure employed by Marshall et al. (1993) towards
determining abundances is, it has some important shortcomings. For instance,
the method requires knowledge of the curves of growth associated with the lines,
which in turn depend upon opacities, which themselves generally depend upon
the physical conditions of the line-emitting gas. The equivalent width of an
absorption feature is, for example, proportional to the elemental abundance
if the clouds are small enough and if other conditions are such that a line
is optically thin. However, it is equivalent to the square root of the abundance
if the cloud geometry and other conditions are such that the line is saturated.
Unfortunately, since most of the physical conditions like this are only poorly
constrained in AGNs, estimates of abundances in AGNs have important systematic
errors in addition to the well-understood random ones. In fact, Marshall et
al. (1993) speculate that their abundance estimates might be incorrect due to
the potentially incorrect values of the iron opacities they assumed in their
calculation. For these reasons, an independent means of estimating AGN abundances
which bypass the problems with the traditional method would be desirable. In
this appendix, I describe and employ a new method for obtaining abundances in
AGNs which, though currently very crude, does precisely this.

\section{Procedure}

This new method takes advantage of the fact that essentially all metals are
formed in stars. It is for this reason that abundances are a direct function
of the time- and position-dependent stellar birthrates, IMFs, and stellar evolution
characteristics.\rfn{Some models assume continuous infall of low-metallicity
gas in order to describe the metallicities that we observe. In such models,
there would be an additional dependence upon the infall characteristics. However,
there does not appear to be significant quantities of intergalactic gas except
at high redshifts and in dense galaxy clusters. Moreover, in models similar
to that proposed in Taylor (1998), infall is not required. For these reasons,
in this appendix I assume that galaxies form their own metals.} In this appendix
let us assume that these functions are closely coupled to the mean stellar luminosity
density averaged over the life of the galaxy (as opposed to the luminosity density
that might exist during a starburst). This assumption is similar to that made
in Appendix \ref{darkma}. It specifically implies

\begin{equation}
\label{loglinearused}
[\textrm{X}/\textrm{H}]=[\textrm{X}/\textrm{H}]_{0}+[\textrm{X}/\textrm{H}]_{1}\log \left( \rho _{\rm l}/\rho _{\rm l\odot }\right) .
\end{equation}
 In the above equation, $  \rho _{\rm l}  $ is intrinsic luminosity density
and the other three variables on the right hand side are free parameters. In
this appendix, we will attempt to extrapolate the abundance gradients in the
Milky Way to other galaxies. In this case, provided we neglect the fact that
the sun is more enriched than its neighbors, $  \rho _{\rm l\odot }  $ can
be taken as the luminosity density in the solar vicinity, $  [\textrm{X}/\textrm{H}]_{0}  $
can be assumed to be zero, and $  [\textrm{X}/\textrm{H}]_{1}  $ specifies
the abundance gradient. Note that the subscripts ``0'' and ``1'' placed
on $  [\textrm{X}/\textrm{H}]  $ are not to be confused with an occasionally
used means of denoting different observational measurement techniques. 

For a hypothetical bulge-less spiral galaxy, the dynamic range in $  \rho _{\rm l}  $
is relatively small. This is because at $  R\ll R_{0}  $ the luminosity density
is approximately constant. In an AGN, however, $  \rho _{\rm l}  $ continues
to increase at $  R\ll R_{0}  $, so equation (\ref{loglinearused}) would yield
drastically larger inner disk to outer disk metallicity ratios. However, in
very bright regions, one would expect that the star formation rate is limited
by negative feedback. For this reason, the value of $  [\textrm{X}/\textrm{H}]_{1}  $
in active galaxies may be lower than the value in the solar vicinity. To determine
whether or not this is the case, $  [\textrm{X}/\textrm{H}]_{1}  $ could in
principle be directly computed using current galactic evolutionary models. However,
such a calculation is far beyond the scope of this appendix and would probably
be poorly constrained anyway due to various free parameters. 

Here I simply assume that $  [\textrm{X}/\textrm{H}]_{1}  $ is the same for
all galaxies and estimate it for the Milky Way from the radial metallicity gradients
observed in H {\footnotesize II} regions, supernova remnants, planetary nebulae,
and stars. Assuming $  \rho _{\rm l}=\rho _{\rm l0}e^{-R/R_{0}}  $ for the
Galaxy, we have
\begin{equation}
\label{abundanceequation}
[\textrm{X}/\textrm{H}]_{1}=-\frac{R_{0}/\textrm{kpc}}{\log e}\frac{d[\textrm{X}/\textrm{H}]}{d(R/\textrm{kpc})}.
\end{equation}

\section{Results}

Equation (\ref{abundanceequation}) yields the luminosity-dependent abundances
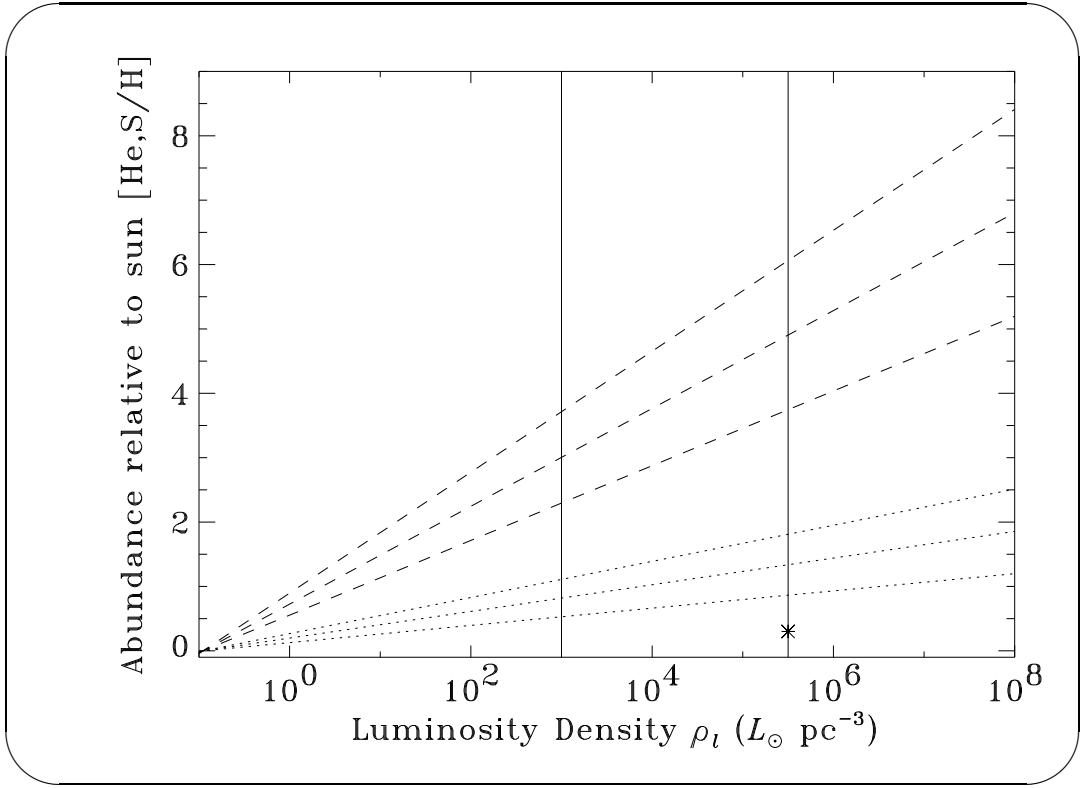
of He and S shown in \Fig{../R136/abundanceofHeandSinAGNs}{Middle dotted line:
abundance estimate of He as a function of mean local luminosity density according
to equation (\ref{loglinearused}) and Table 1 of Peimbert (1992). Outside dotted
lines show 1$  \sigma   $ observational errors ignoring the potentially large
systematic uncertainties of equation (\ref{loglinearused}). Dashed lines: abundance
and 1$  \sigma   $ errors of S. Solid (vertical) lines: lower and upper boundaries
of the luminosity densities that may be appropriate for NGC 5548. Regions to
the left of the left horizontal line have stellar densities that are much lower
than those near AGNs, while regions to the right of the right horizontal line
are not be realized due to mixing and other factors. Asterisk: the actual abundance
employed in the computer runs shown in Chapter 4, assuming the density luminosity
at $  r=1.0\, \textrm{pc}  $ in model 1.}. Because stellar orbits are suspected
of being highly eccentric below $  \sim   $1 pc (see, e.g., Appendix B)\pfn{
( \S~3.x??? needs to be here but seems not!)}, mixing should be dramatically
enhanced in this range, which would reduce any abundance gradients in this very
small region. For this reason, the upper limit to the luminosity density relevant
to equation (\ref{loglinearused}) was adopted to be that at $  r=1.0\, \textrm{pc}  $
in model 1 (which is described in Chapters 3 and 4). The lower limit to the
luminosity density was assumed to be the luminosity density at $  r=10\, \textrm{pc}  $,
approximately where the radial covering function attains its maximum value in
model 2. In these calculations, a mass to light ratio of 10 $  M_{\odot }/L_{\odot }  $
was adopted. This number is also not well constrained, and other ratios are
clearly possible. The abundance gradients $  d[\textrm{X}/\textrm{H}]/(dR/\textrm{kpc})  $
were obtained by statistically averaging results shown in Peimbert (1992), Friel
\& Janes (1993) (for Fe only), and Panagia \& Tosi (1981) (for Fe only). For
simplicity, solar abundances at Galactocentric radius $  R=R_{\odot }=7.8\, \textrm{kpc}  $
(not to be confused with the solar radius) were assumed, giving $  [\textrm{X}/\textrm{H}]_{0}=0  $.
The scale radius $  R_{0}  $ was taken to be $  4.5\textrm{ kpc}  $, while
the axially averaged luminosity density (see Appendix C) at $  R=R_{\odot }  $
was taken to be $  \rho _{l\odot }=3\times 0.039\, L_{\odot }\, \textrm{pc}^{-3}  $.

Figure \ref{../R136/abundanceofHeandSinAGNs} suggests, based upon what we know
about the evolution of the Galaxy, that metallicities in AGNs could be extremely
high. This is not to say that they are definitely this high, as there are far
too many fundamental systematic uncertainties (ignored physical effects) associated
with equation (\ref{loglinearused}). In fact, this is the reason that it was
decided to employ near solar abundances in our models. But very high AGN metallicities
would certainly not be in conflict with the metallicity gradients that are observed
in the Galaxy, especially if the IMF is top-heavy in these extreme environments.

\Fig{../R136/abundanceofCandNinAGNs}{Dotted lines: abundance and $  1\sigma   $
error estimates of C. Dashed lines: abundances and $  1\sigma   $ error estimates
of N. Other lines and data points are as in Figure \ref{../R136/abundanceofHeandSinAGNs}.}
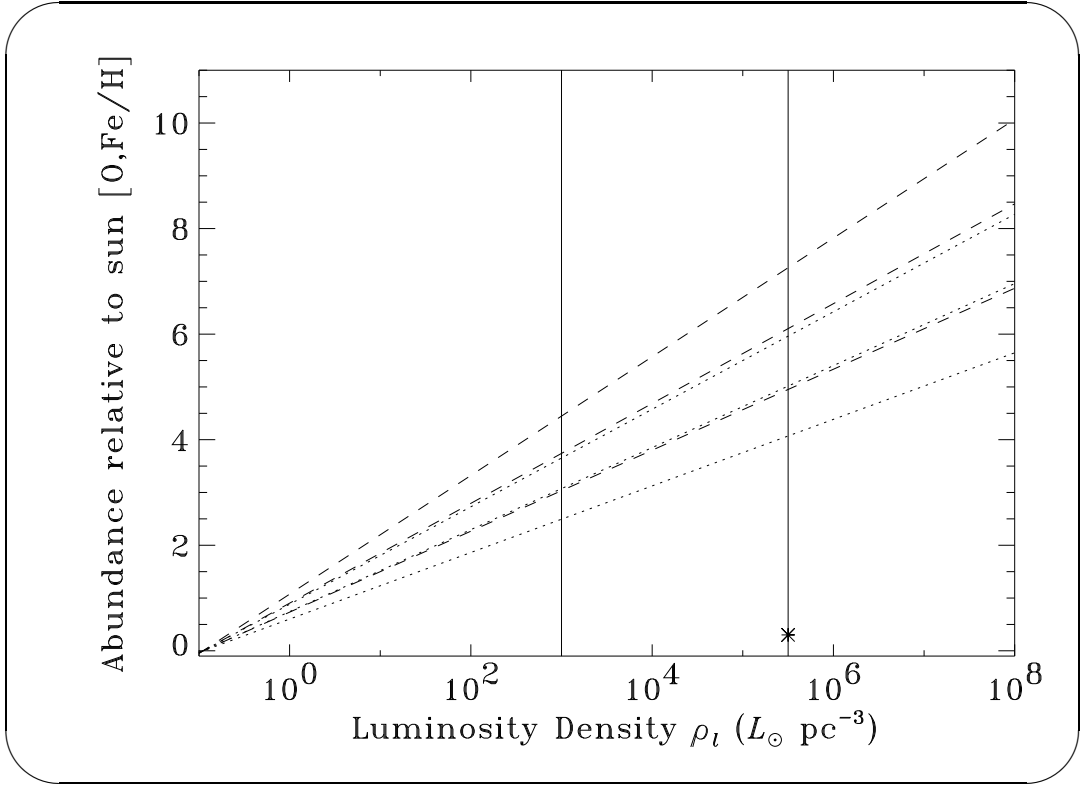
shows the abundance estimates for C and N, while \Fig{../R136/abundanceofOandFeinAGNs}{Dotted
lines: abundance and $  1\sigma   $ error estimates of O. Dashed lines: abundances
and $  1\sigma   $ error estimates of Fe. Other lines and data points are as
in Figure \ref{../R136/abundanceofHeandSinAGNs}.} shows the abundance estimates
of O and Fe. These plots suggest that C, N, O, and Fe abundances in AGNs may
indeed be extremely high. 

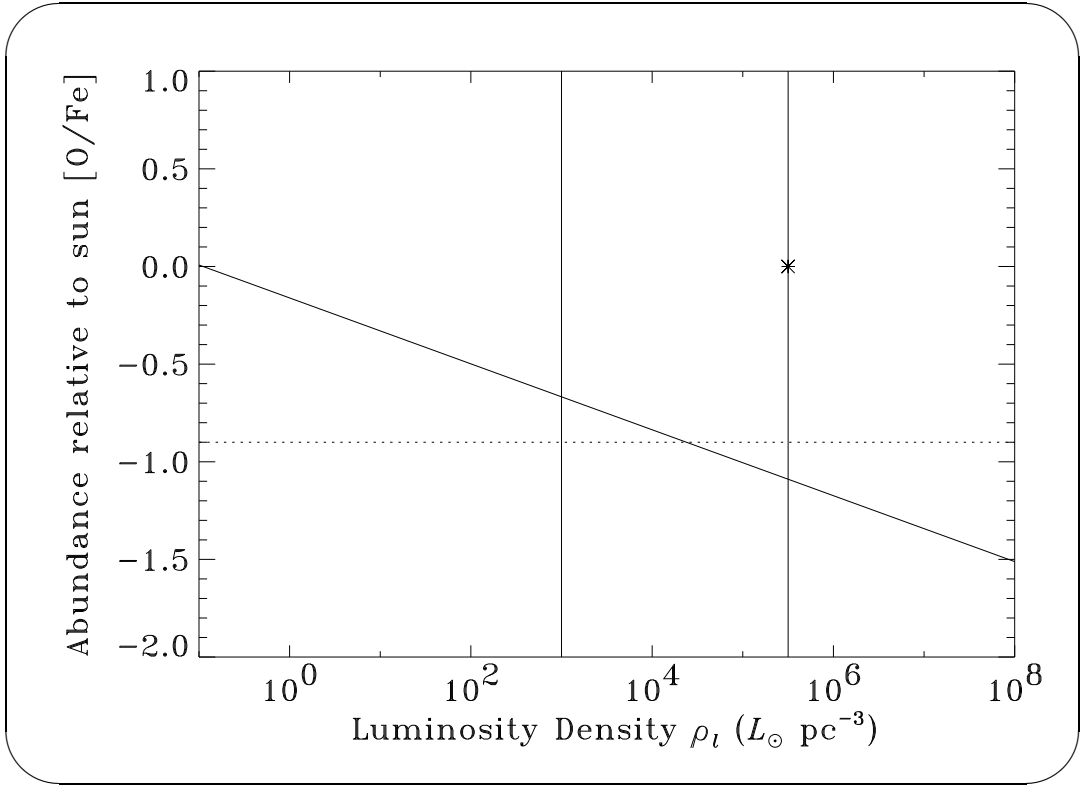
\Fig{../R136/abundanceofOoverFeinAGNs}{Solid line: projected {[}O/Fe{]} as a
function of luminosity density. shows just {[}O/Fe{]} without errors. Dotted
horizontal line: upper limit of {[}O/Fe{]} in the AGN NGC 1068 according to
Marshall et al. (1993). Other lines and datum point are labelled as in Figure
\ref{../R136/abundanceofHeandSinAGNs}.} shows the projected values of just the
{[}O/Fe{]} ratio. They suggest that A(Fe/O) should be enhanced in AGNs by an
order of magnitude or more. 

Taking this ``what-if'' scenario associated with equation (\ref{loglinearused})
one step farther, the lower limit to the luminosity density in NGC 1068 relevant
to equation (\ref{loglinearused}) can be very crudely estimated from the intersection
point of the lines in \ref{../R136/abundanceofOoverFeinAGNs}, yielding $  \rho _{\rm l}\gsim 2\times 10^{4}\, L_{\odot }\, \textrm{pc}^{-3}  $.
This luminosity density is well within reasonable limits. The beauty of what
we have done here is that there has been no need to resort to ``exotic'' physics,
such as drastic changes in iron opacities or sub-solar AGN O abundances (Marshall
et al. 1993).

\section{Discussion}

So far, we have explored the empirical side of metallicity gradients. We have
found that the Galactic metallicity gradients are compatible with the observed
AGN abundances. An obvious question to ask is whether or not these results are
fully consistent with various theoretical predictions. Interestingly, they may
not be. 

It is generally believed that Fe is predominantly produced by Type Ia supernovae
and that O is predominantly produced by Type II supernovae. This belief arises
in part from the observed supernovae spectral line ratios. It also arises from
current models (e.g., Tsujimoto et al. 1995), in which the majority of Fe produced
in Type II supernovae is imploded rather than exploded. This results in traditional
Type II supernovae models yielding a flat or even decreasing Fe yield as progenitor
mass increases. The same is not expected for O, which is synthesized at stellar
radii larger than that of the Fe. 

The potential problem stems from the fact that, for the reasons discussed in
Appendix C, IMFs probably become relatively top-heavy IMF toward galactic centers.
Since the progenitor masses of Type IIs are higher than the progenitor masses
of Type Ias, one may expect higher, rather than lower, O/Fe abundance ratios
towards the Galactic center. 

This suggests that one or more of the following is true:

\begin{enumerate}
\item The ratio of Type Ia to Type II supernovae increases towards the Galactic center
despite indications of IMF gradients. This could be due to, for example, the
binary fraction decreasing with Galactocentric radius. 
\item The Fe yield estimates of high-mass Type II supernovae are too low.
\item The present measurements of the Fe/O abundance Galactic gradient are incorrect,
the Marshall et al. (1993) result is an anomaly, and/or the IMF gradients and
mechanisms for metal production assumed in this chapter (e.g., eq. {[}\ref{loglinearused}{]})
are incorrect.
\end{enumerate}
Though possibility (1) is perhaps most likely, possibility (2) should not be
ruled out. This is because possibility (2) primarily involves the tweaking of
theoretical parameters not well established in the first place, like the radii
of the implosion/explosion boundaries in current one-dimensional models, which
may be smaller for high-mass supernovae progenitors than current estimates.
Some of the indications that current theoretical supernovae models may be inadequate
is the $  ^{138}  $Ba overabundance observed in SN 1987A, which is a factor
of 2.5 greater than theoretical upper limits (Mazzali \& Chugai 1995). Also,
recent Type Ic ``hypernovae'' SN 1998bw (Iwamoto et al. 1998) produced 0.7$  M_{\odot }  $
of Fe precursor $  ^{56}  $Ni. This yield exceeds traditional theoretical upper
limits by about one order of magnitude.

Finally, it is worth discussing an additional implication of equation (\ref{loglinearused})
towards AGN luminosities. Equation (\ref{loglinearused}) predicts that the
metallicity in observable AGNs increases with redshift. This is simply because
higher luminosity AGNs probably have higher maximum effective values of $  \rho _{\rm l}  $.
Physically, this makes sense according to the results of Appendixes B and C
which imply that the stars in the high-luminosity AGNs are more massive than
those in the Milky Way. This should result in much more efficient ISM enrichment.
Note that the lifetimes of massive stars (which are the underlying time scales
for enrichment of these systems) are extremely short compared to the age of
the universe. Therefore, since higher redshifts correspond to higher luminosities
for a flux-limited survey, AGN metallicities should (despite some claims to
the contrary) appear to increase with redshift. According to results of Hamann
\& Ferland (1992), this appears to indeed be the case.

\chapter{Does the Intercloud Medium Pressure Affect the Stellar Wind Shape?\label{cloudshape}
}

\def\figdirprefix{\home/scon/}For simplicity, the stellar winds in the models
we present in Chapter 4 are assumed to be spherically symmetric. In this appendix,
I briefly discuss the validity of this assumption if there is an intercloud
medium. 

The problem of a stellar wind moving through a homogeneous medium at high velocity
has been already addressed by, e.g., Brighenti \& D'Ercole (1995). The situation
is quite similar to that of colliding winds in binaries (e.g., Cooke, Fabian,
\& Pringle 1978) and, to a lesser extent, comet tails (e.g., Beard 1981). There
are probably five basic physical parameters of relevance: the terminal stellar
wind velocity $  v_{\infty }  $ on the ``shocked side'' of the star\rfn{Since
the continuum heating could affect the wind strength, this quantity could be
a function of the angle between the velocity and radius vectors.}, the velocity
of the star through the intercloud medium $  v_{*}  $, the mass loss rate of
the star $  \dot{M}  $, the intercloud density mass density $  \rho _{\rm {HIM}}  $,
and the speed of sound of the intercloud medium $  v_{\rm {s}}  $. 

If $  v_{*}\gg v_{\rm {s}}  $, these parameters can probably be orthogonalized
into a mere two of essential importance: the Mach number $  M\equiv v_{*}/v_{\rm {s}}  $
and the stand off distance of the shock to the star $  l  $. This stand off
distance in front of the stars is the place where the wind momentum flux balances
the intercepted intercloud momentum flux, 

\begin{equation}
\label{standoffdistance}
l=\left( \frac{\dot{M}v_{\infty }}{4\pi \rho _{\rm {HIM}}v^{2}_{*}}\right) ^{1/2}.
\end{equation}
 In the general vicinity of the star, the approximate shape of the shock front
for an $  M\gg 1  $ wind should be spherical directly in front of the star
and conical farther away such that the angle of the shock surface normal to
the stellar velocity vector is $  \leq \tan ^{-1}(1/M)  $.

Of interest here is just the question of spherical symmetry. In particular,
if $  l\gg R_{\rm w}  $, the line-emitting sections of the stellar winds are
clearly emitted from a region inside the shock front that is still spherically
symmetric. Otherwise, they may not be, and the models in Chapter 4 should probably
be re-run using a more sophisticated code capable of accounting for the asymmetries.

A key variable of equation (\ref{standoffdistance}), $  \rho _{\rm {HIM}}  $,
has not yet been reliably measured. Nevertheless, one can obtain potential upper
limits to $  \rho _{\rm {HIM}}  $ in a variety of ways. One upper limit, for
instance, is the value of $  \rho _{\rm {HIM}}  $ which would result in the
heating through drag (as the stars travel through the intercloud medium) being
comparable to or higher than the ordinary heating from the continuum. As shown
in Figure \ref{model2/shockpower}, these two sources of power become comparable
when the intercloud medium pressure is similar to the expected cloud edge pressure.
Another upper limit to $  \rho _{\rm {HIM}}  $ is that which would result in
Fe K-shell photoelectric absorption, which does not appear to be significant
in most AGNs. Mathews \& Ferland (1987) showed that if a BLR size of 10$  ^{19}  $
cm is assumed and the HIM is in pressure equilibrium with BLR clouds that have
edge densities of 10$  ^{9.5}  $ cm$  ^{-3}  $ , the intercloud medium temperature
required for lack of the Fe K-shell absorption edges is greater than 10$  ^{8.7}  $
K. A more modern estimate might assume a BLR radius of 10 light days (2.6$  \times   $10$  ^{16}  $
cm) and a BLR cloud edge density of 10$  ^{10}  $ cm$  ^{-3}  $, which would
yield an HIM temperature of above 10$  ^{7.2}  $ K. Since this temperature
is near the Compton temperature expected in AGNs, the upper limit is similar
to that obtained with an intercloud medium that has the same pressure as the
wind/cloud edges.

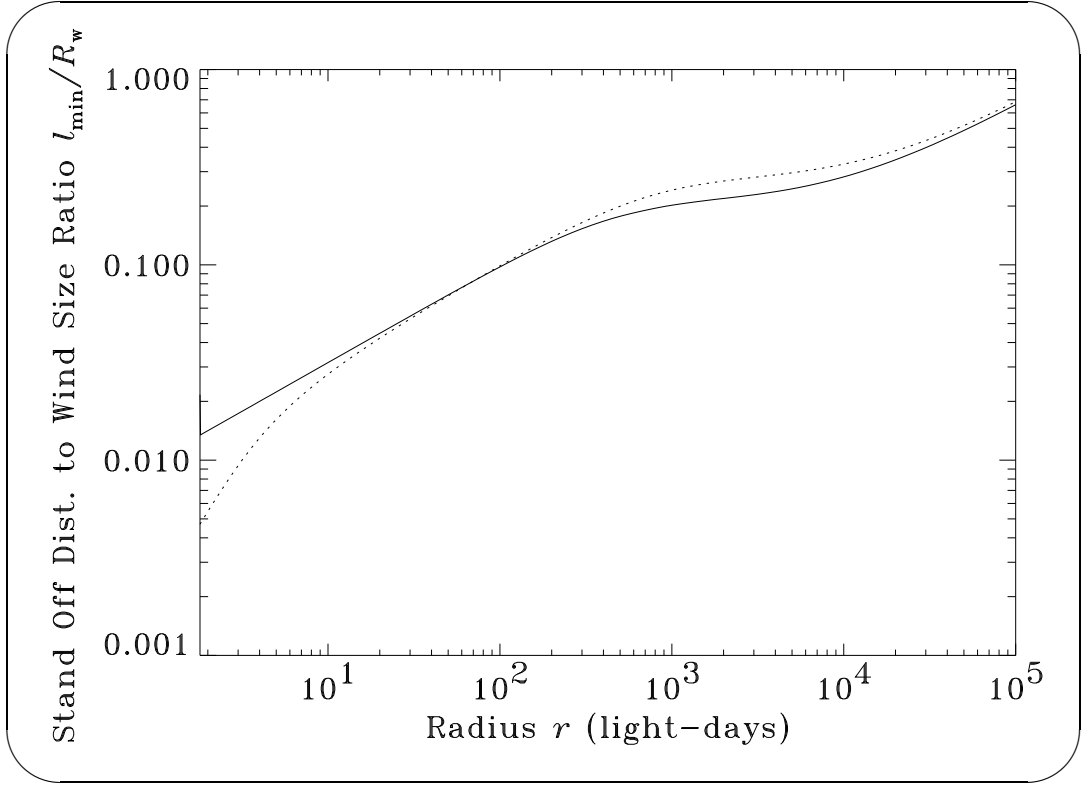
This upper limit gives the lower limit to $  l  $ shown in \Fig{model2/standoff}{Solid
line: minimum stand off distance $  l_{\rm {min}}  $ to wind size $  R_{\rm {w}}  $
ratio assuming $  T_{\rm {HIM}}=10^{7}\, \textrm{K}  $ and equal HIM and wind
edge pressures for model 2. Dotted line: ratio for model 1. If $  l\gg R_{\rm {w}}  $, the
line-emitting regions of the clouds should be unaffected by ram pressure. These
results imply (yet by no means prove) that this might not be the case.} assuming $  T_{\rm {HIM}}=10^{7}\, \textrm{K}  $
and a mean molecular weight per HIM particle of 1.3$  m_{\rm p}  $. This upper
limit to $  \rho _{\rm {HIM}}  $ results in $  l_{\rm {min}}<R_{\rm {w}}  $
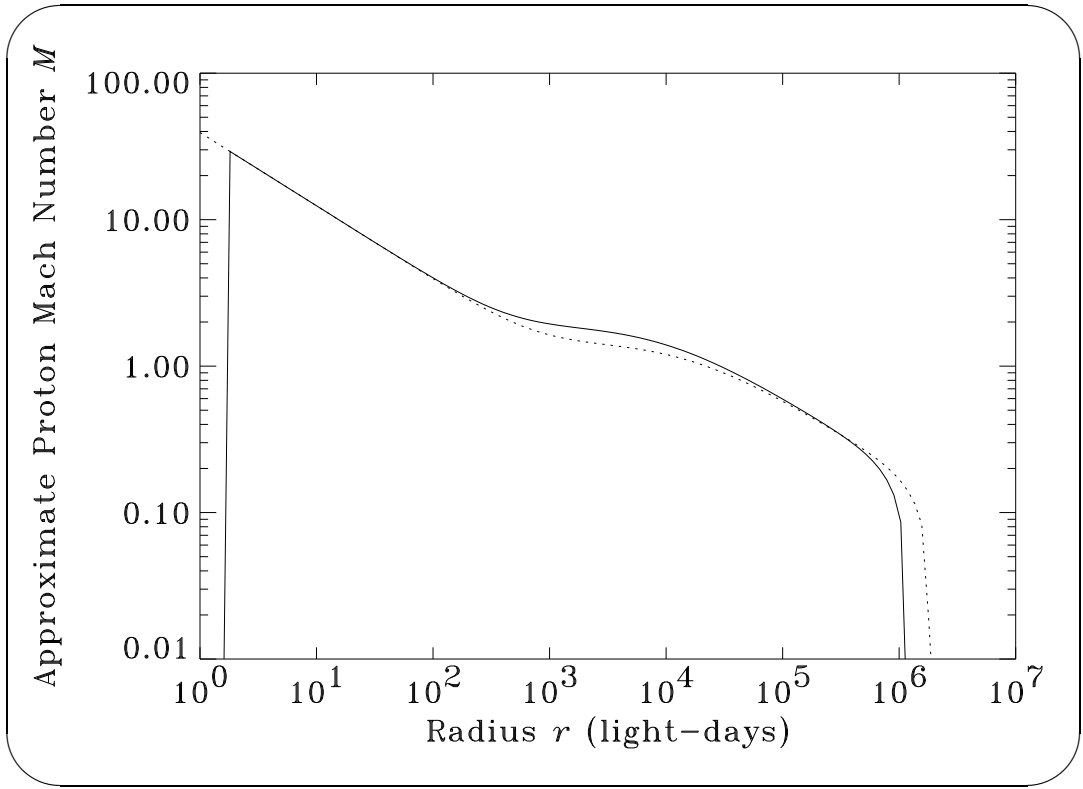
throughout the BLR. Incidentally, as \Fig{model2/machnumber}{Solid line: approximate
Mach number $  M  $ for the intercloud protons of model 2 if the intercloud
medium temperature is 10$  ^{7}  $ K. Dotted line: the approximate Mach number
for protons of model 1 under the same assumptions.} shows, the stars in this
model move supersonically throughout the BLR. These results suggest that nonspherical
effects due to the HIM could be important, particularly in low-$  v_{\infty }  $
models, such as the ones of AN94 and AN97.

However, as argued in Kazanas (1989), a feature of the wind models is that they,
unlike the pressure-equilibrium cloud models, do not require the existence of
any intercloud medium in the first place. Moreover, as is discussed in Appendix
\ref{shiftsap}, nonspherical models would require at least one additional parameter
to characterize the unknown physical properties of the intercloud medium in
the BLR. For these reasons, it was decided to not include nonspherical effects
in the code for the runs shown in Chapter 4. But this does not by any means
imply that this should not be done in future wind models, particularly if a
measurement of the BLR intercloud medium density is obtained which yields $  l\la R_{\rm w}  $.

\chapter{Local Delays\label{localdelays} 
}

\infootbegplain
\noindent \def\labep#1{\label{localdelays#1}}\def\figdirprefix{\home/completedpapers/lags-1/}In Chapter 2
we assumed that the emission characteristics of illuminated
clouds are purely a function of the instant
continuum flux to which they are
exposed.  In this appendix, I analyze the validity of this assumption.
I find that this universally adopted assumption may be wrong, and that
the history of
exposure accounting for ``local delays" due to finite cloud equilibrium times
may also be relevant.  In such cases, I show that the mean response time
is a function of the recent average value of the continuum.  I also show
that if instantaneous or linear response is incorrectly assumed, local delays
and nonlinear response can make a system appear larger than its actual size.
Finally, I show that local delays can be a source of asymmetry about the
peak of the cross-correlation function.
\par
\medskip
\noindent {}\setcounter{section}{1}{}\addtocounter{section}{-1}\section{Background} 
\par
\medskip
Procedures for computing a linearized response function of the
time-dependent line emission given off from an ensemble of clouds illuminated
by a time-dependent source are well known (e.g., Blandford \& McKee 1982).
They assume that the contribution toward line emission from a specific source
is purely a function of the radius from the central object and the
instantaneous
continuum flux to which it is subjected.  This requires that the processes
relevant
to its line emission attain equilibrium much more quickly than the other time
scales involved.  The explicit time-dependent response of individual clouds,
where, e.g., the line emission efficiency in a cloud lags the continuum flux
it experiences, has not yet been accounted for in previous works concerning
AGN variability.
\par
\medskip
Accounting for finite equilibrium times, however, can yield interesting
results for most of the AGN cloud models that have been proposed.  Consider,
for instance, a cloud model in which the cloud area is a decreasing function
of the cloud pressure, which is externally regulated by the pressure of an
intercloud medium.  Rees, Netzer, \& Ferland (1989) additionally assumed
$P\propto r^{-s}$, where $P$ is the pressure throughout the cloud and $r$ is the distance from
the black hole.  Let us consider the analogous case where the pressure is
regulated by the local ionizing continuum flux $F_{{\scriptsize\rm c}}$ and only indirectly through
$r$, namely $P\propto F^{s/2}_{{\scriptsize\rm c}}$.  Such a dependence implies that a change in the continuum
luminosity invokes a change in the cloud pressure as well.  As we shall see,
``reactive" cloud models like this one offer both theoretical and empirical
advantages over static ones.  Note that the clouds would not react
instantaneously; a minimum for the characteristic time scale for internal
pressure equilibrium to be asymptotically obtained is the sound crossing time
of the clouds.  As noted in Netzer (1990), this time scale can be similar
to the continuum variation time scales, suggesting that clouds of this model
rarely might be in actual pressure equilibrium.  Therefore, even though the
outermost layer emitting a line can be a small fraction of the cloud as a
whole, clouds of this model should to some extent ``remember" their prior
pressures and areas.
\par
\medskip
Because line emission from clouds is a strong function of the area,
pressure, and pressure ionization parameter $\Xi ~($defined here as the ionizing
photon to gas pressure), the line efficiency of a cloud has a nontrivial time
dependence.  For instance, consider the case where the continuum flux local to
a cloud suddenly increases.  If $s<2,$ the pressure ionization parameter of the
cloud would at first follow the increase in the continuum flux, but would then
decrease as the pressure begins to approach its new equilibrium value.
Relative to Ly$\alpha $, the flux in a line like N ${\scriptsize\rm v~\lambda }1240,$ which is probably a
relatively high ionization transition in stable cloud sections (Taylor 1994),
would initially rise, but then decay as the ionization parameter decreases.
The response function that one would obtain upon a linear fitting would have
structure not only at the range of lags corresponding to the light crossing
times of the emission region, but also at lags greater than these by the
pressure equilibrium times in the clouds.
\par
\medskip
In such a case, the previous works on AGN variability, which have all
assumed that a response function at a given lag is proportional to the density
of clouds along the corresponding ``iso-delay" surface, are inapplicable.
Specifically, the results based upon equation (2.13) of Blandford \& McKee
(1982), which was derived under the assumption that the equilibrium time
scales of the cloud properties are all much less than the light crossing time
(hereafter, the ``fast cloud" assumption), are now suspect.  This is an
important point because a great deal of effort has been expended to obtain and
analyze variability data using the approach of Blandford \& McKee (1982).
\par
\medskip
In \S ~\ref{localdelays}.2 of this appendix we find that
there are several cloud
properties affecting line emission that could be strong functions of the local
continuum flux with equilibrium times large enough to violate the fast cloud
assumption.  Because, for these cases, the popular formalism of Blandford \&
McKee (1982) is invalid, a new and more general
formalism for analyzing variability data will be developed in
\S ~\ref{localdelays}.3.  This new
formalism is compatible with models that have clouds with finite equilibrium
times and nonlinear responses.  Readers not interested
in the mathematical derivation of the time-dependent line profile with the new
formalism may wish to skip to \S ~\ref{localdelays}.4, where the new theory is
applied to some simple models. A summary is provided in
\S ~\ref{localdelays}.5.
\par
\medskip
\noindent {}\setcounter{section}{2}{}\addtocounter{section}{-1}\section{Motivation}
In order for the formalism of Blandford \& McKee (1982) to be {\it in}valid for
a given
cloud model, two conditions must be satisfied for at least one of the cloud
properties in the model.  The first of these conditions is that the line
emissivity be a moderately strong function of the cloud property and that the
cloud property in turn be a moderately strong function of the local continuum
flux a cloud experiences.  The second condition is that the equilibrium time
scale of the cloud property be near one of the other characteristic time
scales of the system.  If the equilibrium time scale is near or greater than
the line emission region light crossing time, the response function will be
affected.  Conversely, if the equilibrium time is near the time for
clouds to cross the emission region, the time-averaged line profile can be
affected.  Determining the precise way in which the response functions and
profiles are
affected requires a detailed and highly model-dependent analysis.  Before
going through such an analysis, let us first discuss some of the cloud line
emission model properties which apparently meet the above two conditions.
\par
\medskip
Table \ref{localdelaystimescales} lists some of the processes responsible
for reactive
cloud properties in several of the models that have been proposed and the
equilibrium time scales associated with them.  Also shown is whether the
slowness of equilibrium affects the response functions, line profiles, or line
ratios.  The first entry is for a two-phase pressure-equilibrium model
(e.g., Wolfe 1974; Krolik, McKee, \& Tarter 1981).  Assuming in this case that
the cloud pressure is regulated by pressure of the intercloud medium, the
delay in the cloud pressure response to the continuum is limited by the
intercloud temperature equilibrium time scale.  For the model parameters
described in Table \ref{localdelaystimescales}, this is (only) $\simeq 43$ days.  If
the dependence of the
intercloud temperature upon the local continuum flux is strong enough, the
responding pressure will affect the response functions for the parameters
assumed in Table \ref{localdelaystimescales} in a highly line-dependent
fashion, giving the line ratios
a complicated time dependence.  Furthermore, if the cloud identities are
preserved (as in Rees, Netzer, \& Ferland 1989), the slowness of the cloud area
and column density reactions will also affect the response functions
respectively in a line-independent and weakly line-dependent fashion.  The
time-averaged line profiles for this model are not affected by the finite
pressure equilibrium time, which is too small compared to the cloud crossing
time~$(\sim 2$ years for the parameters shown in Table \ref{localdelaystimescales})
to be affected. However,
if the intercloud temperature dependence is moderately strong, this model,
like several others that are not immune to the various processes analyzed in
Table \ref{localdelaystimescales}, requires use of a new formalism.  Such a
formalism will be developed in \S ~\ref{localdelays}.3.
\par
\begin{table}
\renewcommand{\footnotesize}{\normalsize}
\begin{minipage}{\textwidth}
\small\baselineskip=0.5truecm
\begin{center}
\renewcommand{\footnoterule}{\rm} \newdimen\digitwidth \setbox0=\hbox{\rm0}
\digitwidth=\wd0 \catcode`!=\active \def!{\kern\digitwidth}
\caption{Equilibrium Processes of Some Reactive Cloud Model Parameters and Their Effects}
\label{localdelaystimescales}
\begin{tabular*}{\textwidth}{cccc} 
\\ \hline \hline
\par
\noindent Name of Limiting               & Cloud      & Parameter     & Observational \\
\par
\noindent Process                        & Parameters & Equilibrium   & Parameters \\
\par
& Affected   & Times         & Affected \\
\hline
\par
\noindent Inter-cloud Cooling$^{1}$           & $A, N_{{\scriptsize\rm c}}, P_{l}$  & 43 days       &RF, LR (strong)\\
Thermal Evaporation$^{2}$           & $A, N_{{\scriptsize\rm c}}, P_{l}$  & 20 years & RF, LP, LR (strong)\\
\par
\noindent Pressure-limited Evaporation$^{3}$  & $A, N_{{\scriptsize\rm c}}$      & 2.0 days      & RF, LR (weak)\\
\par
\noindent Pressure-limited Evaporation$^{3,4}$& $P_{l}$         & $\sim 0.37$ days    &RF, LR (strong)\\
\par
\noindent Stellar Wind Expansion$^{5}$        & $A, N_{{\scriptsize\rm c}}, P_{l}$  & 20 days       &RF, LR (strong)\\
\par
\noindent Stellar Photospheric Heating$^{6}$  &$A, N_{{\scriptsize\rm c}}, P_{l}$&$10^{-2}-10^{11}$ days&RF, LP, LR (strong)\\
\par
\noindent Magnetic Confinement$^{7}$          & $A, N_{{\scriptsize\rm c}}, P_{l}$  & 4.3 days  &RF{}\def\thefollowingdummymacroisapersonalnotetomyselfgeneratedbymyconverterprogramthatshouldnotbeprintedpleaseignoreit{, LP}, LR
(strong)\\ \hline \hline \end{tabular*}
\end{center}
NOTE---``RF," ``LP," and ``LR" are respective abbreviations for ``response
functions," ``line profiles," and ``line ratios."  These results are for
clouds at a fiducial radius from the continuum source $r_{0}$ of 10 light-days (the
light crossing time scale), a fiducial local continuum flux of $10^{44}/(4\pi r^{2}_{0})$
ergs cm$^{-2}$, a fiducial velocity of 4000 km s$^{-1}$, a fiducial cloud hydrogen
density of $10^{11}$ cm$^{-3}$, and a fiducial mean column density of $10^{22.5}$ cm$^{-2}$.
\par
\medskip
\noindent $^{1}$Only for models with clouds in pressure equilibrium with a hot
inter-cloud medium (e.g., Krolik, McKee, \& Tarter 1981).  Calculation assumes
an inter-cloud temperature of $10^{7}$ K, which implies that the dominant source
of cooling is thermal Bremsstrahlung, which in turn implies flux-dependent
(reactive) cloud parameters.
\par
\medskip
\noindent $^{2}$Adopted from results in Krinsky and Puetter (1992), but after scaling to the
column density assumed here.  Line ratios are only strongly affected for
pressure-stratified clouds.
\par
\medskip
\noindent $^{3}$Only for pressure-stratified cloud models, see Taylor (1994).
\par
\medskip
\noindent $^{4}$Only for lines emitted uniformly from the inverse-Str\"omgren region.
\par
\medskip
\noindent $^{5}$Adopted from parameters assumed in Schaaf \& Schmutzler (1992).
\par
\medskip
\noindent $^{6}$Adapted from Harpaz \& Rappaport (1991) and Antona \& Ergma (1993).
\par
\medskip
\noindent $^{7}$As in Rees (1987), but assuming the field responds to the continuum flux on
the Alv\'en wave cloud crossing time.\pfn{check that reference
\par
 }
\par
\end{minipage}
\end{table}

\medskip
Note that the physical processes considered in Table
\ref{localdelaystimescales} were drawn
from the set of processes invoked by the various cloud models that have
been proposed.  In principle, all of these could be incorrect.  Therefore,
Table \ref{localdelaystimescales} is necessarily incomplete.  For this reason,
the analysis of
time-dependent cloud response could be important even if all of the processes
in Table \ref{localdelaystimescales} somehow accommodated the fast cloud
assumption.\pfn{As will be shown
in Paper IV, such analysis does in fact constrain the mechanisms that are
permissible in AGN models.}
\bigskip
\par
\infootbegplain
\noindent {}\setcounter{section}{3}{}\addtocounter{section}{-1}\section{Theory of Response}
\medskip
\noindent \par In this section we shall extend the formalism in Blandford \& McKee (1982)
so that information can be obtained from variability data about models which
violate two fundamental assumptions made in Blandford \& McKee (1982): (1)
instantaneous response and (2) linear response.
\par
\medskip
\noindent {}\setcounter{section}{3}{}\setcounter{subsection}{1}{}\addtocounter{subsection}{-1}\subsection{Locally Delayed Response}
\par
\medskip
Before one can understand the overall, global response of systems that
can violate the fast cloud assumption, one must first understand the response
of the individual clouds that make up such systems.  In this section a general
method of determining the time dependence of an arbitrary cloud property is
derived.  In \S ~\ref{localdelays}.3.2, this method will be used to obtain the
global response of systems that can violate the fast cloud assumption.
\par
\medskip
The character of the response of an AGN cloud depends critically upon the
relative magnitude of two time scales.  One of these is the variation time
scale of some condition externally imposed upon the cloud, such as the local
continuum flux or intercloud pressure.  Another is the characteristic
equilibrium time scale of a physical property of the cloud, such as its
temperature or size, in response to the variations of the external conditions.
As an example, let us consider the case of a cloud with a physical property
that is an increasing function of the local continuum flux.  Let us also
assume, for this example, that the continuum source, itself, is
time-independent, but
that the cloud is in a periodic orbit about the black hole.  This situation is
a simple variation on those in which the continuum does vary.  If we assume
that the orbital period is significantly greater than the equilibrium time
scale of the property, the physical property would lag the time-averaged
continuum flux to which the cloud is exposed as it orbits the black hole.
The line
emission in such a cloud would depend not only on its position, but also on
another variable which indicates its orbital phase.  Under certain conditions
(see Appendix \ref{equilibriumiscrossing}), it can be shown that this variable
can be the cloud velocity
vector.  For instance, if the line emissivity of a cloud is an increasing
function of just the physical property, then the line emission from the cloud
would be greatest not when it is closest to the black hole, but slightly
farther away, after the cloud has acquired a small outward velocity.  As the
orbital time becomes even larger, we approach the ``fast cloud" regime.  In
this regime we can assume that the physical property in the cloud reacts fast
enough that it is purely a function of the local flux or, in this case, of the
distance from the black hole such that the phase lag is zero.
\par
\medskip
A second case to consider is one wherein the variation time scale of the
local continuum flux of a cloud is significantly smaller than the equilibrium
time of the property.  Here the physical property of the gas would lag the
orbital
motion by a significant phase.  This implies that a line with a strong enough
dependence upon the lagged property could attain maximum flux when the cloud
has a relatively high outward radial velocity.  In this case, as the variation
time scale of the input continuum flux becomes even smaller, we approach the
``slow cloud" regime, where we can simply assume that the relevant physical
property is a constant throughout the orbit.
\par
\medskip
The third possible case to consider is one in which the local continuum
variation time scale is intermediate and similar to the
equilibrium time
scale. Understanding this case requires a more quantitative approach than the
other two cases.  Let us call the generic cloud property of interest $y(t)$
where ${\it t}$ is the time measured in the reference frame of the observer.  The
analysis which follows is quite general and $y(t)$ could represent properties
such as the mean cloud area, pressure, or column density.  Similarly, let $x(t)$
be a generic input, such as the local continuum flux near the cloud, of which
$y(t)$ is assumed to be a function.  Let $y^\prime (x)$ be the asymptotic functional
dependence of $y$ upon $x$ once sufficient time has elapsed for equilibrium to be
established, where the prime denotes the functional dependence in the fast
cloud regime.  Furthermore, let us assume that there exists a characteristic
time scale $\tau _{y}$ for $y$ to respond to changes in {\it x}.  Such a characteristic time
will be equal to the ratio of the extent to which $y$ is out of equilibrium to
the rate at which the non-instantaneously responding component of $y$ actually
attains its equilibrium value.  This gives
\par
\s
\tau _{y}={y^\prime {\hbox{\bf(}}x(t){\hbox{\bf)}}-y(t)\over \dot{y}(t)-\dot{x}(t)\tilde{\hat\Psi}_{y|x}(\infty )<y>/<x>},
\labep{3.10}\e
\noindent where $\tilde{\hat\Psi}_{y|x}(\infty )$ is a free parameter that is the instantaneous component of the
``gain" of $y$ with respect to ${\it x}$, and $<x>$ is the average or ``bias" of ${\it x}$, etc.
The gain itself is an operator (defined by eq. [{}\ref{localdelays3.4}]) that
yields the
dimensionless ratio of the amplitudes of small variations of an output about
its mean with respect to that of some input.  It is merely the Fourier
transform of the linearized response function (see Appendix B).  The implicit
assumption here that $\tau _{y}(x)$ is approximately constant could be invalid under
the following conditions: the variations of $x$ are large enough, the initial
conditions are far enough from equilibrium, or the equilibrium time has an
explicit dependence upon the sign of $\dot{x}(t)$.  In these cases the physics
associated with the response time is not properly described by only one
parameter.  Otherwise, equation ({}\ref{localdelays3.10}) completely characterizes
the system
given the prior inputs $x(t^\prime \le t)$ and the other system characteristics $y^\prime (x), \tau _{y}$,
and $\tilde{\hat\Psi}_{y|x}(\infty )$.
\par
\medskip
Though we will assume that $y^\prime (x)$ is a nonlinear function, in certain
instances we shall find it highly instructive to consider the case in which
the variations in $x$ are small enough that $y^\prime (x)$ is accurately described by a
first-order Taylor expansion.  Performing such linearization of equation
({}\ref{localdelays3.10})~(with eqs.~[{}\ref{localdelays3.1}]-[{}\ref{localdelays3.5}])
yields in the frequency domain
\par
\s
\tilde{\hat\Psi}_{y|x}(\omega )={\eta (y|x)+i \tilde{\hat\Psi}_{y|x}(\infty )\tau _{y}\omega \over 1+i\tau _{y}\omega },
\labep{3.11}\e
\noindent where $\omega _{y}\equiv 1/\tau _{y}$ and $\eta (y|x)\equiv \tilde{\hat\Psi}_{y|x}(0)$ is the ``asymptotic gain" of $y$ with respect
to ${\it x~}($see also eq. [{}\ref{localdelays3.3}]).  Equation ({}\ref{localdelays3.11}) can
be used to formulate a
more precise definition of the fast and slow cloud regimes, which respectively
occur for $\omega \ll \omega _{y}$ and $\omega \gg \omega _{y}$, where the transfer function becomes a trivial
function of $\omega ~($flat in $\log -\log $ coordinates).
\par
\medskip
Equation ({}\ref{localdelays3.11}) yields in the time domain
\par
\s
\hat{\Psi }_{y|x}(\tau )=\tilde{\hat\Psi}_{y|x}(\infty )\delta (\tau )+\Theta (\tau )[\eta (y|x)-\tilde{\hat\Psi}_{y|x}(\infty )]\omega _{y}e^{-\tau \omega _{y}},
\labep{3.12.5.5}\e
\noindent where $\Theta $ is the step function.  This response function tells us (namely via eq.
[{}\ref{localdelays3.13}]) the contribution in the linear regime toward the output
(e.g., the
cloud area) made by an input (e.g., the local continuum flux [measured in the
reference frame of the cloud]) at a prior time.  The first term in equation
({}\ref{localdelays3.12.5.5}) is the component of $y$ that mirrors the variations in
${\it x}$ without
delay, while the second term is the component of $y$ that responds on the time
scale $\tau _{y}$.
\par
\medskip
For the important case in which the instantaneous component of the gain
is zero, equations ({}\ref{localdelays3.11})-({}\ref{localdelays3.12.5.5}) yield the
results indicated earlier
in this section: in the fast cloud regime they yield an output that mirrors
the input variations, while in the slow cloud regime they yield an output that
is constant.
\par
\medskip
\noindent {}\setcounter{section}{3}{}\setcounter{subsection}{2}{}\addtocounter{subsection}{-1}\subsection{The Line Profile}
\par
\medskip
Now that we have prescribed a general way of accounting for individual
cloud properties that exhibit hysteresis-like behavior, we can derive the more
observable properties of AGN models which have finite (rather than zero)
equilibrium times.  Of particular interest here is the angle-dependent
apparent luminosity $L^{{\scriptsize\rm cl}}_{l}$ emitted in line $l$ of a cloud with position vector
from the black hole ${\bf r}$ and velocity vector ${\bf v}$.  An expression for this that is
general enough for the models that will be analyzed in this\pfn{series of
papers} appendix and which takes local delays into account is
\par
\s
L^{{\scriptsize\rm cl}}_{l}(t,{\bf r},{\bf v},\hat{{\bf s}})=F_{{\scriptsize\rm c}}(t,{\bf r})A\epsilon _{l}[1+\epsilon _{\hbox{{\scriptsize\rm A}{\scriptsize\it l}}}\hat{{\bf r}}\cdot \hat{{\bf s}}]+L^{{\scriptsize\rm s}}_{l},
\labep{3.16.5}\e
\noindent where $F_{{\scriptsize\rm c}}(t,{\bf r})$ is the ionizing continuum flux at ${\bf r}$, {\it A} is the cloud area, $\epsilon _{l}$
is the dimensionless emission efficiency for line $l,~\epsilon _{\hbox{{\scriptsize\rm A}{\scriptsize\it l}}}$ is the first-moment
correction to the efficiency for an anisotropically emitting cloud, ${\bf D}\equiv {\bf r}+{\bf s}$ is
the position vector of the observer, and $L^{{\scriptsize\rm s}}_{l}$ is the cloud luminosity in line $l$
due to resonance scattering.  Each of the cloud parameters
in equation ({}\ref{localdelays3.16.5}) that has an equilibrium time near or
greater
than $r/c~($hereafter, the ``spatial time" scale) must be evaluated using the
appropriate form of equations ({}\ref{localdelays3.12.5.5}) and
({}\ref{localdelays3.13}). Before a model
conforming to equation ({}\ref{localdelays3.16.5}) can have predictive power, not
only must
the continuum light curve be measured, but also estimates of the
time scales and the asymptotic functional dependence of each cloud parameter
upon the local continuum flux must be made.
\par
\medskip
Once a specific expression for the observed line flux from an individual
cloud is assumed, the macroscopic characteristics of the global system
composed of several clouds are easy to calculate.  Neglecting absorption, the
flux per cloud observable at ${\bf D}$ is $F^{{\scriptsize\rm cl}}_{l}(t,{\bf r},{\bf v};{\bf D})\propto L^{{\scriptsize\rm cl}}_{l}(t-s/c,{\bf r},{\bf v},\hat{{\bf s}})/s^{2}$.  With
this terminology, the time-dependent line profile becomes (see Appendix
\ref{equilibriumiscrossing})
\par
\s
F_{l}(t,v_{D}) = \int d^{3}rd^{3}vf({\bf r},{\bf v})F^{{\scriptsize\rm cl}}_{l}(t,{\bf r},{\bf v};{\bf D})\delta (v_{D}+{\bf v}\cdot \hat{{\bf D}}),
\labep{3.23}\e
\noindent where $v_{D}$ is the equivalent tangential velocity and $f$ is the distribution
function.  Under the conditions
specified earlier, the above equation permits computation of the line profile
for any class of AGN cloud line emission models which can be described by
equation ({}\ref{localdelays3.16.5}).
\par
\medskip
Though equation ({}\ref{localdelays3.23}) provides a means of computing the
nonlinear line profile response when the input history is known, applying it
can be computationally expensive (though not as much as the
trajectory-dependent
sum method considered in Appendix \ref{equilibriumiscrossing}).  This is
because it requires
modeling the cloud properties such as the $F^{{\scriptsize\rm cl}}_{l}$ function, which from a
numerical perspective is an array with dimensions ${\bf r}$ and ${\bf v}$ that evolves with
time, though probably only weakly in the ${\bf v}$ dimensions.  If local delays are
important, evaluation of $F^{{\scriptsize\rm cl}}_{l}$ at each point in time requires integrating over
history according to the appropriate forms of equation ({}\ref{localdelays3.13}).
Since the
positional integral in the above equation can be interpreted as an integral
over history, the expression for the line profile is a double
integration over lag.  This is in contrast to the analogous expression for
the line profile given by equation (2.12) of Blandford \& McKee (1982), which
involves only a single integration over lag.
\par
\medskip
\noindent {}\setcounter{section}{3}{}\setcounter{subsection}{3}{}\addtocounter{subsection}{-1}\subsection{The Line Transfer Function and Linear Approximation of the Line
Profile}By linearizing equation ({}\ref{localdelays3.16.5})~(see
Appendix
B), one of the integrations in lag in equation ({}\ref{localdelays3.23}) can be
eliminated,
and the computer time required to obtain the time-dependent line profile of a
model can be significantly reduced.  For several key cases,
we find that these benefits outweigh the inaccuracy of linear models.
\par
\medskip
The first step in linearizing the line flux emitted from an individual
cloud is to obtain its gain about the bias continuum flux.  This in turn
requires determining the transfer function (eq. [{}\ref{localdelays3.11}]) of each
flux-dependent cloud parameter affecting the line emission in equation
({}\ref{localdelays3.16.5}).  Because the gain is calculated by considering small
perturbations about the mean of the input, the gain of each of these
parameters can be computed using the time-averaged local continuum flux, which
is dependent only upon position.  In terms of the gains of these cloud
parameters, equation ({}\ref{localdelays3.16.5}) yields for an individual cloud
\par
\s
\tilde{\hat\Psi}_{L^{{\scriptsize\rm cl}}_{l}|F_{{\scriptsize\rm c}}}(\omega ;{\bf r},{\bf v},\hat{{\bf s}})=1+\tilde{\hat\Psi}_{A|F_{{\scriptsize\rm c}}}(\omega )+\tilde{\hat\Psi}_{\epsilon _{l}|F_{{\scriptsize\rm c}}}(\omega )+{<\epsilon _{\hbox{{\scriptsize\rm A}{\scriptsize\it l}}}>\hat{{\bf r}}\cdot \hat{{\bf s}}\over 1+<\epsilon _{\hbox{{\scriptsize\rm A}{\scriptsize\it l}}}>\hat{{\bf r}}\cdot \hat{{\bf s}}}\tilde{\hat\Psi}_{\epsilon _{\hbox{{\scriptsize\rm A}{\scriptsize\it l}}}|F_{{\scriptsize\rm c}}}(\omega )+\tilde{\hat\Psi}_{L^{{\scriptsize\rm s}}_{l}|F_{{\scriptsize\rm c}}}(\omega ),
\labep{3.24}\e
\noindent where we will implicitly assume that the continuum flux is evaluated locally
(at ${\bf r})$.  Each of the above terms is proportional to the gain of one of the
reactive cloud parameters.  The third term itself is the sum of three highly
model-dependent terms if $\epsilon _{l}=\epsilon _{l}(\Xi ,P,N_{{\scriptsize\rm c}})$ sufficiently parameterizes the cloud
emission for a given model.  Note that even if the
various cloud properties such as the area are described without approximation
by a nontrivial linear response function, the output line response in a cloud
is nonlinear nonetheless.  This is because the above equation provides only an
approximation to the response valid for small perturbations about a mean.
Such nonlinearity is a general property of reactive cloud models.
\par
\medskip
With each cloud response linearized, the remaining time dependence in
the system line flux equation is due purely from the $F_{{\scriptsize\rm c}}$ factor, so the global
transfer function of the profile is
\par
\medskip
\noindent $$\tilde{\hat\Psi}_{F_{l}|F_{{\scriptsize\rm c}}}(\omega ;v_{D})=\int d^{3}r d^{3}v{<F^{{\scriptsize\rm cl}}_{l}(t,{\bf r},{\bf v};{\bf D})>_{t}\over <F_{l}>}f({\bf r},{\bf v})\tilde{\hat\Psi}_{F^{{\scriptsize\rm cl}}_{l}|F_{{\scriptsize\rm c}}}(\omega ;{\bf r},{\bf v},\hat{{\bf s}})\times $$
\s
e^{-i\omega (\hat{{\bf r}}-\hat{{\bf D}})\cdot {\bf r}/c}\delta (v_{D}+{\bf v}\cdot \hat{{\bf D}}).
\labep{3.25}\e
\noindent The line flux gain of an individual cloud appearing in this equation is equal
to the gain of the individual line luminosity (eq. [{}\ref{localdelays3.24}])
if, for the time being, we neglect absorption.  Note that unlike the Fourier
transform of the expression given for the response function by Blandford \&
McKee (1982), the above equation has a factor of the cloud gain that can be
nonzero at nonzero frequency.  In the time domain, equation
({}\ref{localdelays3.25})
gives (via eq. [{}\ref{localdelays3.13}]) the linear approximation to the line
profile flux,
\par
\s
F_{l}(t,v_{D})=<F_{l}(t,v_{D})>_{t}\left[1+\int d\tau \left({F_{{\scriptsize\rm c}}(\tau -t)\over <F_{{\scriptsize\rm c}}>}-1\right)\hat{\Psi }_{F_{l}|F_{{\scriptsize\rm c}}}(\tau ;v_{D})\right] \pm  \sigma _{F_{l}},
\labep{3.25.5}\e
\noindent where $\sigma _{F_{l}}$ represents the input-dependent error.  Unlike
equation ({}\ref{localdelays3.23}), this equation does
not have an implicit
nested integration in lag.  Applications of it to the models considered in
this appendix \pfn{paper}that account for the dependence of $F^{{\scriptsize\rm cl}}_{l}$ upon the
direction
of the velocity vector are provided elsewhere (\pfn{Paper IV;}Taylor 1994).
\par
\medskip
It is worth repeating that a condition for the linearized response
function to be descriptive is that at a given radius the continuum variations
are small enough that the second order derivatives can be
neglected.  Since large-scale variations in the continuum luminosity are known
to occur, the system would be somewhat contrived to consistently obey this
condition.  For instance, in the slow cloud regime the linearized form of
equation ({}\ref{localdelays3.24}) will generally be inaccurate at low enough mean
cloud ionization parameters when the emitting ion is in partial fractional
abundance.  In this case, the second derivative of $\epsilon _{l}$ with respect to $F_{{\scriptsize\rm c}}$ would
not only be large and positive for most lines, but would also be a sensitive
function of input level.  If the asymptotic mean cloud ionization parameter is
a decreasing function of the local flux
$(s<2~[$\S ~\ref{localdelays}.1]), overestimates for the size of the line emission
region when using fully linear models are implied.  Even for wind cloud models
(e.g., Kazanas 1989) in the fast cloud regime with $s=2,$ where the effective
ionization parameter can be taken to be constant, nonlinearities would still
arise from the dependence of the cloud area upon flux for which a constant
$\eta (A|F_{{\scriptsize\rm c}})$ term cannot account.  In any of these types of situations, the Fourier
transform of the oscillatory component of the line flux would not be
proportional to that of the continuum, and forms of equation
({}\ref{localdelays3.25.5}) would not accurately describe the variability that
would be observed.
\par
\medskip
In such cases, the ``optimal" response function that best fits real
data becomes a function of both the specific data set as well as the fitting
criterion (see also eqs. [{}\ref{localdelays3.7}]-[{}\ref{localdelays3.8}]).  Therefore,
its utility
in measuring any of the epoch-independent features of AGN and AGN models is
somewhat questionable.  This is in contrast with the linearized response
function (eq. [{}\ref{localdelays3.25}]), which is dependent upon only the mean of
the
continuum flux.  Ideally, a fitting criteria would exist that would reliably
yield this input-independent, poorly fitting linearized response function
rather than the optimal one.  However, there are alternative parameters and
parameterized functions for analyses of variability data that completely
bypass this problem.  One alternative is the cross-correlation function.
However, even in the linear regime this is also a strong function of the
excitation characteristics.  (See also \S ~\ref{localdelays}.4.)  A more
promising alternative
is to fit nonlinear models (e.g., eq. [{}\ref{localdelays3.12}]; Taylor \& Kazanas
1992) to data.
Using nonlinear models offers the potential of epoch-independent fitting or
measurement within the context of a model of physical AGN properties even when
the continuum variations are large or the line emission is a sensitive
function of the flux.  \pfn{This point will be discussed in more detail in
Paper II.}
\medskip
\noindent {}\setcounter{section}{4}{}\addtocounter{section}{-1}\section{Examples}
\par
\medskip
Let us consider cloud models in which the effective reprocessing
efficiencies are increasing functions of the local continuum flux, with the
equilibrium time scale of the relevant physical properties being slightly
larger than the characteristic light crossing time.  From
\S ~\ref{localdelays}.3, we know that
responses to short and weak pulses of continuum radiation in such a system
could be modeled satisfactorily with a linear ``spatial" response function,
which is the response function of the system were the fast cloud regime
applicable.  This response function has structure on just the light crossing
times of the emission region, as the efficiency and physical conditions of the
clouds in such a system would deviate only slightly from their mean values.
Similarly, responses to long pulses of fixed intensity probably also could
be mimicked with a {\it different} linear response function that had additional
structure at lags beyond the cloud equilibrium times.  However, if either
short pulses and long pulses of constant intensity or long pulses of varying
intensity occurred in such a system, a single linear response function would
not be able to fit all aspects of the variability.  A linear system would
respond either too strongly to the weak pulses or too weakly to the strong
pulses and furthermore would respond either too slowly to the weak pulses or
too rapidly to the energetic pulses.  The first two types of nonlinear
behavior are due to an input-dependent asymptotic gain, while the latter two
are due to nested lags or ``inseparability" of the cloud and spatial response
functions for systems in which the input is a multiplicative factor in the
expression of the output.
\par
\medskip
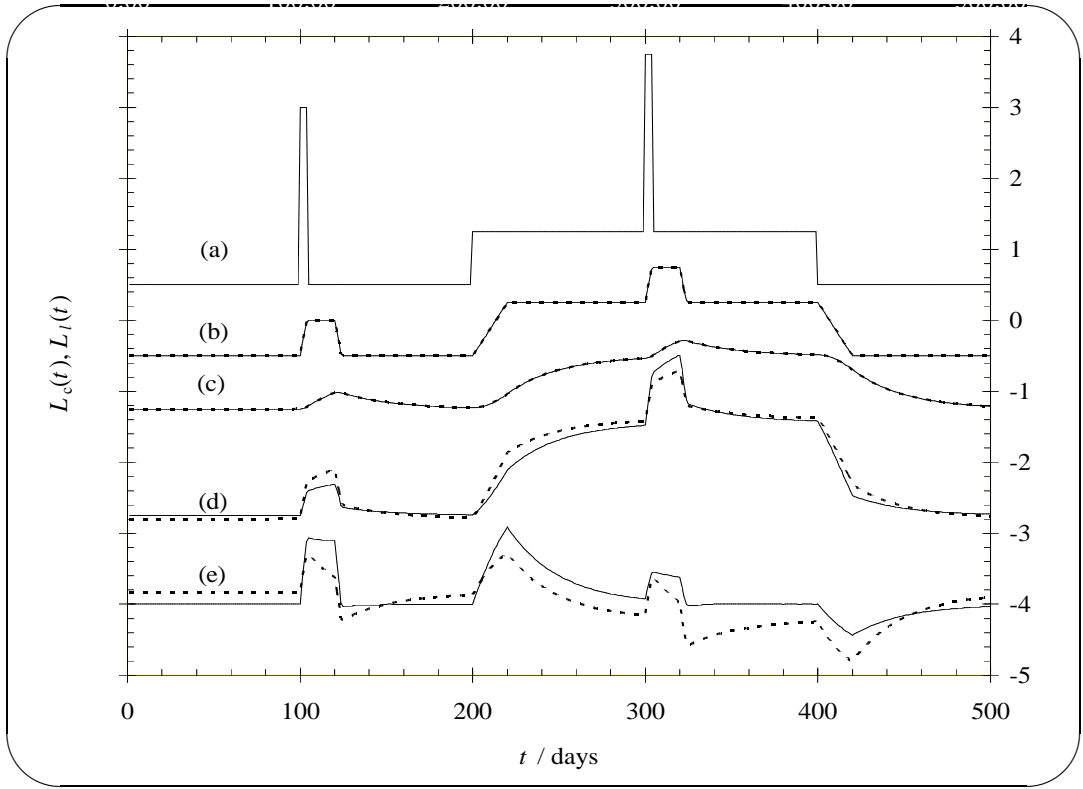
\noindent \fig{localdelays1}{Comparison of the linear approximation to the actual
responses of simple shell-like systems with local delays.  Line {\it a} is the input
continuum that was assumed, which has a luminosity of 0.5 in arbitrary units,
the ``low state," followed by a luminosity of 1.25 units, the ``high state,"
which lasts for 200 days.  Superimposed upon the low and high states are
delta-function-like spikes of area 10 unit-days.  The solid lines ${\it b}-{\it e}$ are the
output line luminosities for the models described in the text
(\S ~\ref{localdelays}.4) offset
respectively by -1, -1.75, -3, and -5 luminosity units while the dotted lines
are approximations of the outputs obtained from linearized response functions.
Though the linearized responses do a reasonable job of matching the actual
responses for most of the models shown here, they fail to exhibit the
differences between weak and strong (time-integrated) excitation.  This is
particularly evident for the model shown in solid line {\it e}.}\clearpage
\par
\medskip
These effects can be seen more clearly by considering a simple shell-like
system in which the light crossing time is slightly shorter than the cloud
equilibrium time.  Specifically, let $f({\bf r},{\bf v})\propto \delta (r-10$~{\rm days}$\cdot c)$, the
asymptotic cloud area function be $A^\prime (F_{{\scriptsize\rm c}})=A_{0}(F_{{\scriptsize\rm c}}/F_{0})^{\alpha }, \tau _{A}=30$~{\rm days},
$\epsilon _{\hbox{{\scriptsize\rm A}{\scriptsize\it l}}}$=$0, \epsilon _{l}=1,$ and the normalization of $f$ determined by the condition that
the mean covering factor be unity, i.e., $<F_{l}>=<F_{{\scriptsize\rm c}}>$.  The ratio of the area
equilibrium time to the light crossing time of this system is 3, which is near
enough to 1 for neither the fast nor the slow cloud regime
(\S ~\ref{localdelays}.3)
to be applicable. Exact outputs obtained upon application of equations
({}\ref{localdelays3.16.5})-({}\ref{localdelays3.23}) for
various values of $\alpha $ of this system are displayed as solid lines $b,~d$, and $e$ of
Figure \ref{localdelays1}, while the input continuum that was assumed is
shown as the solid
line {\it a}.  This input is a ``low state" followed by ``high state" that lasts for
200 days.  Superimposed upon the low and high states are delta-function-like
spikes of area 10 luminosity-unit-days, the responses of which can give a
crude indication of a spatial response function of the system.  The outputs
from
the linearized response functions are shown as dotted lines.  Ideally the
optimal linearized response functions would have been obtained from a fitting
scheme that minimized the discrepancy between the exact outputs.  However, in
this work they were obtained simply from equation ({}\ref{localdelays3.5}),
equations
({}\ref{localdelays3.24})-({}\ref{localdelays3.25.5}), and finally equation
({}\ref{localdelays3.8}), which was derived for
sinusoidal-like inputs but which results in surprisingly good fits for the
input here as well.
\par
\medskip
For the nonreactive $\alpha =0$ case shown in solid line $b$, the gain of the
output line luminosity is unity, and the linearized response function of the
system is just the spatial response function, which is a step function.
Whether in the high or low state, here the amplitude of the response on time
scales larger than the spatial time is the same as that of the input.  The $\alpha =1$
model is shown in solid line {\it d}.  For this reactive model the cloud area
responds linearly to the local continuum flux.  This is shown explicitly in
line {\it c}, which is the time-dependent area of the clouds on the
shell after the spatial delay was removed by artificially setting $f({\bf r},{\bf v})\propto \delta (r)$.
Care must be taken in the interpretation of this response, as the actual size
of this system is infinitely smaller than the size that the formalism of
Blandford \& McKee (1982) would yield, which is $\sim 40$~light-days$\sim c\tau _{\max }/2,$ where
$\Psi _{F_{l}|F_{{\scriptsize\rm c}}}(\tau >\tau _{\max })\simeq 0.$  It is important to understand that although the cloud areas
respond linearly in the $\alpha =1$ model, the output itself is over-responsive
compared to linear, with an asymptotic gain of 2~(eq. [{}\ref{localdelays3.24}]).
This
asymptotic gain is
also approximately the correction factor by which the formalism of Blandford \&
McKee (1982) could overestimate the cloud number density.  (The exact factor
is dependent upon the input, as eq. [{}\ref{localdelays3.8}] indicates, as well as
the fitting
criteria.)  However, by taking into account an asymptotic gain correction
factor that is different from unity, the linear response (dotted line ${\it d})$~does
a surprisingly good job of fitting the actual response (solid line ${\it d})$~of the
system, especially given that the input continuum luminosity function varies
by a factor 7.5.
\par
\medskip
Nonlinear response is more apparent in the ``under-responsive" $\alpha =-1$ model
shown in solid line {\it e}.  Here the cloud area response is given by a nonlinear
input-modified system (eq. [{}\ref{localdelays3.12}]).  A key difference between
the models
shown in solid lines ${\it d}$ and ${\it e}$ is that the gain of an individual cloud area is a
decreasing function of positive frequency for the over-responsive model, but
is an increasing function for the under-responsive model.  Note that because
the areas of the spikes are small, the actual responses to the first spike are
similar in both cases.  The spatial response function (solid line $b)$ does a
crude job of describing both models.  However, low frequency or high
(time-integrated) energy excitation exposes the latent nonlinearities of these
systems.  For instance, the over-responsive system (solid line $d)$ responds
slightly higher to the spike in the high state than the spike in the low
state, while the linearized response function predicts a response that was
the same strength for both spikes.  This aspect of behavior is due to a
nonlinear asymptotic gain.  Even if it were accounted for, the shapes given by
the linearized response function would still not perfectly match the exact
ones.  For instance, the linearized output for the over-responsive system
responds too rapidly to the beginning of the high state.  If its linearized
response function were adjusted to yield a slower response, the linearized
output would then respond too slowly to the beginning of the spike in the low
state.  Ultimately this is due to the area factor and hence the area
equilibrium time playing a less important role in the response to the first
spike of short duration (when the area is relatively constant) than in
response to the energetic high state of long duration (when the area increases
significantly).  These types of problems are particularly evident for the
highly nonlinear under-responsive case.  Because the amplitude of the
asymptotic gain of the area is only 1, the response to both low energy spikes
is square-like.  However, the high state is energetic and long enough to
permit the areas to respond and the asymptotic gain of zero nearly to be
attained, which results in triangle-like responses.  The linearized response
function incorrectly gives triangle-like features in the responses to the
spikes.
\par
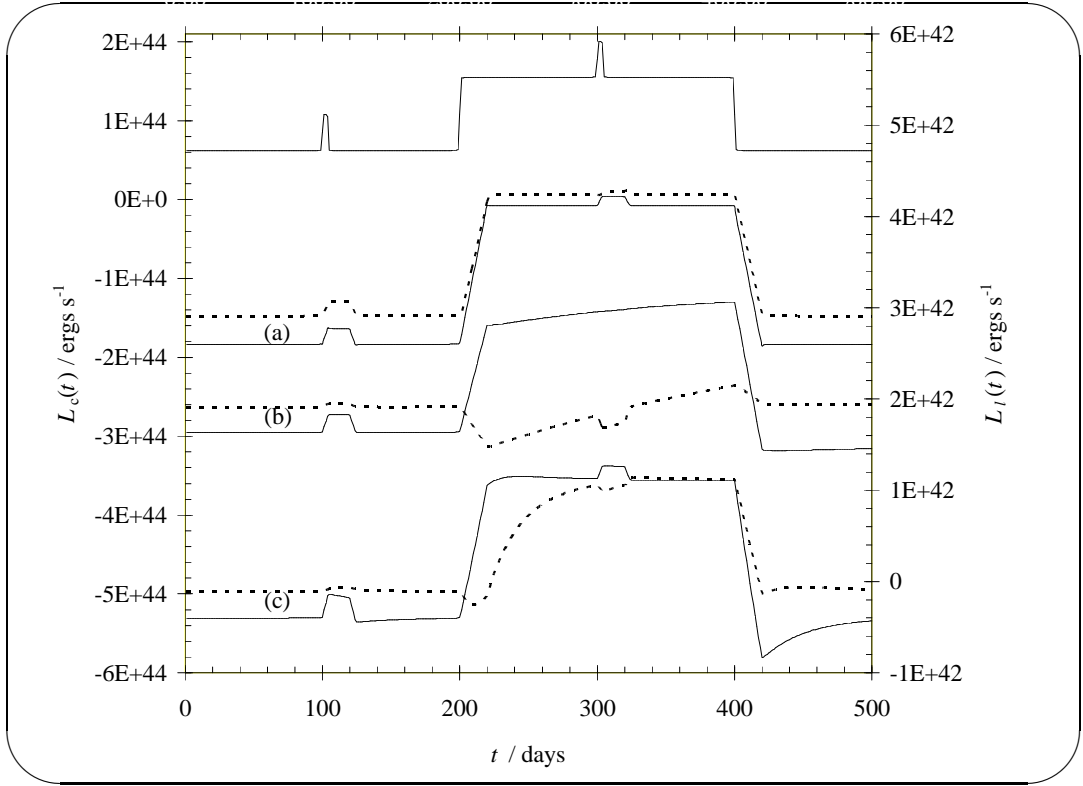
\noindent \fig{localdelays2}{Effect of the pressure equilibrium time upon the
``line-specific" response.  The top line (left axis) is the input continuum
luminosity assumed.  The exact Ly$\alpha ~($solid lines) and C {\scriptsize\rm IV} (dotted lines)
output luminosities (left axis) are also shown for three extremely simple $s=1$
models similar to those shown in Fig. \ref{localdelays1}.  To emphasize the
effect of just the cloud pressures and pressure ionization parameters being
locally delayed, the cloud areas were artificially forced to yield a constant
geometrical covering factor of unity (neglecting absorption) and the cloud
column densities were artificially forced constant at $10^{22}$ cm$^{-2}$.  The initial
column density to pressure ionization parameter ratios assumed were $1.2\times 10^{22}$
cm$^{-2}$.  The line luminosities for models ${\it b}$ and {\it c} are offset
respectively by $-10^{42}$ and $-3\times 10^{42}$ ergs s$^{-1}$.  The pressure-equilibrium times
$\tau _{P_{l}}$ assumed in models ${\it a}, {\it b}$, and {\it c} were respectively 1.1, 500, and 30 days.
For pressure-stratified clouds, this respectively corresponds to wave
propagation speeds of $\sim c_{{\scriptsize\rm s}}~($pressure-limited evaporation$), \sim 2\times 10^{-3}c_{{\scriptsize\rm s}}~($thermal
evaporation), and $\sim 4\times 10^{-2}c_{{\scriptsize\rm s}}~($intermediate evaporation), where $c_{{\scriptsize\rm s}}$ is the sound
speed.  The photoionization code that was used is XSTAR (see, e.g., Kallman
1995).  The spectrum that was assumed is shown in Fig.
\ref{localdelays3}.}\clearpage
\par
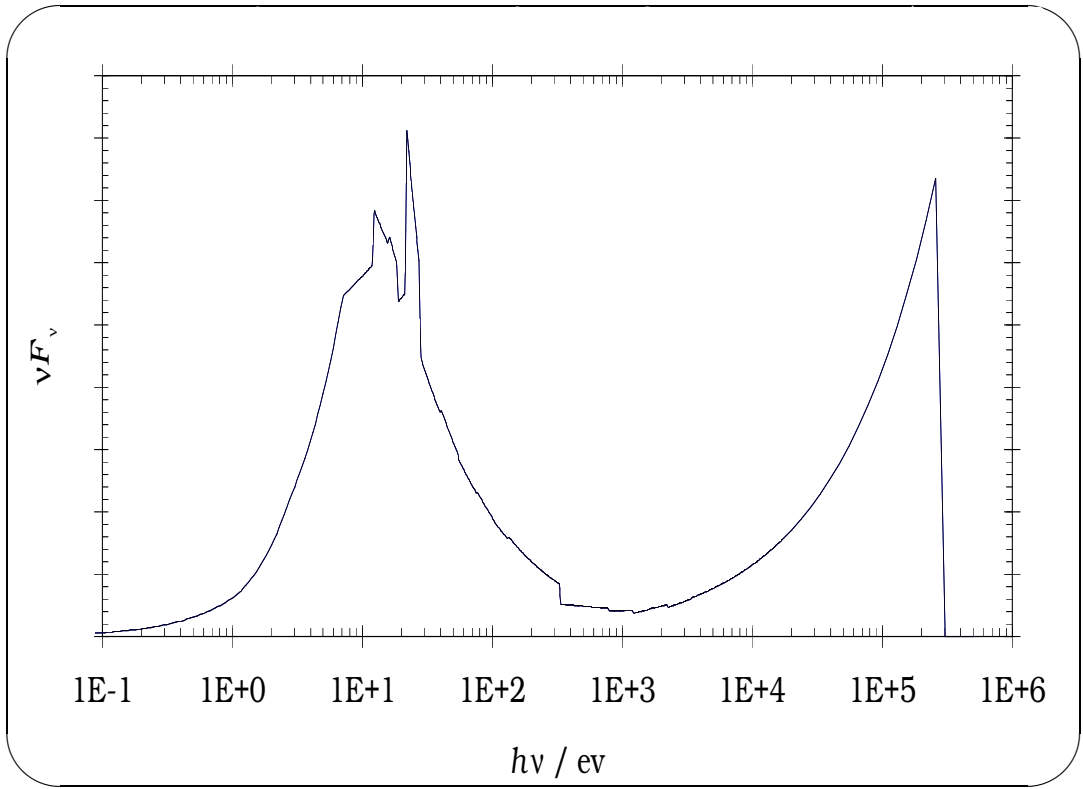
\noindent \fig{localdelays3}{The spectrum that was assumed for the models shown in Fig.
\ref{localdelays2}. It is identical to that used in Krolik et al. (1991).
Note that the $y$-axis is plotted in linear (as opposed to logarithmic)
coordinates.}
\par
\medskip
For the models shown in Figure \ref{localdelays1}, all the cloud property
gain
terms in equation ({}\ref{localdelays3.24}) except that of the area are zero.
However, for
calculations of {\it relative} line strengths, the ionization parameter gain term in
equation ({}\ref{localdelays3.24}) frequently determines the key distinguishing
response characteristics.  This is illustrated by the models shown in Figure
\ref{localdelays2}, where
the response as a function of the pressure equilibrium time scale is shown for
three models nearly identical to those used in Figure \ref{localdelays1}.
However, for
these cases, $P\propto F^{s/2}_{{\scriptsize\rm c}}$, the cloud areas and column densities are forced to be
constant, and the spike strengths are reduced.  For model {\it a}, the pressure
equilibrium time is 1.1 days, which is approximately the mean inverse
Str\"omgren sound crossing time for these clouds.  This is short enough
(compared to the spatial time of 10 days) for the fast cloud regime to be
valid.  Thus, in this case, a single (rather than double) integration in lag
would have sufficed for calculating $L_{l}(t)$.  For model ${\it b}$, the pressure
equilibrium time is 500 days, which is closer to the thermal evaporation time
scale (Table \ref{localdelaystimescales}) of $\sim 2\times 10^{3}$ days.  In the limit in
which the pressure equilibrium
time scale becomes infinite, model ${\it b}$ is identical to a nonreactive $s=0$ model,
which also would not require the computationally expensive double integration.
Finally, for model {\it c} the equilibrium time scale is 30 days.  This intermediate
equilibrium time might be applicable for cloud models in which the evaporation
rate is a strong function of the mean ionization state (see also Taylor 1994).
Because the column density to pressure ionization parameter ratio $(1.2\times 10^{22}$
cm$^{-2})$ was selected to place the model near the ``C {\scriptsize\rm IV}---limited" state in
which the C {\scriptsize\rm IV} gain is $\sim -1,$ the C {\scriptsize\rm IV} line responds very weakly to the initial
spike.  However, it responds strongly on the pressure equilibrium time scale
to the beginning of the extended high state, during which time the pressure
regulation mechanism readjusts the ionization parameter partially back toward
its lower initial equilibrium value.  In contrast, at the ending of the high
state, the C {\scriptsize\rm IV} line drops on the shorter spatial time scale because of the
higher pressures (and lower ionization parameters) of the clouds.  This
example illustrates one of several ways in which the mean response time of a
system can be dependent upon the recent mean continuum luminosity.
\pfn{Response
features like these will be discussed in much more detail in Paper III
(Taylor, in preparation 1996) within the context of more complex models.}
\par
\medskip
The above examples clearly illustrate how the global linearized response
function can have structure not due to the spatial response function.
However, nonlinear effects can even mask the spatial information.  For
example, if the ratio of the local continuum flux to the product of the mean
pressure and column density of the clouds becomes high enough, the emission of
certain lines could be ``recombination-limited," which makes $\tilde{\hat\Psi}_{\epsilon _{l}|F_{{\scriptsize\rm c}}}(\infty )\sim -1$ in
equation ({}\ref{localdelays3.24}).  This occurs during the high state with Ly$\alpha $
for model
${\it b~}($solid line Fig. \ref{localdelays2}).  Therefore, as the model shown in the
solid line {\it c} of
Figure \ref{localdelays1} also illustrates, the response times give, under the
assumption
of spherical symmetry, only {\it upper} limits to the characteristic size.
\par
\medskip
The above examples also help illustrate how local delays can affect the
cross-correlation function.  Consider a simple $\eta (A|F_{{\scriptsize\rm c}})>0$ model, such as that
shown in solid line ${\it d}$ of Figure \ref{localdelays1}.  For a weak enough
high frequency square
wave input occurring above a steady input background, the output would be a
capped triangle wave, which is symmetric about its peak with respect to lag.
The cross-correlation function that one would obtain from such data would be
shifted of approximately a day, yet also be symmetric about the characteristic
lag.
This is because the data from the beginning, or growing phase of the pulse,
which alone would produce a cross-correlation function with a positive slope,
is compensated by the ending or falling phase of the response pulse.
\par
\medskip
Consider, however, the case wherein the input is a moderate-intensity,
low frequency square wave.  Because of local delays, the response would be
higher near the end of the pulse, as in solid line ${\it d}$ of Figure
\ref{localdelays1}a. Since the
cross-correlation function is an amplitude-biased function, the resulting
cross-correlation function would be biased from the data in the falling phase.
Unlike the previous case, this would result in a negatively sloped component
to the cross-correlation function, which is quite common (e.g., Sparke 1993).
\par
\medskip
Note that local delays only give the cross correlation function asymmetry
about their peak above what one would obtain from the global linearized
response function alone.  This is because the cross-correlation function is
the convolution of the response function and the symmetric, input
auto-correlation function.  For this reason, assessing the importance of
asymmetric cross-correlation functions in determining the characteristics of
the local delays of interest here would probably require knowledge of the
best-fitting linear response function.
\par
\medskip
Note that this function would be biased somewhat by the low energy (more
symmetric) pulses.  Therefore, the asymmetry of the ``simulated
cross-correlation function'' that one would obtain from the output of this
response function would not be as great as the one obtained from actual
variability data.  Thus, the difference between the simulated and actual
cross-correlation functions can be asymmetric if local delays are important.
This permits a simple way of testing for local delays.\pfn{More complicated
techniques will be discussed in other papers in this series.}
\par
\infootendplain

{}\setcounter{section}{5}{}\addtocounter{section}{-1}\section{Summary and Conclusions}
\par
\medskip
In this appendix it has been argued that for some reactive cloud
models
the fast cloud assumption is invalid.  In such cases a tight correspondence
between the positional distribution of matter and the global linearized line
flux response function simply does not exist.  This is because the continuum
luminosity at a given time in history affects not only the clouds on a
spherical shell, but also the clouds inside such a shell.  In general,
the response which results is not only nonlinear, but also inseparable,
requiring more than one integration over lag to determine the output flux at a
specified time.  However, in some cases the time-dependence of the line fluxes can be
described using linearized response functions $\Psi _{F_{l}|F_{{\scriptsize\rm c}}}$.  These response
functions have structure at intermediate lags due to the finite size of the
line-emitting region and at lags greater than these due to the finite
equilibrium times of the line-emitting material itself.  For small enough
perturbations the physical cloud properties can be relatively static and a
linear response function can work quite well at describing the responses.
However, when the input continuum variations become extreme enough, such
response functions can fail.  Because of nonlinear asymptotic response, the
integral of a linearized response function of an observable differs from the
time
average of the observable by a correction factor of the asymptotic gain.
Ignoring nonlinear effects can lead to incorrect measurements of the physical
properties of the system, such as sizes that are too large.
\par
\medskip
One of the most fundamental assumptions that has been made in
previous analysis of variability data has been the fast cloud assumption.
With this assumption now in question, this data should be examined again with
models that do not require it.\chapter{The Line Flux of a Cloud When the Equilibrium and Crossing
Times are Comparable\label{equilibriumiscrossing}}     In standard references such as Blandford \& McKee (1982), the
time-dependent line profile is presented as an integral over the cloud phase
space of distribution function and emission flux.  The line emission from
these clouds is implicitly assumed to be independent of the direction of
their velocity vectors. As a result, in order for these models to generate the
line shifts and asymmetries which are observed, the clouds must generally flow
outwards or inwards.  Models with such radial flow can, however, have serious
problems, such as very low accretion efficiencies (see, e.g., Kallman et al.
1993).
\par
\medskip
Before one invests substantial effort to concoct models which address
these
problems, it may be worthwhile to ${\rm re}$-analyze the original, highly intuitive
conclusion that line emission from clouds is independent of the
direction of their velocity vectors.  In fact, a careful examination reveals
that, because of our extreme ignorance of what the clouds in AGNs are, there
are several theoretically permissible reasons that clouds could (at
least in principle) have line emission dependent upon their velocity
direction. Each of these reasons is the result of an uncertainty
in a corresponding assumption made either explicitly or implicitly in
Blandford \& McKee (1982) and similar works.  By examining each of these
assumptions in detail, we clearly outline the boundary of parameter space
in which
the velocity-direction-independent emission result must lie.  Moreover,
by looking just outside these boundaries we could obtain a model which does
not suffer as severely from the types of problems plaguing current AGN
models.
\par
\medskip
In this appendix we will focus on just one of the potentially invalid
assumptions made in previous works, i.e. that the cloud equilibrium times are
negligible compared to the cloud crossing times.  As shown in Table
\ref{localdelaystimescales}, this assumption is invalid for some models.  For
these models, the cloud properties have an additional explicit dependence upon
the prior continuum fluxes $F_{{\scriptsize\rm c}}{\hbox{\bf(}}t-\tau ,{\bf r}(t-\tau ){\hbox{\bf)}}$ and hence the orbital trajectories
of the clouds, which is a different function for each cloud.  This would
complicate modeling
efforts, which would entail integrating over the orbital trajectories of the
clouds.  It could also make the much simpler approach taken by Blandford in
McKee (1982) invalid.  However, we shall find that in certain cases the cloud
luminosity function can be described merely by giving the properties of the
cloud an additional velocity-direction dependence.
\par
\medskip
But let us first obtain the exact solution to this problem.  Ignoring
absorption, the (nonlinear) line flux as a function of time for a cloud is
given by application of equation ({}\ref{localdelays3.12.5.5}) to each of the cloud
parameters in equation ({}\ref{localdelays3.16.5}).  Thus, the
continuum-subtracted,
time-dependent line profile of the ``global" system observed from ${\bf D}$ is
\par
\s
F_{l}(t,v_{D}) =\sum^{N}_{i=1}F^{\hbox{{\scriptsize\rm orb}}}_{li}(t)\delta (v_{D}+{\bf v}_{i}\cdot \hat{{\bf D}}),
\labep{3.18}\e
\noindent where $N$ is the number of clouds and $F^{\hbox{{\scriptsize\rm orb}}}_{li}$ is the flux in line $l$ from cloud {\it i}
computed from the forms of equations
({}\ref{localdelays3.12.5.5})-({}\ref{localdelays3.16.5}) appropriate for the model to
be tested.  Neglecting absorption and non-Doppler line broadening, equation
({}\ref{localdelays3.18}) gives the exact time-dependent line profile for clouds
with arbitrary motions.  However, because the suspected number of clouds in
AGN is high (e.g., Laor et al. $1994; cf$, e.g., Peterson 1994), using it
could prove computationally expensive.
\par
\medskip
Noting the exponential factor in equation ({}\ref{localdelays3.12.5.5}),
integration over the history of local continuum exposure required for
determining the cloud properties at a given time need only be carried out to a
small factor (e.g., $\sim 4)$ of the relevant equilibrium time.  Therefore, if each
cloud property relevant to emission has an equilibrium time scale that is
appreciably less than the emission region crossing time, the radius will not
change drastically over the relevant history interval, and the continuum flux
function can be approximated by its Taylor expansion.  This yields a first
order
correction to the continuum flux function that is proportional to the radial
velocity.  Similar expansions permit estimation of the line flux observed from
an individual cloud and the line shifts that equation ({}\ref{localdelays3.18})
implies.
\par
\medskip
However, if $N$ is independent of the continuum flux and is large enough
that line broadening produces a smooth line profile, a more accurate method
for obtaining the individual cloud line flux that partially accounts for
higher order terms can be obtained simply by taking the statistical average
of the function, which is
\par
\medskip
\noindent $$F^{{\scriptsize\rm cl}}_{l}(t,{\bf r},{\bf v};{\bf D})=\lim_{\delta r,\delta v\rightarrow 0;N\rightarrow \infty } {1\over f({\bf r},{\bf v})\delta ^{3}r\delta ^{3}v} \times \qquad \qquad \qquad \qquad \qquad \qquad \qquad $$
\s
\sum^{N}_{i=1} \int^{{\bf r}+\delta {\bf r},{\bf v}+\delta {\bf v}}_{{\bf r},{\bf v}}d^{3}r^\prime d^{3}v^\prime  \delta {\hbox{\bf(}}{\bf r}^\prime -{\bf r}_{i}(t){\hbox{\bf)}}\delta {\hbox{\bf(}}{\bf v}^\prime -{\bf v}_{i}(t){\hbox{\bf)}}F^{\hbox{{\scriptsize\rm orb}}}_{li}(t).
\labep{3.21}\e
\noindent The dependence of $F^{{\scriptsize\rm cl}}_{l}(t,{\bf r},{\bf v};{\bf D})$ upon time is due to the variation of the
continuum flux to which the cloud is locally exposed.  The dependence upon
position is due to traditional model elements such as changes in the mean
cloud density as a function of average heating.  Finally, the dependence upon
the velocity vector accounts for the intrinsic dependence as well as that due
to the
history of heating being important when the equilibrium time scale is not
completely negligible compared to the emission region crossing time.  Note that for models where
the position and velocity variables impose the integrals of motion of a
trajectory of a cloud, the above condition that the equilibrium times are
small
compared to the crossing times is unnecessary and the flux is an exact
function of only the time, position, and velocity variables.
\par
\medskip
In this case, an analog of equation ({}\ref{localdelays3.21}) can be used to
replace the knowledge of the individual cloud trajectories with the
time-independent phase space distribution function $f({\bf r},{\bf v})$.  Though this function, when combined with equation
({}\ref{localdelays3.21}), permits equation ({}\ref{localdelays3.23}) to be used to
obtain an approximation of the time-dependent line profile, it offers little
advantage over using just equation ({}\ref{localdelays3.18}) because it still
entails explicit time-dependent orbital modeling.
\par
\medskip
However, in the linear regime (eqs.
[{}\ref{localdelays3.25}]-[{}\ref{localdelays3.25.5}]), only the time-average of
equation ({}\ref{localdelays3.21}) is required to obtain the time-dependent line
profile.  Once this (velocity-dependence) has been computed for a model, the
linear approximations to the observable characteristics can be obtained from
equation ({}\ref{localdelays3.25.5}) for various continuum light curves without
explicit time-dependent orbital modeling. Therefore, in the linear regime, the
orbital history of the cloud line emission flux can be approximated as a
simple function of the direction and magnitude of the cloud velocity vector
$\hat{{\bf v}}$.
\par
\medskip
\noindent \chapter{Response
of Nonlinear Systems in the Linear Regime\label{linearization}}
\par
In Blandford \& McKee (1982), linear systems were analyzed in the linear
regime.  In this appendix a formalism is developed for analyzing {\it non}linear
systems in the linear regime.  Though the solution to this problem is a
straightforward and probably necessary prerequisite for any comprehensive
understanding of variability in AGN, it was not correctly obtained or applied
in other works regarding AGN variability.  We shall find that within the
linear regime the analysis in Blandford \& McKee (1982) is inadequate for
general nonlinear systems.
\par
\medskip
Let us consider the generic system described in \S ~\ref{localdelays}.3.1.
Using the notation of \S ~\ref{localdelays}.3.1 for $y(t), y^\prime (x), x(t)$, and $t$,
there is no reason that $\delta y^\prime (x)\propto \delta x$ should generally hold, and one
may be forced to employ a fully nonlinear analysis method to accurately
describe the system.  However, let us assume here that the variations in $x$ are
sufficiently smaller than its mean, in which case, provided $y^\prime (x)$ is a smooth
function, it can be approximated with
\par
\s
y^\prime =<y>\left[1+\eta (y|x)\left({x\over <x>}-1\right)\right] \pm  \sigma _{y},
\labep{3.1}\e
\noindent where the dimensionless, ``asymptotic gain" $\eta $ of changes in $y$ for small and
slow changes in $x$ is defined by
\par
\s
\eta (y|x)\equiv \lim_{\delta x,\dot{x}\rightarrow 0} {\delta y/y\over \delta x/x} ={<x>\over <y>} {\partial y^\prime \over \partial x}|_{x=<x>},
\labep{3.3}\e
\noindent where $\sigma _{y}$ is the error due to nonzero second order derivatives in $y^\prime $.  The asymptotic
gain has elsewhere been termed the ``responsivity" (Krolik et al. 1991; Goad,
O'Brien, \& Gondhalekar 1993).  Here it is an operator to distinguish between
the various gains with the different ``output" and ``input" functions that
will be required, though note that it is independent of the normalizations of
these functions.
\par
\medskip
Let us extend the definition of gain by allowing a dependence upon the
type of input signal.  Consider a time-dependent local ionizing continuum flux
or a Fourier component of it such as $x(t)-x_{0}=x_{1}\cos (\omega t)$.  A dimensionless
frequency-dependent gain or transfer function $\tilde{\hat\Psi}$ of $y({\it t})$ with respect to $x(t)$
can then be defined as
\par
\s
\tilde{\hat\Psi}_{y|x}(\omega ) \equiv  \lim_{\delta x\rightarrow 0} {\delta y/y\over \delta x/x}|_{\ddot{x}/x_{1}=-\omega ^{2}}.
\labep{3.4}\e
\noindent Letting $\tilde{y}(\omega )$ denote the Fourier transform of $y({\it t})$, etc., gives
\par
\s
\tilde{\hat\Psi}_{y|x}(\omega )={<x>\over <y>} {\partial \tilde{y}(\omega )\over \partial \tilde{x}(\omega )}
\labep{3.5}\e
\noindent and an analog of the Fourier transform of equation ({}\ref{localdelays3.1}) for a
frequency-dependent gain,
\par
\s
\tilde{y}(\omega )\simeq <y>\left[\delta (\omega )+\tilde{\hat\Psi}_{y|x}(\omega )\left({\tilde{x}(\omega )\over <x>}-\delta (\omega )\right)\right].
\labep{3.6}\e
\noindent With
this
notation, $|\tilde{\hat\Psi}_{y|x}(\omega )|$ is the dimensionless ratio of the amplitudes of variations
of $y$ to $x$, while -\rm\/Im$[\ln {\hbox{\bf(}}\tilde{\hat\Psi}_{y|x}(\omega ){\hbox{\bf)}}]/\omega $ is the  delay in response.{}\rfn{The sign
in the Fourier transform used here, though different from that in several
references (e.g., Blandford \& McKee 1982), minimizes differences with the
Laplace transform, which offers certain advantages in dealing with this type
of problem.}  Similarly, the asymptotic gain is Re$[\tilde{\hat\Psi}_{y|x}(0)]$, while the
``instantaneous component of the gain" is Re$[\tilde{\hat\Psi}_{y|x}(\infty )]$.
\par
\medskip
Equation ({}\ref{localdelays3.6})
yields in the time domain
\par
\s
y(t)= <y>\left[1+\int d\tau \left({x(t-\tau )\over <x>}-1\right)\hat{\Psi }_{y|x}(\tau )\right] \pm  \sigma _{y},
\labep{3.13}\e
\noindent where we define the inverse Fourier transform of the gain of the output $y({\it t})$
with respect to the input $x(t)$ as the normalized linearized response function,
which is
\par
\s
\Psi _{y|x}(\tau )\equiv {<y>\over <x>}\hat{\Psi }_{y|x}(\tau ).
\labep{3.12.5}\e
\noindent Here the lack of a caret denotes that the normalized response function has
scaling other than that given to it by the inverse Fourier transform.  Note
that the integral of the linearized response function is just the asymptotic
gain, which is only unity for actual linear systems.
\par
\medskip
Linearization is not always advantageous.  In some cases, including
those obeying equation ({}\ref{localdelays3.10}), the exact solution can easily be
obtained from
\par
\s
y(t) = \int d\tau  y^\prime {\hbox{\bf(}}x(t-\tau ){\hbox{\bf)}}\Psi _{y|y^\prime }(\tau ),
\labep{3.12}\e
\noindent where
\par
\s
\Psi _{y|y^\prime }(\tau )=\hat{\Psi }_{y|y^\prime }(\tau )=\hat{\Psi }_{y|x}(\tau )/\eta (y|x)
\labep{3.11.5}\e
\noindent is the response function of an ``input-modified" system.
\par
\medskip
However, linearization can be quite useful, in some cases it allows
complex systems to be accurately described by a single equivalent response
function, which can drastically reduce the simulation time.  Such
is the case wherein one is interested in obtaining an
observable quantity of a system with many clouds, wherein the convolution of $y$
with a spatial linear response function must be evaluated.  For instance,
consider a hypothetical system wherein a physical cloud property $z^\prime (y)\propto y^{\eta (z|y)}$
lags property $y$ on a time scale $\omega ^{-1}_{z}$, while property $y^\prime (x)\propto x^{\eta (y|x)}$ has a
``direct" lag of $\omega ^{-1}_{y}$.  The exact expression for $z({\it t})$ has two nested integrals
over lags.  However, upon linearization the transfer function of $z$ for small
variations in $x$ is
\par
\s
\tilde{\hat\Psi}_{z|x}(\omega )= {\eta (z|y)\over 1+i\omega /\omega _{z}} {\eta (y|x)\over 1+i\omega /\omega _{y}},
\labep{3.14}\e
\noindent or alternatively
\par
\medskip
\noindent $$\hat{\Psi }_{z|x}(\tau )=\Theta (\tau )\eta (z|y)\eta (y|x)\omega _{z}\omega _{y}\left({e^{-\omega _{z}\tau }\over \omega _{y}-\omega _{z}}+{e^{-\omega _{y}\tau }\over \omega _{z}-\omega _{y}}\right) \qquad \{\omega _{y}\neq \omega _{z}\}\qquad \qquad $$
\s
=\Theta (\tau )\eta (z|y)\eta (y|x)\omega ^{2}_{z}\tau e^{-\omega _{z}\tau }~~\qquad \qquad \qquad \qquad \{\omega _{y}=\omega _{z}\},
\labep{3.15}\e
\noindent which when applied (in eq. [{}\ref{localdelays3.13}]) requires only a single
integral
over lag.  For future reference, note that when $\omega _{z}\gg \omega _{y}$ or $\omega _{z}\ll \omega _{y}$, the two gain
factors are ``separable" from one another, i.e. for a restricted range of
excitation frequencies one of the gain factors can be treated as a constant.
\par
\medskip
In this section it has been shown (eq. [{}\ref{localdelays3.11.5}]) that there
is
a ``correction factor" of $\eta (y|x)$ in the expression for the ``gain-corrected
response function."  Previous works (e.g., Blandford \& McKee 1982) assumed
that the systems themselves are linear, which is equivalent to assuming
correction factors of unity.  Some of the problems with making this assumption
are pointed out in Goad, O'Brien, \& Gondhalekar (1993) as well as
\S ~\ref{localdelays}.4
of this work.  Note that the correction factors differ from unity in
equation ({}\ref{localdelays3.3}) in nonlinear systems even if the perturbations
are
arbitrarily small and equation ({}\ref{localdelays3.13}) accurately describes the
system.
\par
\medskip
Partially accounting for even higher order corrections due to
nonlinearity is also possible within the linear regime and, in fact, is
important for accurate interpretation of fits of linear models to nonlinear
systems.  If the variations are not infinitesimal, the above
equations do not necessarily yield the ``optimal" fit that would have obtained
using real variability data.  For instance, consider the case wherein
$y^\prime =y_{0}(x/x_{0})^{\alpha }$.  The better-fitting optimal average for a sinusoidal-like input
is the first Fourier coefficient of $y^\prime $, 
\par
\s
<y>={1\over 2\pi }\int^{2\pi }_{0}d(\omega t)y^\prime {\hbox{\bf(}}x_{0}+x_{1}\cos (\omega t){\hbox{\bf)}}\neq y^\prime (<x>).
\labep{3.7}\e
\noindent Similarly, the observable asymptotic gain is approximately
\par
\s
\eta (y|x)={1\over \pi }\int^{2\pi }_{0}d(\omega t)\cos (\omega t){y^\prime {\hbox{\bf(}}x_{0} +x_{1}\cos (\omega t){\hbox{\bf)}}\over <y>}\neq \alpha ,
\labep{3.8}\e
\noindent where the
inequalities can be removed only for the $x_{1}\ll x_{0}$ case.
\par
\noindent 
\par
\noindent 
\par
\noindent 
\par
\infootendplain


\chapter{Velocity Deconvolution Techniques\label{veldtemp} 
}

\def\labep#1{\label{veldtemp#1}}
\def\figdirprefix{\home/completedpapers/}
\renewcommand\intextsep{40pt}
\infootbegplain
\noindent \title{Line Emission in Active Galactic Nuclei II: Fitting Orbital Cloud
Models with Variability Data}
\par
\noindent \author{Jason A. Taylor\altaffilmark{1}}
\par
\noindent \altaffiltext{1}{Also Department of Physics, University of Maryland, College
Park, MD  20742, USA}
\par
\noindent \affil{NASA/Goddard Space Flight Center}
\affil{Code 661, Greenbelt, MD  20771, USA; taylor@lhea1.gsfc.nasa.gov}\keywords{galaxies: Seyfert---quasars: emission lines---quasars: general}
\par
{}\setcounter{section}{1}{}\addtocounter{section}{-1}\section{Assumptions and Background}
As discussed in Chapter 2, the response functions shown in Chapter 4
were made under the questionable assumption that the wind size equilibrium
times are small compared to the light crossing times.  As shown in Appendix
\ref{localdelays}, without this assumption the response is more complex and
longer than what would be inferred from the spatial distribution of
clouds alone.   This is unfortunate, as the original motivation behind
reverberation mapping was to constrain this spatial distribution of clouds.
\par
\medskip
In this appendix, I show that under certain conditions the
spatial distribution of clouds can nevertheless be measured {\it even if the clouds
have response times comparable to the light crossing times}.  Thus, even if
the
cloud reaction times are of the order of the cloud crossing times, the
original
goal of reverberation mapping may be obtainable after all, at least for
certain classes of models.
\par
\medskip
I also show that the response functions of
properties
of {\it individual} clouds, i.e., the ``cloud response functions," can be
measured.  Knowledge of these cloud response functions would impose very
specific and severe constraints upon the
confinement mechanisms that are invoked in various cloud models and help
resolve the question of what the clouds are if the cloud concept is indeed
valid.
\par
\medskip
\noindent {}\setcounter{section}{2}{}\addtocounter{section}{-1}\section{Separating the Spatial and Cloud Transfer Functions}
\par
\medskip
\noindent In this section, we first define a mathematical transformation that
can be applied to line profiles.  In \S ~\ref{veldtemp}.2.2, we use this
transformation
to derive expressions for the individual cloud response functions. We
find that for certain values of the characteristic cloud equilibrium time
scales, the qualitative response features can be obtained without explicit
use of both the cloud and spatial response functions.  In
\S ~\ref{veldtemp}.2.3 the precise conditions for this simplification
are calculated.
\par
\medskip
There are a wide variety of cloud models that have been proposed.
The results of this appendix, however, are valid only for a small subset
of them.  In particular, in the remainder of this appendix, we assume
that the clouds are in orbital motions inside a centrally symmetric
gravitational field; the results of this appendix do not generally apply
to models violating this assumption.
\par
\medskip
\noindent {}\setcounter{section}{2}{}\setcounter{subsection}{1}{}\addtocounter{subsection}{-1}\subsection{Obtaining the Line Intensity}
\par
\medskip
The primary observable of an AGN is its spectrum.  One of the simplest
ways of analyzing a spectrum is to break it up into an underlying continuum
and several line profiles that are functions of an equivalent line-of-sight
velocity $v_{{\scriptsize\rm D}}$.  These profiles can then be compared to theoretical ones to
gauge the viability of a model, or, more quantitatively, measure the goodness
of a model.
\par
\medskip
Unfortunately, the line profile is somewhat inadequate for our purpose
here.  In this appendix we find that in order to determine some of the key
response characteristics of the flux of a line, it is useful to work with a
variable that is more closely coupled to the radius than is $v_{{\scriptsize\rm D}}$.
\par
\medskip
It is well known that the integral of the spectrum is the bolometric
luminosity or flux.  For this reason, we can consider the spectrum to be the
``wavelength representation" of the bolometric luminosity.  Similarly, we can
consider the line profiles $F_{l}(t,v_{{\scriptsize\rm D}})$ as a function of time $t$ to be a
representation of the continuum-subtracted flux in $v_{{\scriptsize\rm D}}$-space.  The response
functions are yet other representations of the flux.
\par
\medskip
In this appendix we employ a new representation of the line flux.
Consider the integral transform
\par
\s
F_{l}(t,v_{{\scriptsize\rm D}})=\int d\sigma I_{l}(t,\sigma )\phi _{l}(\sigma ,v_{{\scriptsize\rm D}}),
\labep{3.26}\e
\noindent where $F_{l}$ is the continuum-subtracted line flux, $\phi _{l}(\sigma ,v_{{\scriptsize\rm D}})$ is the kernel of the
integral transform, and $I_{l}$ is a new function herein termed the ``line
intensity," which, incidentally, has no relation to the variable commonly used
in radiative transfer calculations.  The above equation maps the
representation of
the line flux in $\sigma $-space $I_{l}(\sigma ,v_{{\scriptsize\rm D}})$ to the representation of the line flux in
$v_{{\scriptsize\rm D}}$-space.  Were $\phi _{l}(\sigma ,v_{{\scriptsize\rm D}})$ constrained to form an orthogonal and complete
basis, the mappings between $I_{l}$ and $F_{l}$ would be unique.  However, for
generality, $\phi _{l}(\sigma ,v_{{\scriptsize\rm D}})$ is permitted to be overcomplete here.  This does
not pose serious problems because only discrete forms of $I_{l}$ are applied
to equation ({}\ref{veldtemp3.26}).  Thus, the resolution in $\sigma $-space can always be
limited such that the mapping from $F_{l}$ to $I_{l}$ is undercomplete and
well-constrained.  In this respect, the only condition that we initially
impose upon $\phi _{l}(\sigma ,v_{{\scriptsize\rm D}})$ is that its integral with respect to $v_{{\scriptsize\rm D}}$ be unity.
\par
\medskip
What particular advantage does the $\sigma $-representation offer over the
$v_{{\scriptsize\rm D}}$-representation?  Let us assume that a
unique mapping between distance from the black hole ${\it r}$ and $\sigma $ exists with the
function $r(\sigma )$.  This permits the radial integration in the time-dependent
line profile (eq. [\ref{localdelays}.5]) to be removed.
Doing this yields
\par
\s
I_{l}(t,\sigma )=|{\partial r\over \partial \sigma }|r^{2} \int d\Omega _{r}{d^{3}vf({\bf r},{\bf v})\delta (v_{{\scriptsize\rm D}}+{\bf v}\cdot \hat{{\bf D}})\over \phi _{l}(\sigma ,v_{{\scriptsize\rm D}})}F^{{\scriptsize\rm cl}}_{l}(t,{\bf r},{\bf v},{\bf D}),
\labep{3.27}\e
\noindent where the dependence of $r$ upon $\sigma $ is implicit, $\Omega _{r}$ is the solid angle of the
spatial vector ${\bf r}$ from the black hole, $F^{{\scriptsize\rm cl}}_{l}(t,{\bf r},{\bf v},{\bf D})$ is the flux from a cloud
at ${\bf r}$ with velocity ${\bf v}$ in line $l$ observable from ${\bf D}$, and $f$ is the probability
distribution of clouds in phase space.  The above expression is an integral
over spatial angle, velocity, and (in its most general form) the history of
the local continuum flux via $F^{{\scriptsize\rm cl}}_{l}$.  The kernel of the transform $\phi _{l}(\sigma ,v_{{\scriptsize\rm D}})$ can
be defined to be the average of the local line profile composed of clouds on
the shell with radius $r(\sigma )$.  With this definition, the fraction in the
integral of equation ({}\ref{veldtemp3.27}) becomes equal to the mean density of
clouds on
this shell.  Thus, equation ({}\ref{veldtemp3.27}) can be viewed simply as a
summation of
the line flux received from clouds on the shell of radius $r(\sigma )$.  In fact, $\sigma $
can be interpreted as a velocity dispersion parameter while the local profile
$\phi _{l}(\sigma ,v_{{\scriptsize\rm D}})$ can be defined such that $3\sigma ^{2}\sim ~<v^{2}>~\simeq GM_{{\scriptsize\rm h}}/{\it r}$, where $M_{{\scriptsize\rm h}}$ is the black
hole mass and $G$ is the gravitational constant.
\par
\medskip
Note that the clouds contributing toward the line intensity at a
specific velocity dispersion $\sigma $ are restricted to lie on a shell that is a
fixed
distance from the black hole.  This is in contrast with the line profile at a
specific equivalent line-of-sight velocity $v_{{\scriptsize\rm D}}$, which is an integral over all
radii.  In the next subsection, we see how this feature can permit
measurement within the context of a model of the cloud response functions
from the intensity light curve.
\par
\medskip
However, let us first review two key prerequisites for applying equation
({}\ref{veldtemp3.26}) to measure (constrain) the line intensities $I_{l}$ from a given
spectrum: \begin{enumerate}\item  knowledge of $r(\sigma )$ and \item knowledge of
the local line profiles $\phi _{l}(\sigma ,v_{{\scriptsize\rm D}})$.\end{enumerate} In Appendix
\ref{veldtempa}, it is shown how both $r(\sigma )$ and
$\phi _{l}(\sigma ,v_{{\scriptsize\rm D}})$
can be obtained for certain simple orbital models.  The methods discussed in
Appendix \ref{veldtempa} are {\it in}applicable for non-orbital models.
Therefore, for
non-orbital models, reliable measurements of $I_{l}$ from a spectrum could be
difficult if not impossible.  For the orbital models considered in
this appendix, however, accurate measurement of $I_{l}$ from a given spectrum
should
be achievable at several different points in dispersion space.  Thus, if we
restrict ourselves to orbital cloud models, the ``intensity light curves" can
be measured from time-resolved spectra, provided that they are available.
Incidentally, this particular theoretical feature of orbital models is
definitely not a reason for invoking them.  It merely permits a particular
type of analysis to be performed that would otherwise be much more difficult.
\par
\medskip
\noindent {}\setcounter{section}{2}{}\setcounter{subsection}{2}{}\addtocounter{subsection}{-1}\subsection{Obtaining the Cloud Transfer Functions}
\par
\medskip
This appendix concerns response functions of several different
time-dependent quantities.  For this reason, we employ a notation that
explicitly denotes what is the specific ``input" and ``output" of a response
function.  In particular, let us define the ``linearized response function"
of an arbitrary output variable $y$ as a function of lag $\tau $ to be $\hat{\Psi }_{y|x}(\tau )$ when
the input variable is ${\it x}$ and the variations are small enough about their means
that the system can be considered to be linear (see Appendix
\ref{linearization}).  With this notation, the dimensionless
Fourier transform of the
linearized response function of the line profile at a given line-of-sight
velocity $v_{{\scriptsize\rm D}}$ with respect to the observed continuum flux $F_{{\scriptsize\rm c}}$ is
\par
\s
\tilde{\hat\Psi}_{F_{l}|F_{{\scriptsize\rm c}}}(\omega ,v_{{\scriptsize\rm D}}).
\e
\noindent An expression for this quantity, which is also called the two-dimensional
transfer function or ``global gain" of the line profile, is provided by
equation (\ref{localdelays}.7).  It contains an integral over all radii.  For this reason, it has
nontrivial structure (is not flat in $\log -\log $ coordinates) over a large range
of lags, even with $v_{{\scriptsize\rm D}}$ held constant.
\par
\medskip
Let us now calculate the transfer function of the line intensity with
respect to the continuum flux.  Applying the formalism of Appendix
\ref{linearization} yields
\par
\noindent $$\tilde{\hat\Psi}_{I_{l}|F_{{\scriptsize\rm c}}}(\omega ,\sigma )=|{\partial r\over \partial \sigma }|r^{2} \int d\Omega _{r}d^{3}v{<F^{{\scriptsize\rm cl}}_{l}(t,{\bf r},{\bf v},{\bf D})>_{t}\over <I_{l}(t,\sigma )>_{t}}\times \qquad \qquad \qquad \qquad \qquad \qquad \qquad \qquad $$
\s
f({\bf r},{\bf v})\tilde{\hat\Psi}_{F^{{\scriptsize\rm cl}}_{l}|F_{{\scriptsize\rm c}}}(\omega )e^{-i\omega (\hat{{\bf r}}-\hat{{\bf D}})\cdot {\bf r}/c},
\labep{3.28}\e
\noindent where the dependence of $r$ upon $\sigma $ is again implicit.  The above equation
states that gain of the intensity at a given excitation frequency and
dispersion $\sigma $ is proportional to the time-averaged, angle-dependent line flux
of the clouds at $r(\sigma )$; the local line flux gain of such clouds; and a factor
that is a strong function of the angle of the position vector of the clouds.
Once evaluated, the expression provides an approximation in the linear regime
of $I_{l}(t,\sigma )$ given $F_{{\scriptsize\rm c}}(t)$.  For reasons discussed in Appendix
\ref{localdelays}, the expression is unique for a given model and
mean continuum flux.
\par
\medskip
While equation ({}\ref{veldtemp3.28}) appears complex, it is actually simple
for several models.  Many plausible models, for instance, have a
time-averaged cloud line flux $F^{{\scriptsize\rm cl}}_{l}$ that is a function of $\hat{{\bf r}}\cdot \hat{{\bf s}}$.  Fortunately,
this alone does not result in the {\it gain} of the cloud line flux having a
similar dependence.  This is because the gain of the cloud line flux is
determined primarily from the mean of the local continuum flux, which is a
function only of $r~($not $\hat{{\bf r}})$ for many simple yet useful models; the
angle-dependent beaming anisotropy factor simply regulates what fraction of
the nonlinear flux of a cloud located at ${\bf r}$ is directed towards the observer.
On the other hand, some models produce line shifts and asymmetries due to
aspherical distribution functions (like those for bulk radial outflow) or
with $\hat{{\bf v}}$-dependent cloud pressures (like the one described in Appendix
\ref{shiftsap}).  At any rate, for the former class of models in which the
cloud gain is independent of $\hat{{\bf r}}$ and ${\bf v}$, we can make a neat simplification to
equation ({}\ref{veldtemp3.28}).  The gain factor can be moved outside the spatial
solid angle and velocity integrals in equation ({}\ref{veldtemp3.28}).  It can
then be written as
\par
\s
\tilde{\hat\Psi}_{F^{{\scriptsize\rm cl}}_{l}|F_{{\scriptsize\rm c}}}(\omega )={\tilde{\hat\Psi}_{I_{l}|F_{{\scriptsize\rm c}}}(\omega ,\sigma )\over \tilde{\hat\Psi}_{I^{{\scriptsize\rm sp}}_{l}|F_{{\scriptsize\rm c}}}(\omega ,\sigma )},
\labep{3.28.7}\e
\noindent where we define $I^{{\scriptsize\rm sp}}_{l}$ to be the ``spatial intensity" of the instantaneous
component of the cloud responses, i.e., $I_{l}$ if each cloud were to respond
linearly with $\hat{\Psi }_{F^{{\scriptsize\rm cl}}_{l}|F_{{\scriptsize\rm c}}}(\tau )$ simply $\delta (\tau )$ for all clouds.  As is made
clear shortly, the spatial intensity contains the geometrical information
about a line's emission in a model.
\par
\medskip
If $\tilde{\hat\Psi}_{I^{{\scriptsize\rm sp}}_{l}|F_{{\scriptsize\rm c}}}$ could be restricted (computed) to some extent theoretically,
the above equation would relate the unique and potentially observable
intensity gain to the mean of the linearized transfer functions of clouds on
the shell with radius $r(\sigma )$.  However, for simple spherical models each of
these clouds on this shell responds identically; the response function of
the shell is the same as that of each cloud on the shell.  Thus, under the
above conditions, equation
({}\ref{veldtemp3.28.7}) can permit measurement of the linearized transfer and
response functions of an {\it individual} cloud.  More precisely, if we account for
the finite resolution in velocity dispersion space to which $I_{l}$ has been
fit (or deconvolved) via equation ({}\ref{veldtemp3.26}) and the systematic
errors in the knowledge of the local line profiles, equation
({}\ref{veldtemp3.28.7}) permits measurement of the individual cloud transfer
functions averaged over a non-zero (but potentially small) range in radii.
This is fortuitous.  Provided firm theoretical constraints can be imposed upon
$\tilde{\hat\Psi}_{I^{{\scriptsize\rm sp}}_{l}|F_{{\scriptsize\rm c}}}$, the original dilemma demonstrated in Taylor (1996) regarding
``contamination'' of the observed response functions by the individual cloud
response functions may be tractable after all.
\par
\medskip
Fortunately for us, $\tilde{\hat\Psi}_{I^{{\scriptsize\rm sp}}_{l}|F_{{\scriptsize\rm c}}}$ probably has only a mild model
dependence.  This can be shown by considering a spherically symmetric model
without occulting material.  Applying $r(\sigma )=r_{0}(\sigma _{0}/\sigma )^{-2}~($where $r_{0}\sigma ^{2}_{0}\simeq M_{{\scriptsize\rm h}}G/3)$ to
equation ({}\ref{veldtemp3.28}) yields the linearized response function $\hat{\Psi }$ of the
spatial intensity
\par
\s
\hat{\Psi }_{I^{{\scriptsize\rm sp}}_{l}|F_{{\scriptsize\rm c}}}(\tau ,\sigma )=\Theta (\tau )\Theta (2\tau ^{{\scriptsize\rm sp}}-\tau ){1+\epsilon _{\hbox{{\scriptsize\rm A}{\scriptsize\it l}}}(1-\tau /\tau ^{{\scriptsize\rm sp}})\over 2\tau ^{{\scriptsize\rm sp}}},
\labep{3.29}\e
\noindent where $\epsilon _{\hbox{{\scriptsize\rm A}{\scriptsize\it l}}}$ is the beaming factor in line $l~($defined such that $\epsilon _{\hbox{{\scriptsize\rm A}{\scriptsize\it l}}}=-1$
implies fully beamed line emission and $\epsilon _{\hbox{{\scriptsize\rm A}{\scriptsize\it l}}}=0$ implies isotropic line emission)
and $\tau ^{{\scriptsize\rm sp}}$ is the characteristic ``spatial lag" defined as $r(\sigma )$/c.  Note that,
regardless
of the precise value of $\epsilon _{\hbox{{\scriptsize\rm A}{\scriptsize\it l}}}$, the above expression for $\hat{\Psi }_{I^{{\scriptsize\rm sp}}_{l}|F_{{\scriptsize\rm c}}}$ does not have
nontrivial structure over a large range of lags.  Rather, it is a trapezoidal
function having structure at a characteristic delay of $\tau ^{{\scriptsize\rm sp}}$.  Similarly,
$\tilde{\hat\Psi}_{I^{{\scriptsize\rm sp}}_{l}|F_{{\scriptsize\rm c}}}$
approaches zero at frequencies significantly above the ``spatial frequency"
defined here by $\omega ^{{\scriptsize\rm sp}}\equiv 1/\tau ^{{\scriptsize\rm sp}}$ and is unity at all frequencies significantly below
the spatial frequency.  Note that for models in which absorption is important
(such as those with an occulting accretion disk), the near clouds are
relatively brighter, in which case the effects of beaming are to some extent
masked.  For such models the spatial response function would probably be
smoother than equation ({}\ref{veldtemp3.29}) and the spatial gain would be even
more constrained to have nontrivial structure only near the spatial
frequency.  Thus even moderate
systematic uncertainties of the models under consideration would probably not
foil measurement of the individual linearized cloud response functions.
\par
\medskip
Equation ({}\ref{veldtemp3.28.7}) tells us that for some models the intensity
gain is simply the product of the spatial intensity and cloud gains.  This
situation is thus similar to that with equation
(\ref{localdelays3.14}), which has the following two
general characteristics that can be applied here as well:
\par
\noindent \begin{enumerate}\item If the line flux from clouds at a given radius
responds linearly then the intensity at the corresponding velocity
dispersion also responds linearly.  This is
because the spatial intensity necessarily responds linearly.  Conversely, any
nonlinearity in the intensity can be ascribed to nonlinear response of clouds
at a particular radius bin.  An advantage of this from a modeling perspective
is that with predictive photoionization codes one could impose relatively
tight constraints upon the cloud parameter space as a function of radius.
This is in contrast with modeling the gain of the line profile, which at a
specific wavelength is a function of clouds spanning a large range in mean
continuum flux.  On the other hand, a potential drawback of this is that the
intensities at some dispersions would be much more nonlinear than those of
the associated profiles because of this lack of radial dilution of cloud
response.  Therefore, measuring the individual cloud intensity transfer
functions could prove difficult at certain dispersions.  In this case, the
individual cloud transfer functions and equation ({}\ref{veldtemp3.28.7}) have
limited utility in model fitting.  Fitting of the physical model parameters
using (nonlinear) equation ({}\ref{veldtemp3.27}) could still be done to
capitalize upon this richness of the intensity representation.  However, as
discussed in Appendix \ref{localdelays} regarding
the fitting of nonlinear time-dependent profiles, doing this would be
relatively expensive computationally.\item A second characteristic equation ({}\ref{veldtemp3.28.7}) shares with
equation (\ref{localdelays3.14}) is that the qualitative
response features of the intensity can be determined by considering the
ratios of the characteristic time scales of the spatial and linearized cloud
response functions.  For instance, when the ratio differs enough from unity,
the cloud line flux and spatial intensity gain factors can be ``separable"
from one another, and one of the two gain factors can be treated as constant.
In this case the time-dependent behaviors at a given excitation frequency
depend
upon only one of the two time scales of the system.\end{enumerate}
\par
\medskip
In the next section we see that the concept of separability can be
partially extended to the time domain, where it implies simplified
relationships between the three linearized response functions.  In this case,
equation ({}\ref{veldtemp3.28.7}) and its application in measuring $\tilde{\hat\Psi}_{F^{{\scriptsize\rm cl}}_{l}|F_{{\scriptsize\rm c}}}(\omega )$
simplifies dramatically.  Let us now turn our attention to the precise
conditions necessary for separability to occur.
\par
\medskip
\noindent {}\setcounter{section}{2}{}\setcounter{subsection}{3}{}\addtocounter{subsection}{-1}\subsection{Conditions for Separability of the Cloud and Spatial Intensity
Gains}
\par
\medskip
Physically, the question of interest is, ``Given the various properties
of a particular AGN line emission model under consideration in which the
clouds take some finite time to react to changes in the continuum flux they
experience, which components of the profiles would respond in simple ways?''
If, for example, the sampling rate of our data set is only high enough to map
out the response from clouds contributing to the narrower profile components,
will the emission making up the core of the C {\scriptsize\rm IV} profile come from clouds
that can be treated as responding instantaneously, or will the time-dependent
responses of these clouds affect what we observed in our data?  If the clouds
respond fast enough, the time-dependence of their responses could perhaps be
ignored in the narrow emission region but not the broad emission region.
This subsection addresses such questions quantitatively.
\par
\medskip
Let us consider the case in which just one time constant $\tau ^{{\scriptsize\rm cl}}_{l}=1/\omega ^{{\scriptsize\rm cl}}_{l}$
characterizes the ``width" of the flux response function in a line
$l$ of clouds at a given radius.  Thus $\omega ^{{\scriptsize\rm cl}}_{l}$ is the characteristic response
frequency of the clouds, $\omega ^{{\scriptsize\rm sp}}_{l}$ is the characteristic response frequency of the
spatial intensity function, and $|\omega |$ is the excitation frequency of the
system.  Both the spatial and cloud gain factors in equation
({}\ref{veldtemp3.28.7}) must be in one of the three frequency regimes
(slow, fast, or intermediate) discussed in Appendix \ref{localdelays}.  For the global combined system, {\it provided} variability of the continuum
source both exists and is measurable on all time scales, we have 3 possible
frequency regimes for each gain factor.  This yields a total of $3^{2}=9$
different possible regimes.  Of these 9 regimes, only 5 have at least one
characteristic frequency that is similar (intermediate) to the excitation
frequency $|\omega |$.  Of these 5 remaining regimes, one is inseparable and
occurs for dispersions and continuum excitation frequencies such that
$|\omega |\sim \omega ^{{\scriptsize\rm cl}}_{l}\sim \omega ^{{\scriptsize\rm sp}}$.  In this regime, neither gain factor is approximately constant
for excitation frequencies near $|\omega |$.  That leaves us with 5-1=4 potentially
useful regimes in which the cloud response is separable from the spatial
response.
\par
\medskip
The characteristic response times of the clouds are probably dependent
upon their position within the BLR and NLR.
Let us assume that the equilibrium time scale of each cloud can be
approximated with a power law in
mean flux, such as $\tau ^{{\scriptsize\rm cl}}_{l}=\tau ^{{\scriptsize\rm cl}}_{l0}(r/r_{0})^{\eta (\tau ^{{\scriptsize\rm cl}}_{l}|r)}$,  where the ``asymptotic gain" $\eta $
of $\tau _{l}$ with respect to $r$ is defined as $\partial \log \tau _{l}/\partial $log{\it r~}for $r=r_{0}~($see also eq.
[\ref{localdelays3.3}]).  Note that only $\eta (\tau ^{{\scriptsize\rm cl}}_{l}|r)=0$
corresponds to flux-independent cloud equilibrium times.  Furthermore,
we have $\tau ^{{\scriptsize\rm cl}}_{l}=\tau ^{{\scriptsize\rm sp}}$ at a ``critical radius" of
\par
\s
r_{l{\scriptsize\rm c}}\simeq r_{0}\left({c\tau ^{{\scriptsize\rm cl}}_{l0}\over r_{0}}\right)^{1/|1-\eta (\tau ^{{\scriptsize\rm cl}}_{l}|r)|}
\labep{3.30}\e
\noindent corresponding to a ``critical velocity dispersion" of
\par
\s
\sigma _{l{\scriptsize\rm c}}\simeq \sigma _{0}\left({r_{0}\over c\tau ^{{\scriptsize\rm cl}}_{l0}}\right)^{1/|2[1-\eta (\tau ^{{\scriptsize\rm cl}}_{l}|r)]|} .
\labep{3.31}\e
\noindent The above equation tells us that the cloud response time scale differs
significantly from the spatial time scale when $\eta (\tau ^{{\scriptsize\rm cl}}_{l}|r)=1$ and $\tau ^{{\scriptsize\rm cl}}_{l0}$ differs
substantially by $\tau ^{{\scriptsize\rm sp}}_{0}$ or when $\eta (\tau ^{{\scriptsize\rm cl}}_{l}|r)\neq 1$ for intensity components with
dispersions much greater or less than $\sigma _{l{\scriptsize\rm c}}$.  In these cases separability can
be attained.  Note that this result is contingent upon $\eta (\tau ^{{\scriptsize\rm cl}}_{l}|r)$ being
constant; there are systems that do not have a critical
velocity dispersion because of the dependence of $\eta (\tau ^{{\scriptsize\rm cl}}_{l}|r)$ upon
radius.
\par
\medskip
The four different separable regimes can then be characterized as
follows:
\par
\noindent \begin{enumerate}\item The ``fast spatial" regime.  In this regime, $|\omega |\ll \omega ^{{\scriptsize\rm sp}}$,
yet $|\omega |\sim \omega ^{{\scriptsize\rm cl}}_{l}$, which requires $\tau ^{{\scriptsize\rm cl}}_{l}\gg \tau ^{{\scriptsize\rm sp}}$.  Let us assume
that each gain factor at a given radius has nontrivial structure only within
one decade of frequency.  The condition $|\omega |\ll \omega ^{{\scriptsize\rm sp}}$ can then be expressed as
$|\omega |<\omega _{l{\scriptsize\rm c}-}$, where
\par
\s
\omega ^{-1}_{l{\scriptsize\rm c\mp }}\simeq \tau _{l{\scriptsize\rm c\pm }}\simeq \tau _{l{\scriptsize\rm c}}10^{\pm 1/|2[1-\eta (\tau ^{{\scriptsize\rm cl}}_{l}|r)]|}
\labep{3.32}\e
\noindent gives the bounding excitation ``separable frequencies" outside of which the
fast or slow approximation becomes valid for only one of the gain
factors.  For $\eta (\tau ^{{\scriptsize\rm cl}}_{l}|r)<1,$ this
regime can occur for components of the intensity with $\sigma >\sigma _{l{\scriptsize\rm c}+}$, where
\par
\s
\sigma _{l{\scriptsize\rm c\pm }}\simeq \sigma _{l{\scriptsize\rm c}}10^{\pm 1/|4[1-\eta (\tau ^{{\scriptsize\rm cl}}_{l}|r)]|}
\labep{3.33}\e
\noindent are the ``separable dispersions."  Similarly, for $\eta (\tau ^{{\scriptsize\rm cl}}_{l}|r)>1,$ the fast
spatial regime can occur for $\sigma <\sigma _{l{\scriptsize\rm c}-}$.  In Figure
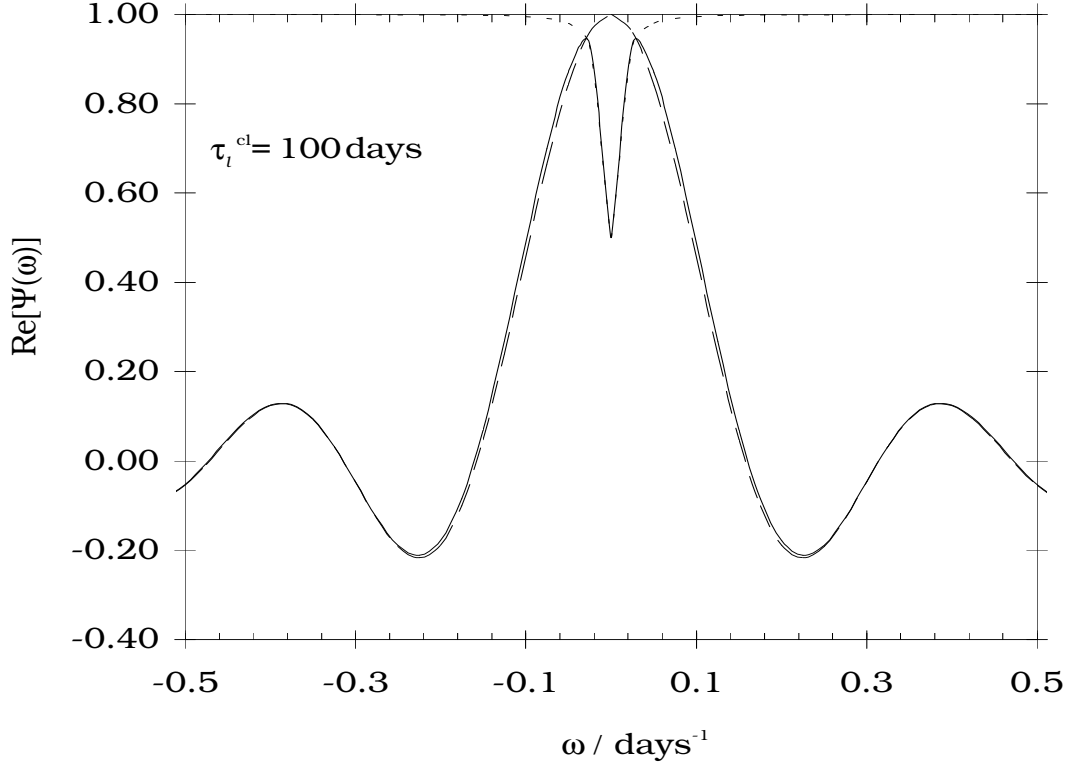
\ref{veldtemp/f1}\fignobox{veldtemp/f1}{Real component of the transfer
function
(the gain) of the $\sigma =4000$ km s$^{-1}$ intensity component (solid line) as a
function of excitation
frequency $\omega $ assuming cloud equilibrium times of $\tau ^{{\scriptsize\rm cl}}_{l}=100$ days.  Also shown
are the gains of the transfer functions of the individual clouds (dotted
line) and the spatial gain (dashed line).  This is for a hypothetical
optically thin line using a very simple model, with a black hole of mass
$M_{{\scriptsize\rm h}}=10^{8}~M_{\odot } ($selected to yield a spatial time scale of $\tau ^{{\scriptsize\rm sp}}=GM_{{\scriptsize\rm h}}/[3c\sigma ^{2}]=10$ days).
For $|\omega |<\omega _{l{\scriptsize\rm c}-}\simeq (30$ days$)^{-1}$, the spatial transfer function is approximately its
asymptotic value.  This defines the fast spatial regime.  Conversely, for
$|\omega |\gtrsim \omega _{l{\scriptsize\rm c}+}\simeq (30$ days$)^{-1}$, the cloud transfer function is approximately its
instantaneous value.  This defines the slow cloud regime.}, the
spatial, cloud, and global (directly observable) intensity gains of an
optically thin line of
a spherically symmetric linear system are shown at a velocity dispersion of
$\sigma =4000$ km s$^{-1}$.  The line emitted by the clouds of this hypothetical model
responds on an equilibrium time scale of $\tau ^{{\scriptsize\rm cl}}_{l}=100$~days, with $\eta (F^{{\scriptsize\rm cl}}_{l}|F_{{\scriptsize\rm c}})=0.5$
and $\tilde{\hat\Psi}_{F^{{\scriptsize\rm cl}}_{l}|F_{{\scriptsize\rm c}}}(\infty )=1.0.$~(Such numbers might exist, for example, in a model with
a linearly-responding line and large clouds that eventually evaporate when
exposed to additional continuum heating.)  The black hole mass is
$10^{8}~M_{\odot }$, which was selected such
that $\tau ^{{\scriptsize\rm sp}}{\scriptsize\rm =}GM_{{\scriptsize\rm h}}/(3c\sigma ^{2})=10$~days. The time constant ratios of this hypothetical
model differ enough that the excitation frequency range of the intermediate
regime is small.  The fast spatial regime occurs here for
$|\omega |<\omega _{l{\scriptsize\rm c}-}\simeq (30$~days$)^{-1}$, where the spatial intensity gain factor is
approximately its asymptotic value of unity.  In Figure
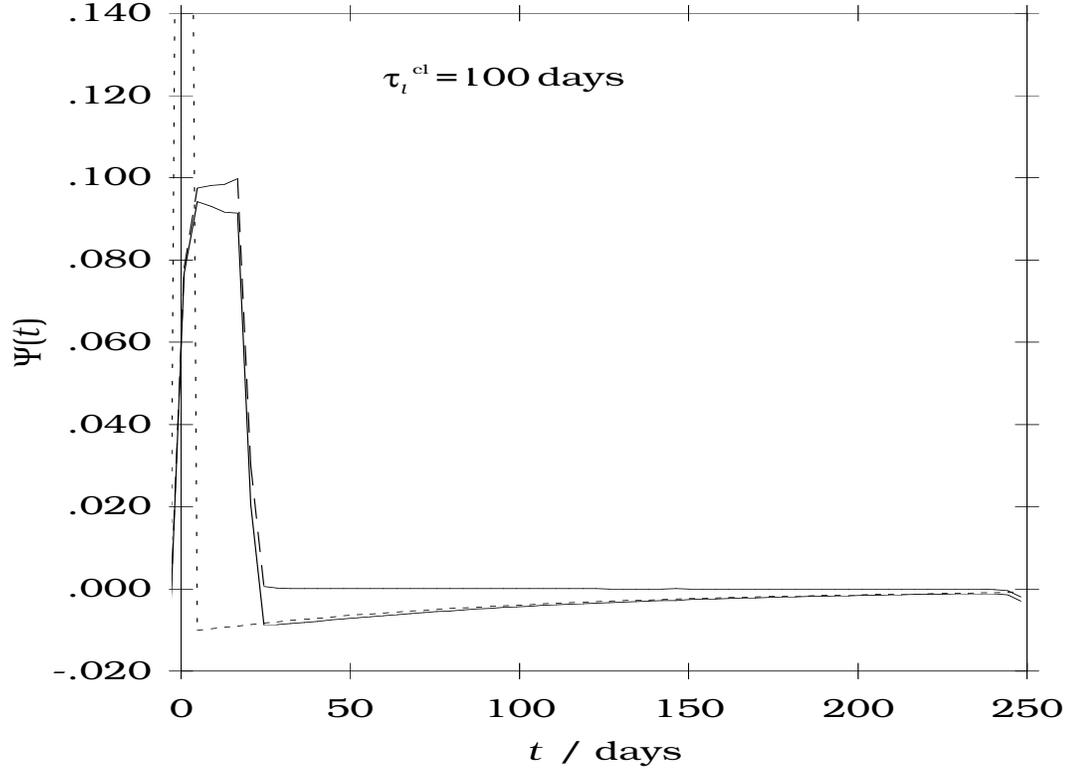
\ref{veldtemp/f2}\fignobox{veldtemp/f2}{Time domain equivalents (response
functions) of the system
shown in Fig. \ref{veldtemp/f1}.  An analog of the fast
spatial regime occurs in the time domain on time resolution scales greater
(poorer) than $\tau _{l{\scriptsize\rm c}+}=20$
days, while an analog of the slow cloud regime occurs on time resolution
scales less (better) than $\tau _{l{\scriptsize\rm c}-}\simeq 30$ days.  Note that the cloud response
function is off the scale near zero lag.  This is because $\tilde{\hat\Psi}_{F^{{\scriptsize\rm cl}}_{l}|F_{{\scriptsize\rm c}}}(\infty )$ is
nonzero.  Also, the solid (and
easiest to measure) line has structure on a time scale much greater than the
spatial response time scale, which indicates either a highly arranged
distribution of matter or (via eq. [{}\ref{veldtemp3.29}]) locally delayed
response.
Unless specific deconvolution techniques were employed to avoid it, the
empirical intensity component would have a nonzero width in dispersion space
due to finite sampling resolution.  This would result in a steep, rather than
infinite, slope near $\tau =20$ days for such a hypothetical system.  There is some
numerical error at high and low values of $\tau $.}, the response functions of
this system are shown.  For lags greater than $\tau _{l{\scriptsize\rm c}+}=20$~days, the global line
intensity response function is equal to the cloud response function.  By
ignoring oscillations with frequencies above $\omega _{l{\scriptsize\rm c}-}$ and subtracting off either
narrow or broad components of the line profile, this fast spatial regime is
in principle always achievable for simple enough systems (that, e.g., are
moderately spherically symmetric).  However, as forewarned above, doing this
in practice could be difficult or even impossible if, e.g., $\eta (\tau ^{{\scriptsize\rm cl}}_{l}|r)$ is
close enough to unity and $\sigma _{l{\scriptsize\rm c}+}$ is too high or $\sigma _{l{\scriptsize\rm c}-}$ is too low compared to the
mean velocity dispersion in the observed line.\item The ``slow cloud" regime.  In this case, $|\omega |\gg \omega ^{{\scriptsize\rm cl}}_{l}$ yet $|\omega |\sim \omega ^{{\scriptsize\rm sp}}$, which
also requires $\tau ^{{\scriptsize\rm cl}}_{l}\gg \tau ^{{\scriptsize\rm sp}}$.  This regime is the same as that illustrated in
Figure \ref{veldtemp/f1}, but for $|\omega |>\omega _{l{\scriptsize\rm c}+}\simeq (30$~days$)^{-1}$, wherein the cloud
gain factor is approximately its instantaneous value.  Because this
instantaneous gain is nonzero, the global intensity response function is
approximately the product of the instantaneous cloud gain and the spatial
response function for lags below $\tau _{l{\scriptsize\rm c}-}\simeq 30$~days.\item The ``slow spatial" regime.  In this case, $|\omega |\gg \omega ^{{\scriptsize\rm sp}}$ yet $|\omega |\sim \omega ^{{\scriptsize\rm cl}}_{l}$, which
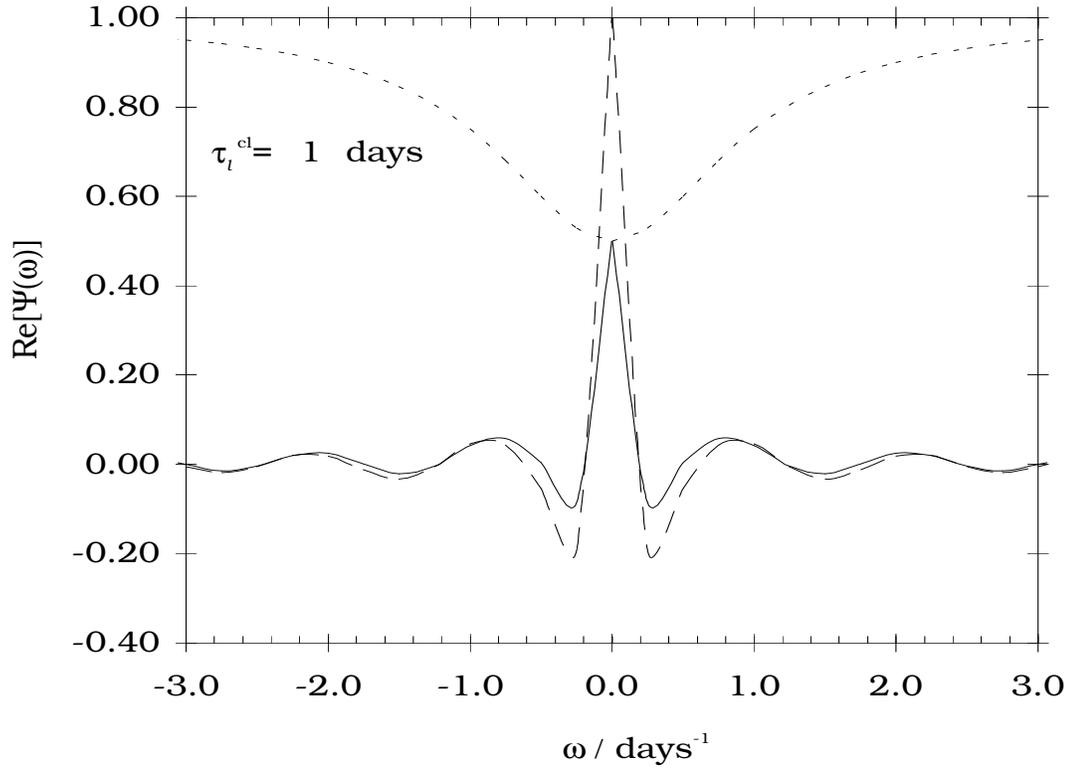
requires $\tau ^{{\scriptsize\rm cl}}_{l}\ll \tau ^{{\scriptsize\rm sp}}$.  In Figure \ref{veldtemp/f3}\fignobox{veldtemp/f3}{Real
component of the transfer function of the $\sigma =4000$ km s$^{-1}$ intensity component
assuming cloud equilibrium times of $\tau ^{{\scriptsize\rm cl}}_{l}=1$ day.  Other parameters are
unchanged from those assumed in Fig. \ref{veldtemp/f1}.  For $|\omega |\gg (3$ days$)^{-1}$,
the intensity gain oscillates about zero.  This is due to the spatial gain
factor being near zero at high frequencies.  This defines the fast spatial
regime.  For $|\omega |<\omega _{l{\scriptsize\rm c}-}\simeq (3$ days$)^{-1}$, the cloud gain factor is approximately
constant.  This defines the fast cloud regime.} the gains are shown for a
system like that of Figures \ref{veldtemp/f1} and \ref{veldtemp/f2} but for
a hypothetical model in which $\tau ^{{\scriptsize\rm cl}}_{l}=1$~day and $\tau ^{{\scriptsize\rm sp}}{\scriptsize\rm =}10$~days.  For $|\omega |\gg (3$~days$)^{-1}$
the intensity gain oscillates about zero because the
instantaneous spatial intensity gain is zero for this model.  This
defines the fast spatial regime.  In the time domain we see this regime on
resolutions better than $\tau _{l{\scriptsize\rm c}-}\simeq 20$~days, where the derivative of the
global intensity response function is equal to the cloud response function
divided by the spatial response function at zero lag (see Fig.
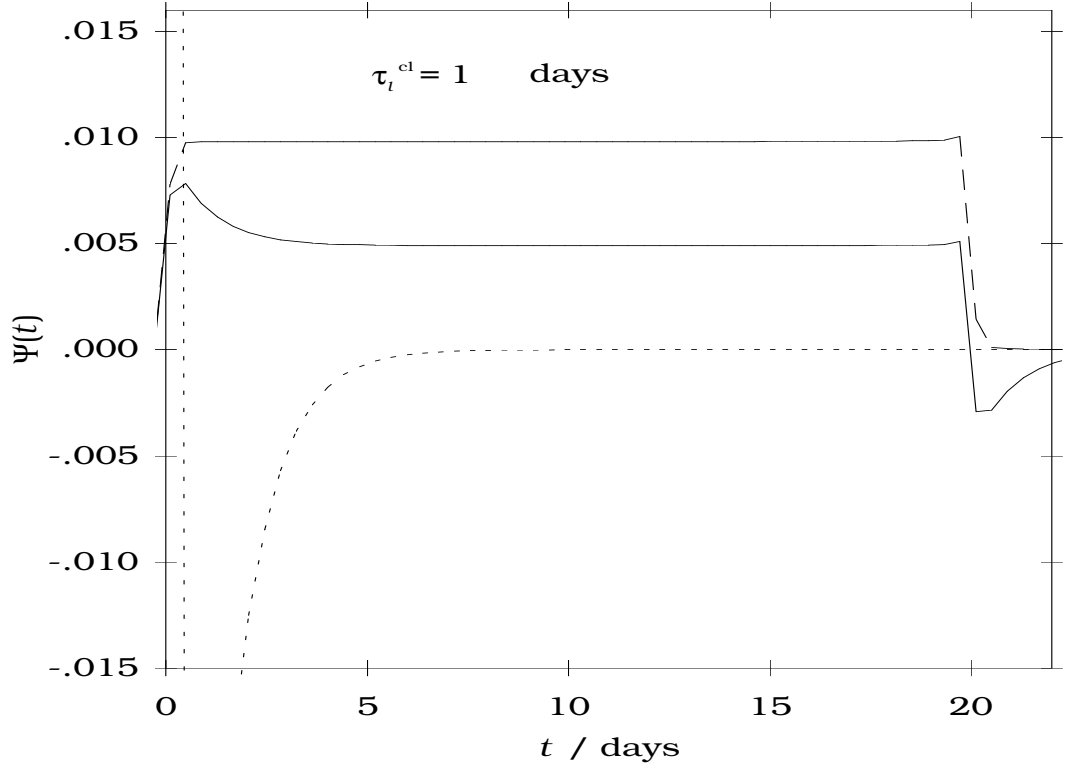
\ref{veldtemp/f4}\fignobox{veldtemp/f4}{Time domain equivalents of Fig.
\ref{veldtemp/f3}.  Note that, as in Fig. \ref{veldtemp/f2}, the solid line
has structure on two distinct time scales.}.)\item The ``fast cloud" regime.  In this regime, $|\omega |\ll \omega ^{{\scriptsize\rm cl}}_{l}$ yet $|\omega |\sim \omega ^{{\scriptsize\rm sp}}$, which
also requires $\tau ^{{\scriptsize\rm cl}}_{l}\ll \tau ^{{\scriptsize\rm sp}}$.  In the time domain this regime corresponds to lags
greater than $\tau _{l{\scriptsize\rm c}+}$, where, as in the slow cloud case, the global response
function is (within a time resolution of $\tau _{l{\scriptsize\rm c}+})$ approximately the spatial
response function multiplied by the asymptotic cloud gain.  This regime is
illustrated in Figures \ref{veldtemp/f3} and \ref{veldtemp/f4}, where
$\omega _{l{\scriptsize\rm c}-}\simeq (3$~days$)^{-1}$ and $\tau _{l{\scriptsize\rm c}+}\simeq 3$~days. If $\eta (\tau ^{{\scriptsize\rm cl}}_{l}|r)\sim 0,$ this fast cloud regime can
be realized if the broader profile components with dispersion velocities
above $\sigma _{l{\scriptsize\rm c}-}$ are subtracted out.  Incidentally, in some prior work in which the
fast cloud regime was apparently assumed, only narrow components were
subtracted out (e.g., Krolik et al. 1991).\end{enumerate}
\par
\medskip
\noindent {}\setcounter{section}{3}{}\addtocounter{section}{-1}\section{Applications and Model Fitting}
\medskip
The previous section contains mathematical expressions for the cloud and
spatial transfer functions.  It raises two questions.  First, how,
specifically, might such equations be used to make real measurements of the
cloud and spatial response functions?  Second, what specific benefits would
this offer over the simpler methods of data analysis that have been employed
in the past?
\par
\medskip
Analysis of variability data permits two potential benefits:
\par
\medskip
\noindent \begin{itemize}\item It permits a reduction in the uncertainties of the
values of the physical parameters that are used to fit a given object as
defined within the context of the specific models that are fit.  Note
that these
physical parameters have no meaning outside the specific models to which they
are associated.  Thus, any truly model-independent analysis of data would
offer no new {\it physical} information about the system being observed.\item It
permits a reduction of the uncertainty of the overall viabilities (i.e.,
goodnesses) of the models being fit provided the actual effective number
of these parameters that have been granted to a model is small enough
compared to that of the data.\end{itemize} Therefore, if local delays or
nonlinearities are important, not only will the linear models that have
traditionally been used in the past fail when they are not given excessive
freedom, but also the values of the parameters obtained
upon fitting such models will have little physical meaning.  Since current
theoretical and observational constraints are suggestive that local delays
{\it are} important, we unfortunately need to employ models that can account for
these effects in order to gain physical information from variability data.
This subsection takes a step in this direction by presenting some models
which are capable of accounting for local delays or nonlinearities.
\par
\medskip
From an experimental perspective, there is a simple way the equations of
the prior section can be used.  First, the various functions in the equations
can be parameterized.  Second, differences between the two sides of each
equation can be assigned to difference variables.  Third, the model
parameters can be numerically adjusted within the constraints dictated by
their error limits until the difference variables have been minimized.
Before using any of the various software programs that have been written to
perform this type of task, one must first select a compatible
parameterization scheme for the model to be tested.  The numerous parameters
associated with certain models (such as those represented by eq.
[\ref{localdelays3.16.5}]) might be
poorly constrained from the limited data and available information.
In particular, if the only input knowledge for such a model is a linearized
velocity-resolved response function, the various model parameters might not
be uniquely determined, even if the observational errors were negligible.
This is analogous to the shadow of an object not uniquely determining its
topology unless, for example, the object is assumed to be both
two-dimensional and viewed perpendicularly.  For this reason, let us first
consider the fitting of relatively simple models.
\par
\medskip
\noindent {}\setcounter{section}{3}{}\setcounter{subsection}{1}{}\addtocounter{subsection}{-1}\subsection{A Simple Linear Example Model}
\par
\medskip
One of the simplest nontrivial models one could construct assumes
instantaneous response, linearity (which under most circumstances implies
that the effective cloud area {\it A} and line efficiency $\epsilon _{l}$ are independent of
flux and hence radius), spherical symmetries in $f$ with respect to ${\bf r}$ and ${\bf v}$,
and $\epsilon _{\hbox{{\scriptsize\rm A}{\scriptsize\it l}}}=0$.  This last assumption would probably be applicable for
lines suspected of having low optical depths, such as C {\scriptsize\rm III}$]~\lambda 1909$.  One way this model could be
parameterized is by letting $f$ be an interpolation of a two-dimensional
grid in $r$ and {\it v} with a resolution dictated by the quality of the data used.
In this case, $f$ is constrained from the ionizing continuum flux $F_{{\scriptsize\rm c}}(t)$ and the
observed spectrum $F_{\lambda }(t)$.  Specifically, it is straightforward to show that
\par
\s
A\epsilon _{l}f(r,v)={r\over \pi c} {d^{2}\Psi _{F_{l}|F_{{\scriptsize\rm c}}}(2r/c,v)\over dr dv}.
\labep{3.33.2}\e
\noindent In this equation, $\Psi _{F_{l}|F_{{\scriptsize\rm c}}}(\tau ,v_{{\scriptsize\rm D}})$ can be obtained (i.e., measured) from the
linearized response function $\Psi _{F_{\lambda }|F_{{\scriptsize\rm c}}}$ of the spectrum $F_{\lambda }$ with respect to the
ionizing flux $F_{{\scriptsize\rm c}}$, the convolution equation, and
\par
\s
\sum^{}_{l} {c\over \lambda _{l}} F_{l}{\hbox{\bf(}}t,c(\lambda -\lambda _{l})/\lambda _{l}{\hbox{\bf)}}\simeq F_{\lambda }(t)-F_{{\scriptsize\rm c\lambda }}(t),
\labep{3.33.4}\e
\noindent where $F_{{\scriptsize\rm c\lambda }}$ is the time-dependent spectrum of just the continuum.  Use of the
above equation in the overall fitting routine with $\epsilon _{l}$ being free parameters
is important because profile de-blending, like any other interpretation of
data, is necessarily model dependent, at least to some extent.
\par
\medskip
In Blandford \& McKee (1982) a clear distinction is made between model
fitting and blind mathematical fitting of a $\Psi _{F_{l}|F_{{\scriptsize\rm c}}}(\tau ,v_{{\scriptsize\rm D}})$ function.  However,
note that equation ({}\ref{veldtemp3.33.2}) implies that a blind mathematical
fitting of a
response function to variability data of a given line is almost equivalent to
model fitting of the arbitrary ${\it A\epsilon }_{l}f$ function above.  In fact, other than
the required transformation of variables (described in this case by eq.
[{}\ref{veldtemp3.33.2}]) and their associated uncertainties, there is only one
difference
between model fitting using an arbitrary but linear model and fitting with an
arbitrary linearized response function: with the latter method no physical
significance is explicitly attached to the fitted function.  Exactly what is
considered here to be the ``fundamental" set of parameters of a model selected
for fitting is unimportant, provided that the measurement errors are also
transformed appropriately.
\par
\medskip
However, this is not the case when more than one line is available.
Because {\it A} and $\epsilon _{l}$ must be constants in the above linear model, the response
functions for each line are proportional to one another.  Thus, provided the
local continuum flux determines the cloud properties, {\it the mere fact that the
response functions obtained by Krolik et al. (1991) appear to be
line-dependent empirically tells us that the linearity assumption (in
addition to this model) is formally incorrect}.  This tells us that the use of
nonlinear models is probably necessary for fully self-consistent
interpretations of AGN variability data.
\par
\medskip
\noindent {}\setcounter{section}{3}{}\setcounter{subsection}{2}{}\addtocounter{subsection}{-1}\subsection{A Simple Nonlinear Example Model}
\par
\medskip
As a second example of a model that could be used in fitting, let us
assume nonlinearly responding clouds with $F^{{\scriptsize\rm cl}}_{l}\propto F^{\eta _{l}}_{{\scriptsize\rm c}}$, where $\eta _{l}$ is an
abbreviation for $\eta (F^{{\scriptsize\rm cl}}_{l}|F_{{\scriptsize\rm c}})$.  Let us also assume that the other model
parameters are the same as for the prior example model.  This second example
system has instantaneous (without delay) nonlinearities.  Its exact nonlinear
profiles are
\par
\s
F_{l}(t,v_{{\scriptsize\rm D}})=F^{{\scriptsize\rm cl}}_{l0} \int d\tau \left({F_{{\scriptsize\rm c}}(t-\tau ,{\bf D})\over <F_{{\scriptsize\rm c}}>}\right)^{\eta _{l}} \hat{\Psi }_{F_{l}|F^{{\scriptsize\rm cl}}_{l}}(\tau ,v_{{\scriptsize\rm D}}),
\labep{3.33.4.1}\e
\noindent where
\par
\s
\hat{\Psi }_{F_{l}|F^{{\scriptsize\rm cl}}_{l}}(\tau ,v_{{\scriptsize\rm D}})=2\pi r_{0}c\Theta (\tau ) \int^{\infty }_{c\tau /2} dr\left({r\over r_{0}}\right)^{1-2\eta _{l}} n(r)\phi _{l}(\sigma ,v_{{\scriptsize\rm D}})\hbox{ }
\labep{3.33.4.2}\e
\noindent are ``input-modified response functions"
similar to the one in Maoz (1992).
\par
\medskip
If $\eta _{l}$ is forced to be independent of radius, it can be measured by
fitting equation ({}\ref{veldtemp3.33.4.1}) to spectra.  This method of obtaining
the mean
nonlinearities in the lines should be significantly more accurate than that
in, e.g., Pogge \& Peterson (1992), where the spatial response function was
crudely approximated to be a delta function in lag.
\par
\medskip
The dependence of the kernel of an input-modified response
function is a function of only lag $\tau $ via $\theta _{rs}\equiv \cos ^{-1}(\hat{{\bf r}}\cdot \hat{{\bf s}})=\cos ^{-1}(1-c\tau /r)$.  If the
dependence of $F^{{\scriptsize\rm cl}}_{l}$ upon $\theta _{rs}$ and $\hat{{\bf v}}$ is known then the local line profiles can
be numerically fit from the velocity-resolved response functions.  For
instance, in this second example model we assume isotropic cloud line fluxes
which yields (via eq. [{}\ref{veldtemp3.33.4.2}])
\par
\s
\phi _{l}(\sigma ,v_{{\scriptsize\rm D}})={d\Psi _{F_{l}|F^{{\scriptsize\rm cl}}_{{\scriptsize\rm c}}}(\tau ,v_{{\scriptsize\rm D}})/d\tau \over d\Psi _{F_{l}|F^{{\scriptsize\rm cl}}_{{\scriptsize\rm c}}}(\tau )/d\tau }|_{\tau =2GM_{{\scriptsize\rm h}}/(3c\sigma ^{2})},
\labep{3.34}\e
\noindent where the denominator is the integral of the velocity-resolved response
function.  With the above equations, the other physical properties such as
$n(r), r_{0}, F^{{\scriptsize\rm cl}}_{l0}$, and $M_{{\scriptsize\rm h}}$ can be measured within the context of this model.
\par
\medskip
Equation ({}\ref{veldtemp3.34}) is actually of little practical utility since
the local profiles can be obtained in a more direct fashion by numerically
fitting equations ({}\ref{veldtemp3.33.4.1})-({}\ref{veldtemp3.33.4.2}) to variability
data.  Equation ({}\ref{veldtemp3.34}), however, illustrates an important point.
The equation would
apply
even if $\eta _{l}$ were a function of radius and the responses were very nonlinear.
One assumption it does require is instantaneous responses from the clouds.
Recall that the first example model illustrated that if there are
line-dependent response functions than at least one of the cloud line
emission functions is non-linear for models in which the emission is a
function of only the local continuum flux.  Here we see that if there is also
a line dependence in the local profiles then either the assumption about the
dependence of $F^{{\scriptsize\rm cl}}_{l}$ upon $\hat{{\bf r}}$ or $\hat{{\bf v}}$ is invalid or local delays are important.  In
the latter case, models like the second example model would fit the data
poorly, yield low goodnesses, and give large measurement errors.
\par
\medskip
\noindent {}\setcounter{section}{3}{}\setcounter{subsection}{3}{}\addtocounter{subsection}{-1}\subsection{A Simple Example Model with Local Delays}
\par
\medskip
As a third example illustrating how local delays and nonlinear response
could be accounted for, let us assume the cloud areas and pressures have
non-zero equilibrium times of $\tau _{P}(\sigma )=\tau _{P0}(\sigma _{0}/\sigma )^{2\eta (\tau _{P}|r)}$ and
$\tau _{A}(\sigma )=\tau _{A0}(\sigma _{0}/\sigma )^{2\eta (\tau _{A}|r)}$.  In the simplest case, these local delays can be
parameterized with a linear model, with individual cloud
line flux gains of
\par
\s
\tilde{\hat\Psi}_{F^{{\scriptsize\rm cl}}_{l}|F_{{\scriptsize\rm c}}}(\omega ,{\bf r},{\bf v},\hat{{\bf s}})=1 + {\eta (A|F_{{\scriptsize\rm c}})\over 1+i\omega \tau _{A}(\sigma )} + {\eta (\epsilon _{l}|F_{{\scriptsize\rm c}})\over 1+i\omega \tau _{P}(\sigma )}.
\labep{3.37}\e
\noindent Let us again assume for this third example model that $\epsilon _{\hbox{{\scriptsize\rm A}{\scriptsize\it l}}}=0$ and
that
\par
\s
f(r,v)=n(r) {e^{-v^{2}/2\sigma ^{2}}\over (2\pi \sigma ^{2})^{3/2}},
\labep{3.38}\e
\noindent where $n$ is the number density of clouds, with $\sigma ^{2}=(r_{0}/r)\sigma ^{2}_{0}$ and $\sigma ^{2}_{0}=GM_{{\scriptsize\rm h}}/(3r_{0})$
so that $<{\bf v}^{2}>~\simeq GM_{{\scriptsize\rm h}}/r$ in accordance with a Maxwellian distribution of clouds.
As mentioned in Appendix \ref{veldtempa}, equation ({}\ref{veldtemp3.38}) is
probably not as
``square-shaped" in velocity space as certain theoretical models would
suggest.  However, it has the advantage of being built into most fitting
packages.  Equation ({}\ref{veldtemp3.27}) implies
\par
\s
\phi _{l}(\sigma ,v_{{\scriptsize\rm D}})= {e^{-v^{2}_{{\scriptsize\rm D}}/(2\sigma ^{2})}\over \sqrt{2\pi } \sigma }.
\labep{3.39}\e
\noindent This in turn permits measurement, without imposing instantaneous local
response, of $I_{l}(t,\sigma )$ and $\Psi _{I_{l}|F_{{\scriptsize\rm c}}}(\tau ,\sigma )$ when fit with equation
({}\ref{veldtemp3.26}), equation ({}\ref{veldtemp3.33.4}), and the convolution
equation.  In fact, for this model, equation ({}\ref{veldtemp3.28.7}) gives in the
linear regime
\par
\s
\tilde{\hat\Psi}_{F^{{\scriptsize\rm cl}}_{l}|F_{{\scriptsize\rm c}}}(\omega ,\sigma )=\omega \tau ^{{\scriptsize\rm sp}}\tilde{\hat\Psi}_{I_{l}|F_{{\scriptsize\rm c}}}(\omega ,\sigma ){e^{i\omega \tau ^{{\scriptsize\rm sp}}}\over \sin (\omega \tau ^{{\scriptsize\rm sp}})},
\labep{3.40}\e
\noindent which can be fit to equation ({}\ref{veldtemp3.37}) to measure $\tau _{P0}$, $\eta (\tau _{P}|r)$,
$\tau _{A0}$, $\eta (\tau _{A}|r)$, $\eta (A|F_{{\scriptsize\rm c}})$, and $\eta (\epsilon _{l}|F_{{\scriptsize\rm c}})$ from time-resolved spectra.  For more
accurate modeling and lower measurement errors at the expense only of
computation time, equation ({}\ref{veldtemp3.37}) could be replaced by a nonlinear
analog.  The point in either case is that the physical parameters of the
model, including the cloud pressure equilibrium times, are indeed measurable
quantities provided theoretical constraints can be imposed upon the angular
dependence of $F^{{\scriptsize\rm cl}}_{l}$.
\par
\medskip
Note that this third example model does not necessarily have more
degrees of freedom than the first two.  This is because upon including the
above six parameters, one would accordingly reduce the grid resolution of the
fitted functions like $n(r)$ such that the effective number of parameters
(e.g., the order of its polynomial) permitted in
the model remained the same.  Incidentally, this reduction would be small
when the quality of data is high and the response functions can be
interpolated from several points.  Performing the computationally expensive
nested integrals (such as those implicit in the line flux equation of models
with local lags) instead of the linear
approximations (eq. [{}\ref{veldtemp3.37}]) admittedly provides a means of
accounting for both nonlinear and inseparable behavior if sufficient computer
time is available.  In the linear regime, however, including parameters to
account for finite response times in a cloud would not change the prediction
error of the model at all.
\par
\medskip
In this context, as stated above, the real benefit of model fitting with
$f$ and cloud lag parameters over ``blind" mathematical de-convolution is
that the fitted quantities have physical meaning.  For instance, features in
$\Psi _{F_{l}|F_{{\scriptsize\rm c}}}$ beyond lags of $4GM_{{\scriptsize\rm h}}/v^{2}_{{\scriptsize\rm D}}c$ would be physically meaningful in the third
example model provided that the equilibrium times scale of at least one cloud
property is sufficiently long.  Similarly, additional information, such as
that from fully nonlinear photoionization codes like CLOUDY, XSTAR, ION, etc.
can be included (albeit with substantial systematic error) along with the
line emission data to lower the errors of the fitted parameters.  This would,
for example, help exclude so-called ``unphysical de-convolutions."  Also, if one is fitting
the underlying physical conditions of the gas emitting the lines rather than
the line emissivities themselves, line blending aids and constrains our
knowledge of what we are fitting (as opposed to hindering it).  With abstract
mathematical de-convolution, valuable information resources like these must
be thrown away. (Invoking physical arguments to interpret an abstract
de-convolution is a crude means of model fitting.  This point, along with
others on the advantages that model fitting has over ``arbitrary" function
fitting, is discussed in more detail elsewhere.)  Though linear response
functions or, for the apparently nonlinear cases, even multidimensional
``time-dependent" response functions might be able to fit variability data
well, they probably do not offer as much to our understanding of AGNs as does
model fitting.
\par
\medskip
\noindent {}\setcounter{section}{4}{}\addtocounter{section}{-1}\section{Conclusion \& Discussion}
This appendix discusses the analysis of AGN spectral variability data
using line emission models that have ``reactive'' clouds in orbital
motions.  For such models, the line intensity
representation of the line profiles is useful for obtaining the cloud
response characteristics.  These response characteristics are simplest when
their transfer functions are ``separable'' from the spatial transfer functions
of the intensities.  A total of four different separable regimes are
possible.  For spherical systems without occulting material, the observed
linearized response function of the intensity is a trapezoidal function only
in the fast and slow cloud regimes.  If the characteristic cloud equilibrium
time or frequency is a weak enough function of the mean local continuum flux,
the fast cloud regime would occur at excitation frequencies significantly
lower than the characteristic equilibrium frequency for components of the
intensity with low enough velocity dispersions.  In the same such system, the
slow cloud regime would occur at excitation frequencies significantly above
the characteristic equilibrium frequency at high enough velocity dispersions.
Even outside these simplifying regimes of
separability, nature permits, under certain assumptions, measurement (via,
e.g., eq. [{}\ref{veldtemp3.28.7}]) of the linearized response functions of clouds
at a given position.  However, these types of measurements would require
variability data with much larger duration to sampling period ratios than
would be necessary in the fast cloud regime.
\par
\medskip
Abstract mathematical deconvolution discussed in
Blandford \& McKee (1982) is probably not the ideal method of analyzing
variability
data. If its purpose were merely to predict variability data, then other
fully nonlinear methods, such as those employed with neural nets to predict,
e.g., stock prices, would probably fare better.  If, on the other hand, the
ultimate goal is to increase our knowledge of the physics of the AGNs, then
the model fitting discussed in Blandford \& McKee (1982) would probably fare
better; only model fitting permits measurement of several parameters within
the context of specific models while formally accounting for their physical
plausibility.  For instance, the example model of \S ~\ref{veldtemp}.3.2
can be used to
measure the mean asymptotic gain $\eta (F_{l}|F_{{\scriptsize\rm c}})$ of a line without approximating the
spatial response function to be a delta function in lag as was done in
previous works (Krolik et al. 1991; Pogge \& Peterson 1992).  Another set of
parameters is the linearized response function of the intensity in a line
with respect to the continuum flux $\Psi _{I_{l}|F_{{\scriptsize\rm c}}}$.  When information from a
photoionization code is included in the fitting of a model, this function
would probably be of more utility than $\Psi _{F_{\lambda }|F_{{\scriptsize\rm c}}}$.  This is because the range of
cloud parameters sampled by the intensity at a given velocity dispersion is
much smaller than that sampled by the linearized spectral response function
at a given wavelength.
\par
\medskip
It is the knowledge of the values of the various physical parameters
that can help us understand AGNs.  For instance, knowledge of the radii in a
fitting of a model that incorporates fully nonlinear results (e.g., $\epsilon _{l})$ from
a photoionization code where contributions to the reduced $\chi ^{2}$ are large would
help indicate the parameter space wherein more physics is necessary and
uniform
pressure cloud models fail.  Knowledge of which lines have behavior that is
inconsistent with the black hole mass inferred from other lines, after
nonlinear behavior is taken into account, would give an independent upper
limit to how important alternative line broadening mechanisms (e.g., electron
scattering) could be in orbital models.  Knowledge of the sign of
$\eta (A_{\hbox{{\scriptsize\rm geom}}}|F_{{\scriptsize\rm c}})$, where $A_{\hbox{{\scriptsize\rm geom}}}$ is the geometrical size of the clouds, would help
answer the very important question of whether or not the source of clouds in
AGNs, which has been debated since their introduction, is in part due to the
continuum flux or just some passive element of the overall AGN environment.
\par
\medskip
Perhaps the most important knowledge that can be gained by proper model
fitting is that related to the {\it cloud} response functions $\Psi _{F^{{\scriptsize\rm cl}}_{l}|F_{{\scriptsize\rm c}}}$. For
instance, consider that the emission efficiency in a line can drop
significantly when the column density to ionization parameter ratio becomes
low enough that the ``back" portions of a cloud are too highly ionized for
significant emission by an ion.  Partly for this reason, intensity line
ratios could be used to obtain measurements of mean column densities as a
function of radius (along with the pressures or densities).  This in turn
could permit measurement of cloud characteristics such as the time-averaged
size of the clouds as a function of radius.  With cloud response functions,
the {\it time}-dependence of these various cloud properties can also be determined,
which would permit measurement of the cloud evaporation and
pressure-equilibrium wave propagation speeds (even if they are nearly zero).
Such knowledge would impose new and important constraints upon the various
cloud confinement mechanisms that have been proposed.
\par
\medskip
Rather than be distressed that nature is more complicated without the
fast cloud assumption, we should consider ourselves very fortunate,
for cloud response functions would allow us to directly probe the structure
of an individual cloud within the context of a model, and, if the cloud
concept is valid, help answer the question ``What are the clouds?"  Answering
this question, rather than ``Where are the clouds?" is probably more
important for understanding AGNs.
\medskip
\noindent \chapter{Obtaining the Local Line
Profiles\label{veldtempa}}
\par
According to Appendix \ref{veldtemp}, {\it a priori} knowledge of both the
local line profiles $\phi _{l}$ and the distance from the black hole $r$ as a function
of velocity dispersion parameter $\sigma $ is necessary before measurements of the
intensities can be made from a spectrum.  This appendix discusses issues
relating to how these two quantities might be constrained.
\par
\medskip
As the example model discussed in \S ~\ref{veldtemp}.3.2 illustrates,
the local line profiles should be particularly straightforward to measure if
the following assumptions are valid:
\par
\noindent \begin{enumerate}
\item the local line shifts (which velocity-dependent emission can cause) are
negligible,
\item the velocity dispersion function $r(\sigma )$ is known,
\item a velocity-resolved input-modified response function can be measured for
at least one line,
\item the approximate dependence of $F^{{\scriptsize\rm cl}}_{l}$ upon $\theta _{rs}\equiv \cos ^{-1}(\hat{{\bf r}}\cdot \hat{{\bf s}})$ is known, and
\item the clouds respond instantaneously (i.e., without ``hysteresis").
\end{enumerate}
\par
\noindent Because of the normalization condition imposed upon $\phi _{l}$,
condition (1) would yield the local profiles to be identical for each line.
When this is not the case, the line flux can also be projected into velocity
shift space.  Condition (2) is satisfied
by
$r(\sigma )=GM_{{\scriptsize\rm h}}/(3\sigma ^{2})$, which should be applicable in the BLR of the
orbital models of interest here.  Condition (3) holds provided the sampling
resolution is high enough and the local continuum flux determines the
properties of the clouds (even if the system is nonlinear). Condition (4) can
be partially met if we apply photoionization codes to determine the
approximate beaming factors $\epsilon _{\hbox{{\scriptsize\rm A}{\scriptsize\it l}}}$, we assume a spherically symmetric cloud
density function, and we assume a specific absorption model or simply assume
that absorption is unimportant.  The validity of condition (5) is
questionable, but could at least be tested if the conditions (1)-(4) are met
by comparing the local profiles of various lines (\S ~\ref{veldtemp}.3.2).
\par
\medskip
For the more general cases in which any of the above conditions are not
met, further constraints upon the local profiles can be made by again
exploiting the
orbital motions assumption, which states that the clouds are accelerated
purely by gravitational interactions.  From this assumption, the only
distinguishing characteristic relevant to the distribution function is the
particle mass.  This is the same primary assumption that is used to determine
stellar distribution functions.  Therefore, provided the other
characteristics of the clouds (e.g., the areas) do not affect the dynamics,
the distribution function of the clouds (and local profiles that are
functions of it) are proportional to those of stars with similar masses.
Techniques used to constrain stellar distribution functions can then be
employed.
\par
\medskip
One common assumption made in working with stellar distribution
functions is that the collision rate is zero, or $df/dt=0.$ In this case,
provided the relaxation time is short enough compared to the age of the
galactic nucleus, the distribution function is purely a function of the
isolating integrals of motion about the mean potential, such as the energy
and angular momentum.  For simple models that obey complete spherical
symmetry, $f$ can only be a one-dimensional function of energy per unit mass {\it E}.
This energy is a simple function of the cloud radius, velocity, and cloud
number density function $n(r)$, which is proportional to the mass density $\rho (r)$
in the above circumstances.  Therefore, because $f(E)$ is a one-dimensional
function, it is a unique function of $n(r)~($see, e.g., discussion in Binney \&
Tremain 1987).  For this reason, the local profiles can be calculated from
the cloud number density $n(r)$ if the mass to cloud number density ratio is
known.
\par
\medskip
For spherically symmetric models the non-velocity-resolved response
function and the time-averaged profile of a line are functions of only the
mean continuum flux and $A\epsilon _{l}(1+\epsilon _{\hbox{{\scriptsize\rm A}{\scriptsize\it l}}})n(r)$, where {\it A} is the cloud area function
and $\epsilon _{l}$ is the dimensionless line efficiency.  Therefore, {\it provided} the mass to
cloud density and the dependencies of ${\it A}, \epsilon _{l}$, and $\epsilon _{\hbox{{\scriptsize\rm A}{\scriptsize\it l}}}$ upon the local continuum
flux are calculable, the non-velocity-resolved response function {\it or} the
time-averaged profile of a line can be used to determine $f(E)$.  With
velocity-resolved response functions, $f(E)$ is therefore over-constrained if
instantaneous response is assumed.  On the other hand, if local delays are
assumed the cloud response functions can be measured under these
circumstances.
\par
\medskip
An analogous situation exists in galactic dynamics, with the
one-dimensional angular brightness function taking the place of the response
function.  In this case, $f(E)$ can be deconvolved only if the mass to light
ratio $M/L$ is known {\it a priori}.  However, with velocity information, both the
radial dependence of the $M/L$ ratio and $f(E)$ can be measured.  Reverberation
mapping is actually worse off in that not all of the parameters are
measurable unless theoretical constraints are imposed upon some of the
parameters; we can chose to measure either $A, \epsilon _{l}$, and dependence of $F^{{\scriptsize\rm cl}}_{l}$ upon
angle by placing theoretical constraints upon the cloud response functions
(e.g., nonlinear instantaneous response) {\it or} we can choose to measure the
cloud response function parameters such as the pressure equilibrium time by
placing theoretical constraints upon ${\it A}, \epsilon _{l}$, and $\epsilon _{\hbox{{\scriptsize\rm A}{\scriptsize\it l}}}$.  For the more complex
and realistic models, $\sigma \propto r^{-1/2}$ merely becomes a very powerful constraint to be
employed in doing numerical model fitting to equation ({}\ref{veldtemp3.28}).
\par
\medskip
As is typical of relaxed stellar systems, the local profiles that result
from applying the above procedure under the assumption of instantaneous
response resemble Gaussians at least to the first order.  If $f$ is additionally a function of angular momentum, the
local line profiles can have a position-dependent shift.  However, the form
of the un-shifted local profiles would probably remain somewhat
Gaussian-like, in accordance with the $df/dt=0$ assumption.  Note that such
model-independence of the local profiles probably does not extend to the
density functions, as it is primarily constrained by the macroscopic boundary
conditions imposed upon $f$ rather than the relaxation time scale.  In this
case, the local line profiles can probably be assumed to be similar for all
AGNs with clouds in orbital motions, with the density functions providing the
primary object-to-object freedom in {\it f}.
\par
\medskip
More careful considerations show that the local line profiles of $df/dt=0$
systems must be more ``square-shaped" than Gaussian.  Physically, this occurs
because of the constraints imposed upon {\it f}.  For instance, with $df/dt=0,$ the
cloud escape rate must be zero, with zero clouds having kinetic energies
above the potential depth (unlike a Maxwellian distribution).  In contrast,
the observed profiles of lines with relatively constant emission efficiencies
(lines similar to, e.g., Ly$\alpha $, which is probably emitted over a large range of
radii) are ``logarithmic." Because of this fundamental difference in form,
measurement of $I_{l}$ within the context of orbital models at several different
points in $\sigma $-space should be possible via application of equation
({}\ref{veldtemp3.26}) to a given spectrum in which the profile of interest is
resolved.
\par
\medskip
However, for some models, $df/dt\neq 0.$  This might occur if the cloud
creation or physical collision rate is high enough, as is expected in very
dense stellar systems. It might also occur at small radii for models
requiring a $f({\bf 0},{\bf v})=0$ boundary condition.  It is straightforward to show that
if $f({\bf r},{\bf v})$ scales to a power of $r$ that is greater than $-{1\over 2}$ at small $r$, then
$df/dt\neq 0, f$ is not positive-definite, or $f$ is not a function of energy alone.
Consideration of such $df/dt\neq 0$ orbital cloud models are outside the scope of
this work.
\par
\noindent 
\par
\infootendplain

\renewcommand\intextsep{10pt}

\chapter{Absorption Reverberation Mapping\label{adpbal} 
}

\def\figdirprefix{\home/completedpapers/adpbal/}
 \def\labep#1{\label{adpbal#1}}
 \infootbegplain
\noindent 
\par
\noindent {}\setcounter{section}{1}{}\addtocounter{section}{-1}\section{Background}
\par
An estimated $10\%$ of all AGNs have line absorption features with
strengths and widths approximately as large as their emission lines.
Moreover, line profile observations at high resolution indicate blueshifted
absorption features in about $50\%$ of Seyfert galaxies.  Despite these facts,
most of the theoretical work on AGNs to date, including the derivations shown
in Chapter 2, neglects absorption.
\par
\medskip
In this appendix I discuss how absorption spectral features could
be used to constrain absorption models.  Although time-{\it averaged} line
absorption features severely constrain absorption line QUASAR (BALQSO)
models, here I will focus upon only analysis of time-{\it dependent} absorption
data, or ``absorption reverberation mapping.''  It is well known that
reverberation mapping using emission lines has been quite useful for
constraining AGN line emission models.  Unfortunately, studies analogous to
these but for absorption spectral features have not yet been performed.  In
this appendix I show that absorption reverberation mapping appears to have at
least as much potential for constraining absorption models as does emission
mapping for AGN line emission models.
\par
\medskip
Though it is conceivable that the absorption mechanisms in BALQSOs are
completely different in separate objects, the differences (e.g., not being
radio loud, having a smooth dependence of the absorption characteristics upon
luminosity, etc.) between BALQSOs and normal AGNs suggest that the absorption
mechanisms in all AGNs may be inherently similar. Therefore, constraining
models of AGNs with only modest absorption should help us understand BALQSOs
as well.  Moreover, because certain AGN models (e.g., Murray et al. 1995)
make predictions about both the emission and absorption features, these
constraints could also help us understand how line emission operates in AGNs.
\par
\medskip
The nearest BALQSO is NGC 4151.  Partly for this reason most of the
calculations in this appendix are scaled for this source.  Observations of NGC
4151 have been obtained at the highest sampling frequency thus far; the 1993
intensive ultraviolet {\it IUE} reverberation mapping campaign of NGC 4151 (Edelson
et al. 1993) had a sampling interval of only $\sim 70$ minutes.  As I show in
\S ~\ref{adpbal}.4.1, this is the approximate time resolution that would be
necessary in order to perform effective absorption reverberation mapping.  I
also show that absorption reverberation mapping requires velocity resolution
at least as high as that of {\it HST}.  For these reasons, currently available
observations are inadequate for performing absorption reverberation mapping,
and this appendix is an attempt to see what results the analysis of AGN data
might yield in the future.
\par
\medskip
Broad absorption line quasar (BALQSO) absorption trough response time
scales can be significantly less than the associated emission response time
scales.  Thus, accurate measurement of absorption delays in AGNs requires
higher sampling frequencies than those used in emission line reverberation
studies.  For instance, data obtained of NGC 5548 from the {\it HST} reverberation
mapping campaign only permits measurement of an upper limit to the absorption
response delay.  This is shown in Figure \ref{5548}.
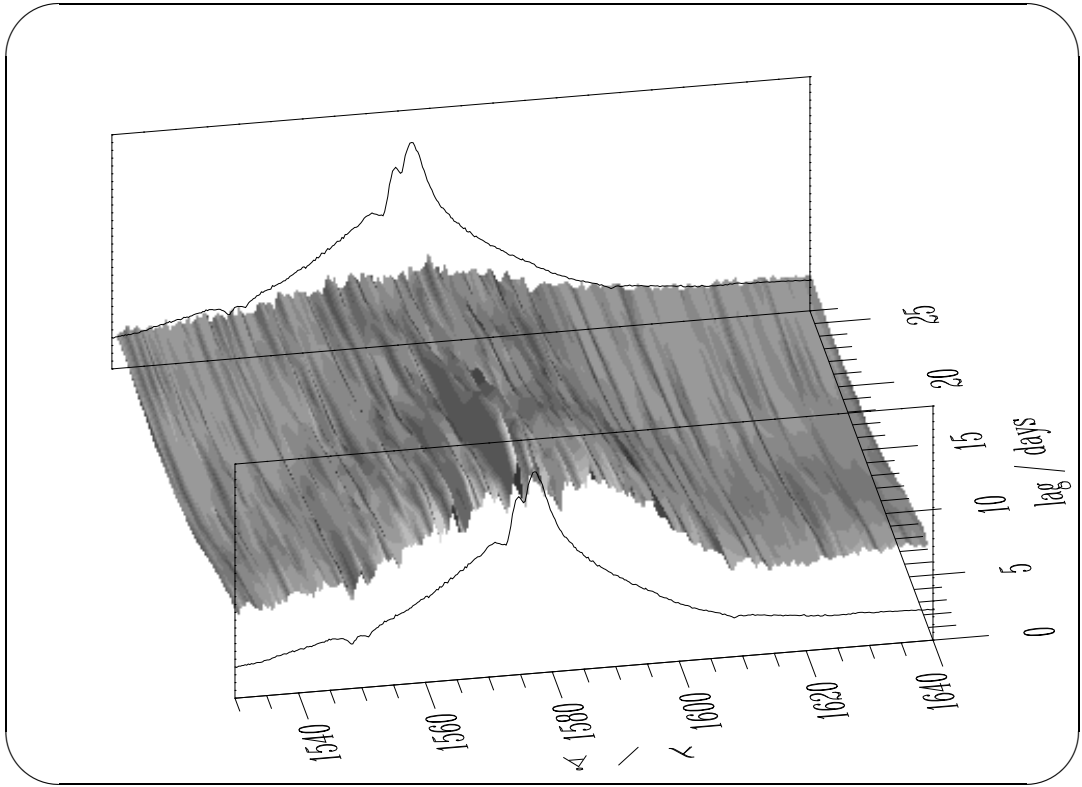
\fig{5548}{The cross covariance function of
the $1135-1180 \hbox{\rm\AA}$ de-redshifted continuum with respect to the spectra of NGC
5548 using the {\it HST} variability campaign data. A weak blueshifted C {\scriptsize\rm IV}
absorption doublet feature is evident even in this non-BAL Seyfert.  The cross
covariances of the absorption features have a width in delay space that is
approximately equal to the sampling rate of $\sim 3$ days.  This result is expected
for both accelerating and decelerating outflow models obeying equation
({}\ref{adpbal1}) in \S \ref{adpbalestimates}.
Therefore, despite the excellent wavelength resolution of {\it HST}, this data set
is
not particularly useful for mapping the absorption trough response functions
and directly distinguishing between the accelerating and decelerating classes
of absorption models.}\clearpage
\par
\medskip
\noindent {}\setcounter{section}{2}{}\addtocounter{section}{-1}\section{Estimates of Possible BALQSO Trough Response
Characteristics\label{adpbalestimates}}
\par
\medskip
Unlike the delays of emission lines, delays in changes of absorption
trough depths in response to changes in continuum levels are not proportional
to the distance of the line-affecting gas from the continuum source.
Several papers have stated that delays in absorption trough depths cannot
reveal any information about the positions of the absorption clouds except
indirectly via the implied lower limits to the recombination time scales
(e.g.; Voit, Shull, \& Begelman 1987; Barlow 1994).  This conception is
essentially valid for absorption of the {\it continuum} component of the spectrum.
It is invalid, however, for absorption of the {\it line emission}
component of the spectrum for the BALQSO models that are currently favored.
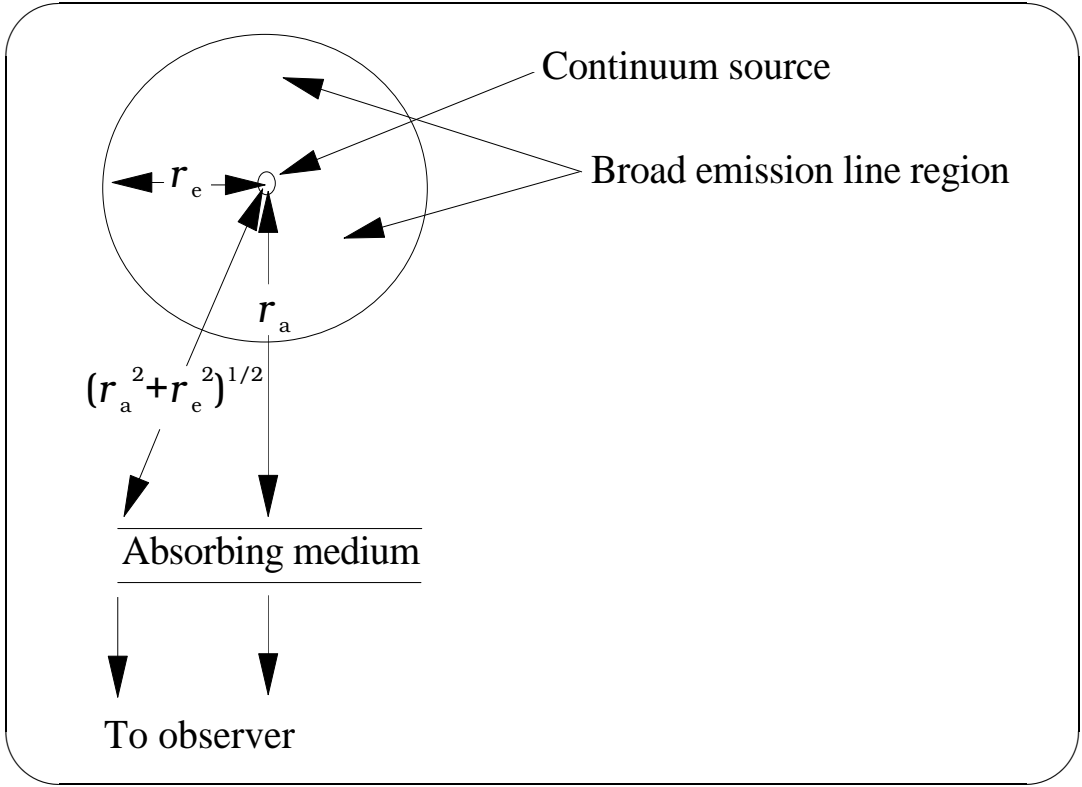
This is shown in the following.\fig{figure1}{Generic BALQSO model.  The
slab of the absorbing medium shown has an outflow speed of $v_{r}$.  Though the
cross-sectional area of the slab absorbing continuum radiation at a wavelength
of $\sim \lambda _{l}(1-v_{r}/c)$ is relatively small, the cross-sectional area of the slab
absorbing line emission radiation is ${\it r}^{2}_{{\scriptsize\rm e}}$.}
\par
\medskip
BALQSO absorption trough components generally appear to be blueshifted
with respect to the emission components.  In most BALQSO models, these
blueshifts occur because the absorbing material is assumed to be outflowing in
radial motion away from the continuum source and toward the observer.  Thus,
BALQSO models are similar to P Cygni stellar wind models.  The BALQSO model
is shown in Figure \ref{figure1}.  The emission region shown has a non-zero
scale size of $r_{{\scriptsize\rm e}}\simeq c/\tau _{{\scriptsize\rm e}}$, where {\it c} is the speed of light and $\tau _{{\scriptsize\rm e}}$ is a mean
response time scale of the emission region that is presumably measurable from
emission variability analysis.  (A more general and precise analysis
accounting for a large range of emission radii, for instance, could be
performed
but is unnecessary for our purposes here.)  Though the continuum absorption
region is assumed to be much smaller than the emission region, note that the
cross-sectional area of the absorption region is the same as that of the line
emission region.  The differential path length gives the relative time delay
for the observed absorption trough component to respond to changes in the
continuum flux due to the dependences of the optical depths of the absorbing
clouds upon the ionization parameter,
\par
\s
\tau _{{\scriptsize\rm a}}\simeq {r_{{\scriptsize\rm a}}\over c}\left(\sqrt{1+r^{2}_{{\scriptsize\rm e}}/r^{2}_{{\scriptsize\rm a}}}-1\right)+\tau _{{\scriptsize\rm r}}.
\labep{1}\e
\noindent In this equation, $r_{{\scriptsize\rm a}}$ is the distance of the absorbing gas from the continuum
source and $\tau _{{\scriptsize\rm r}}$ is the ionizational equilibrium time scale, which includes the
recombination time.  Given a functional dependence of the outflow velocity $v_{r}$
upon radius $r_{{\scriptsize\rm a}}$, equation ({}\ref{adpbal1}) gives the delay as a function of the
equivalent line-of-sight velocity $v\simeq -v_{r}\simeq c(\lambda -\lambda _{l})/\lambda _{l}$, where $\lambda $ is the wavelength
and $\lambda _{l}$ is the wavelength of the line center.  Since absorption of line
emission requires $r_{{\scriptsize\rm a}}>r_{{\scriptsize\rm e}}$, for the $\tau _{{\scriptsize\rm r}}\ll \tau _{{\scriptsize\rm a}}$ case equation ({}\ref{adpbal1}) implies
$\tau _{{\scriptsize\rm a}}<\tau _{{\scriptsize\rm e}}$.  Therefore, for the purposes of determining the velocity dependence
of the approximate minimum absorption response time scale, time dependence
of the line emission variation can be neglected for $\tau _{{\scriptsize\rm r}}\ll \tau _{{\scriptsize\rm a}}$ models.
\par
\medskip
Equation ({}\ref{adpbal1}) indicates that there are three different classes of
models distinguishable via their velocity dependence of the absorption trough
response minimum{}\rfn{Here we are distinguishing between the minimum absorption
response time scale and the actual response time scale that occurs upon a
specific change in the continuum luminosity.  The difference is important when
the luminosity is such that the ``responsivity" or gain of the transmission
coefficient is near zero at either the bluest or reddest trough component of
the absorption.  Equation ({}\ref{adpbal1}) is valid only when the fastest
absorption response is calculated from a data set in which the variation in
the
continuum amplitude is high enough that the gains of both the reddest and
bluest absorption trough components are not always near zero.} variation time
scale: \begin{itemize}\item ``{\it Accelerating} outflow models" in which
the
outermost absorbing clouds, for some reason, are moving
the fastest. \item
``{\it Decelerating} outflow models" in which the innermost absorbing clouds are
moving the fastest. \item ``{\it Detached} models" in which the gas density is low
enough that
the recombination time provides the dominant contribution towards the total
delay.\end{itemize}  Let us discuss each of these model classes in turn.
\par
\medskip
For accelerating models, $v_{r}$ increases as $r_{{\scriptsize\rm a}}$ increases.  Therefore, if
$\tau _{{\scriptsize\rm r}}\ll \tau _{{\scriptsize\rm a}}$, equation ({}\ref{adpbal1}) implies that $\tau _{{\scriptsize\rm a}}$ increases as {\it v} increases for
this model, with the blue side of the absorption trough having a smaller
minimum response time scale than the red side of the absorption trough.
\par
For the decelerating class of models, $\tau _{{\scriptsize\rm a}}$ decreases as {\it v} increases.  In
this case the red side of the absorption trough should be able to respond
first.
\par
\medskip
A further division between classes of outflow models can also be made.
If the acceleration mechanism of the outflow is such that the terminal outflow
velocity is comparable to the gravitational escape velocity, the radial
absorption slab velocity -{\it v} could, at least for some models, be used to
estimate the relative distance to the continuum source.  This permits
estimates of $r_{{\scriptsize\rm a}}$ for such models.  In particular, for such
gravitationally regulated outflow models,
\par
\s
r_{{\scriptsize\rm a}}\sim r_{{\scriptsize\rm e}}(v_{{\scriptsize\rm e}}/-v)^{2},
\labep{1.5}\e
\noindent where $v_{{\scriptsize\rm e}}$ is a velocity scale of the underlying emission line.  From crude
estimates of the absorption ionization parameter $U_{{\scriptsize\rm a}}$ based upon absorption
optical depth ratios, estimates of the hydrogen density $n_{{\scriptsize\rm H}}$ and
corresponding recombination times $\tau _{{\scriptsize\rm r}}$ for certain
gravitationally regulated outflow models can be made.  The densities in AGN
emission line regions are suspected to be high (e.g., $n_{{\scriptsize\rm H}}\sim 10^{11}$ cm$^{-3})$.
Therefore, because BALQSO absorption trough widths are comparable to the FWHM
of the emission profiles, equation ({}\ref{adpbal1.5}) implies that the absorption
recombination times are small in gravitationally regulated outflow systems.
Conversely, recombination could be the dominant contributor toward delay $\tau _{{\scriptsize\rm a}}$
in ``detached outflow systems."  Such detached systems should be different
from the gravitationally regulated outflow systems in that the absorption of
the continuum radiation emitted near a resonance line and the absorption of
the line emission itself should respond on the same time scale (the
recombination time).
\par
\medskip
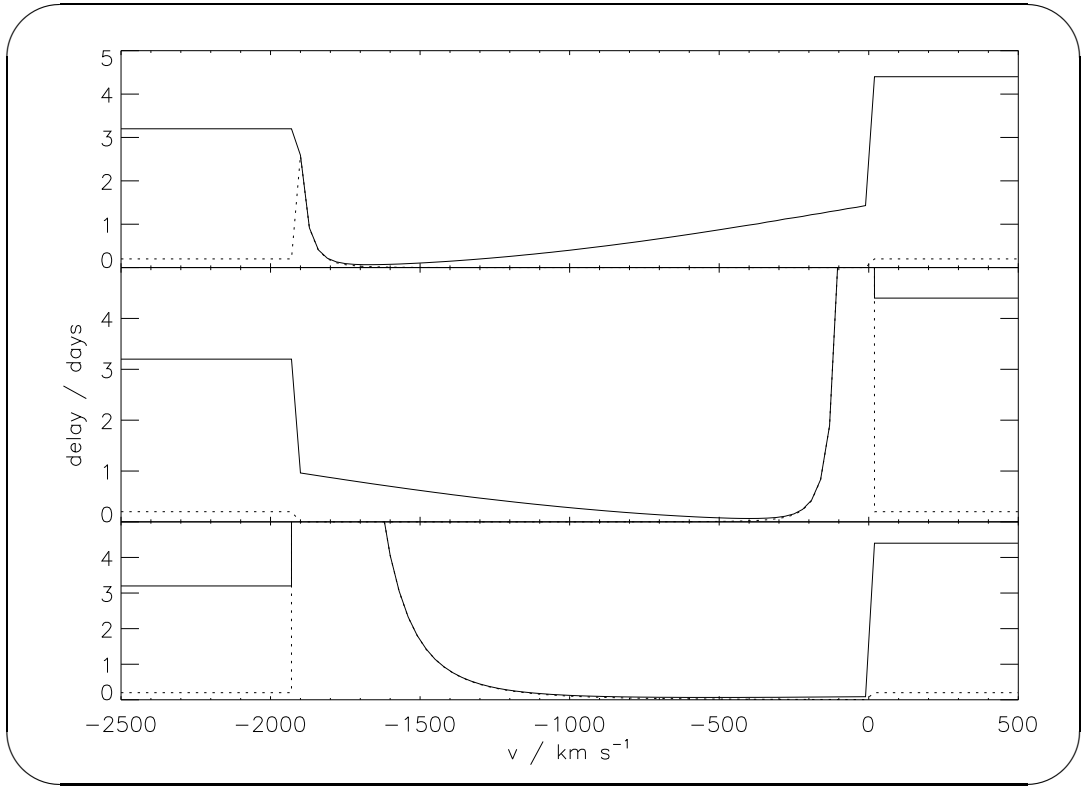
\noindent \fig{figure3}{Approximate minimum delays
in the response of the mean line and continuum absorption components to
changes in continuum flux for three models.  The top graph gives the predicted
delays for an accelerated outflow NGC 4151 absorption model.  The solid line
represents the approximate mean delay of the line absorption component, while
the dotted line represents the approximate mean delay of the continuum
absorption component.  For this model, the continuum absorption delay (which
is approximately the recombination time) is negligible at most velocities.
The delay of the absorption component of the line emission is smallest on the
blueward side of the absorption trough.  The middle graph gives the predicted
delays for the decelerating outflow class of BALQSO models.  It predicts the
opposite response behavior, with the blueward side of the absorption trough
responding the fastest.  Finally, the lower graph gives the predicted
absorption trough response delays for detached outflow models.  The gas
density assumed at the innermost edge of  the absorption region was $n_{{\scriptsize\rm H}}=6.6\times 10^{4}$
cm$^{-3}$.  For this model, both line and continuum absorption components are
predicted to respond on the recombination time scale, with the redward side of
the absorption trough responding
fastest.}\clearpage
\par
\medskip
These results are summarized by the calculations shown in Figure
\ref{figure3}.  The top graph gives the absorption delay/velocity results that
equations ({}\ref{adpbal1})-({}\ref{adpbal1.5}) yield for accelerating outflow, the
middle graph gives the results for decelerating outflow, while the lower graph
gives the results for detached accelerating outflow.  These calculations are
intended for NGC 4151{}\rfn{Results for NGC 3516 are similar.  The predicted
absorption response time scale is of order $\lesssim 0.6$ days for both the accelerating
and decelerating absorption models for the $\tau _{{\scriptsize\rm r}}\ll \tau _{{\scriptsize\rm a}}$ case.  The $\sim 2$ day sampling
period of the data analyzed in Koratkar et al. (1996) is therefore too large
to make accurate estimates of the line absorption response functions for
either accelerating or decelerating outflow models.}; in accordance with
results published in Ulrich \& Horne (1996), they assume a characteristic
emission delay of 4.4 days for the core emission component, 3.2 days for the
blue emission wing, and $v_{{\scriptsize\rm e}}=2000$ km s$^{-1}$ for the underlying emission scale
velocity.  The calculations also assume a position-independent absorption
ionization parameter of ${\it U}_{{\scriptsize\rm a}}=0.01.$  This value of ${\it U}_{{\scriptsize\rm a}}$ was selected to be as
consistent as possible with values quoted in the literature, which include
$U_{{\scriptsize\rm a}}\simeq 0.001~($Turnshek $1994), 0.01\lesssim U_{{\scriptsize\rm a}}\lesssim 0.07~($Hamann, Korista, \& Morris 1993), and
$U_{{\scriptsize\rm a}}\gtrsim 0.05~($Barlow 1995).  Results similar to those shown in Figure \ref{figure3}
are obtained with different absorption ionization parameters and with
absorption gas pressures that are constant along the line of sight.  In the
detached accelerating outflow model, the density was assumed to drop as
$v^{-1}_{r}r^{-2}_{{\scriptsize\rm a}}$.  Therefore, in this model the redward side of the absorption trough
responds before the blueward side.  One prediction of the detached model that
is different from the other two model classes is that it predicts the
absorption of both the continuum and line emission components to be delayed by
the same amounts (the recombination time).
\par
\medskip
\noindent {}\setcounter{section}{3}{}\addtocounter{section}{-1}\section{Crude Predictions for Popular BALQSO Models}
\par
\medskip
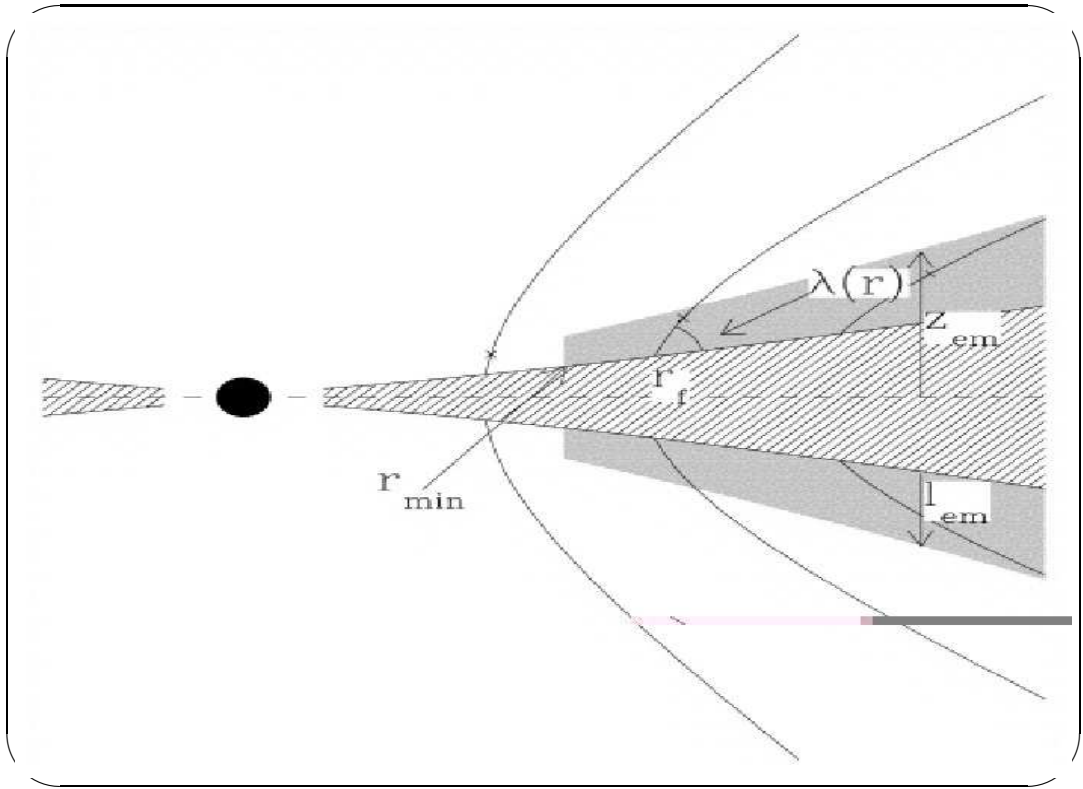
\noindent \fig{figure2}{The Murray, Chiang,
Grossman, and Voit
AGN/QSO model. At high inclination angles, absorption of line emission occurs.
In such BALQSO cases, the absorbing gas farthest from the broad emission
line region has the slowest line-of-sight velocity.}
\par
\medskip
In a recent review article (Weymann 1995), three BALQSO models were
discussed: the model of Murray et al. (1995), the model of Scoville \& Norman
(1995), and the model of Arav, Li, \& Begelman (1994).  In this section, we
discuss each of these models in turn.  We then attempt to estimate how
these models should compare with the models shown in Figure
\ref{figure3}.
\par
\medskip
\noindent {}\setcounter{section}{3}{}\setcounter{subsection}{1}{}\addtocounter{subsection}{-1}\subsection{Murray, Chiang, Grossman, and Voit Disk Wind Model}
\par
\medskip
The Murray et al. (1995) BALQSO model is shown pictorially
in Figure \ref{figure2}.  This model assumes that the accretion disk
surrounding the black hole has a radiatively accelerated outflowing wind with
a shear velocity of the order of the radial velocity.  In this model, $\sim 10\%$ of
AGNs
are BALQSOs because the opening angle of the wind is $\sim 81^{\circ }=(1-0.1)\times 90^{\circ }$.  The
mass loss flux is predicted to adjust itself such that the terminal wind
velocity of a given stream line is comparable to the gravitational escape
velocity (Murray, et al. 1995).  Because of the non-radial shape of
the streamlines, however, the absorbing gas farthest from the central source
has the
slowest tangential velocity.  Therefore, this model is in many ways similar to
the decelerating model shown in the middle of Figure \ref{figure3}.  Thus,
the red side of the absorption trough should be able to respond faster than
the blue side of the absorption trough in this model.  However, this model is
somewhat complex; more accurate calculations would be required before firmer
predictions of this model could be made.
\par
\medskip
Incidentally, this model should have absorption response
characteristics similar to other models in which the innermost clouds move
faster than the outermost clouds and where the velocities are comparable to
the escape velocities.  In particular, this model should be similar to the
model discussed in Appendix \ref{shiftsap}.
\par
\medskip
\noindent {}\setcounter{section}{3}{}\setcounter{subsection}{2}{}\addtocounter{subsection}{-1}\subsection{Scoville and Norman Contrail Model}
\par
\medskip
In the BALQSO model of Scoville \& Norman (1995), ``contrails," or
remnants of stellar winds, are slowly accelerated outward via radiation
pressure.  In this model, remnants with the highest radial velocity also have
the highest mean $r_{{\scriptsize\rm a}}$.  Moreover, the radial velocities are also associated with
the escape velocities.  With regard to the velocity dependence of the
absorption trough component, this model falls into the accelerating
class and the absorption trough response delay should be similar to that
shown in the top of Figure \ref{figure3}.  Therefore, the bluest absorption
trough components should have the shortest minimum response time scale and
$\tau _{{\scriptsize\rm r}}\ll \tau _{{\scriptsize\rm a}}$.
\par
For this model, there is a non-zero radial velocity dispersion at a given
$r_{{\scriptsize\rm a}}$.  For a BALQSO with an absorption equivalent width of 1000 km s$^{-1}$ and a
maximum possible thermal/inter-contrail turbulent velocity gradient of 50
km s$^{-1}$, a lower limit to the number of separate optically thick contrails of
maximal emission blocking region area $\sim r^{2}_{{\scriptsize\rm e}}$ is $\sim 1000/50=20\gg 1.$  Therefore, the
probability that the essential red/blue absorption response behavior for
this model is drastically different from that shown in the middle graph of
Figure \ref{figure3} is probably small.
\par
\medskip
\noindent {}\setcounter{section}{3}{}\setcounter{subsection}{3}{}\addtocounter{subsection}{-1}\subsection{Arav, Li, and Begelman Clumpy Outflow Model}
\par
\medskip
In this model, clumps of absorbing material are assumed to be outflowing
in radial motion away from the continuum source.  The densities are relatively
low in this case because radiative acceleration is assumed to be
high enough that the gravitation acceleration can be neglected (Arav et al.
1994).  As a result, the terminal velocities of the winds are much
higher than the gravitational escape velocities.  This implies much larger
values of $r_{{\scriptsize\rm a}}$ than in the other two classes of models.  Because of constraints
upon the ionization parameter, this in turn implies lower densities for the
absorbing gas than for the other models.  The densities assumed in
the detached model of Figure \ref{figure3}, however, are still approximately
$10^{4}$
times lower than those assumed by Arav, Li, and Begelman (1994).  Therefore,
depending upon the particular values of parameters assumed, this model could
behave either like the detached model or like the accelerating model shown in
Figure \ref{figure3}.
\par
\medskip
\noindent {}\setcounter{section}{4}{}\addtocounter{section}{-1}\section{Requirements of Velocity and Temporal Resolution to Discriminate Between
Popular BALQSO Models}
\par
\medskip
Before any attempt is made to perform absorption reverberation, suitable
data must be obtained first.  In this section I make some crude estimates of
the probably velocity and temporal resolution that would be required.
\par
\medskip
\noindent {}\setcounter{section}{4}{}\setcounter{subsection}{1}{}\addtocounter{subsection}{-1}\subsection{Velocity Resolution}
\par
\medskip
In principle, the absolute minimum number of velocity components
that would permit constraints to be imposed upon the BALQSO models is
{\rm2: a} redward absorption trough component and a blueward absorption
trough component.  The NEWSIPS {\it IUE} SWP spectral resolution is one
pixel per $1.6 \hbox{\rm\AA}$ at $1550 \hbox{\rm\AA}$. GHRS observations of NGC 4151 indicate that
the C {\scriptsize\rm IV} $\lambda  1548.2/1550.8$ broad line absorption trough is $\gtrsim  11 \hbox{\rm\AA}$ wide
(Weymann 1996).  This implies $11/1.6\sim 7$ independent measurable {\it IUE}
velocity absorption trough components.  A more conservative (and
probably more realistic) estimate assuming half this resolution would
yield $\sim 3.4$ different velocity components.  With random and
fixed-pattern noise it could be even less.
\par
\medskip
Since $3.4\ge 2,$ it could be argued that {\it IUE} resolution might be
sufficient to distinguish between the three classes of models discussed in
this appendix.  However, in order to
produce an accurate, smooth velocity-resolved map of the linearized
absorption trough response function, the much higher resolution of {\it HST} would
probably be required.
\par
\medskip
\noindent {}\setcounter{section}{4}{}\setcounter{subsection}{2}{}\addtocounter{subsection}{-1}\subsection{Estimates of the Expected Sampling Rate and Temporal resolution}
\par
\medskip
In a system that responds to input excitations with characteristic delay
of $\tau _{{\scriptsize\rm a}}$, a conservative maximum effective sampling period required to place
empirical lower limits upon $\tau _{{\scriptsize\rm a}}$ is $\lesssim \tau _{{\scriptsize\rm a}}/3.$  Similarly, in order to place upper
limits upon $\tau _{{\scriptsize\rm a}}$, the data set must span a minimum time of $\sim 3\tau _{{\scriptsize\rm a}}$.  In this
section we show that for both the accelerating and decelerating classes of
BALQSO models these two conditions can be met by data that are sampled at
rates as high as or higher than that obtained in the 1993 {\it IUE} continuous
monitoring campaign of NGC 4151 .
\par
\medskip
If the sampling time is less than the time required for the flux within
the trough to change more than the $1\sigma $ uncertainty, the effective absorption
trough flux sampling rate is
\par
\s
\tau _{{\scriptsize\rm a}1\sigma }\simeq  {F_{{\scriptsize\rm a}}\over |\dot{F}_{{\scriptsize\rm a}}|} {\sigma _{F_{{\scriptsize\rm a}}}\over F_{{\scriptsize\rm a}}},
\labep{2}\e
\noindent where $F_{{\scriptsize\rm a}}$ is the flux in or near the absorption trough, $\sigma _{F_{{\scriptsize\rm a}}}$ is its $1\sigma $
uncertainty, and $|\dot{F}_{{\scriptsize\rm a}}|$ is the amplitude of its typical derivative with respect
to time.  The $1275 \hbox{\rm\AA}$ continuum flux change rate was $\lesssim 10\%$ in several hours
(Crenshaw et al. 1996). Therefore, if the asymptotic dependence of the
transmission coefficient were to respond linearly upon variations in the $1275
\hbox{\rm\AA}$ continuum flux, the effective sampling rate would be $\tau _{{\scriptsize\rm a}1\sigma }\sim 3$ hours.
\par
\medskip
For a more realistic estimate of $\tau _{{\scriptsize\rm a}1\sigma }$ in a nonlinear model in which the
density of absorbing gas does not change due to variations in the continuum
flux, we have
\par
\s
\tau _{{\scriptsize\rm a}1\sigma }\simeq  {U_{{\scriptsize\rm a}}\partial T\over T \partial U_{{\scriptsize\rm a}}} {F_{1275}\partial Q\over Q\partial F_{1275}} {F_{1275}\over |\dot{F}_{1275}|} {F_{{\scriptsize\rm a}}\over \sigma _{F_{{\scriptsize\rm a}}}},
\labep{3}\e
\noindent where $T$ is the transmission coefficient through the absorbing medium, $F_{1275}$ is
the continuum flux integrated from $1260-1290 \hbox{\rm\AA}, Q$ is the photon
luminosity from 13.6 eV to 14 keV, and $|\dot{F}_{1275}|$ is the amplitude of a typical
derivative of $F_{1275}$ with respect to time.  Equation ({}\ref{adpbal3}) implies that
the time scales are very large in saturated absorption systems composed of
``black" clouds where $T\sim 0.$  This could conceivably increase $\tau _{{\scriptsize\rm a}1\sigma }$ to a
significant fraction of the theoretical line absorption response time scale,
in which case, the shape of the absorption response functions would be
difficult to determine. However, models that assume mass-flux
conservation (including the decelerating and accelerating
outflow models shown in Fig. \ref{figure3}) have an ionization parameter
that scales inversely with radial velocity.  As a result, the amplitude of
the first factor in equation ({}\ref{adpbal3}) is predicted to be greater than
unity for these models near the trough edges.   Therefore, since the second
factor in equation ({}\ref{adpbal3}) is almost certainly greater than unity (e.g.,
Edelson et al. $1996), \tau _{{\scriptsize\rm a}1\sigma }\lesssim 3$ hours for both the decelerating and accelerating
classes of outflow models.
\par
\medskip
These estimates that $T\neq 0$ are bolstered by the GHRS observations of NGC
4151 shown in Weyman (1996).  The relative absorption amplitudes of
individual absorption features changed significantly over the different
epochs sampled, especially near the blueward side of the absorption trough.
NGC 4151 does not appear unique in this respect: using {\it IUE} observations of
NGC 3516, Koratkar et al. (1996) found that the gain or ``responsivity" of
the C {\scriptsize\rm IV} absorption trough in NGC 3516 varies between -1 and 5 depending upon
velocity and luminosity.  Thus, $\sim 3$ hours is probably a crude yet realistic
estimate of the expected effective time sampling rate within the absorption
trough for NGC 4151.
\par
\medskip
As shown in Figure \ref{figure3}, the predicted response delay time
for both the accelerating and decelerating models is between 0.2 and 1.0 days
in $\sim 75\%$ of the absorption trough.  Thus, in $\sim 75\%$ of the absorption trough the
effective sampling period will be between $\sim 1.6$ and $\sim 8.0$ times less than the
response time scale predicted by the accelerating and decelerating class of
models. This should permit upper limits to the absorption response time
scales to be obtained.  These time scales are between $\sim 50$ and $\sim 10$ times less
than the data set duration.  Therefore, measuring upper limits to the
absorption response time scales should also be possible for both the
accelerating and decelerating models.  However, for detached models with
densities similar to those assumed in the lower graph in Figure
\ref{figure3}, only a measurement of the lower limit of the absorption
response time scale will be possible at velocities below -1500 km s$^{-1}$.
\par
\medskip

{}\setcounter{section}{6}{}\addtocounter{section}{-1}\section{Summary}
\par
\medskip
\noindent The three classes of AGN absorption line models considered in this appendix
have outflowing material that is either accelerating (e.g., the contrail
model of Scoville \& Norman [1995]), decelerating (e.g., the wind disk model
of Murray et al. [1995]; the clumpy outflow model of Arav et al. [1994]), or
``detached" and far from the broad emission line region (e.g., the clumpy
outflow model but with lower gas densities).  All three of these types of
models are able to yield the blueshifted absorption that is observed in AGNs.
But the absorption components respond to continuum variations very
differently in each case. Decelerating outflow models appear to predict that
the red side of the absorption trough should respond before the blue side.
On the other hand, accelerating models appear to predict that the blue side
of the absorption trough should respond first.  If the clouds in the clumpy
outflow model are far enough away from the continuum source, one would
additionally observe relatively small spreads in response delay time
parameter space.  None of the three classes of models predicts a
velocity-independent absorption response time scale.  For these reasons,
future analysis of data obtained from {\it HST} and other observatories should rule
out at least one and possibly two of the three classes of currently popular
broad absorption line models.
\par
\infootendplain

\chapter{An Introduction to the Cometary Star Model \label{shiftsap} 
}

\def\labep#1{\label{shiftsap#1}}
\def\figdirprefix{\home/scon/thesispics/}
\infootbegplain
The models we presented in Chapter 4 employ spherical winds.  As
mentioned in Appendix \ref{cloudshape}, the decision was made to exclude
nonspherical cloud models due to their perceived complexity.  According to
equation (\ref{standoffdistance}),  the assumption of spherical symmetry is
only valid for ``relatively small'' values of the intercloud medium mass
density $\rho _{\hbox{{\scriptsize\rm HIM}}}$.  Since this parameter has not been reliably measured, the
question of spherical symmetry is not necessarily closed.  For this reason, I
have considered a different model which has nonspherical winds.  I term this
model the ``cometary star model.''  In this appendix I briefly highlight some
of its features.
\par
\medskip
The cometary star model was first described in Taylor (1994).  It is
similar in many ways to the red giant cloud line emission model proposed in
Kazanas (1989).  However, it attempts to account for the ram pressure the
stars would experience due to their supersonic motion through the intercloud
medium.  One of the features of the cometary star model is its ability to
self-consistently fit the line profile shift and asymmetric response
characteristics that have been observed.
\par
\medskip
In each cloud of the cometary cloud model, the pressure is assumed to
decrease with distance from the shock front.  My calculations of the cometary
star model have employed a two-zone approximation in which each cometary
cloud is broken up into two zones of different pressures:
\begin{enumerate}\item a high pressure zone near the shock front, hereafter
called the cloud ``head," and \item a low pressure zone, hereafter called the
cloud ``tail," that points away from the cloud velocity
vector.\end{enumerate}  From a modeling perspective, the main difference
between the cometary star model and the red giant wind model discussed in the
body of this dissertation is the introduction of a new parameter.  This
parameter is the head to tail pressure ratio.
\par
\medskip
A more accurate calculation might employ smooth density gradients within
each cloud.  In order to estimate the approximate errors associated with the
simpler two-zone approximation, two special, high resolution runs with XSTAR
were performed.  These special runs assumed an edge ionization parameter of
$\Xi =0.05~(U=0.37)$, a column density of $N_{{\scriptsize\rm H}}=5.5\times 10^{22}$ cm$^{-2}$, and a head/tail
pressure ratio of 2.0.  The velocity-weighted difference between the shifts
of the crude two-zone approximation and the more accurate
radially-dependent-density runs was only $17.7\%$ for the 26 different lines
that were computed.  Thus, at least for clouds with these parameters, the
two-zone approximation employed in this appendix is probably sufficient for
our purposes.
\par
\medskip
On the other hand, it is possible that these parameters are giving
relatively low errors because the ionized region in these clouds is
relatively small.  With more realistic low$-N_{{\scriptsize\rm H}}/\Xi $ clouds (which the Baldwin
effect implies), the errors with the two-zone approximation would probably be
greater.  Therefore, $\sim 18\%$ should probably be considered as a {\it lower} limit to
the systematic errors of the results presented in this appendix.
\par
\medskip
The results of the two-zone approximation applied to the cometary star
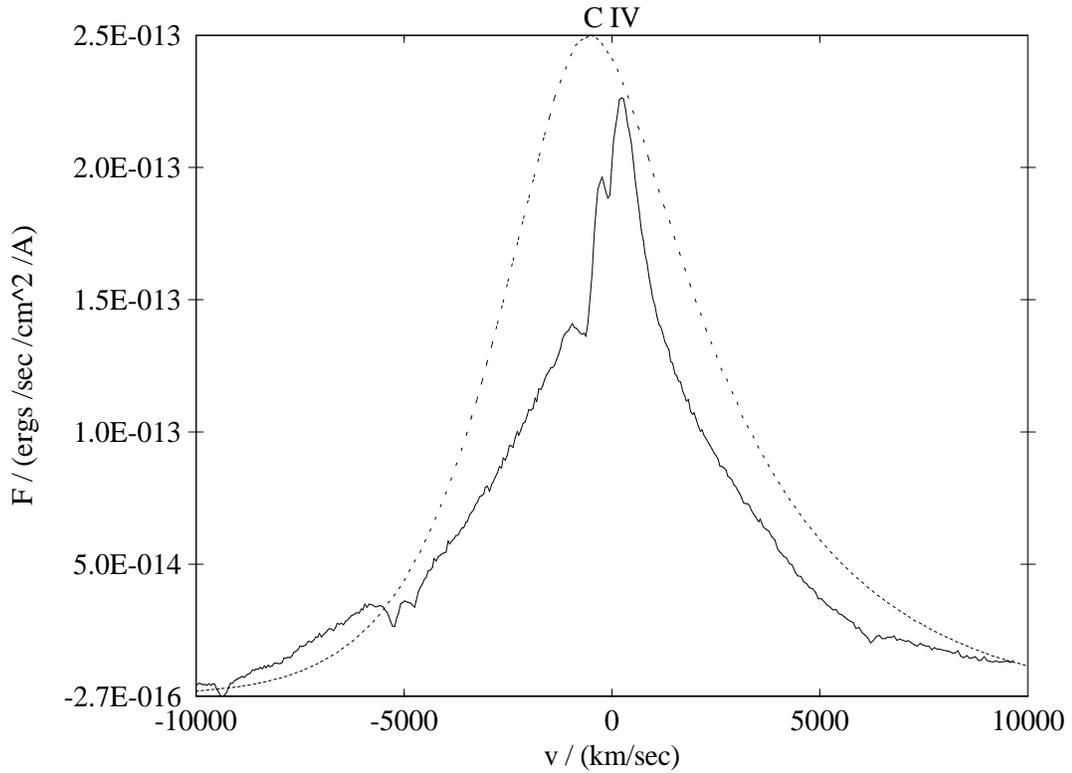
model are fascinating. \fignobox{c4lp}{Dotted line: synthetic (theoretical) C
{\scriptsize\rm IV} line profile assuming a pressure difference of 2.0 in the leading/trailing
cloud edges.  This calculation assumes a simple virial emission cloud model
with an occulting broad line region accretion disk.  Solid line: the
time-averaged continuum-subtracted C {\scriptsize\rm IV} line profile of NGC 5548, shown for
comparison.  Note that the observed line profile has a blueshifted absorption
feature.  Without this absorption feature, the peak of the line profile
(which is slightly redshifted) would probably be
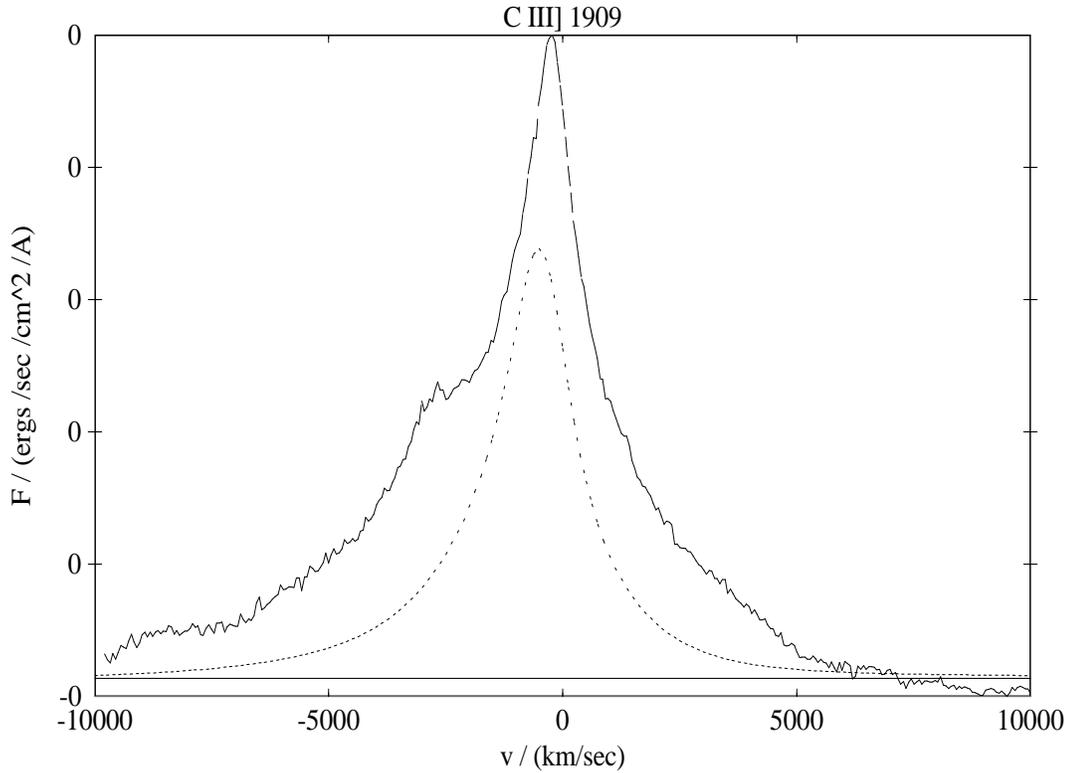
blueshifted.}\fignobox{c3lp}{Dotted line: C {\scriptsize\rm III}] line profile for the same
model as that shown in Figure \ref{c4lp}.  Solid line: C   {\scriptsize\rm III}] profile of NGC
5548.} Figures \ref{c4lp} and \ref{c3lp} show the C {\scriptsize\rm IV} and C {\scriptsize\rm III}] line
profiles for the cometary star model under the two-zone approximation
assuming a face-on accretion disk and a head to tail pressure ratio parameter
of 2.0.  Though there are slight blue shifts of the line peaks, at first
glance these synthetic profiles appear relatively innocuous.
\par
\medskip
A more careful analysis, however, reveals that the high velocity
dispersion components of these synthetic line profiles are also shifted.
These shifts indicate a correlation between mean line-of-sight velocity and
distance from the central black hole for this model, which assumes that the
clouds are in orbital motions.  This correlation is revealed in the bisector
shift plot (see, e.g., Kulander \& Jefferies 1966) shown in
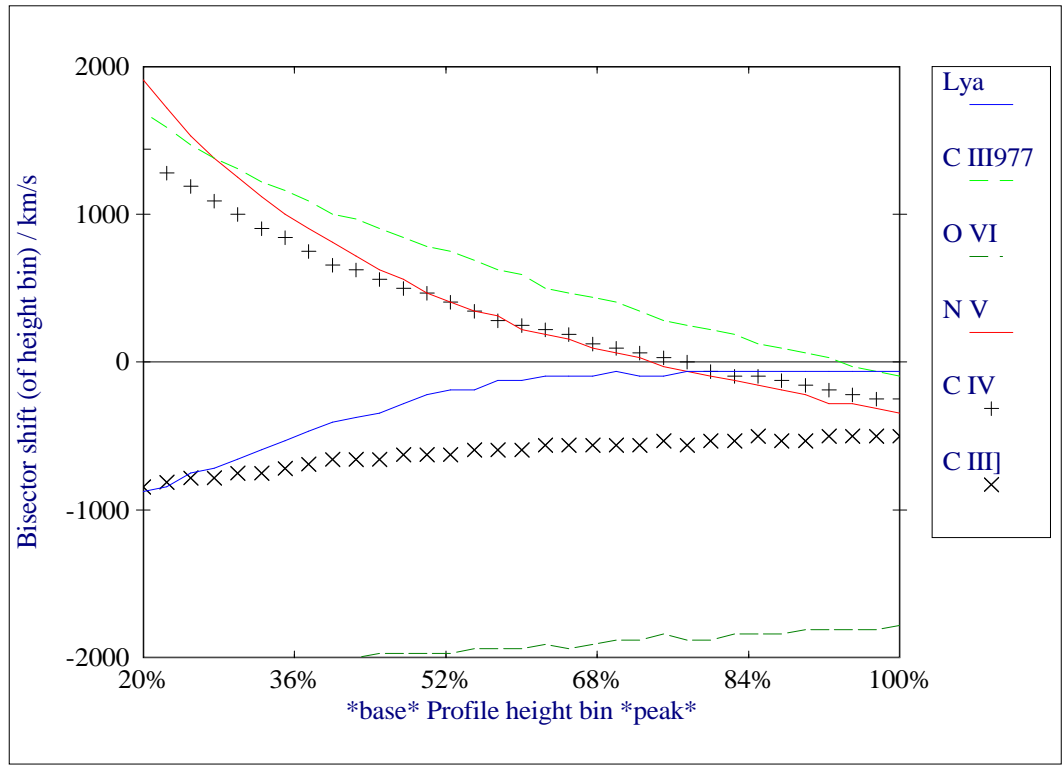
\Fignobox{shifts}{Bisector shift plot for several lines.  Parameters are as in
Figures \ref{c4lp} and \ref{c3lp}.  Bisector shift plots show the shift of
the peak of the square-shaped truncated profile verses truncation height.
For this model, the peaks of the C {\scriptsize\rm IV} and C {\scriptsize\rm III}] lines are shifted towards
the blue even though the bases are shifted in opposite
directions.}.  Note that the broader components (near the base) of the
synthetic C {\scriptsize\rm IV} line profile are shifted to the red, while the narrower
components are shifted towards the blue.  These shifts generally agree with
empirical results (e.g., Brotherton et al. 1994). In particular, they are
able to explain the apparent redshifted C {\scriptsize\rm IV} profile base (see also Figure
\ref{model2/civ}).\rfn{It has been speculated that the emission redward
of the anticipated C {\scriptsize\rm IV} profile base in NGC 5548 may be an Fe II line such as
$\lambda 1608$ or highly redshifted He II $\lambda 1640$~(Korista et al. 1995).  Though these
hypotheses cannot be completely ruled out, they seem unlikely.  Only three Fe
{\scriptsize\rm II} lines in the $1600\hbox{\rm\AA}--1650\hbox{\rm\AA}$ range with spontaneous transition rates above
$10^{3}$ s$^{-1}$ exist in the Morton (1991) tables: $\lambda 1621, \lambda 1618,$ and $\lambda 1608.$  Of
these, only Fe {\scriptsize\rm II} $\lambda 1608$ is a transition to a ground state.  Regardless of the
emissivity functions of these lines, most AGN models would require
that their underlying profile shapes be similar to that of the other BLR AGN
lines---semi-logarithmic.  Thus, if the emission were due primarily from just
these three lines, there should be very prominent peaks at each of these
wavelengths.  There does, in fact, appear to be a peak at $\lambda 1608.$  However, it
is relatively {\it narrow} and less than $\sim 15\%$ above the nearby spectral region,
which incidentally is quite smooth and flat.  Thus, if the emission is due to
Fe {\scriptsize\rm II} $\lambda 1608$ alone, both its strength and profile shape are extremely
abnormal.  So the Fe {\scriptsize\rm II} $\lambda 1608$ postulate is actually two separate postulates
built into one, neither of which is expected {\it a prior}.  If the emission is not
due to Fe II but rather to highly blueshifted He {\scriptsize\rm II} $\lambda 1640,$ the He {\scriptsize\rm II} profile
is highly asymmetric.  A potential problem with this hypothesis is that it
fails to explain why the base of the C {\scriptsize\rm IV} red wing does not begin to rise
below $|v_{z}|\simeq 10,000$ km s$^{-1}$, where the blue wing flux appears to fall near zero.
Also, this hypothesis simply shifts the ``blame" from one line to another.
Therefore, from a ``model fitting" perspective, it offers little advantage
over the assumption that the emission is simply redshifted C {\scriptsize\rm IV}.}
\par
\medskip
Shifts occur in the cometary star model not because of any non-zero mass
flux imposed upon the model, but rather because of differences in the line
luminosities between the inbound and outbound clouds.  In order to understand
how such differences in emissivity are possible, it is necessary to first
understand how the line emissivities depend upon the ionization parameter and
pressure.  The ram pressure from the intercloud medium is assumed to make
the ionization states of the inbound clouds lower in the inverse Str\"omgren
region than the tail-illuminated outbound ones.  The situation is illustrated
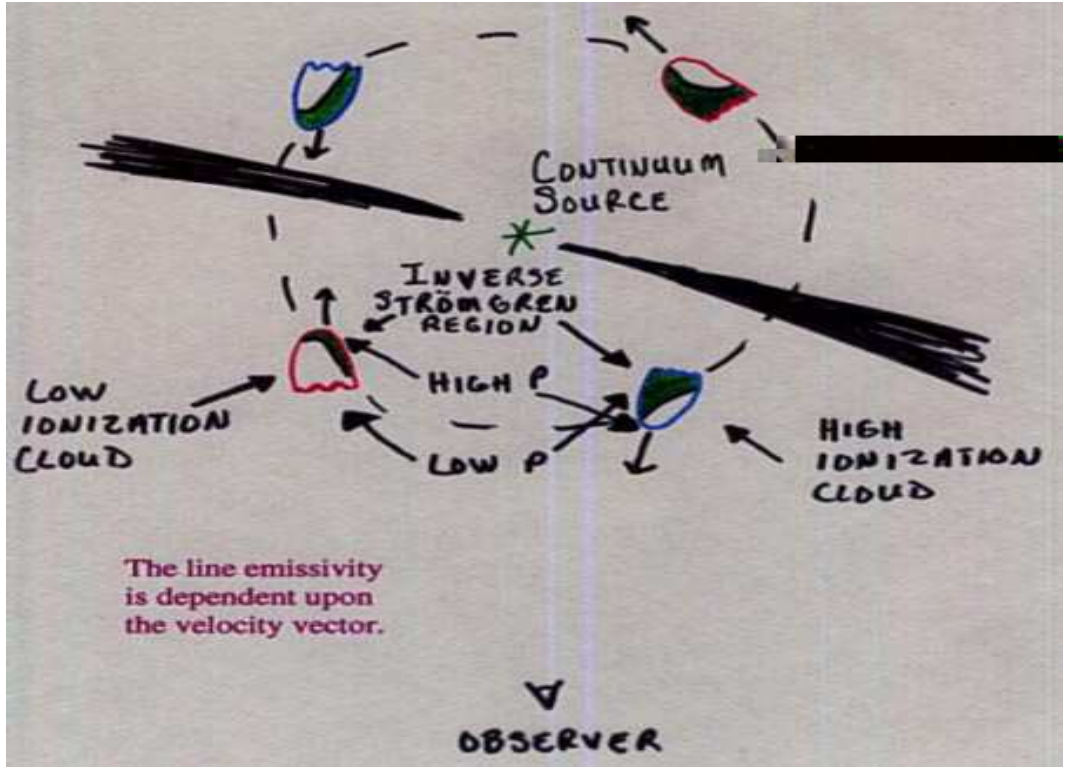
in Figures \ref{write} and \ref{ionversesradius}.\fignobox{write}{Pictorial
representation of clouds in the cometary star model.  Clouds on the far side
of the accretion disk are invisible to the observer.  Clouds on the near side
going towards the observer produce blueshifted line emission.  Clouds on the
near side going away from the observer produce redshifted line emission.
Each of the four clouds shown is assumed to have the same $r$ and local
continuum flux. The total line emission from lines requiring relatively low
ionization is stronger for the inbound clouds than the outbound clouds.  For
lines in which the beaming factor does not change significantly, the
resultant profile is redshifted.  Conversely, lines requiring relatively high
ionization in which the beaming factor does not change significantly are
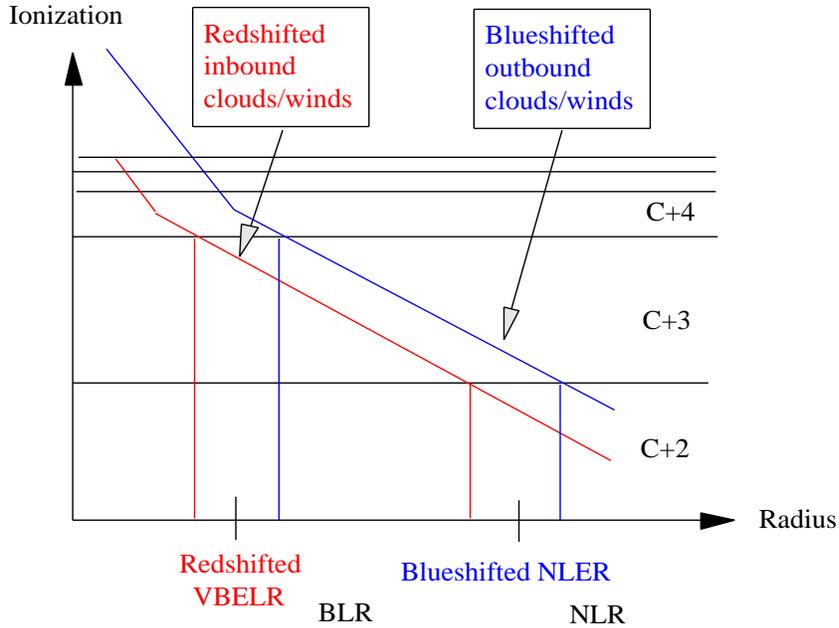
blueshifted.}\fignobox{ionversesradius}{Effective ionization parameter
verses radius from the central continuum source for inbound clouds
going towards the black hole and outbound clouds going away from the black
hole.  The two black horizontal lines indicate the ionization parameters
where the dominant fractional ionizational abundance of carbon changes from
C$^{+2}$ to C$^{+3}$ and from C$^{+3}$ to C$^{+4}$.  Here we assume $s<2,$ where $P\propto r^{-s}, P$ is the
mean cloud pressure, and $r$ is the radius from the central black hole.  Thus
the ionization decreases with radius.  The (upper) diagonal blue line is the
effective ionization parameter for outbound clouds as a function of radius.
The (lower) diagonal red line is similar but is for inbound clouds.  For C
{\scriptsize\rm IV},
there is an outermost radius (shown as the right vertical blue line) beyond
which the fractional ionic abundance of C$^{+3}$ is too low for significant
emission or self-absorption to be possible.  In the region denoted by the
rightmost red and blue vertical lines, only the outbound clouds, which in the
cometary star model have higher mean ionization parameters, should be able to
emit substantial line radiation. Thus the profile bases of normal
intermediate-ionization lines like C {\scriptsize\rm IV} are redshifted, while the profile
peaks are blueshifted.}  Since the line emissivity from a cloud is a strong
function of the ionization state, it is also a strong function of the
velocity direction.  The reason this velocity-dependent emission produces
shifts is because of an occulting accretion disk, which is assumed to block
all of the far clouds.
\par
\medskip
A unique feature of the cometary star model is that each line can have a
different shift without there being any special ion-dependent cloud motions.
Previous models appear to require {\it ion-dependent cloud motions} in order to
explain the ion-dependent line shifts that have been observed.  Such
ion-dependent cloud motions plaguing previous models would appear to be
physically unsatisfactory for several reasons.  One of these reasons is that
the line shifts do not appear to be a monotonic function of the ionization
state associated with the transition (Tytler \& Fan 1992).  In other words,
previous models have difficulty explaining ``out of order" lines, like Ly$\alpha $
and C {\scriptsize\rm III}].  In the cometary star model, the relevant variables are the gains
of the line emissivities with respect to both the ionization parameter and
pressure.  Since Ly$\alpha $ is unusual in that it is a recombination line, it is
shifted much less than the other lines.  On the other hand, C {\scriptsize\rm III}] is unusual
in that it has a very negative pressure gain, and is naturally enhanced in
the outbound clouds.  Thus, in the cometary star model the base of this line
is blueshifted much more than that of other lines without there being any
special ion-dependent cloud motions.
\par
\medskip
These shifts would have a direct effect upon the velocity dependence of
the variability characteristics.  For the C {\scriptsize\rm IV} line profile, which is
predicted to have a redshifted base, the red-ward wing would respond more
rapidly than the blue-ward wing.\rfn{Incidentally, due to the high
values of $\Xi /N_{c}$ in the broad line region, this does {\it not} imply that it would
also respond more strongly.}  For the C {\scriptsize\rm III}] profile, which has a blueshifted
profile base, the situation is reversed.  The blue-ward wing of this line
would respond more rapidly than the red-ward wing.  Previous models attempt
to explain asymmetric response by invoking either outflow or inflow which
would apparently predict the same red/blue response asymmetries for all
lines.  Thus, a good way to test the cometary star model against other models
is to see whether or not the C {\scriptsize\rm III}] line has a reversed red/blue response
asymmetry.
\par
\medskip
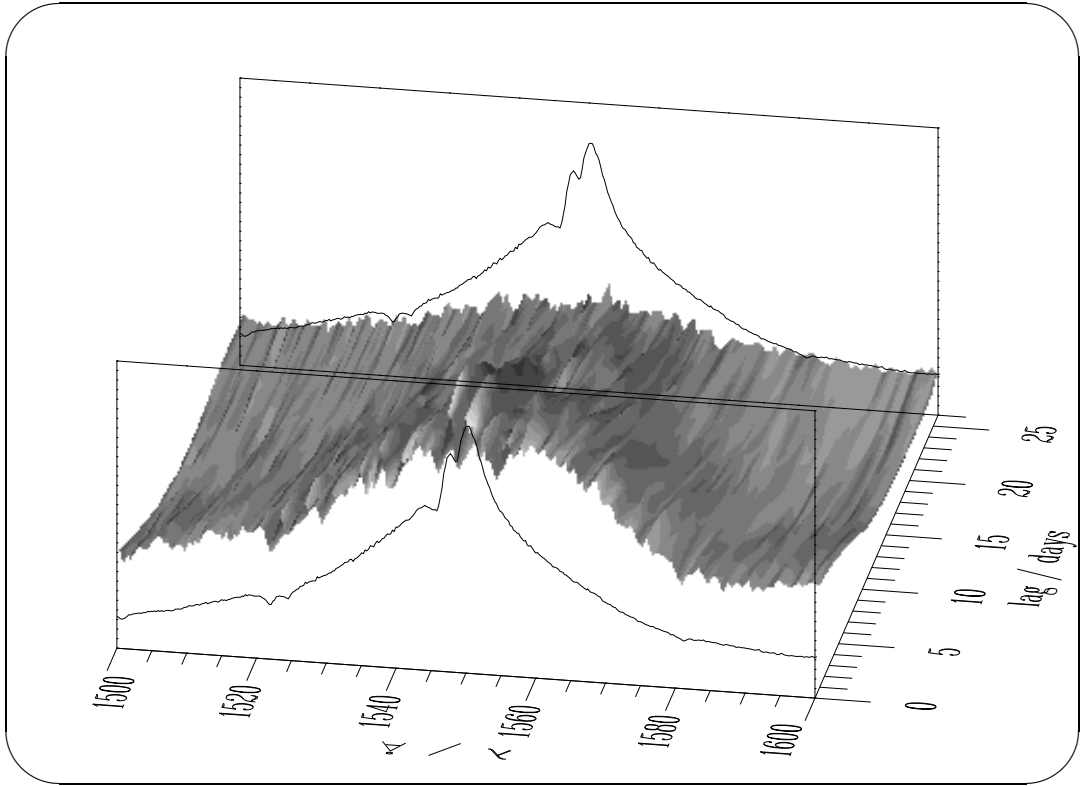
\Fig{c4ccov}{The frequency-dependent cross covariance between the
1135-1180  de-redshifted continuum and the C {\scriptsize\rm IV} spectral region.} shows the
cross covariance between the 1135-1180  de-redshifted continuum for the C {\scriptsize\rm IV}
spectral region of NGC 5548 using the {\it HST} variability campaign data (Korista
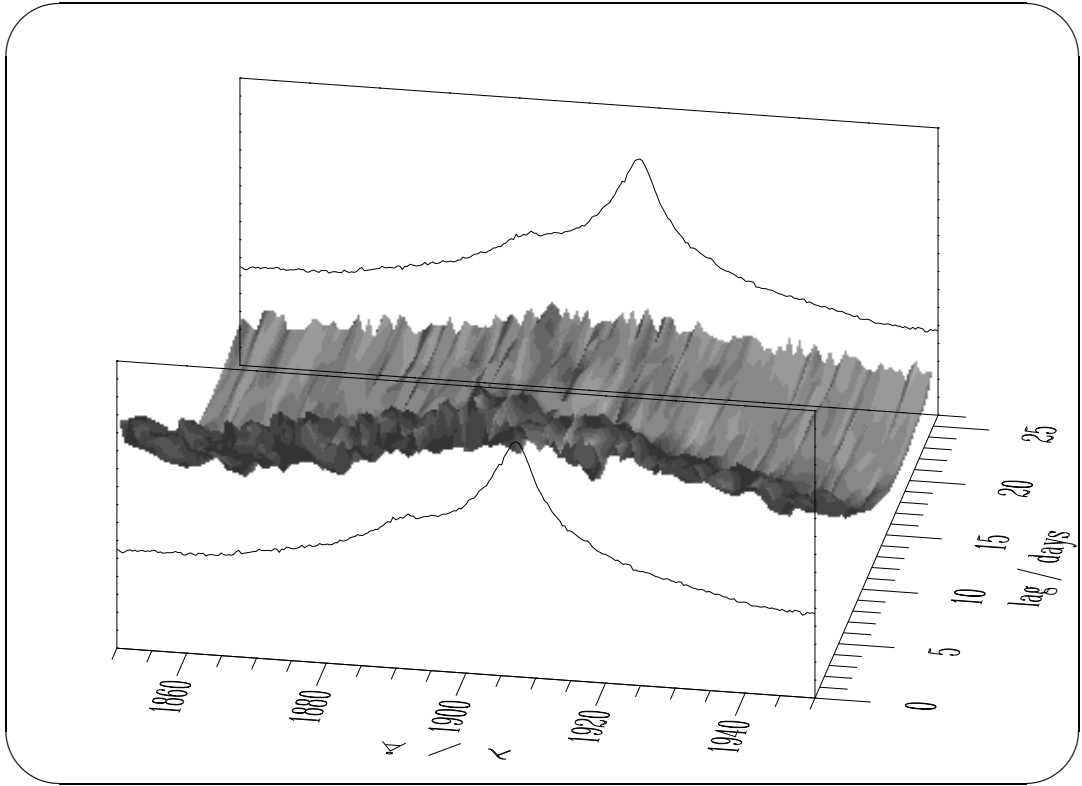
et al. 1995). \Fig{c3ccov}{The frequency-dependent cross covariance between
the 1135-1180  de-redshifted continuum and the C {\scriptsize\rm III}]/Si {\scriptsize\rm III}]/Al {\scriptsize\rm III}
spectral line region.} shows the equivalent C {\scriptsize\rm III}]/Si {\scriptsize\rm III}]/Al {\scriptsize\rm III} spectral
line region.  At lags of less than a few days, there is significant response
due to the underlying continuum, which was not subtracted.  Superimposed upon
this response is an additional response that is proportional to the absolute
line flux presumably emitted from the clouds.  At a lag of 25 days, the
response is nearly symmetric for each wing.  At lags below 5 days, however,
the blue wing of C {\scriptsize\rm IV} appears to have a slightly stronger response amplitude
than the red wing of C {\scriptsize\rm IV}.  Though it is difficult to determine with
certainty, Figure \ref{c4ccov} appears to indicate that C {\scriptsize\rm III}] indeed
exhibits the reverse behavior.  These effects are small and the analysis
provided here is very crude.  Nevertheless, similar results have been found in
the more complete studies of the NGC 5548 data.  While it would be a
straightforward task to concoct models with ion-dependent cloud motions to
explain just these particular observations, perhaps the simplest
interpretation is that the broad components of several lines are often
shifted and that the red-ward C {\scriptsize\rm IV} and blue-ward C {\scriptsize\rm III}] emissions are indeed
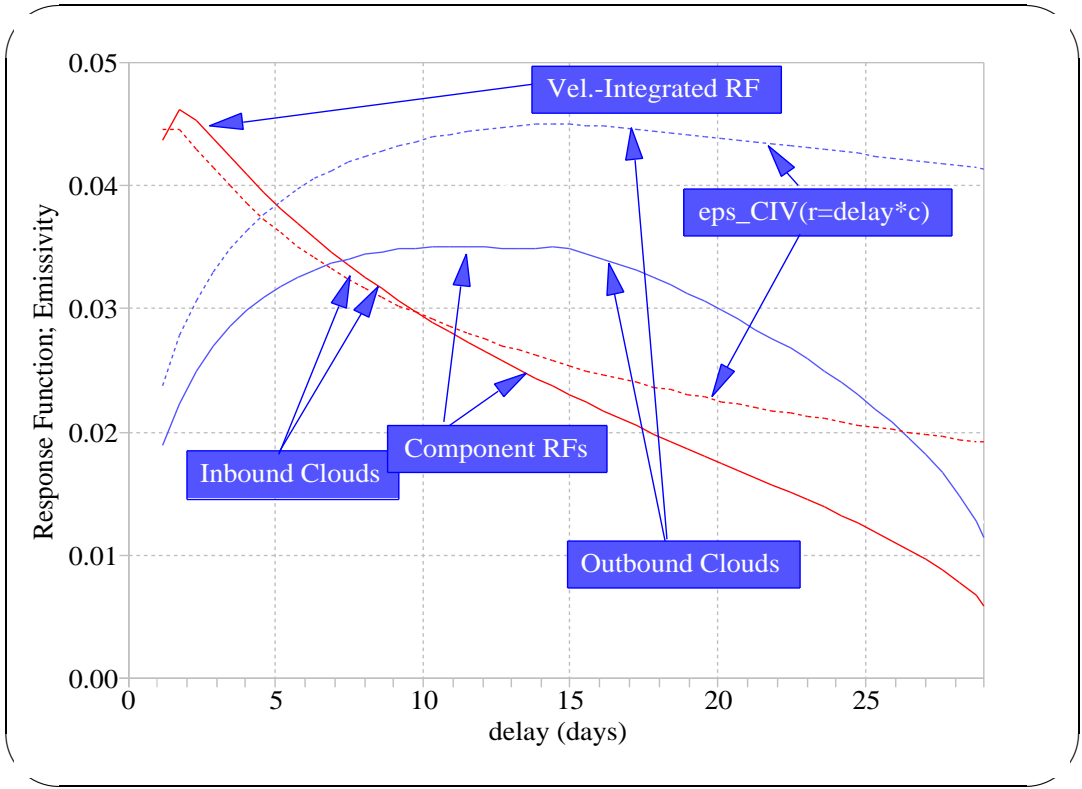
closer on the average to the central black hole.  As \Fig{response}{Solid
lines: response functions for a simplified cometary star model.  Dotted
lines: approximated C {\scriptsize\rm IV} line emissivity.  The head/tail pressure ratio was
assumed to be 2.0 in these calculations.} shows, this is precisely what was
naturally predicted by the cometary star model.
\par
\medskip
Surveys indicate that the peaks of C {\scriptsize\rm III}] are shifted to the blue more
than the peaks of C {\scriptsize\rm IV}.  NGC 5548 is no exception to this trend.  Though the
line shifts shown on the previous page match this ordering, many of the
models I calculated that fared better regarding line ratios did not match
this ordering.
\par
\medskip
One possible explanation for this shortcoming of some of the models is
that self-absorption was not accounted for in my simulations. Self-absorption
from clouds along the line of sight that are farther away from the central
black hole than the other clouds emitting a line should be detectable because
of the required high geometrical covering factors and the high signal to
noise ratios near the line peaks.  As Figure \ref{ionversesradius} shows, for
$s<2$ models there is an outermost radius beyond which the fractional ionic
abundance is too low for significant self-absorption to be possible in most
lines.  For this reason, the self-absorption ``components" of most line peaks
should be blueshifted.  Because the self-absorbing clouds must be along the
line of sight, they should appear to have non-causal local response functions
(zero delay, yet low velocity dispersions in a spherically symmetric orbital
system).
\par
\medskip
Both of these features appear to exist in the lines of the high-quality
NGC 5548 {\it HST} data (including the C {\scriptsize\rm IV} 1548.20/1550.77 doublet resolved in
the plot on the previous page).  Note that this self-absorption results in an
effective countering redshift of the profile peak.
\par
\medskip
This behavior, however, does not apply to all lines.  In particular, C
{\scriptsize\rm III}]~$\lambda 1909$ has such a low absorption oscillator strength $(10^{-6}$ that of C {\scriptsize\rm IV})
that it should be immune from such self-absorption.  Therefore, its profile
shape should not be influenced by self-absorption and the blueshift of its
observed profile peak should be higher than that of most other lines in AGNs
that have high covering factors.  Hopefully, future modeling efforts will
quantitatively test this as well as other apparent predictions of the
cometary star model.
\infootendplain
\end{appendix}
\def\rf{\bibitem{}}
\newenvironment{changemargin}[2]{\begin{list}{}{
         \setlength{\topsep}{0pt}\setlength{\leftmargin}{0pt}
         \setlength{\rightmargin}{0pt}
         \setlength{\listparindent}{\parindent}
         \setlength{\itemindent}{\parindent}
         \setlength{\parsep}{0pt plus 1pt}
         \addtolength{\leftmargin}{#1}\addtolength{\rightmargin}{#2}
         }\item }{\end{list}}
\authorindex
\medskip
{\centering \qquad \it Pages on which citations occur are listed as boldfaced numbers.}
\medskip
\begin{changemargin}{1.0in}{0in}
\addtolength{\parindent}{-1.0in}
\input \jobname.ref
\end{changemargin}
\addtolength{\parindent}{1.0in}
\phraseindex
\printindex
\pagestyle {empty}
\clearpage\mbox{ \/
}
\clearpage

\end{document}